\documentclass[12pt,a4wide]{report}
\usepackage{preamble}
\begin{document}

\begin{titlepage}
   \begin{center}
       \vspace*{1cm}

       {
         \LARGE
         \textbf{Gravitational Scattering of Compact Bodies from Worldline Quantum Field Theory}
       }

            
       \vspace{1.3cm}

       \textbf{Dissertation}
       \\
       zur Erlangung des akademischen Grades
       \\
       doctor rerum naturalium
       \\
       (Dr. rer. nat.)
       \\
       im Fach Physik
       \\
       Spezialisierung: Theoretische Physik
       \\
       \vspace{1.3cm}
       eingereicht an der
       \\
       Mathematisch-Naturwissenschaftlichen Fakult\"at
       \\
       der Humboldt-Universit\"at zu Berlin
       \\
       \vspace{1.3cm}
       von
       \\
       \textbf{M. Sc. Gustav Uhre Jakobsen}
       \\
       Geboren am 05.04.1995 in Kopenhagen
       \\
       \vfill
              {\raggedright
       Pr\"asidentin der Humboldt-Universit\"at zu Berlin:
       \\
       Prof. Dr. Julia von Blumenthal
       \\
       \vspace{.2cm}
       Dekanin der Mathematisch-Naturwissenschaftlichen Fakult\"at:
       \\
       Prof. Dr. Caren Tischendorf
       \\
       \vspace{1.3cm}
       Gutachter:
       1. Prof. Dr. Paolo Di Vecchia
       \\
       \hspace{2.0cm}
       2. Prof. Dr. Jan Plefka
       \\
       \hspace{2.0cm}
       3. Prof. Dr. Pierre Vanhove
       \\
       \vspace{.2cm}
       Tag der m\"undlichen Pr\"ufung: 05.07.2023
       \\
       }            
   \end{center}
\end{titlepage}

\chapter*{Abstract}
In this work the worldline quantum field theory (WQFT) approach to computing observables of the classical general relativistic two-body system is presented.
Compact bodies such as black holes or neutron stars are described in an effective field theory setting by point-like particles with worldline fields.
The WQFT treats all worldline fields on an equal footing with the gravitational space-time field and is formally defined as the tree-level contributions of a path-integral on those fields.
Novel results of the WQFT include the gravitational bremsstrahlung at second post-Minkowskian order and the impulse and spin kick at third post-Minkowskian order all at quadratic order in spins.

We first introduce the effective field theory description of compact bodies in terms of point-like particles and the post-Minkowskian expansion of unbound scattering of these bodies.
The compact bodies are generally described by several worldline fields including their trajectory and spin degrees of freedom and the inclusion of spin is analyzed with particular focus on its supersymmetric description in terms of anti-commuting Grassmann variables.

Next, the WQFT is presented with a comprehensive discussion of its in-in Schwinger-Keldysh formulation, its Feynman rules and graph generation and its on-shell one-point functions which are directly related to the scattering observables of unbound motion.
Here, we present the second post-Minkowskian quadratic-in-spin contributions to its free energy (the WQFT eikonal) which is equivalent to the on-shell action and from which the impulse and spin kick may be derived to the corresponding order.

The computation of scattering observables generally requires the evaluation of multi-loop integrals and for the computation of observables at the third post-Minkowskian order we analyze the required two-loop integrals.
Our discussion uses retarded propagators in contrast to the Feynman ones generally used in quantum field theory and these impose causal boundary conditions of the observables as prescribed by the in-in formalism.

Finally we turn to results of the WQFT starting with the gravitational bremsstrahlung of the scattering of two spinning bodies.
This waveform is discussed together with its radiative information of linear and angular momentum fluxes.
The leading order total radiated angular momentum is derived together with post-Newtonian expansions of the angular and power spectra of the energy flux.

Lastly we present the results of the conservative and radiative impulse and spin kick at third post-Minkowskian order and quadratic order in spins together with the mapping of the unbound results to a conservative (bound) Hamiltonian at the corresponding perturbative order.
These novel spinning results obey a generalized Bini-Damour radiation-reaction relation and their conservative parts can be parameterized in terms of a single scalar.

\chapter*{Zusammenfassung}
In dieser Arbeit wird der Ansatz der Weltlinienquantenfeldtheorie (WQFT) zur Berechnung von Observablen des klassischen allgemeinen relativistischen Zweik\"orpersystems vorgestellt.
Kompakte K\"orper wie Schwarze L\"ocher oder Neutronensterne werden im Rahmen einer effek-tiven Feldtheorie mit Weltlinienfeldern beschrieben.
Die WQFT behandelt alle Weltlinienfel-der gleichberechtigt mit dem Gravitationsfeld und ist definiert als die \textit{tree-level}-Beitr\"age eines Pfadintegrals auf diesen Feldern.
Zu den neuen Ergebnissen der WQFT geh\"oren die Gra-vitationsbremsstrahlung zweiter post-Minkowski'schen Ordnung sowie der Impuls und der Spin-Kick dritter post-Minkowski'schen Ordnung, alle quadratischer Ordnung in Spins.

Zuerst wird die effektive feldtheoretische Beschreibung von kompakten K\"orpern mit Weltlinien und die post-Minkowski'schen Approximation der Streuung dieser K\"orpern vorge-stellt. 
Die kompakten K\"orper werden im Allgemeinen durch mehrere Weltlinienfelder beschrie-ben und die Einbeziehung des Spins wird mit besonderem Augenmerk auf ihre supersymmetrische Beschreibung in Form von antikommutierenden Grassmann-Variablen analysiert.

Anschließend wird die WQFT mit einer Diskussion ihrer in-in Schwinger-Keldysh-Formu-lierung, ihrer Feynman-Regeln und Graphengenerierung sowie ihrer \textit{on-shell} Einpunktfunktionen vorgestellt.
Hier stellen wir den zweiten post-Minkowski'schen quadratischen-in-Spin-Beitrag zu seiner freien Energie (das WQFT-Eikonal) vor, aus dem der Impuls und der Spin-Kick in der entsprechenden Ordnung abgeleitet werden k\"onnen.

Die Berechnung von Streuobservablen erfordert im Allgemeinen die Auswertung von Multi-Loop-Integralen, und wir analysieren die Zwei-Loop-Integrale, die in der dritten post-Minkowski'schen Ordnung der Weltlinienobservablen auftreten.
Im Gegensatz zu den in der Quantenfeldtheorie \"ublicherweise verwendeten Feynman-Propagatoren werden hier retardierte Propagatoren verwendet, die kausale Randbedingungen vorschreiben.

Schlie{\ss}lich wenden wir uns den Ergebnissen der WQFT zu und beginnen mit der gravitativen Bremsstrahlung bei der Streuung zweier rotierender K\"orper.
Diese Wellenform wird zusammen mit der Strahlungsinformation der Linear- und Drehimpulsfl\"usse diskutiert.
Der gesamte abgestrahlte Drehimpuls f\"uhrender post-Minkowski'schen Ordnung wird zusammen mit der post-Newtonian'schen Approximation des Energieflusses abgeleitet.

Wir pr\"asentieren dann die Ergebnisse des konservativen und strahlenden Impulses und des Spin-Kicks bei dritter post-Minkowski'scher Ordnung und quadratischer Ordnung in Spins zusammen mit der Abbildung der ungebundenen Ergebnisse auf einen konservativen (gebundenen) Hamiltonian bei der entsprechenden perturbativen Ordnung.
Diese neuartigen Spinergebnisse folgen einer verallgemeinerten Bini-Damour-Strahlungsreaktionsbeziehung.
Zus\"atzlich k\"onnen ihre konservativen Beitr\"age von einer Skalar parametisiert werden.

\chapter*{List of Publications}

This thesis is mainly based on the content of the following peer-reviewed publications:

\vspace{.4cm}
\noindent
\cite{Jakobsen:2021smu}
G. U. Jakobsen, G. Mogull, J. Plefka and J. Steinhoff,
``\textit{Classical Gravitational Bremsstrahlung from a Worldline Quantum Field Theory}'',
Phys. Rev. Lett. 126 (2021) 201103,
arXiv: 2101.12688

\vspace{.4cm}
\noindent
\cite{Jakobsen:2021lvp}
G. U. Jakobsen, G. Mogull, J. Plefka and J. Steinhoff,
``\textit{Gravitational Bremsstrahlung and Hidden Supersymmetry of Spinning Bodies}'',
Phys. Rev. Lett. 128 (2022) 011101,
arXiv: 2106.10256

\vspace{.4cm}
\noindent
\cite{Jakobsen:2021zvh}
G. U. Jakobsen, G. Mogull, J. Plefka and J. Steinhoff,
``\textit{SUSY in the Sky with Gravitons}'',
JHEP 01 (2022) 027,
arXiv: 2109.04465

\vspace{.4cm}
\noindent
\cite{Jakobsen:2022fcj}
G. U. Jakobsen and G. Mogull,
``\textit{Conservative and Radiative Dynamics of Spinning Bodies at Third Post-Minkowskian Order Using Worldline Quantum Field Theory}'',
Phys. Rev. Lett. 128 (2022) 141102
arXiv: 2201.07778

\vspace{.4cm}
\noindent
\cite{Jakobsen:2022psy}
G. U. Jakobsen, G. Mogull, J. Plefka and B. Sauer,
``\textit{All Things Retarded: Radiation-Reaction in Worldline Quantum Field Theory}'',
JHEP 10 (2022) 128
arXiv: 2207.00569

\vspace{.4cm}
\noindent
\cite{Jakobsen:2022zsx}
G. U. Jakobsen and G. Mogull,
``\textit{Linear Response, Hamiltonian and Radiative Spinning Two-Body Dynamics}'',
Phys. Rev. D 107 (2023) 044033, 
arXiv: 2210.06451

\vfill
\noindent
\textbf{Declaration of Independent Work:}
\\
I declare that I have completed the thesis independently using only the aids and tools specified. I have not applied for a doctor’s degree in the doctoral subject elsewhere and do not hold a corresponding doctor’s degree. I have taken due note of the Faculty of Mathematics and Natural Sciences PhD Regulations, published in the Official Gazette of Humboldt-Universität zu Berlin no. 42/2018 on 11/07/2018.

\chapter*{Acknowledgements}
First, I would like to thank my supervisor Jan Plefka for his support and encouragement and for numerous illuminating discussions on physics, and I am grateful to be part of his group at the Humboldt University of Berlin with important values of independent curiosity-driven research and an aim towards collaboration across fields and groups.

Next, I would like to thank my colleagues Gustav Mogull, Jan Plefka, Benjamin Sauer and Jan Steinhoff for fruitful and exciting research and collaboration during my years as a PhD student.

I am thankful for having been part of the research training group ``Rethinking Quantum Field Theory'' with its broad scope on modern methods and applications of quantum field theory.
Here I am thankful to my fellow PhD students for creating a stimulating environment and more generally to my colleagues at the Humboldt University of Berlin for an open and motivating workspace.

Likewise I would like to thank Alessandra Buonanno for being part of her group at the Max Planck Institute for Gravitational Physics in Potsdam (Albert Einstein Institute) where the broad expertise in gravitational wave physics and related topics has inspired my research and research interests.
There, too, I would like to thank my fellow PhD students and colleagues for always welcoming me and for many enlightening discussions.

At both the Humboldt University of Berlin and the Albert Einstein Institute in Potsdam I would like to thank the administrative staff including Sylvia and Brit for helping me out with all practical matters.

I am thankful for having taken part in several international workshops, schools and conferences with an encouraging and friendly community and most notably for the chance to participate in the workshop ``High-Precision Gravitational Waves'' in Santa Barbara in April 2022.

I am thankful to Paolo Di Vecchia and Pierre Vanhove who agreed to be referees for this thesis.
And along with Paolo Di Vecchia and Pierre Vanhove I am thankful to Alessandra Buonanno, Peter Uwer and Arno Rauschenbeutel for being members of my thesis defense committee.

Finally I would like to thank my family Marianne, Henrik and Leo and my girlfriend Lea for supporting me with respect to my scientific work and also more generally for being there anytime.

This work is funded by the Deutsche Forschungsgemeinschaft (DFG, German Research Foundation) Projektnummer 417533893/GRK2575 “Rethinking Quantum Field Theory”.

\tableofcontents

\chapter{Introduction}
\label{INT}
The problem of motion of particles in the theory of general relativity (GR) is still, after more than a hundred years after its formulation in 1914~\cite{Einstein:1915ca}, an active area of research.
This includes in particular the understanding of the general relativistic bound system of two compact bodies which is essential for the computation of precise theoretical predictions of gravitational wave signals which since the first direct observation in 2015~\cite{LIGOScientific:2016aoc,LIGOScientific:2016wyt} by the LIGO interferometers are now regularly observed by gravitational wave detectors.
The problem of motion of particles in relativistic theories including both GR and electromagnetism has a long rich history with early work by Einstein~\cite{Einstein:1935tc,Einstein:1938yz}, Dirac~\cite{Dirac:1938nz}, Feynman~\cite{Wheeler:1945ps,Wheeler:1949hn} and numerous other authors.
An important property of relativistic motion is the self-interaction of bodies which is a consequence of force being carried by space-time fields and results in radiation of energy.
Naturally, this process of radiation of gravitational waves and energy is crucial to the generation of the gravitational wave signals that are being observed on earth today.

The direct observation of gravitational wave signals on earth~\cite{LIGOScientific:2018mvr,LIGOScientific:2020ibl,LIGOScientific:2021usb,LIGOScientific:2021djp,Antelis:2016icm} promises new exciting possibilities for testing our theoretical understanding of gravity~\cite{Will:2014kxa,Yunes:2016jcc,Shankaranarayanan:2022wbx} and offers new data from which we among others can improve our knowledge of neutron stars and black holes and their abundance in the universe~\cite{Flanagan:2007ix,Dietrich:2020eud}.
These gravitational wave signals are sourced by compact binary coalescences composed of either black holes or neutron stars.
Typically, the two bodies initially orbit each other in (quasi-)circular motion with their relative radius slowly shrinking due to the radiation of energy to gravitational waves.
This results in an inspiralling motion with the final merging of the two bodies into one and ringdown of that body.
The existence of gravitational waves is a consequence of GR first pointed out by Einstein~\cite{Einstein:1916cc}.
Early indirect evidence of gravitational waves was due to the discovery of the Hulse-Taylor pulsar in 1974~\cite{1975ApJ...195L..51H} whose orbital frequency increased (and radial separation decreased) in agreement with GR.

The central theoretical challenge for the prediction of gravitational wave signals (waveforms) relevant for detection in gravitational observatories is the description of the general relativistic two-body system of compact binary coalescences.
A successful analytic approach for the derivation of complete waveforms is the effective-one-body formalism invented by Buonanno and Damour with first waveforms in 2000~\cite{Buonanno:2000ef} where one importantly combines our knowledge of non-perturbative black hole space-times with analytic perturbative data.
Alongside numerical integration of the Einstein field equations of black hole coalescences with first complete results by Pretorius in 2005~\cite{Pretorius:2005gq} waveforms are known for a variety of initial parameters of the compact coalescing bodies~\cite{Boyle:2019kee}.
The initial state of the two-body system where the bodies are still widely separated can conveniently be described by analytic perturbative approaches where the post-Newtonian (PN)~\cite{Blanchet:2013haa} expansion takes advantage of the typically small relative velocity of the two bodies.
This expansion of the gravitaional dynamics which formally may be taken in $1/c$ with $c$ the speed of light dates back to the birth of GR with the Einstein-Infeld-Hoffmann potential first derived in 1917 describing the leading order corrections to the Newtonian potential~\cite{droste,Einstein:1938yz}.
The PN approach is ideally suited to describe the initial inspiralling motion of the compact bound coalescences where it is a dual expansion in both the relative (large) separation $r$ and relative (low) velocity $v$ of the two bodies which for the approximate Keplerian orbits are related through the virial theorem $\frac{GM}{c^2r}\sim \frac{v^2}{c^2}\ll1$ with Newton's constant $G$ and total mass of the system $M$.
Post-Newtonian results are a key ingredient to the effective-one-body formalism which takes as input perturbative analytic results which are then resummed into the effective-one-body Hamiltonian.
With the fourth post-Newtonian order well established~\cite{Schafer:2018kuf} current state of the art results focus on the fifth order~\cite{Levi:2022rrq}.

The consistent and efficient description of compact bodies is a requisite for all theoretical approaches to the gravitational two-body problem.
In the perturbative setting where the bodies are widely separated compared with their radii such a description is achieved with the worldline effective field theory (WEFT) proposed by Goldberger and Rothstein in 2006~\cite{Goldberger:2004jt}.
Here, the compact bodies are described by point-like particles with effective curvature couplings describing their finite size effects and thus, as an example, distinguishing black holes and neutron stars from each other.
The simplest leading order description is spherically symmetric point particles obeying the geodesic equation.
A particular success of this approach is the consistent inclusion of all order spin effects to linear order in the curvature by Levi and Steinhoff~\cite{Levi:2015msa} with other interesting developments the description of horizon absorption of gravitational waves and consequent increase in mass~\cite{Goldberger:2005cd,Saketh:2022xjb}.
The effective field theory (EFT) setting~\cite{Porto:2016pyg,Levi:2018nxp,Goldberger:2022ebt,Goldberger:2022rqf} offers a consistent framework for including finite size effects of the bodies and keeping track of the power counting of the significance of different contributions.
It is interesting to note that both the effective-one-body formalism and the (worldline) effective field theory of compact bodies draws inspiration from quantum field theory (QFT)~\cite{Buonanno:1998gg,Goldberger:2004jt}.

An alternative approach to the WEFT is given by the classical limit of quantum field theory.
While no fundamental quantum theory of gravity exists much work on quantum gravity within the framework of QFT has been done including initial work by Feynman~\cite{Feynman:1963ax}, De Witt~\cite{DeWitt:1967yk,DeWitt:1967uc} and Veltman and t'Hooft~\cite{tHooft:1974toh}.
It was then Donoghue who first in 1994~\cite{Donoghue:1994dn} consistently computed low-energy, long-range quantum gravity corrections to the Newtonian potential within an effective field theory framework.
An important realization was that loops in the perturbative expansion of QFT do not only contain quantum dynamics but also classical physics~\cite{Iwasaki:1971vb,Holstein:2004dn}.
Another invention came, then, when these ideas were applied to the classical limit of (on-shell, quantum) scattering amplitudes~\cite{Bjerrum-Bohr:2013bxa,Bjerrum-Bohr:2018xdl,Cheung:2018wkq}.
The effectiveness of this approach comes from its focus on on-shell building blocks and takes advantages of modern amplitudes techniques including double-copy, on-shell recursion relations and generalized unitarity~\cite{Elvang:2013cua,Bern:2019crd}.
In the effective quantum gravity approach compact bodies are described by massive quantum (space-time) fields with the simplest example being spinless bodies described by the Klein-Gordon scalar field.
In the following we will refer to this approach as the QFT-amplitudes approach (see reviews~\cite{Bjerrum-Bohr:2022blt,Kosower:2022yvp,Buonanno:2022pgc}).

Interestingly, through the use of a worldline quantum field theory (WQFT)~\cite{Mogull:2020sak} the two approaches, QFT-amplitudes and WEFT (in particular the post-Minkowskian EFT~\cite{Kalin:2020mvi}), were seen to be closely related.
This synthesis of the two approaches was achieved by integrating out the quantum scalar fields of the QFT-amplitudes approach by the introduction, instead, of classical worldline fields.
In practice, and from the classical perspective, the WQFT formalism is an efficient and systematic method for solving the classical equations of motion relevant to the worldline effective description of compact bodies.
In its most basic application, the WQFT approach solves the equations of motion in a Feynman diagrammatic perturbative expansion where both the gravitational field and worldline fields propagate on an equal footing within the diagrams.
Like the QFT-amplitudes approach, the main focus is on gauge invariant observables derived from on-shell WQFT amplitudes.
The WQFT approach resembles several other approaches including the Post-Minkowskian EFT~\cite{Kalin:2020mvi,Mougiakakos:2021ckm}, heavy mass EFT~\cite{Damgaard:2019lfh,Brandhuber:2021kpo} and velocity cuts of scattering amplitudes~\cite{Bjerrum-Bohr:2021din,Bjerrum-Bohr:2021wwt}.
In a series of papers~\cite{Jakobsen:2021smu,Jakobsen:2021lvp,Jakobsen:2021zvh,Jakobsen:2022fcj,Jakobsen:2022psy,Jakobsen:2022zsx} including the present author, the WQFT was further developed and applied to the perturbative expansion of unbound two-body scattering dynamics.
The description and presentation of the WQFT and its applications is the main goal of this thesis.

The WQFT is especially suited for describing the unbound scattering of compact bodies in the post-Minkowskian expansion (which is generally the case also for the QFT-amplitudes approaches).
The post-Minkowskian (PM) expansion is a weak field expansion of the gravitational field with formal expansion parameter Newton's constant $G$.
In relation to the relativistic two-body problem it is especially suited for the perturbative expansion of unbound scattering with large impact parameter and small scattering angle.
Here, the relative velocity of the two bodies may well be relativistic and in contrast to the PN expansion the PM approach does not restrict the velocity to being small.
Early work on the PM expansion includes Refs.~\cite{Bertotti:1956pxu,Kerr:1959zlt,Portilla:1979xx,Westpfahl:1979gu,Portilla:1980uz,Bel:1981be} with the leading PM order gravitational bremsstrahlung computed in the 1970s by Kovacs, Thorne and Crowley~\cite{Kovacs1,Crowley:1977us,Kovacs:1977uw,Kovacs:1978eu} and the second PM order scattering angle by Westpfahl in 1985~\cite{Westpfahl:1985tsl} in both cases with the simplest point-like compact bodies without spin and finite size effects.
A new breakthrough, then, came in 2019 with the computation of the conservative Hamiltonian and scattering angle at the third post-Minkowskian order by Bern\textit{ et al.}~\cite{Bern:2019nnu}.
Since then, the field has developed rapidly with the first complete radiative results for the 3PM impulse in 2021~\cite{Herrmann:2021tct} (building on~\cite{Damour:2020tta,DiVecchia:2021ndb,Herrmann:2021lqe}) and by now both conservative and radiative results at the fourth post-Minkowskian order~\cite{Bern:2021dqo,Bern:2021yeh,Bern:2022jvn,Dlapa:2021npj,Dlapa:2021vgp,Dlapa:2022lmu,Bjerrum-Bohr:2022ows}.

While the post-Minkowskian expansion of the unbound gravitational scattering of compact bodies is interesting in its own right, it is not at first clear, how it can be applied to the description of bound systems and the consequent derivation of (bound) waveforms.
This problem, however, is circumvented by mapping the unbound scattering data onto variables relevant to the bound motion.
First, for conservative motion, the unbound (local) potential, or Hamiltonian, may simply be mapped to the bound one~\cite{Iwasaki:1971vb,Cheung:2018wkq,Cristofoli:2019neg}.
Second, in many cases gauge invariant unbound observables may be mapped directly to corresponding bound ones~\cite{Kalin:2019rwq,Kalin:2019inp,Cho:2021arx,Saketh:2021sri} with the simplest example given by the mapping of the unbound scattering angle to the bound periastron advance.
Currently, however, both approaches are not yet able to deal with the non-local in time tail effects that appear at the fourth post-Minkowskian order~\cite{Bern:2021yeh,Cho:2021arx}.
The mapping of PM data to the bound system effectively resums the post-Newtonian expansion in the relative velocity of the two bodies.
A particularly interesting application is to include post-Minkowskian data into the effective-one-body model~\cite{Damour:2016gwp,Damour:2017zjx,Antonelli:2019ytb,Damgaard:2021rnk} which has most recently been done with the 4PM data~\cite{Khalil:2022ylj,Damour:2022ybd} which indicates that the PM data might be of importance for eccentric bound motion.

The basic effect of radiation of gravitational waves and thereby loss of linear and angular momentum must be included in the effective field theory approaches and for the WEFT this is achieved with the in-in Schwinger-Keldysh formalism~\cite{Schwinger:1960qe,Keldysh:1964ud} which conveniently incorporates causal, retarded boundary conditions at the level of the action.
Essentially, at the level of the equations of motion, retarded boundary conditions are imposed with the use of retarded propagators which are non-zero only for future times.
The in-in (W)EFT description of the gravitational two-body system was introduced by Galley \textit{et al.}~\cite{Galley:2008ih,Galley:2009px}.
In the WQFT formalism retarded propagators are consistently used to impose causal boundary conditions~\cite{Mogull:2020sak,Jakobsen:2021smu,Jakobsen:2021lvp,Jakobsen:2021zvh,Jakobsen:2022fcj,Jakobsen:2022psy,Jakobsen:2022zsx} and formally introduced at the level of the action with the in-in (WQFT) formalism in Ref.~\cite{Jakobsen:2022psy}.
An approach for deriving full radiative observables within the QFT-amplitudes approach was given by Kosower, Maybee and O'Connell~\cite{Kosower:2018adc} where observables are derived directly from scattering amplitudes and their cuts and bears similarities to the in-in formalism (see also the complementary approach with the eikonal operator of Refs.~\cite{DiVecchia:2021bdo,DiVecchia:2022nna,DiVecchia:2022piu}).

An important property of compact bodies such as black holes and neutron stars is their spin.
Thus in the case of (stationary, asymptotically flat) black holes in pure gravity, their most general form is described by the Kerr metric~\cite{Kerr:1963ud} which is fully determined by its two parameters mass and spin and in that sense similar to fundamental quantum particles.
Historically, the gravitational dynamics of a compact body including its spin (dipolar moment) and quadrupolar moment is given by the Mathisson-Papapetrou-Dixon equations~\cite{Mathisson:1937zz,Papapetrou:1951pa,Dixon:1970zza}.
Relating the quadrupolar moment to its spin results, then, in equations accurate to quadratic order in its spin.
This equation is well understood within the (W)EFT framework with much recent work~\cite{Porto:2005ac,Marsat:2014xea,Levi:2015msa,Vines:2016unv,Liu:2021zxr,Saketh:2022wap} going to higher orders in spin and multipole moments and an improved understanding of the gauge symmetry related to the choice of spin supplementary condition.
In the post-Minkowskian expansion initial exciting results at leading PM order and all orders in spin were given by Vines~\cite{Vines:2017hyw}.
It was then later realized~\cite{Guevara:2018wpp,Guevara:2019fsj,Chung:2018kqs,Chung:2019duq,Arkani-Hamed:2019ymq} that those results are a direct consequence of the minimal (tree-level) three-point scattering amplitude of Arkani-Hamed, Huang and Huang~\cite{Arkani-Hamed:2017jhn}.
In general, the description of spin in the QFT-amplitudes uses spin-$\frac{n}2$ quantum fields which are related to the perturbative description of classical spin to the $n$th power.
The initial success of the results at 1PM to all orders in spin has lead to an eager search for similar results at 2PM including~\cite{Chiodaroli:2021eug,Saketh:2022wap,Bern:2022kto,Aoude:2022trd,Aoude:2022thd,Bautista:2021wfy,Bautista:2022wjf,Bjerrum-Bohr:2023jau,Alessio:2023kgf} where the relevant scattering amplitude is the massive gravitational Compton amplitude (with partial 3PM all order spin results in Ref.~\cite{Alessio:2022kwv}).
While all order spin results are interesting from a fundamental perspective, a perturbative expansion in spin is relevant to the inspiral phase of binary mergers where the spin often is of the same order as the (typically low) velocity.
Perturbative spinning PM results include~\cite{Bern:2020buy,Liu:2021zxr,Haddad:2021znf,Chen:2021kxt,Kosmopoulos:2021zoq,Jakobsen:2021lvp,Jakobsen:2021zvh,Jakobsen:2022fcj,Jakobsen:2022zsx,Riva:2022fru,FebresCordero:2022jts,Damgaard:2022jem,Menezes:2022tcs} (For PN results see e.g. the recent Refs.~\cite{Kim:2022bwv,Mandal:2022ufb}).

In the WQFT formalism spin can be incorporated using the description of spin of standard (W)EFT approaches.
Here, however, it was found advantageous to describe the (classical) spin degrees of freedom in terms of anti-commuting Grassmann variables~\cite{Jakobsen:2021lvp,Jakobsen:2021zvh,Jakobsen:2022fcj,Jakobsen:2022zsx} giving rise to a supersymmetry of the worldline action.
Such supersymmetric descriptions of spin have been considered in various contexts~\cite{Berezin:1976eg,Brink:1976sz,Brink:1976uf,Barducci:1976wc,Galvao:1980cu,Howe:1988ft,Howe:1989vn,vanHolten:1990we,Gibbons:1993ap} with the interesting result in~\cite{Gibbons:1993ap} that their SUSY could be related to the Carter constant~\cite{Carter:1968rr} of the Kerr metric.
It was then another interesting observation in~\cite{Jakobsen:2021zvh} that the $\mN=2$ supersymmetric worldline action~\cite{Bastianelli:2005vk,Bastianelli:2005uy} describes the dynamics of Kerr black holes to quadratic order in their spins.
In general the SUSY worldline action with $\mN$ flavors of Grassmann fields corresponds to a classical spinning particle to $\mN$th order in its spin.
The use of Grassmann variables in the WQFT for describing spin allows for a consistent description in terms of an action with only one additional worldline field in contrast to the case in traditional approaches where at least two fields are necessary.
The SUSY is directly related to the freedom of choosing a spin supplementary condition and captures this freedom which mixes the trajectory and spin degrees of freedom in a natural manner.

The progress of deriving worldline observables such as the scattering angle at the $n$th post-Minkowskian order is especially limited by the complex $(n-1)$-loop integrals that are required at each order.
Most importantly, the mass dependence of the scattering observables is bootstrapped to being polynomial~\cite{Damour:2019lcq} and the relevant loop integrals depend only on a single dimensionless scale, namely the relative velocity of the two bodies.
At the third PM order, the first truly post-Minkowskian integrals appear at two loops which describe an intricate dependence on the relative velocity $v$ including the appearance of the rapidity $\operatorname{arctanh}(v)$ in the observables.
These two-loop integrals have been reproduced in several contexts~\cite{Bern:2019crd,Kalin:2020fhe,Parra-Martinez:2020dzs,Cheung:2020gyp,Herrmann:2021tct,Ruf:2021egk,Bjerrum-Bohr:2021vuf,Bjerrum-Bohr:2021din,Brandhuber:2021eyq,DiVecchia:2021bdo,Riva:2021vnj,Mougiakakos:2022sic,Jakobsen:2022zsx,Jakobsen:2022psy,Jakobsen:2022fcj,DiVecchia:2022piu,Heissenberg:2022tsn,Riva:2022fru,FebresCordero:2022jts} with current state of the art being three loops (4PM) as mentioned above.
In contrast to the Feynman propagator predominantly used in the QFT-amplitudes approach, the WQFT uses retarded propagators~\cite{Jakobsen:2022psy} which automatically enforces causal boundary conditions.
In addition to the worldline observables, one may also consider the scattering waveform, or bremsstrahlung, which requires another type of integrals.
The leading order (W)QFT integrals for the waveform were first computed (in time domain) in Ref.~\cite{Jakobsen:2021lvp} with recent results at next-to-leading order~\cite{Elkhidir:2023dco,Herderschee:2023fxh,Brandhuber:2023hhy,Georgoudis:2023lgf}.
In general we may identify two orthogonal directions for higher precision which are the integration discussed here and the construction of the integrand.
It is interesting that adding perturbative effects such as spin to the observables does not change the integration significantly but instead is a challenge at the integrand level.

Advanced integration techniques of QFT and quantum chromo dynamics are employed in the evaluation of the (classical) loop integrals including IBP-reduction~\cite{Chetyrkin:1981qh,Laporta:2000dsw}, differential equations~\cite{Kotikov:1990kg,Bern:1993kr,Remiddi:1997ny,Gehrmann:1999as,Henn:2013pwa,Henn:2013nsa} and the method of regions~\cite{Beneke:1997zp,Smirnov:2002pj,Smirnov:2012gma,Becher:2014oda} (asymptotic expansion).
These methods allow for a surprisingly direct reduction of the post-Minkowskian problem to the post-Newtonian one.
Thus, the differential equations bootstrap the velocity dependence of the integrals and boundary conditions for these equations are provided in the PN limit of small velocity.
Genuine integration, then, is required only in this PN limit where the PM integrals reduce to integrals encountered in the PN literature.
These methods apply both to the Feynman and retarded propagator prescriptions and the significance of the propagator enters mainly at this final step in the boundary integrals.
This bootstrapping of the PM integrals with differential equations (and IBP-reduction), however, gets increasingly challenging at every new order although the single scale nature of the integrals and the classical limit, which introduces delta function constraints on the energy components of the loop momenta, make the integrals significantly simpler than corresponding multi-scale (quantum) integrals encountered in QFT.

In this work we present the WQFT formalism with focus on its development and applications in the series of papers~\cite{Jakobsen:2021smu,Jakobsen:2021lvp,Jakobsen:2021zvh,Jakobsen:2022fcj,Jakobsen:2022psy,Jakobsen:2022zsx} including the present author.
This includes the novel, state of the art results of Ref.~\cite{Jakobsen:2021lvp} of the leading order spinning bremsstrahlung and of Refs.~\cite{Jakobsen:2022fcj,Jakobsen:2022zsx} of the conservative and radiative contributions to the impulse, spin kick and Hamiltonian at the third PM order all with spin effects to quadratic order in the spins.
The structure of the present work does not, however, follow the chronological order of the papers~\cite{Jakobsen:2021smu,Jakobsen:2021lvp,Jakobsen:2021zvh,Jakobsen:2022fcj,Jakobsen:2022psy,Jakobsen:2022zsx} and instead presents the WQFT and its applications in an independent manner.
The task of describing compact bodies coupled to gravity as point particles is considered a problem of WEFT and presented in \chap~\ref{WEFT} together with the inclusion of spin effects and its SUSY description in terms of Grassmann variables~\cite{Jakobsen:2021zvh,Jakobsen:2022zsx}.
From this perspective, then, the WQFT is presented in \chap~\ref{WQFT} with the main goal of solving in a systematic and efficient manner the classical equations of motion of point-like particles, that is, worldline fields.
This includes the in-in formalism~\cite{Jakobsen:2022psy} in Sec.~\ref{WQFT:CDfPI} and (off-shell) WQFT Feynman rules and graph generation in Sec.~\ref{WQFT:Diagrammatics} and (on-shell) observables and their structure in Sec.~\ref{WQFT:PostMinkowskian}.

In \chap~\ref{sec:TL} we focus, then, on integration techniques for loop-integration, namely IBP-reduction to master integrals, tensor reduction, symmetries, differential equations, method of regions and evaluation of boundary integrals.
The main objective is the derivation of the (3PM, classical) two-loop integrals with retarded propagators~\cite{Jakobsen:2022fcj,Jakobsen:2022psy}.

The remaining \chaps~\ref{GB} and~\ref{sec:Scattering} present spinning PM results computed with the WQFT though their presentation generally is independent of the way they were derived.
In \chap~\ref{GB} the leading PM order results of the gravitational bremsstrahlung~\cite{Jakobsen:2021smu,Jakobsen:2021lvp} at $\mO(G^2,S^2)$ are presented including the derivation of the 2PM radiated angular momentum and post-Newtonian fluxes of energy.
In \chap~\ref{sec:Scattering} results for the full conservative and radiative worldline observables the impulse and spin kick at $\mO(G^3,S^2)$ are presented~\cite{Jakobsen:2022fcj,Jakobsen:2022zsx}.
From the conservative results a two-body Hamiltonian at the same perturbative orders is derived which effectively maps the unbound results to bound dynamics.
For aligned spins the gauge invariant mapping of the scattering angle to the binding energy is considered.
We note that while tidal effects for both the waveform and worldline observables were computed at corresponding orders to the above in Ref.~\cite{Jakobsen:2022psy}, we will mainly focus on the spin effects in this work.

Finally, in \chap~\ref{DIS} we conclude with a discussion and perspectives for further research.
In the appendix~\ref{NOT} we have gathered some formulas and conventions relevant to this thesis.

\chapter{General Relativistic Dynamics of Compact Bodies}
\label{WEFT}
This chapter serves as background to the introduction of the worldline quantum field theory and the several perturbative post-Minkowskian results derived from it.
Thus, in this chapter we set up a theoretical model for the two-body problem of compact bodies and consequent equations of motion which are then to be solved perturbatively within the framework of WQFT.
After briefly considering aspects of general relativity in Sec.~\ref{sec:Gravitational}, we move on to the description of compact bodies in terms of worldlines and the post-Minkowskian expansion in Sec.~\ref{WEFT:WEFT} and finally the inclusion of spin and its SUSY description in terms of anti-commuting Grassmann variables in Sec.~\ref{sec:Spin}.
The main objective is the introduction of the WEFT description of compact bodies which allows for a systematic description of the general relativistic two-body problem of compact objects taking into account finite size effects and multipole moments (including spin) as long as the gravitational field is sufficiently weak compared with the scale of the bodies.

\section{General Relativity and Gravitational Waves}
\label{sec:Gravitational}
Classical gravitational physics is described by Einstein's general theory of relativity.
Basic accounts of general relativity include Refs.~\cite{Dirac:GR,Weinberg:1972kfs,tHooft:2012wiu}.
This section starts out with the fundamentals of general relativity and focuses then on two topics of importance for the present work.
These are the background expansion around flat space-time and gauge fixing and the gauge fixed equations of motion.
For parts of those two last topics some background is also given in Refs.~\cite{Jakobsen:2020ksu,Jakobsen:2020diz}.
We generally use the mostly minus metric, we put the speed of light to unity $c=1$ and we use Einstein's summation convention.

The basic field carrying gravitational force is the metric field $g_\mn(x)$.
It is a space-time field and thus a function of the space-time coordinates $x^\mu$.
The space-time dimension is denoted by $d$ and is kept arbitrary in order to use dimensional regularization in the worldline effective field theory.
The Einstein-Hilbert action describes the gravitational field in vacuum and reads:\checked
\begin{align}
  S_{\rm EH}
  =
  \frac1{16 \pi G}
  \int
  \di^d
  x
  \sqrt{-g}
  R
  \ .
\end{align}
Here, $G$ is Newton's constant and we put the speed of light to unity $c=1$.
The metric determinant is $g=\det(g_\mn)$ and $R$ is the Ricci curvature scalar derived from the Ricci tensor, $R_\mn$, and Riemann curvature, $R^{\mu}_{\ \nu\ab}$, as:\checked
\begin{align}
  R_\mn
  &=
  R^\alpha_{\ \mn\alpha}
  \ ,\qquad
  R
  =
  R_\mn g^\mn
  \ .
\end{align}
Here, $g^\mn$, is the inverse metric satisfying $g^{\ab}g_{\beta\gam}=\del^\alpha_\gam$ with the Kronecker delta $\delta^\alpha_\gam$.
General covariant indices are lowered and raised with the metric and its inverse respectively.

The Riemann curvature tensor in terms of the Christoffel symbol reads:\checked
\begin{align}
  R^\mu_{\ \nu\ab}
  =
  2
  \pat_{
    [\alpha}^{\vphantom{\mu}
    }
    \Gam^\mu_{\ \beta]\nu}
  +2
  \Gam^\mu_{\ \kap[\beta}
    \Gam^{\kap\vphantom{\mu}}_{\ \alpha]\nu}
  \ .
\end{align}
Square brackets on indices indicates the averaged antisymmetric sum (Eqs.~\ref{A:Brackets}).
Partial derivatives with respect to $x^\mu$ are denoted by $\pat_\mu$.
The Christoffel symbol is:\checked
\begin{align}
  \Gam^\mu_{\ \ab}
  =
   g^\mn
  \Big(
  \pat_{(\beta} g_{\alpha)\nu}
  -
  \frac12
  \pat_\nu g_\ab
  \Big)
  \ .
\end{align}
Here, round brackets on the indices denote the averaged symmetric sum (Eqs.~\ref{A:Brackets}).

In general, the action will contain terms beyond the Einstein-Hilbert action which describe other fields and matter.
We denote this part of the action collectively as $S_{\rm matter}$.
The fundamental equations of (classical) general relativity then follow from the variational principle $\delta(S_{\rm EH}+S_{\rm matter})=0$.
They are the Einstein field equations and usually written as:\checked
\begin{align}
  G^\mn
  =
  -8\pi G\,
  T^\mn
  \ ,
\end{align}
with the Einstein tensor\checked
\begin{align}\label{EinsteinTensor}
  G^\mn
  &=
  -
  \frac{
    16\pi G
  }{
    \sqrt{-g}
    }
  \frac{
    \del S_{\rm EH}
  }{
    \del g_\mn
  }
  =
  R^\mn
  -
  \frac12
  R\,
  g^\mn
  \ ,
\end{align}
and energy-momentum tensor\checked
\begin{align}
  T^\mn
  =
  -\frac2{\sqrt{-g}}
  \frac{
    \del S_{\rm matter}
  }{
    \del g_\mn
  }
  \ .
\end{align}
Here, we used the conventions of Eqs.~\eqref{WEFT:MomentumConvention} for functional derivatives.
The general covariance of the gravitational theory implies the Bianchi identity and conservation of energy-momentum:\checked
\begin{align}
  D_\rho G^\rs =
  D_\rho T^\rs
  =
  0
  \ .
\end{align}
Here $D_\alpha$ denotes covariant differentiation,\checked
\bse
\begin{align}
  D_\alpha X^\mu
  &=
  \pat_\alpha X^\mu
  +
  \Gam^\mu_\ab
  X^\beta
  \ ,
  \\
  D_\alpha X_\mu
  &=
  \pat_\alpha X_\mu
  -
  \Gam^\beta_{\alpha\mu}
  X_\beta
  \ ,  
\end{align}
\ese
with some vector field $X^\mu$.
Covariant differentiation obeys the chain and Leibniz rules, $D_\alpha g_\mn=0$ and reduces to partial differentiation $\partial_\alpha$ for scalars.

For the description of spinning objects we will use tetrad fields.
The tetrad field (or vielbein), $e_{a\mu}(x)$, is a square root of the metric field and satisfies:\checked
\bse\label{WEFT:Vielbein}
\begin{align}
  e_{a\mu}(x)
  e^a_{\, \nu}(x)
  &=
  g_\mn(x)
  \ ,
  \\
  e_{a\mu}(x)
  e^{\, \mu}_b(x)
  &=
  \eta_{ab}
  \ .
\end{align}
\ese
It has local flat indices which we denote $a$ and $b$ and general covariant indices.
The local indices are lowered and raised with $\eta^{ab}=\eta_{ab}$ and the general covariant ones with the full metric.
In a sense, the vielbein translates between local flat indices and general covariant indices.
We may define the covariant derivative to be compatible with the local indices.
In that case, the spin connection $\oma^{ab}_\mu$ plays a similar role to the Christoffel symbol for the local indices.
It may be defined from the requirement $D_\mu e^{a\nu}=0$,\checked
\begin{align}
  \label{WEFT:SpinConnection}
  0
  =
  D_\mu
  e^{a\nu}
  =
  \pat_\mu
  e^{a\nu}
  +
  \Gam^\nu_{\mu\sig}
  e^{a\sig}
  +
  \oma_\mu^{ab}
  e_b^{\,\nu}
  \ ,
\end{align}
which implies:\checked
\begin{align}
  \oma^{ab}_\mu
  =
  e_{\, \nu}^a \pat_\mu e^{b\nu}
  +
  e^a_{\, \nu}
  \Gam^\nu_{\mu\sig}
  e^{b\sig}
  \ .
\end{align}
The general rules for covariant derivatives of tensors with mixed combinations of local and general covariant indices follow from the Leibniz rule.
The spin connection $\oma^{ab}_\mu=\oma^{[ab]}_\mu$ is antisymmetric in the upper indices $a$ and $b$.

\subsection{Gravitation Around Flat Space-Time
}
We may expand the metric around some background $b_\mn(x)$ which is advantageous if we consider small perturbations around that background.
In the following we will focus on the case where the background is flat space-time which is relevant to the post-Minkowskian expansion.
We note, however, that the formalism can be generalized to any background.

We choose the following simple parametrization of the metric:\checked
\begin{align}\label{WEFT:Graviton}
  g_\mn(x)
  =
  \eta_\mn
  +
  \kappa
  h_\mn(x)
  \ .
\end{align}
We will often refer to $h_\mn(x)$ as the graviton field since it plays the role of gravitons from the point of view of quantum field theory.
The inclusion of the constant $\kap$ is conventional.
It is defined in terms of the Newton constant, $G$, by:\checked
\begin{align}
  \kap =
  \sqrt{32\pi G}
  \ .
\end{align}
In general, other parameterizations than Eq.~\eqref{WEFT:Graviton} can also be considered in an equivalent fashion (see e.g. Ref.~\cite{Cheung:2020gyp}).

Working with the background expansion of the metric, it is natural to use covariant quantities with respect to that background.
In our case the background is flat space and covariant quantities are simple partial derivatives and Lorentz indices raised and lowered with the flat space background metric.
We will then raise and lower indices on the graviton field, $h_\mn(x)$, on partial derivatives and on tensors to be defined below with the flat space background metric.
When it is not clear from the context, we will indicate the metric (background or full) explicitly.

It is natural to separate the gravitational action into terms quadratic in $h_\mn$ and terms of higher order in $h_\mn$.
Correspondingly, the field equations are separated into linear and higher order terms in $h_\mn$.
We refer to the quadratic part of the action as the kinetic term.
Expanding the Einstein-Hilbert action we find:\checked
\begin{align}\label{WEFT:EHK}
  S_{\rm EH}^{\rm kin}
  =
  \frac12
  \int \di^d x\,
  \pat_\rho h_\mn(x)
  Q^{\rs\,\mn\,\ab}
  \pat_\sig h_\ab(x)
  \ .
\end{align}
The tensor $Q^{\rs\,\mn\,\ab}$ appearing in the kinetic term is,\checked
\begin{align}
  Q^{\rs\,\mn\,\ab}
  =
  \eta^\rs
  \mP^{\mn\ab}
  -
  2
  \mP^{\mn\gam(\rho}
  \mP^{\sig)\del\ab}
  \eta_{\gam\del}
  \ ,
\end{align}
with the tensor $\mP^{\mn\ab}$ given by:\checked
\begin{align}
  \label{WEFT:PTensor}
  \mP^{\mn\ab}
  =
  \eta^{\mu(\alpha}
  \eta^{\beta)\nu}
  -
  \frac12
  \eta^\mn
  \eta^\ab
  \ .
\end{align}
From its definition the tensor $Q^{\rs\,\mn\,\ab}$ is clearly symmetric under $\ab\leftrightarrow\mn$ but it is also symmetric under $\rs\leftrightarrow\mn$ and $\rs\leftrightarrow\ab$.
It obeys, in particular, the identity:\checked
\begin{align}\label{WEFT:QT}
  Q^{\rs\,\ab\,\mn} p_\rho p_\sig p_\alpha=0
  \ ,
\end{align}
for any vector $p_\rho$.
This gives rise to a gauge symmetry of $h_\mn(x)$ corresponding to the general covariance of the full theory.

We define $S^{\rm int}_{\rm EH}$ to include all other terms of $S_{\rm EH}$ than the (quadratic) kinetic terms.
The interaction terms are then of cubic and higher order in $h_\mn$.
We treat the interaction terms on an equal footing to the matter terms $S_\rmm{matter}$ and define the (local, pseudo) energy-momentum tensor, $\tau^\mn(x)$:\checked
\begin{align}
  \label{WEFT:PT}
  \tau^\mn(x)
  =
  -
  \frac{2}{\kap}
  \frac{
    \del
    \big(
    S_{\rm matter}
    +
    S_{\rm EH}^{\rm int}
    \big)
  }{
    \del h_\mn(x)}
  \ .
\end{align}
This is a tensor with respect to the background metric (i.e. Lorentz covariant).
With the present normalization it scales as $\kap^0$.
The interaction part of the gravitational action can be defined as a functional integral of the gravitational (pseudo) energy-momentum tensor:\checked
\begin{align}
  \label{WEFT:EHInteraction}
  S_{\rm EH}^{\rm int}
  =
  -\frac{\kap}{2}
  \int \di^d x
  \int
  D h_\mn(x)
  \tau^\mn_{\rm GR}(x)
  \ .
\end{align}
The functional integral can simply be considered as the inverse operation to the functional differentiation of Eq.~\eqref{WEFT:PT} and the energy-momentum $\tau_{\rm GR}^\mn(x)$ is the part of $\tau^\mn(x)$ due to the Einstein-Hilbert action.

The Einstein field equations may now be written as:\checked
\begin{align}
  Q^{\mn\,\rs\,\ab}
  \pat_\rho
  \pat_\sig
  h_\ab(x)
  =
  -\frac{
    \kap}{2}
  \tau^\mn(x)
  \ .
\end{align}
From the property of the $Q$-tensor in Eq.~\eqref{WEFT:QT} it follows that $\tau^\mn(x)$ is locally conserved:\checked
\begin{align}
  \pat_\mu \tau^\mn(x)=0
  \ .
\end{align}
The energy-momentum tensor $\tau^\mn(x)$ is the source of the graviton field.
It can be written in terms of the general covariant energy-momentum $T^\mn$ and gravitational energy-momentum $\tau_{\rm GR}^\mn$:\checked
\begin{align}
  \tau^\mn
  &=
  T^\mn
  +
  \frac1{\sqrt{-g}}
  \tau_{\rm GR}^\mn
  \\
  &=
  \sqrt{-g} \, T^\mn
  +
  \tau_{\rm GR}^\mn
  \ .
  \nn
\end{align}
The relation between the two expressions are derived using the equations of motion.

\subsection{Gauge Symmetry and Gauge Fixing}
\label{WEFT:GSGF}
The gravitational theory is invariant under general covariant coordinate transformations which gives rise to spin 2 gauge symmetry of the kinetic term of $h_\mn(x)$.
Thus, the kinetic action in Eq.~\eqref{WEFT:EHK} has the symmetry that a (infinitesimal) change $\del h_\mn(x)$ given by\checked
\begin{align}
  \del h_\mn(x)
  =
  2\partial_{(\mu} \eps_{\nu)}(x)
  \ ,
\end{align}
with infinitesimal parameter $\eps_\nu(x)$ does not change the action.
This follows from the property of the $Q^{\ab\,\mn\,\rs}$ tensor in Eq.~\eqref{WEFT:QT}.
In fact, this is the weak field limit of a generic infinitesimal coordinate transformation of $h_\mn(x)$:\checked
\begin{align}\label{WEFT:GCT}
  \del h_\mn(x)
  =
  2
  \pat_{(\mu} \eps_{\nu)}(x)
  +
  2
  h_{\alpha(\mu}(x)
  \pat_{\nu)}
  \eps^{\alpha}(x)
  +
  \eps^\sig(x)
  \pat_\sig
  h_\mn(x)
  \ .
\end{align}
This symmetry is the general covariance of the full gravitational action which mixes terms of different orders in $h_\mn(x)$.

The gauge symmetry must be fixed in order to invert the kinetic operator $Q^{\mn\,\ab\,\rs}\pat_\rho\pat_\sig$.
In classical theory this can be done by imposing that some function of the gravitational field, $G^\sig[h_\mn]$, vanishes.
We refer to $G^\sig$ as the gauge fixing function.
In this work, we will use the simple choice of linear de Donder gauge:\checked
\begin{align}\label{WEFT:LDG}
  G^\rho[h_\mn]
  =
  \mP^{\rs\mn}
  \pat_\sig h_\mn(x)
  \ .
\end{align}
If we impose this function to vanish then the kinetic term in the Einstein field equations simply becomes the wave operator $\pat^2$ with the tensor structure $\mP^{\mn\ab}$:\checked
\begin{align}\label{WEFT:QHP}
  Q^{\mn\,\rs\,\ab}
  \pat_\rho
  \pat_\sig
  h_\ab(x)
  &=
  \mP^{\mn\ab} \pat^2 h_\ab
  \\
  &=
  -\frac{\kap}{2}
  \tau^\mn(x)
  \nn
  \ .
\end{align}
With this gauge constraint, it is clear that the local energy-momentum tensor is conserved.

In quantum field theory the gauge fixing is often done at the level of the action.
In the classical limit we simply add a gauge fixing term to the action of the form:\checked
\begin{align}
  S_{\rm gf}
  =
  \int \di^d x\,
  \eta_{\sig\rho}
  G^\sig G^\rho
  \ .
\end{align}
We define the gauge fixed Einstein-Hilbert action:\checked
\begin{align}\label{WEFT:SGR}
  S_{\rm GR}=S_{\rm EH} + S_{\rm gf}
  \ .
\end{align}
With the gauge function of linear de Donder gauge in Eq.~\eqref{WEFT:LDG} we get the following kinetic term:\checked
\begin{align}
  \label{WEFT:GRKin}
  S_{\rm GR}^{\rm kin}
  =
  \frac12
  \int \di^d x\,
  \pat_\rho h_\mn(x)
    \eta^{\rs}
  \mP^{\mn\ab}
  \pat_\sig h_\ab(x)
  \ ,
\end{align}
and the interaction terms are unchanged, $S_{\rm EH}^{\rm int}=S_{\rm GR}^{\rm int}$.
In general we can also add nonlinear terms to the gauge fixing function and e.g. impose the harmonic gauge $\Gam^\mu_\rs g^\rs=0$.

In the rest of this thesis we will usually use the gauge fixed version of the action $S_{\rm GR}$.
We argue that the equations of motion that follow from this action are equivalent to the ones of the Einstein-Hilbert action together with the gauge constraint $G_\sig=0$.
That is, our claim is that,\checked
\begin{align}\label{WEFT:GRE0}
  \frac{
    \del
    (
    S_{\rm GR}+S_{\rm matter}
    )
  }{
    \del g_\mn(x)
  }
  =0
  \ ,
\end{align}
is equivalent to the Einstein field equations together with the gauge constraint $G_\sig=0$.
This, to some extend, follows from the Bianchi identities.
Thus, the equation of motion that follows from $S_{\rm GR}$ for general $G_\sig(x)$ is (i.e. Eq.~\ref{WEFT:GRE0}),\checked
\begin{align}\label{WEFT:GRE}
  G^\mn(x)
  -
  \frac{\kap^2}{\sqrt{-g}}
  \eta_{\rs}
  \frac{\del G^\sig}{\del g_\mn(x)}
  G^\rho
  =
  -
  \frac{\kap^2}{4}
  T^\mn(x)
  \ ,
\end{align}
with the Einstein tensor $G^\mn(x)$ (Eq.~\ref{EinsteinTensor}).
The Bianchi identities then imply:\checked
\begin{align}
  D_\mu
  \Big(
      \eta_\rs
    \frac{\del G^\sig}{\del g_\mn(x)}
    G^\rho
    \Big)
  =0
  \ .
\end{align}
This equation is satisfied by $G_\sig=0$ and in that case Eq.~\eqref{WEFT:GRE} reduces to the Einstein field equations.
This equation is particularly simple for the linear de Donder gauge choice Eq.~\eqref{WEFT:LDG} where $\delta G^\sig/\delta g_\mn(x)$ is independent of the metric.
In any case, when we use $S_{\rm GR}$ with the linear de Donder gauge function $G^\sig$ then Eq.~\eqref{WEFT:GRE} is exactly Eq.~\eqref{WEFT:QHP}.
Thus, we claim that solving $\del (S_{\rm GR}+S_{\rm matter})=0$ is equivalent to the Einstein field equations $\del (S_{\rm EH}+S_{\rm matter})=0$.

Finally, we discuss freely propagating gravitational waves and their propagation to infinity from a source.
Details beyond what is presented here can e.g. be found in chapter 10 on gravitational radiation of Ref.~\cite{Weinberg:1972kfs}.

First, if the gravitational field is weak the kinetic part of the gravitational action describes the dynamics at leading order.
In vacuum this part of the action gives rise to freely propagating plane waves.
A single such wave takes the form:\checked
\begin{align}
  h_\mn(x)
  =
  \operatorname{Re}
  \Big(
  \eps_\mn(k)
  e^{-ik\cdot x}
  \Big)
  \ ,
\end{align}
with $k$ its momentum (or wave vector) and $\eps_\mn(k)$ its polarization.
From the perspective of quantum field theory these may be thought of as free gravitons in analogy with photons and other massless bosons.
Redundant gauge freedom implies that there are only two independent polarizations which may be chosen as helicity states with helicity $\pm2$.

Second, we consider the gravitational waveform far away from a source with a typical length $l$.
Starting from the gauge fixed Einstein field equations Eq.~\eqref{WEFT:QHP}, we invert the wave operator with the retarded propagator.
In the wave zone ($|\vct{x}|\gg\{l,l^2\oma,1/\oma\}$) far away from the source, the leading order graviton field is:\checked
\begin{align}\label{WEFT:Waveform}
  \kap h_\mn(x)
  =
  \frac{4G}{|\vct{x}|}
  \int_\omega
  \mP_{\mn\ab}\,
  \tau^\ab(k)
  e^{-ik\cdot x}
  +
  \mO(|\vct{x}|^{-2})
  \ .
\end{align}
Here, $\tau_\mn(k)$ is the (local) energy-momentum tensor Eq.~\eqref{WEFT:PT} in momentum space and its momentum is parametrized as $k^\mu=(\oma,\oma\vct{\hat x})$ and, also, we have assumed $d=4$.
Here we use the following conventions for momentum space which are also summarized in the appendix in Eqs.~\eqref{WEFT:MomentumConvention}:\checked
\bse
\begin{align}
  \tau_\mn(k)
  &=
  \int \di^d x
  \,
  \tau_\mn(x)
  \,
  e^{ik\cdot x}
  \ ,\qquad
  \tau_\mn(x)
  =
  \int_k
  \,
  \tau_\mn(k)
  \,
  e^{-ik\cdot x}
  \ ,
\end{align}
with:\checked
\begin{align}
  \int_k
  =
  \int
  \frac{
    \di^d k
  }{
    (2\pi)^{d}
  }
  \ ,
\end{align}
and the one-dimensional integration on frequency (or energy):\checked
\begin{align}
  \int_\oma
  =
  \int
  \frac{\di\oma}{
    2\pi}
  \ .
\end{align}
\ese
The waveform Eq.~\eqref{WEFT:Waveform} is a sum of plane waves with polarizations $\tau_\mn(k)$ and wave vectors $k^\mu=(\oma,\oma \vct{\hat x})$.
It describes all linear and angular momentum carried away to infinity from the source by gravitational waves.
It is convenient to contract the waveform with polarizations which effectively projects out gauge dependence.
The waveform radiated away from a scattering event will be discussed in \chap~\ref{GB}.

\section{Worldline Effective Field Theory and the Gravitational Two-Body System}
\label{WEFT:WEFT}
In this section we focus on the worldline effective field theory (WEFT)~\cite{Goldberger:2004jt,Goldberger:2006bd,Goldberger:2005cd} for describing compact bodies and their gravitational interactions.
This EFT framework and the relevant equations of motion is the topic of the first Sec.~\ref{sec:Effective} and afterwards in Sec.~\ref{WEFT:GS} we focus on the post-Minkowskian expansion of the WEFT two-body system.
Here, our main focus is on the simplest spinless bodies adding spin afterwards in Sec.~\ref{sec:Spin}.

\subsection{Compact Bodies as Point-Like Particles}
\label{sec:Effective}
The basic idea of WEFT is to describe extended massive bodies such as black holes or neutron stars as point particles.
Such a description requires certain assumptions on the bodies and the gravitational field surrounding each body to be satisfied.
Considering a single body, we may take as basic expansion parameter its typical size $|l|$ compared with the typical radius of curvature $|L|$ of the gravitational field surrounding that body.
The scale $|l|$ typically scales as $Gm$ with $m$ the total mass of the body and $|L|$ as the (inverse) Riemann scalar: $R\sim1/|L|^2$.
The WEFT then describes the extended body as an expansion in $|l|/|L|\ll 1$.
This approximation is relevant for compact bodies which will have a small radius and for weak gravitational fields which will have a large radius of curvature.

Point particles are ideally described by worldline fields which describe the body degrees of freedom as a function of time which we will generally denote by $\tau$.
The most basic worldline field is the worldline parametrization $z^\sig(\tau)$.
This field may be considered as a map from the one dimensional space-time of the worldline (namely the time $\tau$) to the $d$ dimensional space-time in which it propagates.
It simply parametrizes the position of the point particle, $z^\mu(\tau)$, in our coordinate system $x^\mu$ as a function of the time $\tau$.
To this simple worldline field we can add other worldline fields which may describe internal degrees of freedom such as multipoles of the body including its spin.
We will include spin effects of the compact bodies in Sec.~\ref{sec:Spin}.

The avatar of a relativistic point particle with (proper) worldline time $\tau$ is described by the Polyakov action (see e.g.~\cite{Mogull:2020sak}):\checked
\begin{align}
  \label{eq:Polyakov}
  S_\rmm{Pol}
  =
  -\frac{m}2
  \int
  \di\tau
  \,
  g_\mn
  \big(
  z(\tau)
  \big)
  \dot z^\mu(\tau)
  \dot z^\nu(\tau)
  \ .
\end{align}
Here, dots on $z^\mu(\tau)$ indicate differentiation with respect to $\tau$.
In the spirit of effective field theory, we then add effective couplings to this action.
They describe finite size effects of the body under consideration and they will generally be proportional to the curvature and scale as some power of $|l|/|L|$.
Not all effective couplings are independent and the identification of all independent such couplings and their relative scaling in terms of the effective expansion parameter is a problem in effective field theory.
As an example of genuine non-minimal couplings on the worldline we have the tidal interactions (see e.g. Refs.~\cite{Kalin:2020lmz,Jakobsen:2022psy}):\checked
\begin{align}
  S_{E^2+B^2}
  =
  m
  \int \di \tau\,
  \Big(
  c_{E^2}
  E_\mn(\tau) E^\mn(\tau)
  +
  c_{B^2}
  B_\mn(\tau) B^\mn(\tau)
  \Big)
  \ .
\end{align}
Here, the worldline couples to the square of the electric and magnetic parts of the Riemann curvature tensor which in $d=4$ are given by:\checked
\bse
\begin{align}
  \label{WEFT:ETensor}
  E_\mn(\tau)
  &=
  R_{\mu\alpha\nu\beta}(z)
  \frac{
    \dot z^\alpha
    \dot z^\beta}{
    \dot z^2}
  \ ,
  \\
  B_\mn(\tau)
  &=
  \frac12
  R_{\mu\alpha}^{\ \ \rs}(z)
  \eps_{\rs\nu\beta}
  \frac{
    \dot z^\alpha
    \dot z^\beta}{
    \dot z^2}
  \ .
\end{align}
\ese
Here, $\eps_{\rs\nu\beta}$ is the $d=4$ Levi-Civita symbol.
The electric and magnetic curvatures are defined on the worldline trajectory $z^\sig(\tau)$ with respect to the instantaneous worldline frame of reference $\dot z^\sig(\tau)$.
In vacuum spacetime they include all information of the full Riemann tensor (evaluated on the worldline) and play an important role in the curvature expansion.
The couplings $C_{E^2}$ and $C_{B^2}$ are usually referred to as the quadrupole Love numbers and due to the quadratic appearance of the curvature they must scale as $|l|^4$.

With worldline effective field theory we have constructed an action describing a compact body viewed as a point particle interacting with the gravitational field at its own coordinate position.
A system of several (compact) bodies is now easily described by adding one copy of the WEFT action for each body together with the (gauge fixed) Einstein-Hilbert action describing the gravitational field in the vacuum between the bodies.
Let us generally write the action for a single worldline as:\checked
\begin{align}
  S_{\rmm{wl},i}
  &=
  S_{\rmm{wl},i}[g_\mn(x),w_i^\sig(\tau),C_{i}]
  \ .
\end{align}
On the right-hand-side we indicate its dependence on the gravitational field, a collection of worldline fields $w_i^\sig(\tau)$ labelled by $w$ and effective coupling constants (body parameters) parametrizing internal structure of the bodies.
The body parameters of the body $i$ are collectively denoted by $C_{i}$ and include its mass $m_i$, tidal numbers $C_{E^2,i}$ and $C_{B^2,i}$, and other possible parameters.
We use $w_i^\sig(\tau)$ to denote all relevant worldline fields collectively.
In the simplest case, this includes only the worldline parametrization, $z^\sig_i(\tau)$.
Later, however, when we add spin it will be described by an additional worldline field, $\psi_i^\sig(\tau)$.
In general, we could imagine any number of worldline fields to be relevant with any combinations of indices and $w_i^\sig(\tau)$ is simply a placeholder for the general case (see e.g. Ref.~\cite{Saketh:2022xjb} where all multipoles of the body are considered).

A system of any number of particles labelled by $i$ which interact gravitationally is then described by the following action,\checked
\begin{align}\label{WEFT:S}
  S
  =
  S_{\rm GR}
  +
  \sum_i
  S_{\rmm{wl},i}
  \ ,
\end{align}
with the gauge fixed Einstein-Hilbert action $S_{\rm GR}$ from Eq.~\eqref{WEFT:SGR}.
The case of interest to us will be the two-body system so that $i\in\{1,2\}$ (For the general case of many-body systems see e.g. Refs~\cite{Loebbert:2020aos,Jones:2022aji}).
With the action, $S$, we are then in the position to describe compact massive bodies and their gravitational interaction as long as the assumptions of WEFT are satisfied.

Let us discuss the equations of motion of this system using the simple example of the Polyakov action for the worldlines and hence ignoring any effective curvature couplings.
In that case the action simply reads:\checked
\begin{align}\label{eq:SpinlessSimple}
  S =
  S_{\rm GR}
  -
  \sum_{i=1,2}
  \frac{m_i}2
  \int d\tau
  \,
  g_\mn
  \big(
  z_i(\tau)
  \big)
  \dot z_i^\mu(\tau)
  \dot z_i^\nu(\tau)
  \ .
\end{align}
The equations of motion follow from the variational principle $\del S=0$.
The Einstein field equations for the gravitational field were discussed in Sec.~\ref{WEFT:GSGF} and with the gauge fixed Einstein-Hilbert action they are:\checked
\bse
\begin{align}
  G^\mn
  &=
  -
  8\pi G
  \,T^\mn
  \ ,
  \\
  G^\sig&=0
  \ .
\end{align}
\ese
Here $G_\sig$ is the gauge fixing function which we take to be the linear de Donder gauge Eq.~\eqref{WEFT:LDG}.
These equations are generic to gravity.
However, we now have a specific expression for the matter part of the action and we derive the corresponding energy-momentum tensor to be:\checked
\bse
\begin{align}
  T^\mn
  &=
  \frac1{\sqrt{-g}}
  \sum_i  \mT^\mn_i
  \ ,
  \\
  \mT^\mn_i
  &=
  \int \di\tau\,
  \delta^d
  \big(
  x-z_i(\tau)
  \big)
  \dot z^\mu_i(\tau)
  \dot z^\nu_i(\tau)
  \ .
\end{align}
\ese
The energy-momentum tensor is a sum of individual contributions from each worldline.
Due to the ($d$-dimensional) delta-function in the second line it is only non-zero when evaluated on the individual worldlines.
Thus, the extended bodies have effectively shrunk into points.
The individual energy-momentum (density) tensors $\mT^\mn_i$ are surprisingly simple and independent of the metric which is due to the linearity of the Polyakov action in the metric.

The classical equation of motion of the worldlines derived from $\del S/\del z_i^\sig(\tau)=0$ is the geodesic equation:\checked
\begin{align}
  0
  =
  \frac{
    \Di \dot z^\sig_i
    }{
    \di
    \tau}
  =
  \ddot
  z^\sig_{i}
  +
  \Gam^\sig_{\mn}[g_\ab(z_{i})]
  \dot z^\mu_{i}
  \dot z^\nu_{i}
  \ .
\end{align}
Here the Christoffel symbol, $\Gamma^\sig_\mn$, is evaluated as a function of the (full) metric $g_\mn(z_i)$ which in turn is evaluated on the worldline.
The linearity of the Polyakov action in the metric also allows us to write the geodesic equation linearly in $g_\mn$:\checked
\begin{align}
  0
  =
  g_\rs \ddot z^\rho
  +
  g_\rs \Gam^\sig_\mn
  \dot z^\mu
  \dot z^\nu
  \ .
\end{align}
This version is useful in the weak field expansion.

The equations of motion for both the gravitational field and the worldlines are ill defined.
This is most apparent in the geodesic equation.
Here, the full metric must be evaluated on the worldline.
However, since it is also sourced by the worldlines, it will generally be divergent at this point.
Concerning the Einstein field equations, the sources are delta-functions localized on the worldlines.
As delta-functions are not functions in the usual sense, divergences may very well appear in the metric too.
See e.g. \cite{Wald:2009ue} for further discussion of this topic.

The appearance of divergences, however, is not surprising in the framework of effective field theory.
In our case, they appear due to the fact that we have shrunken a finite body into a point and we cannot trust our equations of motion very near to the body.
In order, however, to make the equations well-defined we have to use a regularization scheme.
In our case, this will be dimensional regularization (dim. reg.).
Thus we generally work with the gravitational field in $d$ space-time dimensions keeping the worldlines one-dimensional.
In principle, when performing a computation, we can only let $d\to4$ at the end of the computation (or to some other dimension if we are interested in that).

The regularization procedure is well-known from quantum field theory and it is essential in effective field theory.
In QFT, if the theory is renormalizable, the divergences must disappear into renormalizations of the existing couplings.
In effective field theory, however, the divergences will generally require the addition of new effective couplings.
In a sense, the equations of motion can only be expected to have well defined perturbative solutions (perturbative in $|l|/|L|$).
Generally, it would be expected that divergences appear and renormalization introduces finite size effective terms.
This indeed happens in the computation of the (gauge dependent) metric~\cite{Jakobsen:2020ksu,Mougiakakos:2020laz,Goldberger:2004jt} or tail effect~\cite{Galley:2015kus}.

\subsection{Gravitational Scattering and the Post-Minkowskian Expansion}
\label{WEFT:GS}
Scattering events of compact bodies are particularly well described by the worldline effective field theory approach.
In fact, the whole dynamical evolution of the system may very well stay within the assumptions of the point particle approximation which will be the case if they are widely separated from each other compared with their Schwarzschild radii.
That is, if both of $\frac{G m_i}{|b|}\ll 1$ for $i=1,2$ where $|b|$ is a typical length between the bodies.
In this case the gravitational field of the second body will be weak at the point of the first body and vice versa and we may thus use a weak field expansion of the metric.
This perturbative scheme can be identified as a formal expansion in Newton's constant $G$ and is referred to as the post-Minkowskian expansion.
The dimensionless expansion parameters, however, are $\frac{Gm_i}{|b|}$ which are the WEFT expansion parameters of each body and the two expansions are thus intimately related.

A scattering event is particularly clean in the sense that the initial and final worldline states can be described in terms of asymptotic states defined on flat space-time which are independent of the general covariance of gravity.
The knowledge of the asymptotic initial and final states of the worldlines together with the waveform encode all gauge invariant information of the scattering event.
In a sense, the asymptotic states describe Minkowskian motion and post-Minkowskian interactions appear in the bulk in between the initial and final states.
We will first analyze the (spinless) asymptotic worldline states and then consider the post-Minkowskian expansion and equations of motion.
Asymptotic states with spin and spinning particles in general relativity are then introduced in Sec.~\ref{sec:Spin}.

\subsubsection{Kinematics of Worldline Asymptotic States}
Asymptotic spinless worldline states are described by the Minkowski space-time Polyakov action:\checked
\begin{align}
  \label{WEFT:WLK}
  S^{\rm kin}_{\rmm{wl},i}
  =
  -
  \frac{\tilde m_i}{2}
  \int
  \di \tau_i\,
  \eta_\mn
  \dot z_i^\mu(\tau_i)
  \dot z_i^\nu(\tau_i)
  \ .
\end{align}
We refer to this part of the worldline action as the kinetic part which is denoted with the superscript ``kin''.
It is the Minkowskian limit of the full (spinless) worldline action $S_{\rmm{wl},i}$ in which the graviton field $h_\mn(x)$ is send to zero.
In this limit the curvature disappears and all effective couplings (except the mass) are irrelevant.
We have put a tilde on the mass, because we have not yet fixed a gauge.
We will refer to the remaining symmetries of the full action as gauge symmetries which include scaling of the worldline time and translation of the worldline coordinate and time.

Solutions to the classical equations of motions of the kinetic action are straight lines which we parametrize as:\checked
\begin{align}\label{WEFT:SL}
  z_i^\mu(\tau)
  =
  b^\mu_i
  +
  \tau \tilde v_i^\mu
  \ .
\end{align}
We consider $b_i^\mu$ and $\tilde v_i^\mu$ as arbitrary four vectors except that we require a timelike motion ${\tilde v_i^2>0}$.
Space-time coordinate translation $\del c^\mu$ and worldline time translations $\del \xi_i$ result in transformations of $b_i^\mu$,\checked
\bse\label{WEFT:GT}
\begin{align}
  \del b_i^\mu
  =
  \del c^\mu
  +
  v^\mu_i \del \xi_i
  \ ,
\end{align}
with $\tilde v^\mu_i$ and $\tilde m_i$ invariant.
In addition, scaling of the worldline time and mass result in the (infinitesimal) transformation:\checked
\begin{align}
  \del
  \tilde
  v_i^\mu
  =
  \alpha
  \tilde v_i^\mu
  \ ,
  \qquad
  \del \tilde m_i=-\alpha \tilde m_i
  \ ,
  \qquad
  \del \tau_i = -\alpha \tau_i
  \ ,
\end{align}
\ese
with $b^\mu_i$ invariant.
The kinetic action is invariant under all of these transformations.

The only gauge invariant combinations of variables are the two four-momenta and the (relative, orthogonal) impact parameter, $b^\mu$:\checked
\bse\label{WEFT:Vectors}
\begin{align}
  p_i^\mu
  &=
  \tilde m_i
  \tilde v_i^\mu
  \ ,
  \\\label{WEFT:IP}
  b^\mu
  &=
  P_{12}^\mn  (b_2-b_1)_\nu
  \ ,
\end{align}
\ese
with the projector $P_{12}^\mn$ which maps four vectors into the subspace orthogonal to both of $p_i^\mu$.
It is given by:\checked
\begin{align}
  P_{12}^\mn
  =
  \eta^\mn
  +
  \frac{
    \hat p_1^\mu
    \hat p_1^\nu
    +
    \hat p_2^\mu
    \hat p_2^\nu
    -
    2
    \gam\,
    \hat p_{1}^{(\mu}
    \hat p_{2}^{\nu)}
  }{
    \gam^2-1}
  \ .
\end{align}
Here, and in the following, we use the notation $\hat Z^\mu=Z^\mu/|Z^\mu|$ with $|Z^\mu|=\sqrt{|Z^2|}$ for any four-vector $Z^\mu$ (see also the appendix Eq.~\ref{A:VA}).
The relative Lorentz factor is $\gam=\hat p_1\cdot \hat p_2$.
The vectors in Eqs.~\eqref{WEFT:Vectors} have eight arbitrary remaining vector components which is consistent with the original 16 components and eight symmetry degrees of freedom.
Lorentz space-time rotations reduce these to the invariant scalar products:\checked
\bse\label{WEFT:Scalars}
\begin{align}
  p_i^2
  &=
  m_i^2
  \ ,
  \\
  \hat p_1\cdot \hat p_2
  &=
  \gam
  \ ,
  \\
  \sqrt{-b^2}
  &=
  |b|
  \ .
\end{align}
\ese
Gauge invariant observables (i.e. invariant under Eqs.~\eqref{WEFT:GT} and possibly Lorentz transformations) can depend only on the corresponding gauge invariant quantities.
Thus a scalar observable can depend only on $m_i$, $\gam$ and $|b|$.

In the following, we will always use proper time for the asymptotic states and we denote the proper time velocity by $v_i^\sig$:\checked
\begin{align}
  v^\sig_i
  =
  \hat p^\sig_i
  \ .
\end{align}
The genuine masses of the bodies are denoted by $m_i=|p_i|$.
We may also pick a gauge where both $b_i^\sig$ are proportional to $b^\sig$.
Indeed, by shifting the coordinate center we may set either to zero choosing e.g. $b_2=b$ and $b_1=0$.
Having picked any gauge we can always re-express gauge invariant quantities in terms of the unconstrained variables using Eqs.~\eqref{WEFT:Vectors}.

The asymptotic (spinless) worldline states of the two-body system are now described by the three gauge invariant vectors $p_i^\mu$ and $b^\mu$.
We may, however, use different combinations of those vector which each describe important physical variables of the system.

The total (worldline) four-momentum $P^\mu$ is given by:\checked
\bse\label{WEFT:CoM}
\begin{align}
  P^\mu
  &=
  p_1^\mu+p_2^\mu
  \ .
\end{align}
This momentum defines the center-of-mass (CoM) inertial frame of reference.
We note that the labelling of this frame as center-of-mass frame is conventional although the concept of ``center-of-mass'' is not unique in relativistic dynamics.
In the CoM frame the bodies have a relative momentum $p^\mu$:\checked
\begin{align}
  p^\mu
  &=
  (\eta^\mu_\nu
  -
  \hat P^\mu
  \hat P_\nu
  )
  p_1^\nu
  =
  -
  (\eta^\mu_\nu
  -
  \hat P^\mu
  \hat P_\nu
  )
  p_2^\nu
  \nn
  \\
  &=
  \frac{m_1 m_2}{E^2}
  \Big(
  (\gam m_1+m_2)
 v_1^\mu
  -
  (\gam m_2+m_1)
  v_2^\mu
  \Big)
  \ .
\end{align}
\ese
Here, $E$ is the total CoM energy given by $E=|P^\mu|$.
Naturally, from their definition, the total and relative momenta are orthogonal $p\cdot P=0$.
The inverse transformations to Eqs.~\eqref{WEFT:CoM} are given by:\checked
\bse
\begin{align}
  p_1^\mu
  &=
  E_1 \hat P^\mu
  +
  p_\infty \hat p^\mu
  =
  (E_1,\vct{p})
  \ ,
  \\
  p_2^\mu
  &=
  E_2 \hat P^\mu
  -
  p_\infty \hat p^\mu
  =
  (E_2,-\vct{p})
  \ .
\end{align}
\ese
The last expression of each line refers to the CoM frame.
Here, $E_i$ are the energies of each body in the CoM frame $E_i=\hat P\cdot p_i$ and $p_\infty$ is the magnitude of the relative momentum $p_\infty=|p|$.

It is also common to define the total mass $M$, the reduced mass $\mu$ and the symmetric mass ratio $\nu$ by:
\begin{align}
  M=m_1+m_2
  \ ,\qquad
  \mu = \frac{m_1 m_2}{M}
  \ ,\qquad
  \nu = \frac{\mu}{M}
  \ .
\end{align}
The relative mass difference $\delta$ is then defined by,
\begin{align}
  \del = \frac{m_2-m_1}{M}
  =
  \sqrt{1-4\nu}
  \ ,
\end{align}
where the last equality holds only when $m_2>m_1$.
We also define the reduced energy $\Gam$:
\begin{align}
  \Gam=
  \frac{E}{M}
  =
  \sqrt{1+2\nu(\gam-1)}
\end{align}

Another important physical quantity is the (total) angular momentum tensor $L^\mn$,\checked
\begin{align}\label{WEFT:Angular}
  L^\mn
  &=
  \sum L_i^\mn
  \ , \qquad
  L_i^\mn
  =
  2b^{[\mu}_i
    p^{\nu]}_i
  \ ,
\end{align}
which is, however, not invariant under translations of the coordinate center.
The corresponding Pauli-Lubanski vector $L^\mu$, however, is invariant and given by,\checked
\begin{align}\label{WEFT:OPB}
  L^\mu
  =
  \frac12
  \eps^{\mn\ab}
  \hat P_\nu
  L_\ab
  \ .
\end{align}
with magnitude $|L^\mu|=|b| p_\infty$.
Its gauge invariance implies that it can be related to the momenta and (orthogonal) impact parameter:\checked
\begin{align}
  L_\mu
  &=
  -
  |b|\, p_\infty\,
  \eps_{\mn\ab}
\hat b^{\nu} \hat p^\alpha \hat P^\beta
=
  -\frac{1}{E}\eps_{\mn\ab}
  b^{\nu} p_{1}^{\alpha} p_{2}^{\beta}
  \ .
\end{align}
We will discuss angular momentum and spin in greater detail in Sec.~\ref{sec:Dynamics}.
When spin is introduced neither the impact parameter $b^\sig$ nor the Pauli-Lubanski vector of orbital angular momentum $L^\sig$ are gauge invariant.

An additional symmetry to the ones discussed above is the exchange of the two worldlines.
Thus, with the generic action Eq.~\eqref{WEFT:S} each worldline is described by the same worldline action with its own set of parameters labelled by $1$ and $2$.
The system is invariant under exchange of these two sets of parameters.
We refer to this as the particle exchange symmetry.
The transformation of asymptotic state variables under this symmetry is easily derived with e.g. $p_1^\sig\leftrightarrow p_2^\sig$ and $b^\sig\to-b^\sig$.

Let us finally introduce a generic frame $V^\mu$ defined by:\checked
\begin{align}
  \label{WEFT:GF}
  V^\mu
  =
  \alpha_1
  v_1^\mu
  +
  \alpha_2
  v_2^\mu
  \ ,
\end{align}
with the $\alpha_i$ chosen so that $V^2=1$ and $V\cdot v_i>0$.
This frame includes as special cases all frames spanned by $v_1^\mu$ and $v_2^\mu$.
If we put $\alpha_i=m_i/E$ we get the CoM frame and for $\alpha_1=0$ we get the frame of $v_2^\mu$ and vice versa.
The unit vector orthogonal to $V^\mu$ in the $v_i$ subspace is:\checked
\begin{align}
  V^\mu_\bot
  =
  \frac{
    2
      v_1^{[\mu} v_2^{\nu]}
  V_\nu
  }{\sqrt{\gam^2-1}}
  \ ,
\end{align}
with normalization $V_\bot^2=-1$ and such that $V_\bot^\mu$ points in the direction of motion of the first body or opposite to the motion of the second body.
Clearly this vector is purely spacial in the generic frame: $V_\bot\cdot V=0$.

In the CoM frame we find that $V_\bot^\mu \to \hat p^\mu$ and in the rest frames of each body we find $V_\bot\to v_{i\bot}^\mu$ with:\checked
\begin{align}
  \label{WEFT:VBot}
  v_{1\bot}^\mu
  =
  \frac{-v_2^\mu+\gam v_1^\mu}{
    \sqrt{\gam^2-1}
  }
  \ ,
  \qquad
  v_{2\bot}^\mu
  =
  \frac{v_1^\mu-\gam v_2^\mu}{
    \sqrt{\gam^2-1}
  }
  \ .
\end{align}
In the generic frame $V^\mu$ in four space-time dimensions we have the spanning set of orthonormal vectors $V^\mu$, $V_\bot^\mu$, $\hat b^\mu$ and $\hat L^\mu$.
The vector $V^\mu$ is timelike and the rest are spacelike.
We may write the ($d=4$) identity as:\checked
\begin{align}
  \eta^\mn
  =
  V^\mu V^\nu
  -
  V_\bot^\mu
  V_\bot^\nu
  -
  \hat b^\mu
  \hat b^\nu
  -
  \hat L^\mu
  \hat L^\nu
  =
  \frac{
  v_2^\mu
  v_{1\bot}^\nu
  -
  v_1^\mu
    v_{2\bot}^\nu
  }{
    \sqrt{\gam^2-1}
    }
  -
  \hat b^\mu
  \hat b^\nu
  -
  \hat L^\mu
  \hat L^\nu
  \ .
  \nn
\end{align}
Such expressions are sometimes useful in order to simplify results.

\subsubsection{Observables and Post-Minkowskian Expansion}
The initial and final worldline states of the scattering process are both described by asymptotic states.
If we for a given asymptotic variable X label its initial and final values by $X_{-\infty}$ and $X_{\infty}$ respectively, we may define worldline observables (kicks) $\Del X$ with the following simple formula:\checked
\begin{align}  \label{WEFT:Kick}
  \Del X
  =
  X_{\infty}-X_{-\infty}
  \ .
\end{align}
For the momenta $p_i^\sig$ we get the impulses $\Del p_i^\sig$ and for the orbital angular momentum $L^\mu$ we get the total change of orbital angular momentum $\Del L^\mu$.

The (post-Minkowskian) interaction terms are defined as the remaining part of the action to the kinetic terms, or equivalently, the part of the action which scales with $\kappa$ to some power $n>0$.
For the gauge fixed Einstein-Hilbert action its interaction term $S_{\rm GR}^{\rm int}$ was given in Eq.~\eqref{WEFT:EHInteraction} and the leading order worldline interaction term is:\checked
\begin{align}
  S_{\rmm{wl},i}^{\rm int}
  =
  -\frac{\kap m_i}2
  \int
  \di \tau\,
  h_\mn\,
  \dot z^\mu_i
  \dot z^\nu_i
  +
  \mO\big(
  |l|/|L|,|S^\mn_i|
  \big)
  \ .
\end{align}
Here, we indicated further terms scaling with the effective expansion parameter $|l|/|L|$ (i.e. the curvature, see Sec.~\ref{sec:Effective}) or the spins of the bodies labelled $S_i^\mn$.
The leading order term is simply the interaction term of the Polyakov action.

The kicks can be obtained perturbatively in the post-Minkowskian expansion from solving the equations of motion.
Each worldline field is expanded as its flat space-time asymptotic motion together with a perturbation:\checked
\bse\label{WEFT:PME}
\begin{align}
  w^\sig_i(\tau)
  =
  w^\sig_{i,\rmm{Min}}(\tau)
  +
  \Del w^\sig_i(\tau)
  \ .
\end{align}
For the worldline parametrization $\Del z^\sig_i(\tau)$, the background motion $w^\sig_{i,\rmm{Min}}(\tau)$ is given by straight-line motion $b_i^\sig+\tau v_i^\sig$ (Eq.~\ref{WEFT:SL}).
We generally refer to $\Del w^\sig_i(\tau)$ as the worldline fluctuations and specifically to $\Del z^\sig_i(\tau)$ as the trajectory fluctuation.
The gravitational field is expanded around flat space-time as in Eq.~\eqref{WEFT:Graviton}:\checked
\begin{align}
  g_\mn(x) =
  \eta_\mn
  +
  \kap h_\mn(x)
  \ .
\end{align}
\ese
The equations of motion are expanded in $\kap$ where the worldline fluctuations $\Del w^\sig_i(\tau)$ are assumed to scale with $\kap$.
The equations of motion of the perturbation fields can be written as:\checked
\bse\label{WEFT:PEOMs}
\begin{align}
  m_i
   \Del\ddot z^\sig_i(\tau)
  &=
   -\frac{\del S_{\rmm{wl},i}^{\rm int}}{
    \del \Del z_i^\sig(\tau)
  }
  \ ,
  \\
  P^{\mn\ab}
  \partial^2
  h_{\ab}(x)
  &=
  -\frac{\kap}{2}
  \tau^\mn(x)
  \ .
\end{align}
\ese
Here, we focused only on the trajectory fluctuation and graviton field with $\tau^\mn(x)$ the local energy-momentum tensor Eq.~\eqref{WEFT:PT}.
The systematic perturbative solution to these equations with the Worldline Quantum Field Theory approach will be discussed in \chap~\ref{WQFT}.
By other classical means, see e.g. Ref.~\cite{Saketh:2021sri}.

The solution of the classical equations of motion with dimensional regularization is uniquely defined as soon as suitable boundary conditions are given.
The boundary conditions may be given as the initial state of the system from which, then, the final one is fully determined.
The boundary conditions allow us to invert the kinetic terms of the perturbation fields in Eq.~\eqref{WEFT:PEOMs} and for background variables defined at past infinity a retarded prescription is required for the propagators.
The kinetic operators of the trajectory fluctuation and graviton are $\di^2/\di\tau^2$ and $\pat^2$ respectively with the first operator being the one-dimensional version of the $d$-dimensional $\pat^2$.
The ($d$-dimensional) retarded propagator $D_{\rm R}(x-y)$ is most easily derived from its momentum space version and given by:\checked
\begin{align}\label{WEFT:RP}
  D_{\rm R}(x-y)
  =
  \int_k
  e^{-ik\cdot(x-y)}
  \frac1{
    (k^0+i\eps)^2
   -
    \vct{k}^2
  }
  \ .
\end{align}
Here, the momentum $k^\mu$ is split into time and spacial parts with respect to some arbitrary (timelike) frame.
The fact that the retarded propagator is independent of this choice of frame is because the split of $k^\mu$ into time and spacial parts in the denominator only matters when the propagator goes on-shell $k^2=0$.
For the one-dimensional case we get the propagator,\checked
\begin{align}\label{WEFT:WLP}
  D_{\rm R}(\tau-\tau')
  =
  \int_\oma
  e^{-i\oma(\tau-\tau')}
  \frac1{(\oma+i\eps)^2}
  =
  -(\tau-\tau')\,\theta(\tau-\tau')
  \ ,
\end{align}
with the Heaviside function $\theta(\tau)$ which makes manifest that this propagator is non-zero only for future times.

The use of the retarded propagator leads to loss of energy of the worldline fields to radiation.
One may, however, also consider the use of time-symmetric propagators which lead to conservative dynamics for the worldline fields.
In that case the motion of the (worldline) point particles may be described by Hamiltonian dynamics.
Such a description is common in the post-Newtonian literature where the retarded propagator is split into a conservative and a radiative piece.

\section{Spinning Compact Bodies with Supersymmetry}
\label{sec:Spin}
The inclusion of spin effects introduces new fields, symmetries and observables.
Several approaches to the description of spin in the (W)EFT context exist~\cite{Porto:2005ac,Marsat:2014xea,Levi:2015msa,Vines:2016unv,Liu:2021zxr,Saketh:2022wap} and it is not clear which is the most advantageous.
Here, we use the the supersymmetric (SUSY) description of spin first introduced in the WEFT context in Refs.~\cite{Jakobsen:2021lvp,Jakobsen:2021zvh} with prior work on that approach discussed in the introduction (Sec.~\ref{INT}).
First, in Sec.~\ref{sec:Dynamics} we focus on the description of the spinning asymptotic states moving in flat space-time and the different gauge choices relevant to the SUSY of the spinning worldline action.
In principle, this description is accurate to all orders in spin, in the sense that perturbative corrections appear only when the coupling to the gravitational field is considered.
This coupling to the curvature, or covariantization, is then considered in the next Sec.~\ref{WEFT:SUSY} with results accurate to quadratic order in the spins.
The content of this section is based mostly on material of the Refs.~\cite{Jakobsen:2022zsx,Jakobsen:2021zvh}.

\subsection{Gauge Symmetries and Kinematics of Spinning Bodies}
\label{sec:Dynamics}
In this section we first analyze the asymptotic states of spinning point-like particles and then, in the final part, we introduce two common gauge choices for the spinning variables: Covariant and canonical gauge.
The inclusion of spin introduces one additional worldline field and a corresponding observable the spin kick.

\subsubsection{Spinning Asymptotic States and Observables}
Asymptotic states of spinless particles were analyzed in Sec.~\ref{WEFT:GS} and we will now generalize those to include spin.
Using Grassmann variables, $\psi^\sig_{i,A}(\tau)$ to describe the spin degrees of freedom, the kinetic action for the spinning asymptotic states (motion in flat space-time) read:\checked
\begin{align}
  \label{WEFT:WLKin}
  S^{\rm kin}_{\rmm{wl},i}
  =
  -
  \frac{m_i}2
  \int
  \di \tau\,
  \Big(
  \dot z_i(\tau)^2
  +
  i\,
  \psi_{i,A}(\tau)
  \cdot
  \dot \psi_{i,A}(\tau)
  \Big)
  \ .
\end{align}
The Grassmann variables $\psi^\sig_{i,A}$ are anti-commuting $\psi^\sig_{i,A}(\tau)\psi^{\sig'}_{i',A'}(\tau)=-\psi^{\sig'}_{i',A'}(\tau)\psi^\sig_{i,A}(\tau)$ worldline vector fields with a flavor index $A=1,...,\mN$ with $\mN$ the number of flavors.
Here, and in the following we assume summation on the flavor index $A$.

The important additional (gauge) symmetry to time and coordinate translation and time scaling is the supersymmetry:\checked
\bse\label{WEFT:STG}
\begin{align}
  \del \psi_{i,A}^\sig(\tau)
  &=
  \eps_{i,A} \dot z^\sig_i(\tau)
  \ ,
  \\
  \del z^\sig_i(\tau)
  &=
  -i\eps_{i,A}
  \psi_{i,A}^\sig(\tau)
  \ ,
\end{align}
\ese
with infinitesimal SUSY parameter $\eps_{i,A}$.
In addition the action is invariant under internal rotations of the Grassmann fields with respect to the flavor index $A$.

The Grassmann variables themselves do not have direct physic relevance but, instead, they may be combined into the spin tensor $\mS_i^\mn(\tau)$ as follows:\checked
\begin{align}
  \mS^\mn_i(\tau)
  =
  -i m_i
  \psi^\mu_{i,A}(\tau)
  \psi^\nu_{i,A}(\tau)
  \ .
\end{align}
The anti-commutativity of the Grassmann variables ensures that the spin tensor is antisymmetric.
In a sense, the Grassmann variable is a square root of the spin tensor.
The spin tensor describes the (physical) relativistic internal angular momentum of each body (see e.g. Refs.~\cite{Weinberg:1972kfs,Manohar:2022dea}).

We assume proper time and parametrize the asymptotic states (which are solutions to $\del S_{\rmm{wl},i}^{\rm kin}=0$) as:\checked
\bse
\begin{align}
  z_i^\mu(\tau)
  &=
  b^\mu_i
  +
  \tau v_i^\mu
  \ ,
  \\
  \psi_{i,A}^\mu(\tau)
  &=
  \Psi_{i,A}^\mu
  \ ,
  \\
  \mS^\mn_i(\tau)
  &=
  S^\mn_i
  \ .
\end{align}
\ese
The third equation is a consequence of the second one with the identification\checked
\begin{align}\label{WEFT:SP}
  S_i^\mn
  =
  -i
  m_i
  \Psi^\mu_{i,A}\Psi^\nu_{i,A}
  \ .
\end{align}
In flat space, the spin tensor is constant.
Proper time implies that $v^2_i=1$ and $p^\mu_i=m_i v^\mu_i$.

Each body is now described by two vectors and a tensor: $p^\mu_i$, $b_i^\mu$ and $S^\mn_i$.
The four-momenta $p_i^\mu$ are still invariant under all gauge transformations but the SUSY transformations do not leave the (orthogonal, relative) impact parameter $b^\sig$ invariant.
The impact parameter was defined in Eq.~\eqref{WEFT:IP} as\checked
\begin{align}
  b^\mu
  =
  P_{12}^{\mn}
  (b_2-b_1)_\nu
  \ ,
\end{align}
with the projector $P_{12}^\mn$ into the space orthogonal to $v_i^\mu$.
Thus, it has to be generalized in order to stay invariant under SUSY transformations.

The SUSY transformations act on the background parameters as follows:\checked
\bse\label{WEFT:ST}
\begin{align}
  \del b^\sig_i
  &=
  -i\eps_{i,A} \psi^\sig_{i,A}
  \ ,
  \\
  \del \psi^\sig_{i,A}
  &=
  \eps_{i,A} v^\sig
  \ ,
  \\
  \del S^\mn_i
  &=
  2
  p_i^{[\mu}
    \del b_i^{\nu]}
  \ ,
\end{align}
\ese
with $\del v_i^\sig=0$.
Again, the third line is a consequence of the first two lines.
The SUSY acts on the spin tensor in the same way as a change of spin supplementary condition (SSC).
For that reason, we also refer to the SUSY as the SSC symmetry.
We easily identify the gauge invariant spin variables by projecting out the velocity direction:\checked
\bse
\label{WEFT:SpinO}
\begin{align}
  \Psi^\mu_{i,A,\bot}
  &=
  (P_i \cdot \Psi_{i,A})^\mu
  \ ,
  \\
  S^\mn_{i,\bot}
  &=
  S_\ab
  P^{\alpha\mu}_i
  P^{\beta\nu}_i
  \ .
\end{align}
\ese
These variables are clearly invariant under SUSY transformations and remove one degree of freedom from $\Psi^\mu_{i,A}$ corresponding to the (one degree of) symmetry.
In four space-time dimensions, we may use the Pauli-Lubanski vector $S^\mu_i$ instead of the tensor $S_{i,\bot}^\mn$:\checked
\bse\label{WEFT:CPB}
\begin{align}\label{WEFT:VPB}
  S^\mu_i
  &=
  \frac12
  \eps^{\mu}_{\ \nu\ab}
  v_{i}^{\nu}
  S_{i}^{\ab}
  \ ,
  \\
  S^\mn_{i,\bot}
  &=
  \eps^{\mn}_{\ \ \ab}
  v_{i}^{\alpha}
  S_{i}^{\beta}
  \ .
\end{align}
\ese
In the following we will often use the Pauli-Lubanski spin vector $S_i^\sig$ which we also refer to as the covariant spin vector.
Its mass scaled version $a^\sig_i$ is also often useful:\checked
\begin{align}
  a^\sig_i = \frac{S^\sig_i}{m_i} \ .
\end{align}

In addition to the spin tensors, the parameters $b_i^\mu$ are also easily generalized to be independent of the SUSY transformations:\checked
\begin{align}
  \label{WEFT:Beta}
  \beta_i^\mu
  =
  b_i^\mu 
  +
  \frac1{m_i}
  v_{i,\nu}
  S_i^{\mn}
  \ .
\end{align}
The parameters $\beta_i^\mu$ are SUSY independent as long as the SUSY shift in $b_i^\mu$ is orthogonal to $v_i^\mu$.
This can always be ensured with an additional shift in the proper time.
We may now simply define the SUSY impact parameter:\checked
\begin{align}
  \beta^\mu
  =
  P_{12}^\mn
  (\beta_2
  -
  \beta_1)_\nu
  \ .
\end{align}
This parameter is invariant under all background gauge symmetries.
It is a generalization of the former impact parameter $b^\sig$ and we will refer to it as the SUSY impact parameter.

The gauge invariant background parameters are then the four-momenta, the orthogonal spin tensors (or Grassmann variables) and the SUSY impact parameter: $p_i^\mu$, $S^\mn_{i,\bot}$ and $\beta^\mu$.
Let us now move on to other physical variables of interest to the scattering system.

The individual orbital momenta of the bodies $L_i^\mn$ and the total $L^\mn$ were all defined in Eqs.~\eqref{WEFT:Angular}.
The total angular momentum of each body, $J^\mn_i$, and of the system as a whole, $J^\mn$, are then:\checked
\bse
\begin{align}
  J^\mn_i
  &=
  L^\mn_i+S_i^\mn
  \\
  J^\mn
  &=
  J_1^\mn+J_2^\mn
\end{align}
\ese
The SUSY transformation Eqs.~\eqref{WEFT:ST} may be interpreted as changing the coordinate center of the (point-like) extended massive bodies.
We have then three coordinate centers, the one of the coordinate system with variation $\del c$ and the centers of each body with variation $\del c_i$.
Changing those affect the different angular momentum tensors, $J_i^\mn$, $L_i^\mn$ and $S_i^\mn$ with $\del b_i=\del c+ \del c_i$:\checked
\bse
\begin{align}
  \del S_i^\mn
  &=
  2 p_i^{[\mu} \del c_i^{\nu]}
  \\
  \del L_i^\mn
  &=
  2 p_i^{[\mu}
    (\del c_i
  +
  \del c)^{\nu]}
    \\
    \del J_i^\mn
    &=
    2 p_i^{[\mu}
      \del c^{\nu]}
\end{align}
\ese
In particular, the total angular momentum $J^\mn$ transforms as:\checked
\begin{align}
  \del J^\mn
  =
  2 P^{[\mu}
    \del c^{\nu]}
  \ .
\end{align}
And the Pauli-Lubanski vector defined as:\checked
\begin{align}
  J^\mu
  =
  \frac12
  \eps^{\mn\ab}
  \hat P_\nu
  J_\ab
  \ ,
\end{align}
is invariant under all gauge symmetries.
This is an alternative gauge invariant variable to the impact parameter $\beta^\mu$.
They are related by:\checked
\bse
\begin{align}
  \label{WEFT:TAM}
  J^\mu
  &=
  \mL^\mu
  +
  2\hat P_\nu
  \sum_i
  S_i^{[\mu}v^{\nu]}_i 
  \ ,
  \\
  \mL^\mu
  &=
  -\frac1{E}
  \eps^{\mn\ab}
  \beta_\nu
  p_\alpha
  P_\beta
  \ .
\end{align}
\ese
The vector $\mL^\mu$ is gauge invariant, too, and corresponds to the total CoM orbital angular momentum with respect to the SUSY $\beta^\mu_i$ parameters.

Observables with spin are now easily defined as in Eq.~\eqref{WEFT:Kick} as the difference of asymptotic variables at future and past infinity.
We have, in principle, five vectorial observables corresponding e.g. to the momenta $p_i^\mu$, the total angular momentum $J^\mu$ and the Pauli-Lubanski spin vectors $S_i^\mu$.
In particular, the new observable when considering spin is the spin kick where we may consider both the spin (vector) kick $\Del S^\mu_i$ or the spin (tensor) kick $\Del S^\mn_{i,\bot}$.
In addition we have the Grassmann kick $\Del \psi^\mu_{i,A}$ from which both spin kicks may be derived (see below Eqs.~\ref{WEFT:SpinKicks}).
Just like the momentum $p_i^2=m^2$ is conserved the internal rotations of $\psi_{i,A}^\mu$ on $A$ implies conservation of $\psi^\mu_{i,A} \psi_{i,A,\mu}$ which in turn implies conservation of the spin length $S^\mn_i S_{i,\mn}$ or $S^\mu_i S_{i,\mu}$.
The SUSY invariant parameters introduced here were indirectly considered in Refs.~\cite{Jakobsen:2022zsx,Jakobsen:2021zvh} where the corresponding symmetries were discussed.

\subsubsection{Covariant and Canonical SSC}
Concerning the SSC (SUSY) symmetry, there are two special gauges that are often used in the literature: The covariant and canonical spin supplementary conditions.
We will first consider the covariant (Tulczyjew-Dixon) SSC which in many respects seem the simplest gauge choice.
The canonical (Pryce-Newton-Wigner) SSC, however, is essential for using Hamiltonian dynamics for the spin.
For additional discussions of the different choices of SSC see e.g. Refs.~\cite{Vines:2016unv,Vines:2017hyw,Barausse:2009aa,Antonelli:2020ybz,Jakobsen:2022zsx}.

The covariant SSC is defined by the following constraints of the spin tensor and Grassmann variable:\checked
\bse\label{WEFT:CV}
\begin{align}
  S^\mn_{i,\rmm{cov}} v_{i,\mu} = 0\ ,
  \\ 
  \Psi_{i,A}^{\rmm{cov}}\cdot v_{i}=0
  \ .
\end{align}
\ese
The two equations follow from each other via Eq.~\eqref{WEFT:SP}.
One realizes from the SUSY transformations Eqs.~\eqref{WEFT:ST} that this choice is always possible.
It has the following simple property that the SUSY impact parameter coincides with the covariant impact parameter and similarly the orthogonal spin tensor coincides with the covariant spin tensor:\checked
\bse
\label{WEFT:CVP}
\begin{align}
  b_{\rm cov}^\mu
  &=
  \beta^\mu
  \ ,
  \\
  S^\mn_{i,\rmm{cov}}
  &=
  S^\mn_{i,\bot}
  \ .
\end{align}
\ese
Here, we use the subscript ``cov'' to indicate that the gauge dependent variables $b^\mu$ and $S^\mn_i$ are evaluated in covariant gauge.
Covariant gauge is then very similar to working with the SUSY independent variables $\beta^\mu$ and $S_{i,\bot}^\mn$.
That is, in this gauge, the SUSY impact parameter coincides with the ``physical'' impact parameter and likewise for the spin tensor.

The canonical SSC is defined with respect to some frame $V^\mu$ (which could be the frame introduced in Eq.~\eqref{WEFT:GF}):\checked
\begin{align}
  S^\mn_{i,\rmm{can}} (v_{i,\mu}+V_\mu)
  &=
  0
  \ ,
  \\
  \Psi_{i,A}^{\rmm{can}}\cdot (v+V)
  &=
  0
  \ .
\end{align}
The SUSY parameter $\eps_{i,A}$ which relates the covariant to the canonical gauge is found to be:\checked
\begin{align}
  \eps_{i,A}
  =
  \frac{\psi_{i,A}^{\rmm{cov}}\cdot V}{
    v_i\cdot(v_i+V)}
  \ .
\end{align}
The inverse transformation may also easily be derived from Eqs.~\eqref{WEFT:ST}.

The relation of the impact parameter and the spin tensor in the canonical gauge to the SUSY parameters is found to be:\checked
\bse
\begin{align}
  b_{i,\rm can}^\mu
  -
  \frac1{m_i}
  V_\nu S^{\mn}_{i,\rmm{can}}
  &=
  \beta^\mu_i
  \ ,
  \\
  S^\mn_{i,\rmm{can}}
  -2
  v^{[\mu}_i
    S^{\nu]\rho}_{i,\rmm{can}}
  V_\rho
  &=
  S^\mn_{i,\bot}
  \ .
\end{align}
\ese
With these equations and Eqs.~\eqref{WEFT:CVP} we can straightforwardly derive relations between the canonical and covariant parameters.

In addition, the canonical spin vector $S_{i,\rmm{can}}^\mu$ plays an important role in Hamiltonian spinning dynamics.
It is defined by:\checked
\begin{align}
  \label{WEFT:NFV}
  S^{\mu}_{i,\rmm{can}}
  &=
  \Lambda^\mu_{\ \nu}(v_i\to V)
  S^\nu_i
  \\
  &=
  S^\mu_i
  -
  \frac{
    (v_i+V)^\mu
  }{
    V\cdot(v_i+V)}
  V\cdot S_i
  \ .
  \nn
\end{align}
Here $S_i^\mu$ is the (Pauli-Lubanski) spin vector of Eq.~\eqref{WEFT:VPB} which we will usually refer to as the covariant spin vector.
The operator $\Lambda^\mu_{\ \nu}(v_i\to V)$ is a Lorentz boost between the frames $v_i^\mu$ and $V^\mu$ and in that sense, the canonical spin vector is identical to the covariant spin vector as seen from the reference frame of the body itself.
While the covariant vector is orthogonal to $v_i^\mu$, the canonical vector $S_{i,\rmm{can}}^\sig$ is orthogonal to the frame $V^\sig$.
The Eq.~\eqref{WEFT:NFV} is easily inverted (e.g. by dotting with $v_i^\mu$ on both sides) and for the covariant vector in terms of the canonical one finds:\checked
\begin{align}
  S_i
  &=
  \Lambda^\mu_{\ \nu}(V\to v_i)
  S^{\nu}_{i,V}
  \\
  &=
  S^{\mu}_{i,\rmm{can}}
  -
  \frac{
    (v_i+V)^\mu
  }{
    v_i\cdot (v_i+V)
  }
  v_i\cdot S^{\mu}_{i,\rmm{can}}
  \ .
  \nn
\end{align}
We note that the labelling of the spin vectors as covariant and canonical is misleading in the sense that both are SUSY invariant.
The canonical vector does, however, depend explicitly on the choice of frame $V^\mu$.
Both vectors have a simple relationship to the corresponding tensors:\checked
\bse
\begin{align}
  S_{i}^\mu
  &=
  \frac12
  \eps^{\mu}_{\ \rs\nu}
  S^\rs_{i,\rmm{cov}}
  v^\nu_i
  \ ,
  \\
  S_{i,\rmm{can}}^\mu
  &=
  \frac12
  \eps^{\mu}_{\ \rs\nu}
  S^\rs_{i,\rmm{can}}
  V^\nu
  \ .
\end{align}
\ese
The first relation is the same as Eq.~\eqref{WEFT:VPB}.

In the present work, the canonical spin vector will be essential in the construction of a two-body Hamiltonian.
In that case we specialize to the CoM frame with $V^\sig=\hat P^\sig$.
In that frame, the total CoM angular momentum is related to the canonical vectors in a simple manner:\checked
\begin{align}\label{WEFT:JTOT}
  J^\mu
  =
  L^\mu_{\rmm{can}}
  +
  S_{1,\rmm{can}}^\mu
  +
  S_{2,\rmm{can}}^\mu
  \ .
\end{align}
Here, $L^\mu_{\rmm{can}}$ is the CoM orbital angular momentum in the canonical gauge.\checked
\begin{align}\label{WEFT:LCN}
  L^\mu_{\rm can}
  &=-\frac1{E}
  \eps^{\mu}_{\ \nu\ab}
  b_{\rmm{can}}^{\nu}
  p^\alpha
  P^\beta
  \ .
\end{align}
The equation~\eqref{WEFT:JTOT} highlights a special case where the canonical vectors are simpler than the covariant ones.
The relation of $J^\mu$ to the covariant variables is essentially given in Eq.~\eqref{WEFT:TAM} when one identifies SUSY and covariant variables.
In the CoM frame we may choose $P^\mu$, $J^\mu$, $p^\mu$ and the two canonical spin vectors $S_{i,\rmm{can}}^\mu$ as independent asymptotic variables.
The first vector $P^\mu$ defines the frame and the latter four four-vectors are all purely spacial in this frame.
For conservative motion both $\Del P^\mu$ and $\Del J^\mu$ vanish and the Hamiltonian may be used to derive the remaining three kicks.

\subsection{SUSY Spinning Worldline Action to $\mO(S^2)$}
\label{WEFT:SUSY}
Having discussed the flat space kinematics and symmetries of point like spinning particles we move on to the theory on curved backgrounds.
In fact, we may try to directly covariantize the flat space action replacing partial derivatives by covariant ones.
Focusing on a single particle, we get:\checked
\begin{align}\label{WEFT:SA1}
  S_{\rmm{wl},S}
  =
  -
  \frac{m}{2}
  \int
  d\tau\,
  g_\mn(z)
  \Big(
  \dot z^\mu
  \dot z^\nu
  +
  i
  \psi^\mu_{A}
  \frac{
    \mathrm{D}\psi_{A}^{\nu}
  }{
    \di \tau}
  \Big)
  \ .
\end{align}
Here, all worldline fields are evaluated as a function of the worldline time $\tau$.
The covariant derivative of $\psi_A^\nu(\tau)$ is:\checked
\begin{align}
  \frac{
    \mathrm{D}\psi_{A}^{\nu}
  }{
    \di \tau}
  =
  \frac{\di \psi_{A}^\nu}{\di \tau}
  +
  \Gam^{\nu}_{\ab}
  \dot z^\alpha
  \psi^\beta_{A}
\end{align}
The action Eq.~\eqref{WEFT:SA1} is indeed general covariant but the SUSY is broken at quadratic order in the spins.
That is, using Eq.~\eqref{WEFT:STG}, we find:\checked
\begin{align}
  \del S_{\rmm{wl},S}
  =
  \mO(\psi^3_A)
  \ .
\end{align}
Thus, only to linear order in spins does the general covariant worldline action $S_{\rmm{wl},S}$ enjoy a SUSY.

Let us note the general pattern in the relation between the Grassmann variables and orders in spin.
In order to describe classical spin to some power $n$ we must have an equivalent amount of flavors $\mN=n$.
Thus, as an example, for linear in spin we only need one flavor and in that case we may ignore the flavor index.
For spin to the $n$th power, the SUSY symmetry is required to the same order, namely including $\psi^{2n}\sim S^n$.

At quadratic order in spins we must therefore adjust the action and SUSY transformations in order to preserve the SUSY symmetry.
At this order, too, we find a term which identically obeys the SUSY and thus may appear with an arbitrary coefficient.
This is an effective coupling and describes finite size effects of the point-like compact bodies.
At first we will simply present the resulting action and then later, below, discuss its derivation.
The result to quadratic order in spins, then, is found to be,\checked
\begin{align}
  \label{GR:SUSYAction}
  S_{\rmm{susy}}
  =
  -m\int\di\tau\,
  \Big(
  \frac12
  g_\mn
  \dot z^\mu
  \dot z^\nu
  +
  i\bar\psi_a
  \frac{
    \Di \psi^a}{
    \di \tau}
  +
  \frac12
  R_{abcd}
  \bar \psi^a
  \psi^b
  \bar \psi^c
  \psi^d
  +
  C_{\rmm{E}}
  E_{ab}
  \bar\psi^a
  \psi^b
  P_{cd}
  \bar \psi^c
  \psi^d
  \Big)
  \ ,
\end{align}
with the following infinitesimal SUSY,
\bse\label{WEFT:LSUSY}
\begin{align}
  \del z^\mu
  &=
  -2\operatorname{Im}
  \bar\eps \psi^\mu
  \ ,
  \\
  \del \psi^a
  &=
  -
  \eps
  e^a_\mu
  \dot z^\mu
  -
  \del z^\mu
  \oma_\mu^{ab}
  \psi_b
  \ ,
\end{align}
\ese
with infinitesimial Grassmann parameter $\eps$ satisfied to the required quadratic order in spins:\checked
\begin{align}
  \del S_{\rm susy}
  =
  \mO(\psi^5)\ .
\end{align}
In this action, all worldline variables are evaluated as a function of the time $\tau$ and gravitational variables as a function of the worldline parametrization.
We use local indices $a$, $b$, $c$, $d$ as defined from a local frame (vielbein) as discussed around Eqs.~\eqref{WEFT:Vielbein}.
The electric curvature $E_{ab}=E_{\mn}e^{\ \mu}_a e^{\ \nu}_b$ is given by (as in Eq.~\ref{WEFT:ETensor}),\checked
\begin{align}
  E_\mn(\tau)
  &=
  R_{\mu\alpha\nu\beta}(z(\tau))
  \frac{
    \dot z^\alpha(\tau)
    \dot z^\beta(\tau)
  }{
    \dot z^2(\tau)
  }
  \ ,
\end{align}
and the projector $P_{ab}=P_{\mn}e^{\ \mu}_a e^{\ \nu}_b$ by:\checked
\begin{align}
  P_\mn(\tau)
  &=
  g_\mn(z(\tau))
  -
  \frac{
    \dot z_\mu(\tau)
    \dot z_\nu(\tau)
  }{
    \dot z^2(\tau)
  }
  \ .
\end{align}
In both equations $\dot z^2=g_\mn \dot z^\mu \dot z^\nu$.
The covariant derivative with the local index was defined in Eq.~\eqref{WEFT:SpinConnection} and reads:\checked
\begin{align}
  \frac{D\psi^a}{\di \tau}
  =
  \dot \psi^a
  +
  \dot z^\sig
  \oma^{ab}_\sig
  \psi_b
  \ .
\end{align}

Since we work to quadratic order in spins only we use two flavors for $\psi_A^\sig(\tau)$ so that $A=1,2$.
Instead of the real $SO(2)$ basis, we have then used a complex $U(1)$ basis defined by:\checked
\bse
\begin{align}
  \psi^\mu
  &=
  \frac1{\sqrt{2}}
  (\psi^\mu_1+i\psi^\mu_2)
  \ ,
  \\
  \bar \psi^\mu
  &=
  \frac1{\sqrt{2}}
  (\psi^\mu_1-i\psi^\mu_2)
  \ .
\end{align}
\ese
The two vectors $\psi^\mu$ and $\bar \psi^\mu$ are Hermitian conjugates of each other:\checked
\begin{align}
  (\psi^\mu)^\dagger
  =
  \bar \psi^\mu
  \ .
\end{align}
Here, a dagger denotes Hermitian conjugation.
This is a consequence of the simpler fact that the fields in the real basis are Hermitian vectors:\checked
\begin{align}
  (\psi_A^\mu)^\dagger =\psi_A^\mu
  \ .
\end{align}
The complex conjugate of a bosonic combination of Grassmann variables is most easily computed using the Hermitian conjugation:\checked
\begin{align}
  (\psi_{A_1}^{\mu_1}
  ...
  \psi_{A_{2n}}^{\mu_{2n}})^*
  =
    (\psi_{A_1}^{\mu_1}
  ...
  \psi_{A_{2n}}^{\mu_{2n}})^\dagger
  &=
  \psi_{A_{2n}}^{\mu_{2n}}
  ...
  \psi_{A_1}^{\mu_1}
  \ .
\end{align}
In particular both the spin tensor in the complex basis,\checked
\begin{align}
  \label{WEFT:SpinTensor}
  \mS^\mn(\tau)
  =
  -2
  i
  m
  \bar \psi^{[\mu}(\tau)
    \psi^{\nu]}(\tau)
  =
  2m \operatorname{Im}\!
  \big[
  \bar \psi^\mu(\tau) \psi^\nu(\tau)
  \big]
  \ ,
\end{align}
and the spin length,\checked
\begin{align}
  \sqrt
  {
    \frac{
      \mS^\mn(\tau)
      \mS_\mn(\tau)
      }{2}
  }
  =
  |\mS(\tau)|
  =
  m\bar\psi^a(\tau)
  \psi_a(\tau)
  \ ,
  \nn
\end{align}
are real.

The three gauge symmetries, scaling of $\tau$, U(1) rotations of $\psi^\mu(\tau)$ and the SUSY imply the conservation of the three following quantities respectively:\checked
\bse
\begin{align}
  g_\mn \dot z^\mu \dot z^\nu
  &=
  1
  +
  R_{abcd}
  \bar \psi^a
  \psi^b
  \bar \psi^c
  \psi^d
  +
  2
  C_{\rmm{E}}
  E_{ab}
  \bar\psi^a
  \psi^b
  P_{cd}
  \bar\psi^c
  \psi^d
  \ ,
  \\
  |S|
  &=
  m\bar\psi^a \psi_a
  \ ,
  \\
  0&=\psi_\mu \dot x^\mu
  \ .
\end{align}
\ese
Here, we have directly gauge fixed our variables to satisfy proper time at infinity and the covariant SSC.
Only with the covariant SSC enforced does the action $S_{\rm susy}$ match directly with traditional approaches~\cite{Jakobsen:2021zvh}.
For the spin length $|S|=|\mS(\tau)|$ we have simply inserted its asymptotic value since it is conserved.

The constraints may, in fact, be used to simplify the action with EFT techniques.
Thus, we may drop terms quadratic in the SSC ($\psi\cdot \dot z$) and effectively assume $\dot z^2$ to be of quadratic order in the spins.
We get:\checked
\begin{align}\label{WEFT:SUSYA}
  S_{\rm spin}
  =
  -m\int\di\tau\,
  \Big(
  \frac12
  g_\mn
  \dot z^\mu
  \dot z^\nu
  +
  i\bar\psi_a
  \frac{
    \Di \psi^a}{
    \di \tau}
  +
  \frac12
  R_{abcd}
  \bar \psi^a
  \psi^b
  \bar \psi^c
  \psi^d
  +
  C_{\rm E}
  \dot z^\mu
  \dot z^\nu
  R_{a\mu b\nu}
  \bar \psi^a
  \psi^b
  \bar\psi\cdot\psi
  \Big)
  \ .
\end{align}
The only difference to $S_{\rm susy}$ is in the finite size term with $C_{\rm E}$.

The actions $S_{\rm susy}$ or $S_{\rm spin}$ may be derived in a variety of ways.
One method is to start from the action of traditional (W)EFT approaches Refs.~\cite{Vines:2016unv,Porto:2016pyg,Levi:2018nxp},\checked
\begin{align}  \nn
  S
  &=
  - \int\!\di\tau
  \bigg[
    \pi_\mu \dot{x}^\mu
    + \frac{1}{2} \mS_{\mu\nu} \Lambda_A{}^\mu  \frac{D \Lambda^{A\nu}}{D\tau} - \lambda (\pi_\mu \pi^\mu - \mathcal{M}^2) - \chi_\mu \mS^{\mu\nu} \bigg( \frac{\pi_\nu}{\sqrt{\pi^2}} + \Lambda_{0\nu} \bigg) \bigg] \ ,
  \\
  \mathcal{M}^2
  &= m^2 - \frac14 R_{\mu\nu\alpha\beta} \mS^{\mu\nu} \mS^{\alpha\beta} +
  C_{E} E_{\mu\nu} \mS^{\mu\rho} P_{\rho\sigma} \mS^{\nu\sigma}
  +\mO(S^3)\ ,
  \label{WEFT:Standard}
\end{align}
and directly re-express the spin tensor in terms of the Grassmann variables using Eq.~\eqref{WEFT:SpinTensor}.
In this first-order action with covariant momentum $\pi^\mu$ and Lagrange multipliers $\lambda$ and $\chi_\mu$ the body fixed frame $\Lambda_A^\mu(\tau)$ is an additional worldline field to $S_\mn$.
However, after re-expressing the spin tensor in terms of $\psi^\mu$ the body fixed frames identically drop out of the action taking advantage of their constraint $\Lambda^{\ \mu}_A \Lambda_{B\mu}=\eta_{AB}$.
In the process of simplifying this traditional action to $S_{\rm susy}$ or $S_{\rm spin}$ one must rewrite it as a second-order action and gauge fix $\lambda$ and $\chi_\mu\to0$.
This comparison effectively verifies that the action~\ref{WEFT:Standard} is equivalent to the actions in terms of Grassmann variables considered here.
The equations of motion of $S_{\rm spin}$ or $S_{\rm susy}$ are thus equivalent to the classical Mathisson-Papapetrou-Dixon (MPD) equations (see e.g. Ref.~\cite{Saketh:2022wap} for the MPD equations).
For more details, see Ref.~\cite{Jakobsen:2021zvh}.

The coefficient $C_{\rm E}$ describes finite size effects and is zero for (Kerr) black holes.
It is related to the more standard coefficient $C_{E S^2}$ of Refs.~\cite{Vines:2016unv,Porto:2016pyg,Levi:2018nxp} through,
\begin{align}
C_{\rm E}=1-C_{ES^2} \ .
\end{align}
Interestingly, for the Kerr black hole case with $C_{\rm E}=0$ the action $S_{\rm susy}$ obeys the SUSY~\eqref{WEFT:LSUSY} to all orders in spin:\checked
\begin{align}\label{WEFT:S2}
  \delta S_{\rm susy}
  \Big|_{C_{\rm E}=0}
  =
  0
  \ .
\end{align}
This is related to the fact that the Kerr black hole action may be derived by coupling of the $\mN=2$ worldline particle to gravity.
In general the SUSY of $S_{\rm susy}$ Eqs.~\eqref{WEFT:LSUSY} and~\eqref{WEFT:S2} may be derived by explicit computation.
For more details, see Ref.~\cite{Jakobsen:2021zvh}.

Finally let us consider the relation of the spin (vector and tensor) kicks to the Grassmann kick.
We derive the spin tensor kick in terms of the Grassmann kick and in turn the spin vector kick in terms of the spin tensor kick:\checked
\bse
\label{WEFT:SpinKicks}
\begin{align}\label{eq:spinKick}
  \Delta S^{\mu\nu}
  &=
  2m
  \operatorname{Im}
  \!\big[
    2\bar\Psi^{[\mu}\Del\psi^{\nu]}
    +
    \Del\bar\psi^\mu\Del\psi^\nu
    \big]
  \ ,
  \\
  \Delta S^\mu
  &=
  \frac1{2m}
        {
    \eps^\mu
        }_{\nu\rho\sigma}
        \left(
        S^{\nu\rho}
        \Delta p^\sigma
        +
        \Delta S^{\nu\rho}
        p^\sigma
        +
        \Delta S^{\nu\rho}\Delta p^\sigma\right)
        \ .
\end{align}
\ese
Thus from the Grassmann kick, the other spin kicks are directly derived.


\chapter{Worldline Quantum Field Theory}
\label{WQFT}
The worldline quantum field theory approach to classical gravitational scattering of compact bodies is an efficient and systematic formalism with which high orders in perturbation theory can be derived and generic physical properties of the compact bodies can be included.
It builds on the worldline effective field theory description of compact bodies importing at the same time several inventions and techniques from quantum field theory.
In its most basic form gravitational scattering observables are computed as on-shell tree-level one-point functions defined by a partition function from an action constructed with worldline effective field theory.
This action includes both the gravitational metric and worldline fields and in the partition function all fields are treated equally with the path integral extending both over bulk and worldline fields.

The worldline quantum field theory was first proposed in Ref.~\cite{Mogull:2020sak} as a link between the two prominent approaches to post-Minkowskian classical gravity, namely PM-EFT~\cite{Kalin:2020mvi} and QFT-amplitudes~\cite{Bjerrum-Bohr:2018xdl}.
From this perspective, the WQFT explains how the (non-trivial) classical limit of the QFT-amplitudes approach simplifies to the worldline effective field theory description of compact bodies also used in the PM-EFT approach.
From this perspective, WQFT brings together important techniques from both approaches.
In subsequent work \cite{Jakobsen:2021smu,Jakobsen:2021lvp,Jakobsen:2021zvh,Jakobsen:2022fcj,Jakobsen:2022psy,Jakobsen:2022zsx} including the present author, the WQFT was further developed to include both spin, tidal and radiation reaction effects and several state of the art results were derived such as the leading order bremsstrahlung and (radiative) worldline observables at the third post-Minkowskian order both with spin and tidal effects.
Additional work with WQFT on double copy~\cite{Shi:2021qsb,Comberiati:2022cpm}, gravitational light bending~\cite{Bastianelli:2021nbs}, classical off-shell currents~\cite{Comberiati:2022ldk} and scattering in electrodynamics~\cite{Wang:2022ntx} has appeared.

While the link of WQFT to the classical limit of the QFT-amplitudes approach to classical gravity is an important aspect, the focus in this chapter will mostly be on the role of the WQFT in solving the classical equations of motion of the worldline fields coupled to gravity and Einstein's field equations.
From this perspective, the WQFT is presented as a highly streamlined formalism for perturbative computations in the worldline effective field theory of compact bodies.
The main application, then, is the computation of high-precision classical gravitational observables.
The classical limit of QFT-amplitudes and quantum corrections to this limit, however, is another interesting perspective of the WQFT.

The main objective of this chapter, then, is the elegant WQFT formalism for solving the classical WEFT equations of motion.
They follow from the variational principle of the action $S$ Eq.~\eqref{WEFT:S} which reads:\checked
\begin{align}
  \label{WQFT:Action}
  S
  =
  S_{\rm GR}[g_\mn]
  +
  \sum_i
  S_{\rmm{wl},i}[g_\mn,w_{i}^\sig]
  \ .
\end{align}
The gauge fixed Einstein-Hilbert action is $S_{\rm GR}$ and the worldline action of the $i$th body is $S_{\rmm{wl},i}$ with a collection of generic worldline fields $w_i^\sig(\tau)$ labelled by $w$.
The main example of a worldline action $S_{\rmm{wl},i}$ is the supersymmetric action $S_{\rmm{spin},i}$ Eq.~\eqref{WEFT:SUSYA} of Sec.~\ref{WEFT:SUSY} which describes spinning compact bodies to quadratic order in their spins.

In the first Sec.~\ref{WQFT:CDfPI} we consider the generic framework of solving classical equations of motion from a path integral.
We cover the important distinction of different boundary value problems and show how causal boundary conditions are achieved with the in-in formalism.
In contrast to the first section, the next two Secs.~\ref{WQFT:Diagrammatics} and~\ref{WQFT:PostMinkowskian} are more practical and focus on the off-shell and on-shell aspects of WQFT respectively.
Thus, in the second section we discuss Feynman rules and graph generation and in the third section we discuss the derivation of observables from on-shell one-point functions and their computation to second post-Minkowskian order.

\section{Classical Dynamics from Path Integrals}
\label{WQFT:CDfPI}
The content of this section can be summarized in the brief statement that tree-level correlation functions satisfy the classical equations of motion defined from the variational principle.
That is, if we consider a generic action $S[X_A]$ which depend on a collection of fields $X_A$ labelled by $A$ then the tree-level contribution to the one-point correlation functions of the fields $X_A$ solve the equations of motion defined by the variational principle $\del S/\del X_A=0$.
This generic example will be discussed in more detail in Sec.~\ref{WQFT:CD}.
Then, in Sec.~\ref{WQFT:Causal} we discuss how causal boundary conditions of the classical solutions are enforced with the in-in Schwinger-Keldysh formalism.

In the following we will usually refer to the tree-level contribution as the classical limit and the equations of motion derived from the variational principle $\del S=0$ as the classical equations of motion.
We note, however, that this terminology generally is misleading as the \textit{physical} classical limit of a given QFT will have contributions from loops as well.
Thus, loop contributions are essential in the QFT-amplitudes approach to classical gravitational scattering~\cite{Iwasaki:1971vb,Holstein:2004dn,Kosower:2018adc}.
In the case of the WQFT, however, the equations of motion derived from the variational principle will describe the physical classical limit of the system of interest.

From the perspective of classical physics, it may seem overly complicated to introduce a path integral in order to solve the classical equations of motion.
For the application to classical theory the path integral is indeed superfluous in the sense that we consider only its stationary phase approximation which gives rise to tree level graphs and the variational principle.
Instead of starting from the path integral it is thus also possible to base the classical applications of the WQFT approach on the principle of least action.
The path integral and the corresponding formalism of QFT do, however, offer a systematic notation and terminology which is useful for perturbative computations.

\subsection{Classical Dynamics from One-Point Functions}
\label{WQFT:CD}
Let us start from the generic action $S[X_A]$ which is a function of several fields, $X_A(x_A)$, labelled by $A$ and functions of coordinates $x_A$.
Here we anticipate the inclusion of space-time fields that are functions of $x^\mu$ and worldline fields that are functions of $\tau$.
The basic idea is to solve the classical equations of motion of this action with one-point correlation functions.
We introduce the partition function $Z[J_A]$ which is given by a path integral as follows:\checked
\begin{align}\label{CD:1}
  Z[J_A]
  =
  \int D[X_A]
  \exp
  \!
  \bigg[
  \frac{i}{\hbar}
  \bigg(
  S[X_A]
  +
  \sum_A \int \di x_{\!A}\,
  J_A(x_A)
  X_A(x_A)
  \bigg)
  \bigg]
  \ .
\end{align}
For bulk fields $\di x_{\!A}=\di^d x$ and for worldline fields $\di x_{\!A}=\di \tau$.
Expectation values of generic functions of the fields, $F[X_A]$, are now defined as:\checked
\begin{align}\label{CD:2}
  \bgev{F(X_A)}
  &=
  \frac{1}{Z_0}
  \int D[X_A]
  F(X_A)
  e^{
    \frac{i}{\hbar}
    S[X_A]
  }
  \Bigg|_{\hbar\to0}
  \\
  &=
  \frac{1}{Z_0}
  F
  \Big(
    \frac{\hbar}{i}
    \frac{\del}{\del J_A}
    \Big)
  Z[J_A]
  \bigg|_{
    \substack{
      J_A\to0
      \\
      \ \hbar\to0}
  }
  \ .
  \nn
\end{align}
Here, the normalization $Z_0$ is given by $Z[J_A=0]$ and ensures that $\ev{1}=1$.
The partition function $Z[J_A]$ is a generator of expectation values.
Correlation functions of the fields $X_A$ are defined by choosing $F(X_A)$ to be a simple product of fields.

Let us discuss the classical limit $\hbar\to0$ of the partition function and expectation values.
As discussed above, we use $\hbar$ as a formal power counting parameter and do not assume any internal variables of the action to scale with $\hbar$.
In the following discussion we will assume a perturbative expansion of the partition function in terms of Feynman diagrams.
The partition function is given, then, by the sum of all (disconnected) vacuum diagrams sourced by the sources $J_A$ with appropriate symmetry factors.
A connected vacuum diagram scales as $\hbar^{-1+L}$ with $L$ the loop order and there is, thus, no well-defined $\hbar\to0$ limit of the partition function.
In contrast, the expectation values have a well-defined $\hbar\to0$ limit.
In particular, $n$-point correlation functions are given as the sum of all (disconnected) diagrams with a total of $n$ external legs and all vacuum diagrams removed due to the normalization $Z_0$.
A connected $n$-point function scales as $\hbar^{-1+L+n}$ and in the $\hbar\to0$ limit we see that only one-point tree-level functions ($n=1$ and $L=0$) contribute.
The disconnected $n$-point correlation functions are thus given as a product of connected one-point correlation functions in this limit.
This discussion is, essentially, the stationary phase approximation of the path integral.
For more detail on basic objects in the path integral formulation of QFT, see e.g. Ref.~\cite{Srednicki:2007qs}.

The above discussion implies that our (classical) expectation values factorize:\checked
\begin{align}
  \label{WQFT:Factor}
  \bgev{F_1[X_A] F_2[X_A]}
  =
  \bgev{F_1[X_A]}
  \bgev{F_2[X_A]}
  \ .
\end{align}
This equation, in particular, explains how the quantum Schwinger-Dyson equation,\checked
\begin{align}
  \bigev{\frac{\del S[X_A]}{\del X_A}}
  =0
  \ ,
\end{align}
becomes the classical equation of motion:\checked
\begin{align}
  \frac{
    \del S\big[\ev{X_A}\big]
  }{\del X_A}
  =0
  \ .
\end{align}
Thus, indeed, the one-point correlation functions satisfy the classical equations of motion in the classical limit $\hbar\to0$.

In passing, we note that one may also define non-trivial higher-point functions in the classical limit that are not the product of simpler one-point functions.
Their generator is $W[J_A]$ given by:\checked
\begin{align}
  \label{WQFT:Connected}
  \frac{i}{\hbar}
  W[J_A]
  =
  \log(Z[J_A])
  \ .
\end{align}
In this work, however, we will focus on the one-point functions.

Let us now focus on the case of interest to us, namely the worldline effective field theory action describing point-like particles coupled to gravity.
We use the generic action $S[g_\mn,w_i]$ of Eq.~\eqref{WQFT:Action} which is a sum of the gauge fixed Einstein-Hilbert action $S_{\rm GR}[g_\mn]$ and a worldline action $S_{\rmm{wl},i}[g_\mn,w_i]$ for each particle $i$.
The gravitational field $g_\mn(x)$ depends on the space-time coordinates and the collection of worldline fields $w^\sig_i(\tau)$ labelled by $w$ depend only on a time variable (one-dimensional space-time).
The path integral for the partition function now reads:\checked
\begin{align}
  \label{WQFT:PI}
  Z_{\rm in-out}[J^\mn,f^{w_i}_\sig]
  &=
  \int D[h_\mn,\Del w_i^\sig]
  \\
  &\times
  \exp
  \!
  \bigg[
  \frac{i}{\hbar}
  \bigg(
  S[g_\mn,w^\sig_i]
  +
  \int \di^d x\,
  J^\mn(x) h_\mn(x)
  +
  \int \di\tau\,
  f^{w_i}_\sig(\tau) \Del w_i^\sig(\tau)
  \bigg)
  \bigg]
  \ .
  \nn
\end{align}
This integral defines the worldline quantum field theory that we will use to derive classical gravitational dynamics.
The integral is defined in the post-Minkowskian expansion with integration on the perturbative fields $h_\mn(x)$ and $\Del w_i^\sig(\tau)$ (defined in Eqs.~\ref{WEFT:PME}).
One may consider non-perturbative definitions as well with integration on the full fields $g_\mn(x)$ and $w^\sig_i(\tau)$ but the perturbative setting is natural from the point of view of effective field theory and will be sufficient for our application.
We use $f^{w_i}_\sig(\tau)$ for the source terms of the worldline fields since in the case of the worldline parametrization they imitate an external force.
We use the subscript ``in-out'' in order to distinguish this path integral from the ``in-in'' path integral to be defined below in Sec.~\ref{WQFT:Causal}.
The two integrals differ with respect to the boundary conditions on the fields.

Correlation functions and expectation values of observables are defined as in Eq.~\eqref{CD:2}.
In terms of the partition function it simply is:\checked
\begin{align}
  \bginout{F[h_\mn,\Del w_i^\sig]}
  &=
  \frac{1}{Z_0^{\rm in-out}}
  F
  \bigg[
    \frac{\hbar}{i}
    \frac{\del\ \ \ }{\del J^\mn}
    \, 
    ,\, 
    \frac{\hbar}{i}
    \frac{\del\ \ \ }{\del f^{w_i}_\sig}
    \bigg]
  Z[J^\mn,f_\sig^{w_i}]
  \bigg|_{\substack{
      J^\mn\to0
      \\
      \,f^{w_i}_\sig\to0
      \\
      \ \ \ \hbar\to0
  }}
  \ .
\end{align}
Again, the normalization $Z_0^{\rm in-out}$ is defined so that $\inout{1}=1$.

The one-point functions of this theory are:\checked
\begin{align}\label{WQFT:WS0}
  g_\mn^{(\rmm{c})}(x)=
  \bginout{
    \eta_\mn+\kap h_\mn(x)
  }
  \ ,
  \qquad
  w^{(\rmm{c})\sig}_{i}(\tau)
  =
  \bginout{
    w_{i,0}^\sig(\tau)
    +
    \Del w^\sig_i(\tau)
  }\ ,
\end{align}
and they obey the classical equations of motion.
As an example we consider the Polyakov action (Eq.~\ref{eq:SpinlessSimple}) where the only worldline field is the worldline parametrization $z_i^\sig(\tau)$ and its classical (in-out) solution is:\checked
\begin{align}\label{WQFT:WS}
  z_{i}^{(\rmm{c})\sig}(\tau)
  =
  b^\mu_i+\tau v_i^\mu
  +
  \bginout{\Del z^\sig_i(\tau)}
  \ .
\end{align}
The gravitational field $g_{\mn}^{(\rmm{c})}(x)$ and the worldline trajectories $z_{i}^{\rmm{(c)}\sig}(\tau)$ satisfy Einstein's field equations and the geodesic equation respectively.
The solution Eq.~\eqref{WQFT:WS} includes all self-interactions of the worldline fields and divergencies are regularized with dimensional regularization as discussed in Sec.~\ref{sec:Effective}.

While the one-point functions of Eqs.~\eqref{WQFT:WS0} and~\eqref{WQFT:WS} satisfy the classical equations of motion it is important to ask which boundary conditions they obey.
The boundary conditions are determined from the kind of propagator used to invert the kinetic operators.
The kinetic operator for the graviton field is $\partial^2$ and for the worldline fluctuations, $\Del z^\sig_i(\tau)$, the one-dimensional $\partial_\tau^2$.
The inversion of these kinetic terms is most easily considered in momentum space.
The convergence of the path-integral Eq.~\eqref{WQFT:PI} requires the use of the Feynman propagator $D_F(k)$ for the graviton field which in momentum space reads:\checked
\begin{align}
  D_F(k)=\frac1{k^2+i\eps}
  \ .
\end{align}
This propagator is different from the retarded propagator which is usually desired for (causal) classical physics.
In momentum space the retarded propagator is given by (\ref{WEFT:RP}):\checked
\begin{align}
  D_R(k)
  =
  \frac1{
    (k_0+i\eps)^2-\vct{k}^2
  }
\end{align}
In contrast to the retarded propagators, the Feynman propagators are time symmetric which is a natural consequence of the in-out path integral.
In the next section we will introduce the in-in path integral which gives rise to retarded propagators.

Finally, let us note that the in-out WQFT suffers from bad behavior of the worldline fields at infinity.
Thus, we would by no means expect the worldline fluctuations $\Del z^\sig_i(\tau)$ to vanish both at past and future infinity.
Thus, the in-out path integral with classical worldline fields seems ill-defined.
As we will see later, we may still compute observables from this theory.
However, the in-in formalism will improve on this point as the perturbation fields are then only assumed to be zero at past infinity.

\subsection{Causal Dynamics from In-In Formalism}
\label{WQFT:Causal}
The in-in Schwinger-Keldysh formalism~\cite{Schwinger:1960qe,Keldysh:1964ud,Jordan:1986ug,Weinberg:2005vy} is designed to compute ``in-in'' expectation values between two incoming states at past infinity rather than expectation values between incoming and outgoing states (transition amplitudes).
In our case it will formally introduce retarded propagators rather than the Feynman propagators of the in-out path integral.
See also Refs.~\cite{Galley:2008ih,Galley:2009px,Foffa:2011np,Kalin:2022hph,Almeida:2022jrv} for applications to the (W)EFT of classical gravity and Ref.~\cite{Jakobsen:2022psy} including the present author.

The in-in partition function may be defined in an equivalent manner to the in-out action with the only difference being the time integration in the action.
Instead of a time integration on the fields from past to future infinity we use a closed time integration path going from past infinity to future infinity and back again:\checked
\begin{align}
  S_{\rm in-in}[X_A]=
  \int_\mC
  \di t
  \,L
  \ .
\end{align}
Here $L$ is the Lagrangian so that $S=\int \di t \,L$.
The contour, $\mC$, may be taken from minus infinity above the real line to plus infinity and then back to minus infinity below the real line.
The value of the fields $X_A(x_A)$ above and below the real line at $t_\pm=t\pm i\eps$ are considered independent.
The boundary conditions on $X_A(x_a)$ are that it should vanish at past infinity in the limit $t\to-\infty$ and that it is identical at the turning point at future infinity $X(t_+)=X(t_-)$ when $t\to\infty$.
Below, we will first consider this action and its partition function in the (1-2) basis and afterwards introduce the Schwinger-Keldysh basis which greatly simplifies the in-in formalism in the classical limit.

\subsubsection{In-in formalism in the (1-2) basis}
Instead of working with the closed time path of integration with fields depending on $t\pm i\eps$ we split every field into two fields, $X_A(t_+)=X_{(1)A}(t)$ and $X_A(t_-)=X_{(2)A}(t)$ and write the closed contour $\mC$ as a sum of two terms:\checked
\begin{align}\label{CD:3}
  S_{\rm in-in}[X_A]
  &=
  \int_{-\infty}^\infty
  L_{(1)}
  -
  \int_{-\infty}^\infty
  L_{(2)}
  \nn
  \\
  &=
  S[X_{(1)A}]-S[X_{(2)A}]
  \ .
\end{align}
The Lagrangians $L_{(i)}$ with $i=1$ or $i=2$ are the Lagrangian $L$ evaluated on the first or second fields.
The terms with the fields $X_{(2)A}$ describe the contribution of the part of the contour from future to past infinity.
In the end the in-in action is simply written as the difference of two copies of the action evaluated on the 1 and 2 fields respectively.

We now have an action with the double amount of degrees of freedom, $X_{(1)A}$ and $X_{(2)A}$, as compared with the in-out action.
These two fields are treated as independent fields except for their boundary conditions.
At future infinity they must coincide,\checked
\begin{align}\label{CD:4}
  X_{(1)A}(t\to\infty)
  =
  X_{(2)A}(t\to\infty)
  \ ,
\end{align}
and at past infinity both fields are required to vanish.
Both conditions follow from their definition in terms of the contour $\mC$.

The interaction terms of each of the $1$ and $2$ fields are independent from each other.
The propagator matrix of the $1$ and $2$ fields, however, mixes the two fields and ensures the boundary condition at future infinity.
It is given by:\checked
\begin{align}\label{CD:Propagator}
  \mD_{(1-2)}(k)=
  \begin{pmatrix}
    D_F(k)
    &
    -W_-(k)
    \\
    -W_+(k)
    &
    -D_D(k)
  \end{pmatrix}
  \ .
\end{align}
We will verify below in Eq.~\eqref{WQFT:BC} that this propagator matrix satisfies the desired boundary conditions.
The Wightman functions $W_\pm(k)$ are given by:\checked
\begin{align}
  W_\pm(k)
  =
  i\dd(k^2)\theta(\pm k_0)
  \ .
\end{align}
They are homogeneous solutions to the wave equation.

Let us briefly pause and recap the different propagators of the wave equation.
The Feynman and retarded propagators in momentum space are:\checked
\begin{align}
  D_F(k)=\frac1{k^2+i\eps}
  \ ,
  \qquad
    D_R(k)
  =
  \frac1{
    (k_0+i\eps)^2-\vct{k}^2
  }
\ .
\end{align}
They are intimately related to the Dyson and advanced propagators:\checked
\begin{align}
  D_D(k)
  =
  (D_F(k))^*
  =
  \frac1{k^2-i\eps}
  \ ,
  \qquad
  D_A(k)
  =
  (D_R(k))^*
  =
  D_R(-k)
  =\frac1{(k_0-i\eps)^2-\vct{k}^2}
\end{align}
In Eq.~\eqref{CD:Propagator}, the field traveling forward in space-time is propagated by the Feynman propagator.
It is then consistent that the field traveling back in time is propagated by the Dyson propagator.
The Feynman, Dyson, retarded and advanced propagators are all inverse operators to the wave operator $k^2$.
In contrast, the Wightman functions $W_\pm(k)$ are annihilated by the wave operator.
In momentum space the retarded and Feynman propagator are related through the Wightman function:\checked
\begin{align}
  D_R(k)
  =
  D_F(k)+W_-(k)
  \ .
\end{align}
In position space the different operators are also related as follows:\checked
\begin{align}\label{CD:Position}
  D_F(x)
  &=
  -\theta(t)W_+(x)
  -\theta(-t)W_-(x)
  \ ,
  &&
  D_R(x)
  =
  \Big(
  W_-(x)-W_+(x)
  \Big)\theta(t)
  \ ,
  \\
  D_D(x)
  &=
  \theta(t) W_-(x)
  +
  \theta(-t)W_+(x)
  \ ,
  &&
  D_A(x)
  =
  \Big(
  W_+(x)-W_-(x)
  \Big)
  \theta(-t)
  \ .
  \nn
\end{align}
Here, we recall the momentum and position space conventions Eqs.~\eqref{WEFT:MomentumConvention}.
For these relations see e.g. the appendix to Ref.~\cite{Kalin:2022hph}.
Clearly, the retarded propagator is only non-zero for $t>0$ on account of the Heaviside function $\theta(t)$.

Due to the Wightman functions being homogeneous solutions to the wave equation, the propagator matrix~\eqref{CD:Propagator} is indeed an inverse to the kinetic term of the two fields:\checked
\begin{align}
  \int_k
  \Big(
  X_{A,1}(-k) k^2 X_{A,1}(k)
  -
  X_{A,2}(-k) k^2 X_{A,2}(k)
  \Big)
  =
  \int_k
  \begin{pmatrix}
    X_{(1)A}(-k) \\ X_{(2)A}(-k)
  \end{pmatrix}
  \cdot
  \mD^{-1}_{(1-2)}(k)
  \cdot
  \begin{pmatrix}
    X_{(1)A}(k) \\ X_{(2)A}(k)
  \end{pmatrix}
  \ .
\end{align}
Examining the propagator matrix in position space using Eqs.~\eqref{CD:Position} we clearly see, that it ensures the boundary condition at future infinity.
In position space it is given by:\checked
\begin{align}
    \mD_{(1-2)}(x)=-
  \begin{pmatrix}
    \theta(t)W_+(x)
    +\theta(-t)W_-(x)
    &
    W_-(x)
    \\
    W_+(x)
    &
    \theta(t)W_-(x)
    +\theta(-t)W_+(x)
  \end{pmatrix}  
\end{align}
In the limit $t\to\infty$ this matrix becomes:\checked
\begin{align}\label{WQFT:BC}
  \mD_{(1-2)}(x)
  \big|_{t\to\infty}
  =
  -
  \begin{pmatrix}
    W_+(x)
    &
    W_-(x)
    \\
    W_+(x)
    &
    W_-(x)
  \end{pmatrix}
  \ .
\end{align}
If we act on two arbitrary sources,\checked
\begin{align}\label{CD:Infty}
  \int \di^d x
  \mD_{(1-2)}(y-x)
  \cdot
  \begin{pmatrix}
    J_{(1)}(x)
    \\
    J_{(2)}(x)
  \end{pmatrix}
  \ ,
\end{align}
that do not necessarily vanish at infinity we get identical values for the fields at infinity and thus the boundary condition Eq.~\eqref{CD:4} is ensured by this propagator matrix.
Starting from an ansatz to the propagator matrix with the Feynman and Dyson propagators on the diagonal we may indeed fix the off-diagonal elements using the $t\to\infty$ limit of Eq.~\eqref{CD:Infty} and requiring the boundary conditions.
Alternatively, the propagator matrix may be derived using an operator based approach (see e.g. Ref.~\cite{Jakobsen:2022psy}).

\subsubsection{Schwinger-Keldysh Basis and In-In Dynamics}
\label{WQFT:Schwinger}
At this point, the in-in formalism seems overly complicated.
We have doubled our field degrees of freedom and our propagator matrix is not diagonal.
It is also not clear, how retarded propagators appear.
However, the formalism simplifies greatly in the classical limit once we use the Schwinger-Keldysh basis with fields $X_{(\pm)A}$ related to the (1-2) fields by:\checked
\begin{align}
  \label{WQFT:Transformation}
  X_{(+)A}
  &=
  \frac12(
  X_{(1)A}+X_{(2)A})
  \ ,
  &&
  X_{(1)A}
  =
  X_{(+)A}+\frac12 X_{(-)A}
  \ ,
  \\
  X_{(-)A}
  &=
  X_{(1)A}-X_{(2)A}
  \ ,
  &&
  X_{(2)A}
  =
  X_{(+)A}-\frac12 X_{(-)A}
  \ .
  \nn
\end{align}
For classical solutions we will find $X_{(1)A}(t)=X_{(2)A}(t)$ and hence $X_{(-)A}=0$.

The propagator matrix in this basis $\mD_{\rm (S-K)}(k)$ can be found by a linear transformation of $\mD_{(1-2)}(k)$ from Eq.~\eqref{CD:Propagator}.
We find:\checked
\begin{align}\label{CD:SKPropagator}
  \mD_{\rm (S-K)}(k)=
  \begin{pmatrix}
    -\frac12(W_+(k)+W_-(k))
    &
    D_R(k)
    \\
    D_A(k)
    &
    0
  \end{pmatrix}
  \ .
\end{align}
The off-diagonals of this matrix are the retarded and advanced propagators.
As we will see only the upper right retarded propagator plays a role for classical dynamics and the two other non-zero entries can be ignored.
The conventions used here are different to Refs.~\cite{Galley:2009px,Kalin:2022hph} where $(\pm)$ indices are raised and lowered with the two-dimensional Levi-Civita symbol.
We simply interpret the $(+)$ and $(-)$ fields as a doubled set of fields with some mixing propagator given by $\mD_{\rm (S-K)}$ acting on vectors $(J_{(+)A},J_{(-)A})^{T}$ ($T$ indicating transpose).

Let us analyze the in-in one point functions and their equations of motion.
We define expectation values using Eq.~\eqref{CD:2}:\checked
\begin{align}
  \bginin{F(X_{(\pm)A})}
  &=
  \frac{1}{Z_0^{\rm in-in}}
  \int D[X_{(\pm)A}]
  F(X_{(\pm)A})
  e^{
    \frac{i}{\hbar}
    S_{\rm in-in}[X_{(\pm)A}]
    }
  \Bigg|_{\hbar\to0}
  \ .
\end{align}
Again the normalization $Z_0^{\rm in-in}$ ensures that $\inin{1}=1$.
The equations of motion follow from the in-in action.
How do they look like in the Schwinger-Keldysh basis?
We insert the $(\pm)$ fields into the in-in action Eq.~\eqref{CD:3} using the relations Eqs.~\eqref{WQFT:Transformation}.
To leading order in the minus fields, we find:\checked
\begin{align}\label{CD:SKAction}
  S_{\rm in-in}[X_{(\pm)A}]
  =
  \sum_A
  \int \di x_{\!A}\,
  X_{(-)A}(x_{\!A})
  \frac{
    \del S
  }{
    \del X_A(x_{\!A})
  }\Big[
    X_{(+)A}
    \Big]
  +
  \mO
  \Big(
  X_{(-)A}^3
  \Big)
  \ .
\end{align}
As we will see, the higher order terms in the minus fields play no role in the classical limit.

Let us first verify that the one-point functions of the minus field vanish.
We define:\checked
\begin{align}
  \bginin{X_{(\pm)A}(x_A)}=X_{(\pm)A}^{\rm (c)}(x_A)
  \ .
\end{align}
The equation of motion of the minus field follows from variation of the action with respect to the plus field and reads:\checked
\begin{align}
  \int \di x_{\!A}\,
  X_{(-)A}^c(x_{\!A})
  \frac{
    \del^2 S[X^c_{(+)A}]}{
    \del X_A(x_{\!A})
    \del X_B(x_{\!B})
    }
    =
    \mO
    \Big(
    (X^c_{(-)A})^3
    \Big)
    \ .
\end{align}
This equation of motion implies that the minus field as a perturbative field must be zero.
That is, then, also the reason why we can neglect higher quadratic and higher terms in the action, as they will never play a role in the equations of motion.
This also implies that we can neglect all parts of the in-in propagator matrix Eq.~\eqref{CD:SKPropagator} except the retarded propagator since the two other non-zero terms only affect the minus field which is zero.
This analysis would change, however, if we included background values for the minus field.
See e.g. Ref.~\cite{Jakobsen:2022psy} for a more detailed discussion of this.

The equation of motion of the plus field follows from variation with respect to the minus field.
Keeping only linear terms in the minus fields, it is clear from the in-in action in the Schwinger-Keldysh basis~\eqref{CD:SKAction} that this results in the desired classical equations of motion for the plus field.
In addition, the plus field is always propagated from a minus source with the retarded propagator which follows from the propagator matrix $\mD_{\rm (S-K)}(k)$ Eq.~\eqref{CD:SKPropagator}.
We have thus arrived at a path integral which in the classical limit gives rise to causal dynamics with retarded propagators.
In fact, the Feynman rules of this theory in the classical limit are almost identical to those of the in-out formalism.
These will be examined in Sec.~\ref{WQFT:Diagrammatics}.

Before moving on, let us specialize to the resulting in-in WQFT in the classical limit.
The in-in action in the Schwinger-Keldysh basis reads:\checked
\begin{align}
  S_{\rm in-in}[g_{(\pm)\mn},w_{(\pm)i}^\sig]
  =
  \int \di^d x\,
  g_{(-)\mn}(x)
  \frac{
    \del S[g_{\mn},w_i^\sig]
  }{
    \del g_{\mn}(x)
  }
  +
  \sum_{w_i}
  \int \di\tau\,
  w_{(-)i}^\sig(\tau)
  \frac{\del
    S[g_\mn,w_i^\sig]
  }{
    \del w_i^\sig(\tau)}
  \ .
\end{align}
Here, we mostly omit the $(+)$ label on the plus fields and we neglected higher order terms of the minus fields.
We generally assume the background parameters of the minus fields to be zero so that $g_{(-)\mn}=h_{(-)\mn}$ and $w_{(-)i}^\sig=\Del w_{(-)i}^\sig$.
One-point functions in the in-in theory are defined by\checked
\begin{align}
  \inin{X}
  =
  \int D[
    h_{(\pm)\mn},\Del w_{(\pm)i}^\sig,
  ]
  X
  e^{
  \frac{i}{\hbar}
  S_{\rm in-in}
  }
  \ ,
\end{align}
with $X$ either the graviton field or any of the worldline perturbation fields $\Del w_i^\sig$.

\section{Diagrammatics of WQFT}\label{WQFT:Diagrammatics}
We have now established that the one-point functions of the WQFT contain all (classical) dynamical information of the worldline and gravitational fields.
Later in Sec.~\ref{WQFT:PostMinkowskian} we will see that the gauge invariant information is contained in their on-shell values.
The focus of this section will be the expansion of the correlation functions in terms of (WQFT) Feynman diagrams.
In the first two sections~\ref{WQFT:FV} and~\ref{WQFT:PropagatorsAnd} we derive the Feynman rules for composing those diagrams and in the third section~\ref{sec:TLWQFT} we show how all Feynman diagrams contributing to a certain correlation function may be obtained from the equations of motion obeyed by that function.
This process builds on an off-shell recursion relation satisfied by the correlation functions which, essentially, is Berends-Giele off-shell recursion~\cite{Berends:1987me}.

We will continue working with the generic action $S[g_\mn,w_i^\sig]$ (Eq.~\ref{WQFT:Action}) composed of the bulk action $S_{\rm GR}$ and the worldline action $S_{\rm{wl},i}$.
The example of main interest, however, will be the SUSY spinning worldline action Eq.~\ref{WEFT:SUSYA} which reads:\checked
\begin{align}
  \label{WQFT:SpinAction}
  &S_{\rmm{spin},i}
  =
  -m_i
  \int
  \di\tau\,
  \bigg[
  \frac12
  g_\mn
  \dot z^\mu_i
  \dot z^\nu_i
  +
  i
  \bar\psi^\mu_i
  \frac{
    D\psi_{i,\mu}
  }{
    D\tau}
  +
  \frac12
  R_{\mn\ab}
  \bar \psi_i^\mu
  \psi_i^\nu
  \bar \psi_i^\alpha
  \psi_i^\beta
  +
  C_{\rmm{E},i}
  R_{\mu\alpha\nu\beta}
  \dot z^\alpha_i
  \dot z^\beta_i
  \bar\psi_i^\mu
  \psi_i^\nu
  \bar \psi_i
  \!\cdot\!
  \psi_i
  \bigg]
  \ .
\end{align}
In this section we will develop the tools for solving the equations of motions of this action coupled to gravity perturbatively using Feynman diagrams.
As discussed in Sec.~\ref{WEFT:SUSY}, these equations are the MPD equations well known in the traditional approach to classical general relativity.

\subsection{WQFT Feynman Vertices}
\label{WQFT:FV}
Feynman vertex rules are generally derived from the interaction part of the action.
The vertices are defined as functional derivatives of these interaction terms with respect to the relevant (perturbative) fields specializing to the desired background afterwards:\checked
\begin{align}\label{WQFT:FRules}
  V[X_1,X_2...X_n]
  =
  \left.
  \frac{
    \delta^n i S_{\rm int}
  }{
    \delta X_n
    ...
    \delta X_2
    \delta X_1
  }
  \right|_{\rm{Min}}
  \ .
\end{align}
Here, the subscript ``Min'' for Minkowski instructs us to insert the Minkowski background value for all fields.
In our example with worldline action $S_{\rmm{spin}, i}$ the fields $X_n$ represent the graviton field $h_\mn(x)$ or the worldline fluctuations $\Del z_i^\sig(\tau)$ or $\Del \psi^\sig_i(\tau)$.
The order of functional derivatives matter when $X_i$ include the Grassmann worldline fields.
The factor of $i$ in front of the action in Eq.~\eqref{WQFT:FRules} is conventional.
In this section we focus on the Feynman vertices of the in-out theory.
However, the vertex rules of the in-in theory are exactly the same as the in-out theory which we will discuss in Sec.~\ref{WQFT:PropagatorsAnd} .

The gravitational bulk action $S_{\rm GR}$ and the worldline action $S_{\rmm{wl},i}$ have very different properties and it is advantageous to define vertex rules for each of those parts individually.
They are defined from the interaction terms of each action as follows:\checked
\bse
\begin{align}
  \label{WQFT:GRFR}
  V_{\rm GR}[h_\mno(k_1)...h_\mnn(k_n)]
  &=
  \left.
  \frac{
    \delta^n iS^{\rm int}_{\rm GR}
  }{
    \delta h_\mno(k_1)
    ...
    \delta h_\mnn(k_n)
  }
  \right|_{\rm{Min}}
  \ ,
  \\
  \label{WQFT:WLVertices}
  V_{\rmm{wl},i}[X_1,X_2...X_n]
  &=
  \left.
  \frac{
    \delta^n iS^{\rm int}_{\rmm{wl},i}
  }{
    \delta X_n
    ...
    \delta X_2
    \delta X_1
  }
  \right|_{\rm{Min}}
  \ .  
\end{align}
\ese
In the first line we have immediately inserted the graviton field in place of $X_n$ because the gravitational action only depends on that field.
In case of the worldline interaction in the second line both worldline fields and the graviton field are relevant.
Also, in the first line we have used the graviton in momentum space instead of position space as we will use momentum space Feynman rules.
Our conventions imply that the momenta $k^\mu_i$ of the Feynman vertices Eq.~\eqref{WQFT:GRFR} correspond to incoming momenta.

The WQFT mixes $d$-dimensional bulk fields (the graviton) and one-dimensional worldline fields.
The bulk vertices derived from $S^{\rm int}_{\rm GR}$ will impose conservation of $d$-dimensional momenta.
Instead, the worldline vertices derived from $S^{\rm int}_{\rmm{wl},i}$ will only impose conservation of energy (with respect to the frame of that body).
This has the important effect that when a graviton interacts on the worldline only its energy is conserved and the integration on its spacial momentum is left unconstrained.
This leads to loop-like integrations.

Apart from the usual dependence on momenta and energies, the WQFT worldline vertices depend on the background parameters of the bodies.
For the worldline parametrization $z^\sig_i(\tau)$ these are $b_i^\mu$ and $v_i^\mu$ and for the spin tensor $\mS^\mn_i(\tau)$ it is $S^\mn_i$.

In the following we will focus on the Feynman vertices derived from the worldline part of the action.
Afterwards we will briefly discuss the bulk Feynman vertices derived from the gauge fixed Einstein-Hilbert action which are a standard result of effective quantum gravity.

\subsubsection{Worldline Vertices with Spin}
We will focus on the worldline Feynman vertices derived from the spinning SUSY action $S_{\rmm{spin},i}$ Eq.~\eqref{WQFT:SpinAction}.
These are relevant for the computation of the leading order waveform and $\mO(G^3,S^2)$ worldline observables considered in later chapters.

The interaction part of the (gauge fixed) spinning SUSY worldline action reads:\checked
\begin{align}
  \label{WQFT:InteractionSpin}
  S_{\rmm{spin}}^{\rm int}
  &=
  -
  \int \di\tau
  \Big[
    \frac{m}2
  h_\mn
  \dot z^\nu
  \dot z^\mu
  -
  \frac1{2}
  \dot z^\mu
  \omega_\mu^{ab}
  \mS_{ab}
  +
  \frac1{8m}
  R_{abcd}
  \mS^{ab}
  \mS^{cd}
  +
  \frac{C_{E}}{2m}
  R_{a\mu b\nu}
  \dot z^\mu
  \dot z^\nu
  \mS^{ac}
  \mS^{b}_{\ c}
  \Big]
  \ .
\end{align}
Here, we have omitted particle labels on the action and worldline variables.
We will do so throughout this section because we are considering the vertices of each body by themselves.
In the end, we can easily restore the particle label on the resulting Feynman vertices and body variables.

In the interaction terms Eq.~\eqref{WQFT:InteractionSpin} we have made use of the fact that they depend on the Grassmann variables only implicitly via the spin tensor, $\mS^\mn(\tau)$.
The spin tensor in terms of the Grassmann field $\psi^\mu(\tau)$ (Eq.~\ref{WEFT:SpinTensor}) is:\checked
\begin{align}
  \mS^{ab}(\tau)
  =
  -2 i
  m
  \bar\psi^{[a}(\tau)
    \psi^{b]}(\tau)
    =
  2m \operatorname{Im}\!
  \big[
  \bar \psi^\mu(\tau) \psi^\nu(\tau)
  \big]
  \ .
\end{align}
Also, in the interaction terms we have explicitly used the local (vielbein) indices $a$, $b$, $c$ and $d$ on the spin tensors instead of covariant ones.
In our post-Minkowskian expansion, the spin tensor $\mS^{ab}(\tau)$ with local indices are considered independent of the graviton field in contrast to $\mS^\mn=\mS^{ab}e_a^\mu e_b^\nu$.
The spin connection $\omega^\mu_{ab}$ was given in Eq.~\eqref{WEFT:SpinConnection} by:\checked
\begin{align}
  \omega_\mu^{ab}
  =
  e^a_{\,\nu}
  \Big(
  \partial_\mu
  e^{b\nu}
  +
  \Gamma^\nu_{\mu\sig}
  e^{b\sig}
  \Big)
  \ .
\end{align}
The expansion in $h_\mn$ of the spin connection, the vielbein and other gravitational variables will briefly be considered at the end of this section in connection with the gravitational bulk Feynman vertices.

Feynman vertices $V_{\rm spin}[X_1,X_2...X_n]$ are now defined by Eq.~\eqref{WQFT:WLVertices} in terms of functional derivatives of the spinning action $S_{\rm spin}^{\rm int}$.
Expressions for the vertex rules can be derived in several different ways. A straightforward approach in which the action is expanded in the perturbative fields is found in Refs.~\cite{Mogull:2020sak,Jakobsen:2021zvh}.
Here, we will present an alternative method where we use a recursive relation satisfied by the vertices in order to add worldline legs to them.

We define reduced vertex rules, $\hat V_{\rm spin}[X_1,X_2...X_n]$, in which we remove the (one-dimensional) energy-conserving delta-function:\checked
\begin{align}
  \label{WQFT:ReducedVertices}
  \dd
  \Big(
  \sum_{m=1...n} \omega_m
  \Big)
  \hat V_{\rm spin}[X_1,X_2...X_n]
  =
  V_{\rm spin}[X_1,X_2...X_n]
  \ .
\end{align}
When $X_m$ is a graviton, $h_\mn(k)$, the corresponding energy is $\omega_m=k\cdot v$.
For worldline fields $w_i^\sig(\oma)$ it is simply their energy $\oma$. 
The recursive relations satisfied by the reduced vertices read:\checked
\bse
\label{WQFT:Recursive}
\begin{align}
  \hat V_{\rm spin}[\Del z^\sig(\oma),X_1...X_n]
  &=
  \Big(
    \frac{
    \pat
  }{
    \pat b_{\sig}
  }
    -i
  \oma
  \frac{
    \pat
  }{
    \pat v_{\sig}
  }
  \Big)
  \hat V_{\rm spin}[X_1...X_n]
  \ ,
  \\
  \hat V_{\rm spin}[\Del \psi^\sig(\oma),X_1...X_n]
  &=
  \frac{\pat}{
    \pat \Psi^\sig
  }
  \hat V_{\rm spin}[X_1...X_n]
  \ .
\end{align}
\ese
These relations tell us that we can add any trajectory fluctuation leg $\Del z^\sig(\oma)$ or Grassmann leg $\Del \psi^\sig(\oma)$ by differentiation of a lower point vertex with respect to the background parameters.
In fact, they tell us that we need only the multipoint graviton vertex rules and from those we can derive vertex rules with any number of worldline legs by simple differentiations.
The multipoint graviton (worldline) vertices are:\checked
\begin{align}
  V_{\rmm{spin}}[h_{\mno}(k_1),h_\mnt(k_2)...h_\mnn(k_n)]
  &=
  \left.
  \frac{
    \delta^n iS^{\rm int}_{\rmm{spin}}
  }{
    \delta h_\mnn(k_n)
    ...
    \delta h_\mnt(k_2)
    \delta h_\mno(k_1)
  }
  \right|_{\rm{Min}}
  \ .    
\end{align}
These vertices, then, are the basic building blocks of all other worldline vertices.

When the energy conserving delta-function of the vertices is stripped off in Eq.~\eqref{WQFT:ReducedVertices}, the remaining (reduced) vertices are not uniquely defined.
This arbitrariness, however, is fixed by demanding that they satisfy the recursive relations Eqs.~\eqref{WQFT:Recursive}.
That it is possible to satisfy those relations, we will see now.

The recursive relations are derived starting from the worldline fields in momentum space:\checked
\bse
\begin{align}
  z^\sig(\oma)
  &=
  b^\sig\dd(\oma)
  -i v^\sig\dd'(\oma)
  +
  \Del z^\sig(\oma)
  \ ,
  \\
  \psi^\sig(\oma)
  &=
  \Psi^\sig
  \dd(\oma)
  +
  \Del\psi^\sig(\oma)
  \ .
\end{align}
\ese
The following chain rule differentiations are derived as a consequence:\checked
\bse
\begin{align}
  \frac{\pat}{
    \pat b^\sig
  }
  &=
  \int_\oma
  \frac{
    \del z^\sig(\oma)
  }{
    \pat b^\rho}
  \frac{
    \del}{
    \del z^\rho(\oma)
  }
  =
  \int_\oma
  \dd(\oma)
  \frac{
    \del}{
    \del z^\sig(\oma)
  }
  \ ,
  \\[10pt]
  \frac{\pat}{
    \pat v^\sig
  }
  &=
  \int_\oma
  \frac{
    \del z^\sig(\oma)
  }{
    \pat v^\rho}
  \frac{
    \del}{
    \del z^\rho(\oma)
  }
  =
  \int_\oma
  -i\dd'(\oma)
  \frac{
    \del}{
    \del z^\sig(\oma)
  }
  \ ,
  \\[10pt]
  \frac{\pat}{
    \pat \Psi^\sig
  }
  &=
  \int_\oma
  \frac{
    \del \psi^\rho(\oma)
  }{
    \pat \Psi^\sig}
  \frac{
    \del}{
    \del \psi^\rho(\oma)
  }
  =
  \int_\oma
  \dd(\oma)
  \frac{
    \del}{
    \del \psi^\sig(\oma)
  }
  \ .
\end{align}
\ese
In the first line we have differentiation with respect to $b^\sig$, then in the second line velocity $v^\mu$ and finally in the third line Grassmann variable $\Psi^\sig$.
The first relation implies that differentiation of a vertex with respect to $b^\mu$ pulls out an external worldline fluctuation leg with $\oma=0$ on that leg.
The second relation implies that differentiation with respect to $v^\mu$ pulls out an external worldline fluctuation leg with all linear contributions in the energy $\oma$ but not the constant term.
The third relation implies that differentiation with respect to $\Psi^\sig$ pulls out a Grassmann fluctuation leg with the energy $\oma=0$.

Now, we note that the interaction piece of the worldline action only depends on $\dot z^\sig(\tau)$, $z^\sig(\tau)$ and $\psi^\sig(\tau)$ and no higher derivatives.
We may thus define the reduced vertex rules $\hat V_{\rm wl}[...]$ to be at most linear in the trajectory fluctuation energies and constant in the Grassmann fluctuation energies.
In that case the three derivatives above are indeed sufficient to reconstruct higher order vertex rules from lower order vertex rules because they relate all necessary information about higher point vertices from lower point vertices.
In this way, we arrive at the recursive relations, Eq.~\eqref{WQFT:Recursive}.
We note that in order to differentiate in these relations we must use unconstrained background variables.
Thus, if we want to use the recursive relations we cannot impose gauge choices on the background parameters until after differentiation has been performed.
That is, we cannot insert $v^2=1$ or $v_\mu \Psi^\mu=0$.

Let us now focus on vertex rules with Grassmann legs.
As a starter we note that vertex rules with Grassmann variables anti-commute:\checked
\begin{align}
  \label{WQFT:Anticommute}
  V_{\rm spin}[...,\Del \psi^\sigo(\omao),...,\Del \psi^\sigt(\omat),...]
  =
  -
  V_{\rm spin}[...,\Del \psi^\sigt(\omat)
    ,...,
    \Del \psi^\sigo(\omao)
    ,...]
  \ .
\end{align}
Here, the respective dots on either side of the equation signify the same collection of fields.

The derivation of vertices with Grassmann legs can be simplified greatly by making use of the fact that the interaction terms depend only on these variables through their bosonic combination the spin tensor $\mS^\mn(\tau)$.
This implies that all vertex rules without Grassmann legs depend on the Grassmann background parameters only through their combination $S^\mn$.
Grassmann legs are then added by differentiation with respect to the background Grassmann variables and, again, we use the chain rule:\checked
\begin{align}
  \frac{
    \pat\ \ 
  }{\pat \Psi^\sig}
  &=
  \frac{\pat S^\mn}{\pat \Psi^\sig\ }
  \frac{\pat\ \ \ }{\pat S^\mn}
  \\
  &=
  2im
  \bar \Psi^\rho
  \frac{\pat\ \ \ }{
    \pat S^{\rs}}
  \ .
  \nn
\end{align}
Naturally, this rule is valid only when applied to functions that depend on $\Psi^\sig$ only through the spin tensor.
The analogous rule with $\bar \Psi^\sig$ is derived by Hermitian conjugation:\checked
\begin{align}
  \frac{
    \pat\ \ 
  }{\pat \bar \Psi^\sig}
  &=
  2im
  \Psi^\rho
  \frac{\pat\ \ \ }{
    \pat S^{\rs}}
  \ .
\end{align}
Note that under the Hermitian conjugation the (usual) left derivative changes into a right derivative and rewriting that in terms of a left derivative introduces an extra sign.

Iterations of differentiations with respect to the Grassmann variables lead to additional cross terms.
As an example, we have:\checked
\begin{align}
  \frac{
    \pat\ \ 
  }{\pat \bar \Psi^\rho}
  \frac{
    \pat\ \ 
  }{\pat \Psi^\sig}
  &=
  2im
  \frac{\pat\ \ \ }{
    \pat S^{\rs}}
  +
  4m^2
  \bar\Psi^\nu
  \Psi^\mu
  \frac{\pat\ \ \ }{
    \pat S^{\mu\rho}}
  \frac{\pat\ \ \ }{
    \pat S^{\nu\sig}}  
  \ .
\end{align}
Subsequent derivatives with respect to the same Grassmann variable do not generate cross terms.
In general, we can re-express vertex rules with Grassmann legs in terms of Feynman rules with external ``spin tensor legs'' which we simply define by partial derivatives with respect to the spin tensor.
To quadratic order in the Grassmann fields, we find:\checked
\bse
\begin{align}
  \label{WQFT:SToG}
  \hat V[\Del \psi^a,...]
  &=
  -2im\bar \Psi^b
  \,
  \hat V[\Del \mS^{ab},...]
  \ ,
  \\
  \hat V[\Del \psi^{a_1},\Del \psi^{a_2},...]
  &=
  4m^2
  \bar \Psi^{b_1}
  \,
  \bar \Psi^{b_2}
  \,
  \hat V[\Del \mS^{a_1b_1},\Del \mS^{a_2b_2},...]
  \ ,
  \\
  \hat V[\Del \bar \psi^{a_1}
    ,\Del \psi^{a_2}
    ,...]
  &=
  2im
  \hat V[\Del \mS^{a_1 a_2},...]
  +
  4m^2
  \Psi^{b_1}
  \bar \Psi^{b_2}
  \hat V[\Del \mS^{a_1b_1},\Del \mS^{a_2b_2},...]
  \ .
\end{align}
\ese
Here, the ellipsis (...) on either side of the equations denote the same collection of fields excluding the Grassmann or spin tensor field.
We have omitted all energy dependence of the Grassmann and spin tensor fields because the reduced vertices are independent of those energies.
The spin tensor Feynman rules are derived from the interaction terms treating $\mS^\mn(\tau)$ as another worldline field or, equivalently, with the recursive relations by differentiation with respect to $S^\mn$.
The Grassmann vertices are thus simply derived from bosonic vertex rules in terms of the graviton, trajectory fluctuation and spin tensor fluctuation.

Finally, we only have to derive the multipoint graviton worldline vertices.
At this point we may simply insert the background values of the worldline fields into the action and thus neglect the worldline perturbation fields.
If we take the (spinless) Polyakov action as an example, we have:\checked
\begin{align}
  -\frac{m}2
  \int\di\tau
  h_{\mn}(b^\sig+v^\sig\tau)v^\mu v^\nu
  \ .
\end{align}
Here, a functional derivative with respect to $h_\mn(k)$ is straightforward.
The Jacobian between momentum and coordinate space reads:\checked
\begin{align}
  \label{WQFT:Jacobian}
  \frac{
    \del h_\mn(x)}{
    \del h_\ab(k)}
  =
  \eta_{\mu(\alpha}
  \eta_{\beta)\nu}
  e^{-ik\cdot x}
  \ .
\end{align}
The only difficulty of including spin, then, is the expansion of gravitational variables like the spin connection and the vielbein in powers of the graviton field.
Equations relevant to that expansion are given below in Eqs.~\eqref{WQFT:Variations}.

In order to get an idea of the structure of the worldline vertices we print the reduced one point graviton emission vertex:\checked
\begin{align}
  \label{WQFT:OnePoint}
  \hat V[h_\mn(k)]
  =
  -i\frac{m\kappa }{2}
  e^{-ik\cdot b}
  \bigg(
  v^\mu v^\nu
  -
  i
  \frac1{m}
  (k \!\cdot\! S)^{(\mu} v^{\nu)} 
  -
  \frac1{2m^2}
  (k \bdot S)^\mu
  (k \bdot S)^\nu
  +\frac{C_{{\rm E}}}{2m^2}
  v^\mu v^\nu
  k\bdot{S}\bdot{S}\bdot k
  \bigg)
  \ .
\end{align}
This relatively simple vertex rule represents the energy-momentum tensor of the background straight-line motion.
For Kerr black holes its structure is known to all orders in spin and given by an exponential structure~\cite{Vines:2017hyw,Guevara:2018wpp,Guevara:2019fsj}.

The vertex rules are presented in terms of unconstrained background variables and with the recursive relations we may add external worldline legs.
The graviton interaction with a trajectory fluctuation becomes:\checked
\begin{align}\label{eq:vertexHZ}
  &\hat V[
    h_\mn(k),
    \Del z^\sig(\oma)]
  \Big|_{C_E\to0}
  =
  -\frac{m\kappa }{2}e^{-ik\cdot b}
  \\\nn
  &\qquad\qquad\qquad
  \times
  \Big( k_\sigma
  v^\mu v^\nu
  +2\omega
  v^{(\mu}\del^{\nu)}_\sigma
  -
  i
  \frac1{m}
  (k\bdot S)^{(\mu}
  \big(
  k_\sigma v^{\nu)}+\omega\del_\sigma^{\nu)}
  \big)
  -
  \frac1{2m^2}
  k_\sigma(k\bdot S)^{\mu}
  (k\bdot S)^{\nu}
  \Big)
  \ .
\end{align}
Here, we specialized to Kerr black holes.
For the vertex rule with one Grassmann fluctuation and one graviton, we get:\checked
\begin{align}
  &
  \hat V[\psi^\sig,h_\mn(k)]  \Big|_{C_E\to0}
  =
  im\kappa e^{-ik\cdot b}
  k_{[\sig}^{\vphantom{(\mu}}
    \del_{\rho]}^{(\mu}
  \left(
  v^{\nu)_{\vphantom{\rho]}}}
    +
    i\frac1{m}
    (S\cdot k)^{\nu)}
  \right)
  \bar \Psi^\rho
  \ .
\end{align}
The corresponding Feynman rule with an external spin leg is then derived with Eq.~\eqref{WQFT:SToG}.
As noted above, all momenta and energies in the above Feynman vertices are incoming.

Using the recursive relations we may for each multipoint graviton vertex derive all order expressions for the corresponding vertices with worldline fluctuation legs.
Thus, in the spinless case we find the same expression as presented in Ref.~\cite{Mogull:2020sak}:\checked
\begin{align}
  &V[h_{\mn}(k),\Del z^\sigo(\omao),\Del z^\sigt(\omat),...,\Del z^\sign(\oman)]=
  -(-i)^{n-1} m\kap e^{-ik\cdot b}
  \\\nn
  &\qquad\qquad\qquad\qquad\qquad\qquad\times
  \bigg[
    \frac12
    v^\mu v^\nu
    k_{\sigma_i} k_{\sigma_j}
    +
    \sum_{i=1}^n
    \oma_i \del^{(\mu}_{\sig_i}v^{\nu)}
    k_{\sigma_j}
    +
    \sum_{i<j}^n
    \oma_i\oma_j
    \del^{(\mu}_{\sig_i}\del^{\nu)}_{\sig_j}
    \bigg]
    \prod_{l\neq i,j}^n
    k_{\sig_l}
  \ .
\end{align}
Here $i\neq j$.
Repeated differentiation with respect to $b^\mu$ add the factors of $k^\mu$ and repeated differentiation with respect to $v^\mu$ hits only the initial two factors of $v^\mu$ and generates the three terms above.
The above formula may easily be generalized to include spin and a similar version could be derived for higher point gravitational interactions.

\subsubsection{Multipoint Graviton Vertices and Expansions in the Graviton Field}
The expansion in $h_\mn(x)$ of gravitational variables such as the curvature, the vielbein or the spin connection allows us to derive multipoint graviton vertex rules.
These are relevant both for the bulk and worldline vertices.
Indeed the derivation of general worldline vertices was reduced to the derivation of multipoint graviton interactions on the worldline.
Here, we will provide a few formulas which are sufficient for the functional variation of the gravitational variables present in this work.

In particular we give formulas for the variation of the inverse metric $g^\mn$, the metric determinant $\sqrt{-g}$ and the vielbein $e_{a \mu}$.
With these formulas variations of any other variables in this work such as the Riemann curvature, the Christoffel symbol or the spin connection can be derived using the chain and Leibniz rules.
In particular the variation with respect to $h_\mn(x)$ of the gauge fixed Einstein-Hilbert action and the SUSY worldline action is possible.
The conversion to momentum space can be achieved with the Jacobian Eq.~\eqref{WQFT:Jacobian}.

The three rules read:\checked
\bse\label{WQFT:Variations}
\begin{align}
  \del
  g^\mn
  &=
  -
  g^{\mu\alpha}
  g^{\nu\beta}
  \del
  g_\ab
  \ ,
  \\
  \del \sqrt{-g}
  &=
  \frac12
  \sqrt{-g}
  g^\mn
  \del g_\mn
  \ ,
  \\
  \del e_{a\mu}
  &=
  \frac12
  e^{\ \sig}_{(a}
  \del g_{\mu)\sig}^{\vphantom{\sig}}
  \ .
\end{align}
\ese
These rules may be derived from linear algebra.
The third equation is not unique to the same extend that the definition of the vielbein Eq.~\eqref{WEFT:Vielbein} does not determine it uniquely.
Here we have chosen a gauge in which we require the vielbein to be symmetric in its two indices $e_{a\mu}=e_{\mu a}$.
This equation mixes local and covariant indices but they get mixed in the post-Minkowskian expansion regardless.
With the symmetry constraint on $e_{a\mu}$ it is simply the (matrix) square root of $g_\mn$ and its variation follows from that fact.

Higher order worldline and bulk graviton interactions may be derived with these rules.
In particular, the graviton bulk interactions are a standard result in effective quantum gravity and are e.g. discussed in Refs.~\cite{DeWitt:1967uc,Donoghue:1994dn,Jakobsen:2020diz,Cheung:2017kzx,Cheung:2020gyp}.

\subsection{Propagators and Flow of Causality}
\label{WQFT:PropagatorsAnd}
Let us now turn our attention to the kinetic term of the action from which we derive the propagators of the WQFT.
We then discuss the Feynman rules for assembling diagrams with special focus on the in-in theory where the introduction of a flow of causality effectively reduces the two Schwinger-Keldysh $(\pm)$ fields into a single field.

We generally work in momentum space where the inversion of the kinetic operators is algebraic.
In the in-out theory we will get Feynman propagators for the graviton field and we adopt a time-symmetric $i\eps$ pole displacement ($i\eps$-prescription) for the worldline propagators following Ref.~\cite{Mogull:2020sak}.
Instead, in the in-in theory we have retarded propagators for all bulk and worldline fields.
These are not time-symmetric and enforce the causal boundary conditions of the in-in theory.

The kinetic terms of the in-out theory are given as follows (Eqs.~\ref{WEFT:GRKin} and~\ref{WEFT:WLKin}):\checked
\bse
\begin{align}
  S^{\rm kin}_{\rm spin}
  &=
  -m
  \int \di\tau
  \bigg(
  \frac12
  \Del \dot z^\mu(\tau)
  \Del \dot z_\mu(\tau)
  +
  i
  \Del \bar \psi^a(\tau)
  \Del \dot \psi_{a}(\tau)
  \bigg)
  \ ,
  \\
  S_{\rm GR}^{\rm kin}
  &=
  \frac12
  \int
  \di^d x\,
  \mP^{\mn\ab}
  \partial_\rho
  h_\mn(x)
  \partial^\rho
  h_\ab(x)
  \ .
\end{align}
\ese
Again, as in the previous section, we omit the particle label on the worldline fields.
The propagators of several worldlines are then derived by reintroducing the particle label.
The in-in kinetic term is derived from the in-out kinetic term with Eq.~\eqref{CD:SKAction}.
We get:\checked
\bse\label{WQFT:InInK}
\begin{align}
  S^{\rm kin}_{\rm in-in,\,spin}
  &=
  - m
  \int\!\!
  \di \tau
  \Big(
  \Del z^\mu_{(-)}
  \,
  \Del \ddot z_{\mu}
  -
  2\operatorname{Im}\!
  \big[
  \Del \bar \psi^a_{(-)}
  \,
  \Del \dot \psi_{a}
  \big]
  \Big)
  \\
  S_{\rm in-in,\,GR}^{\rm kin}
  &=
  -\int\!
  \di^d x\,
  \mP^{\ab\mn}\,
  h_{(-)\mn}(x)\,
  \partial^2\,
  h_{\ab}(x)
  \ .
\end{align}
\ese
Here, we simply omit the $(+)$ subscript on the plus fields.

Let us start with the propagators of the in-out theory where we define the propagator $\Del_{\rm in-out}(k)$ of the field $X(k)$ by the equation:\checked
\begin{align}\label{WQFT:IOD}
  i
  \dd^d(k-k')
  \Del^{-1}_{\rm in-out}(k)
  =
  \frac{
    \del^2 S^{\rm kin}
  }{
    \del X(k)
    \del X^\dagger(k)
  }
  \ .
\end{align}
This gives a formula for $\Del_{\rm in-out}^{-1}(k)$ which is inverted with a suitable $i\eps$-prescription.
When considering the Grassmann variables the order of the two functional derivatives is significant.
For the graviton, we use the Feynman prescription and its propagator is found to be:\checked
\bse\label{WQFT:IOP}
\begin{align}
  \label{WQFT:GPropagator}
  	\begin{tikzpicture}[baseline={(current bounding box.center)}]
	\coordinate (in) at (-1,0);
	\coordinate (out) at (1,0);
	\draw[snake it] (in) -- (out);
        \node[above] at ($(in)!0.5!(out)$) {$k$} ;
        \node[below] at ($(in)!0.5!(out)$) {$\vphantom{k}$} ;
        \node[left] at (in) {$\ab$} ;
        \node[right] at (out) {$\mn$} ;
	\end{tikzpicture}&=
        i
        \frac{
          \mP^{-1}_{\ab\mn}
        }{k^2+i\eps}
        \ .
\end{align}
The tensor $\mP^{-1}_{\mn\ab}$ is the inverse to $\mP^{\mn\ab}$ (Eq.~\ref{WEFT:PTensor}) and in arbitrary dimensions it is given by:\checked
\begin{align}
  \mP^{-1}_{\mn\ab}
  =
  \eta_{\mu(\alpha}
  \eta_{\beta)\nu}
  -
  \frac{1}{d-2}
  \eta_\mn\eta_\ab
  \ .
\end{align}
The graviton propagator Eq.~\eqref{WQFT:GPropagator} is a standard result in effective quantum gravity.
For the in-out worldline propagators we use the average prescription which is the symmetric sum of retarded and advanced propagators:\checked
\label{WQFT:WLPropagators}
\begin{align}
  	\begin{tikzpicture}[baseline={(current bounding box.center)}]
	\coordinate (in) at (-1,0);
	\coordinate (out) at (1,0);
	\draw[worldline] (in) -- (out);
        \node[above] at ($(in)!0.5!(out)$) {$\Del z(\oma)$} ;
        \node[below] at ($(in)!0.5!(out)$) {$\vphantom{z(\omega)}$} ;
        \node[left] at (in) {$\sig$} ;
        \node[right] at (out) {$\rho$} ;
	\end{tikzpicture}&=
        -\frac{
          i\eta_{\rho\sigma}
        }{2m}
        \Big[
        \frac1{
          (\oma+i\eps)^2
        }
        +
        \frac1{
          (\oma-i\eps)^2
        }
        \Big]
        \ ,
        \\
  	\begin{tikzpicture}[baseline={(current bounding box.center)}]
	\coordinate (in) at (-1,0);
	\coordinate (out) at (1,0);
	\draw[worldline] (in) -- (out);
        \node[above] at ($(in)!0.5!(out)$) {$\Del \psi(\omega)$} ;
        \node[below] at ($(in)!0.5!(out)$) {$\vphantom{z(\omega)}$} ;
        \node[left] at (in) {$\sig$} ;
        \node[right] at (out) {$\rho$} ;
	\end{tikzpicture}&=
        -\frac{
          i\eta_{\rho\sigma}
          }{2m}
        \Big[
        \frac1{
          \oma+i\eps
        }
        +
        \frac1{
          \oma-i\eps
        }
        \Big]
        \ .
\end{align}
\ese
These propagators are time symmetric.
We indicate all worldline fields with a solid line and the different species and their energy above the line.
The Grassmann propagator connects $\Del \psi^\sig(\oma)$ and $\Del \bar \psi^\sig(\oma)$.

With the in-out propagators Eqs.~\eqref{WQFT:IOP}, the Feynman Rules of the in-out WQFT are complete.
In principle, we draw all (tree-level) diagrams relevant for the observable under consideration.
They are then evaluated by inserting vertex rules and propagators.
The only challenge is the manipulation of the anti-commuting Grassmann variables.
We will compute observables to $\mO(G^2,S^2)$ in Sec.~\ref{WQFT:Worldline2} using the in-out Feynman rules.

Let us now turn to the propagators and diagrammatics of the in-in theory.
The kinetic terms Eqs.~\eqref{WQFT:InInK} mix the minus and plus fields and the propagator matrix in the Schwinger-Keldysh basis was derived in Eq.~\eqref{CD:SKPropagator}.
We will assume that the minus background parameters are zero and in that case it was argued that only the retarded propagator of that matrix can contribute to (classical) correlation functions.
The propagation from minus to plus is then conveniently represented by an arrow which points in the direction of the retarded propagator.
The in-in propagators will thus have an explicit direction associated to them which is interpreted as the flow of causality.
We define, formally, the in-in propagator $\Del_{\rm in-in}(k)$ of a field $X_{(\pm)}(k)$ by:\checked
\begin{align}\label{WQFT:IID}
 i
  \dd^d(k-k')
  \Del_{\rm in-in}^{-1}(k)
  =
  \frac{
    \del^2 S^{\rm kin}
  }{
    \del X_{(+)}(k)
    \del X^\dagger_{(-)}(k)
  }
  \ .
\end{align}
As with Eq.~\eqref{WQFT:IOD} this gives a formula for $\Del_{\rm in-in}^{-1}(k)$ which in this case must be inverted using the retarded prescription.
In fact, when considering the same field $X(k)$ the two $\Del_{\rm in-out}^{-1}(k)$ and $\Del_{\rm in-in}^{-1}(k)$ are identical and are simply the kinetic operator of that field.
With the convention Eq.~\eqref{WQFT:IID}, the propagators of the in-in theory are given as follows:\checked
\bse\label{WQFT:IIP}
\begin{align}
  	\begin{tikzpicture}[baseline={(current bounding box.center)}]
	\coordinate (in) at (-1,0);
	\coordinate (out) at (1,0);
        \draw[snake it] (in) -- (out);
        \draw[snake it A] (in) -- (out);
        \node[above] at ($(in)!0.5!(out)$) {$\vphantom{j}k$} ;
        \node[below] at ($(in)!0.5!(out)$) {$\vphantom{k}$} ;
        \node[left] at (in) {$\ab$} ;
        \node[right] at (out) {$\mn$} ;
	\end{tikzpicture}&=
        i\frac{
          \mP^{-1}_{\ab\mn}
        }{
          (k^0+i\eps)^2
          -
          \vct{k}^2
        }
        \ ,
        \\
  	\begin{tikzpicture}[baseline={(current bounding box.center)}]
	\coordinate (in) at (-1,0);
	\coordinate (out) at (1,0);
	\draw[worldline] (in) -- (out);
        \draw[worldline A] (in) -- (out);
        \node[above] at ($(in)!0.5!(out)$) {$\Del z(\omega)$} ;
        \node[below] at ($(in)!0.5!(out)$) {$\vphantom{\Del z(\omega)}$} ;
        \node[left] at (in) {$\sig$} ;
        \node[right] at (out) {$\rho\hphantom{\nu}$} ;
	\end{tikzpicture}&=
        -\frac{i\eta_{\rho\sigma}}{m}
        \frac1{
          (\oma+i\eps)^2
        }
        \ ,
        \\
  	\begin{tikzpicture}[baseline={(current bounding box.center)}]
	\coordinate (in) at (-1,0);
	\coordinate (out) at (1,0);
        \draw[worldline] (in) -- (out) ;
        \draw[worldline A] (in) -- (out) ;
        \node[above] at ($(in)!0.5!(out)$) {$\Del \psi(\omega)$} ;
        \node[below] at ($(in)!0.5!(out)$) {$\vphantom{\Del z(\omega)}$} ;
        \node[left] at (in) {$\sig$} ;
        \node[right] at (out) {$\rho\hphantom{\nu}$} ;
	\end{tikzpicture}&=
        -\frac{i\eta_{\rho\sigma}}{m}
        \frac1{
          \oma+i\eps
        }
        \ .
\end{align}
\ese
All in-in propagators have the retarded propagator corresponding to their space-time dimension (one or $d$).
In these expressions, we assume the energy or momentum to be aligned with the flow of causality.
Otherwise, the expression on the right-hand-side changes so that e.g:\checked
\begin{align}
  	\begin{tikzpicture}[baseline={(current bounding box.center)}]
	\coordinate (in) at (-1,0);
	\coordinate (out) at (1,0);
	\draw[worldline] (in) -- (out);
        \draw[worldline A] (in) -- (out);
        \node[above] at ($(in)!0.5!(out)$) {$\Del z(-\omega)$} ;
        \node[below] at ($(in)!0.5!(out)$) {$\vphantom{\Del z(\omega)}$} ;
        \node[left] at (in) {$\sig$} ;
        \node[right] at (out) {$\rho\hphantom{\nu}$} ;
	\end{tikzpicture}&
        =
        -\frac{i\eta_{\rho\sigma}}{m}
        \frac1{
          (-\oma+i\eps)^2
        }
        =
        -\frac{i\eta_{\rho\sigma}}{m}
        \frac1{
          (\oma-i\eps)^2
        }
        \ .
\end{align}

We can now discuss the diagrammatics of the (classical, tree-level) in-in theory.
As a consequence of Eq.~\eqref{CD:SKAction} every in-in vertex has one $(-)$ leg and any number of $(+)$ legs.
With the arrow notation of the propagators this implies that every vertex must have one arrow pointing away from it and the rest pointing into it.
In other words, every vertex is sourced by any number of arrows flowing into it and acts as a new source with one arrow flowing out of it.
The one-graviton vertices act as simple backgrounds with a single outgoing (causality) arrow.
These one-graviton vertices act as the basic sources of all one-point correlation functions.

In general we may draw any (multipoint) Feynman diagram and assign a consistent causality flow to it.
In this way we generate in-in diagrams.
The vertex rules of the in-in theory are insensitive to the causality flow.
That is, their expressions do not depend on the $(+)$ or $(-)$ nature of the fields.
They are thus exactly equivalent to the vertex rules of the in-out theory.
Symmetry factors of one-point in-in diagrams are also equivalent to the corresponding in-out diagrams.

\begin{figure}
  \renewcommand*\thesubfigure{\arabic{subfigure}}
    \centerline{
    \begin{adjustbox}{minipage=\linewidth,scale=1.1}
  \centering
  \begin{subfigure}{.09\textwidth}
    \centering
    \includegraphics{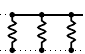}
    \caption{}
  \end{subfigure}
  \begin{subfigure}{.09\textwidth}
    \centering
    \includegraphics{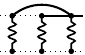}
    \caption{}
  \end{subfigure}
  \begin{subfigure}{.09\textwidth}
    \centering
    \includegraphics{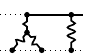}
    \caption{}
  \end{subfigure}
  \begin{subfigure}{.09\textwidth}
    \centering
    \includegraphics{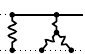}
    \caption{}
  \end{subfigure}
  \begin{subfigure}{.09\textwidth}
    \centering
    \includegraphics{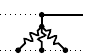}
    \caption{}
  \end{subfigure}
  \begin{subfigure}{.09\textwidth}
    \centering
    \includegraphics{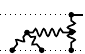}
    \caption{}
  \end{subfigure}
  \begin{subfigure}{.09\textwidth}
    \centering
    \includegraphics{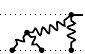}
    \caption{}
  \end{subfigure}
  \begin{subfigure}{.09\textwidth}
    \centering
    \includegraphics{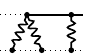}
    \caption{}
  \end{subfigure}
  \begin{subfigure}{.09\textwidth}
    \centering
    \includegraphics{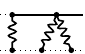}
    \caption{}
  \end{subfigure}
    \begin{subfigure}{.09\textwidth}
    \centering
    \includegraphics{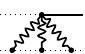}
    \caption{}
  \end{subfigure}
  \end{adjustbox}}
  \caption{
    Diagrams relevant to the probe limit of the worldline observables at the third post-Minkowskian order.
    Every solid line represents either a Grassmann or worldline fluctuation and wiggly lines gravitons.
    Only the graphs (1) - (6) are relevant in the spinless theory.
    This figure is reproduced with minor changes from Ref.~\cite{Jakobsen:2022fcj}.
  }
  \label{WQFT:ProbeGraphs}
\end{figure}

In Fig.~\ref{WQFT:ProbeGraphs} we show some examples of WQFT diagrams with one external worldline leg.
We generally draw gravitons with a wiggly line and worldline fluctuations with a solid line.
The worldline background is then indicated with a dotted line which, however, does not indicate any fluctuations.
We usually draw diagrams such that graviton bulk interactions occur in the middle between the two worldlines with interactions on the first worldline above and interactions on the second worldline below.
In the examples in Fig.~\ref{WQFT:ProbeGraphs} worldline fluctuations only occur on the first worldline (the first body).
While these diagrams are tree-level the dotted background lines make them resemble loop diagrams.

\subsection{Equations of Motion and Off-Shell Recursion}
\label{sec:TLWQFT}
The classical equations of motion satisfied by the one-point correlation functions can be used as an efficient tool for generating all Feynman diagrams relevant to those one-point functions.
This leads to an off-shell recursion relation between the one-point functions where higher orders in perturbation theory are given in terms of lower orders.
This method is analogous to Berends-Giele off-shell recursion~\cite{Berends:1987me}.
In this section we discuss how all Feynman diagrams contributing to the one-point functions are systematically generated from this off-shell recursion.
In addition, we discuss the explicit form of the classical equations of motion in momentum space in terms of the Feynman vertices.

The one-point functions relevant to the example $S_{\rmm{spin},i}$ Eq.~\eqref{WQFT:SpinAction} are $\ev{h_\mn(k)}$, $\ev{z^\sig(\oma)}$ and $\ev{\psi^\sig(\oma)}$.
Let us first, however, focus on the simpler example of the non-spinning Polyakov action.
The system under consideration, then, is the gauge fixed Einstein-Hilbert action and two copies of the Polyakov action:\checked
\begin{align}\label{WQFT:Polyakov}
  S =
  S_{\rm GR}
  -
  \sum_{i=1,2}
  \frac{m_i}2
  \int d\tau
  \,
  g_\mn
  \big(
  z_i(\tau)
  \big)
  \dot z_i^\mu(\tau)
  \dot z_i^\nu(\tau)
  \ .
\end{align}
This action was discussed around Eq.~\eqref{eq:SpinlessSimple} together with the corresponding classical equations of motion.
They are the geodesic equation for each of the bodies $\ev{z^\sig_i(\tau)}$ and the Einstein field equations for $\ev{g_\mn(x)}$.
While these equations often are considered in position space, we will instead work in momentum space with the (momentum space) Feynman rules.

Let us examine the expansion of the classical equations of motion in terms of Feynman vertices.
We start with the worldline fluctuation and denote by $S_{\rmm{wl},i}$ the Polyakov worldline action (the second term of Eq.~\ref{WQFT:Polyakov}).
The (worldline) interaction terms, $S_{\rmm{wl},i}^{\rm int}$, are easily expanded in terms of the corresponding worldline (WQFT) Feynman vertices:\checked
\begin{align}\label{WQFT:FAction}
  iS_{\rmm{wl},i}^{\rm int}
  =
  \sum_{n=0}^\infty
  \frac1{n!}
  V_{\rmm{wl},i}
  \Big[
    \smallint\! h_\mn(k)
    ,
    \smallint\! \Del z^{\sigo}_i(\omao)
    ,...,
    \smallint\! \Del z^\sign_i(\oman)
    \Big]
  \ .
\end{align}
Here, and in the following, we use a small integration sign inside of the vertex rules to indicate an integration on the corresponding field.
With this convention, the right-hand-side of Eq.~\eqref{WQFT:FAction} is written in terms of explicit integrations as follows:\checked
\begin{align}
  &
  V_{\rmm{wl},i}
  \Big[
    \smallint\! h_\mn(k)
    ,
    \smallint\! \Del z^{\sigo}_i(\omao)
    ,...,
    \smallint\! \Del z^\sign_i(\oman)
    \Big]
  =
  \int_{k,\omao,...,\oman}
  \nn
  \\
  &\qquad\times
    V_{\rmm{wl},i}
  \Big[
     h_\mn(k)
    ,
     \Del z^{\sigo}_i(\omao)
    ,...,
     \Del z^\sign_i(\oman)
     \Big]
     h_\mn(k)
     \Del z^{\sigo}_i(\omao)
     ...
     \Del z^\sign_i(\oman)
     \ .
     \label{WQFT:IntegrationNotation}
\end{align}
This compact notation can admittedly lead to some confusing.
On the right-hand-side of this equation the fields inside of the vertex rules indicate the different legs of the corresponding vertex rule.
The vertex is then contracted with the collection of fields to the right of it and we integrate on all momenta (and energies).
Thus, the fields inside of the square brackets on the right-hand-side are just place-holders indicating the relevant vertex rule.
In the following, we will also use this notation in a mixed version with an example,\checked
\begin{align}
  &
  V_{\rmm{wl},i}
  \Big[
    h_\mn(k)
    ,
    \smallint\! \Del z^{\sigo}_i(\omao)
    ,...,
    \smallint\! \Del z^\sign_i(\oman)
    \Big]
  =
  \int_{\omao,...,\oman}
  \nn
  \\
  &\qquad\times
    V_{\rmm{wl},i}
  \Big[
     h_\mn(k)
    ,
     \Del z^{\sigo}_i(\omao)
    ,...,
     \Del z^\sign_i(\oman)
     \Big]
     \Del z^{\sigo}_i(\omao)
     ...
     \Del z^\sign_i(\oman)
     \ ,
\end{align}
where the graviton leg on both sides of this equation is free.
The expansion of the interaction terms in Eq.~\eqref{WQFT:FAction} follows from the general definition Eq.~\eqref{WQFT:FRules} of the Feynman vertices and the expansion of the interaction terms of the gauge fixed Einstein-Hilbert action is analogous to Eq.~\eqref{WQFT:FAction} but written in terms of the bulk vertices instead.

Starting from the interaction terms Eq.~\eqref{WQFT:FAction} it is now straightforward to derive the geodesic equation of $\Del z_i^\sig(\oma)$ in terms of Feynman vertices.
We combine the interaction terms together with the kinetic terms and use the variational principle $\del S=0$.
The equation of motion becomes:\checked
\bse\label{WQFT:FEOM}
\begin{align}\label{WQFT:GE}
  im_i\omega^2
  \bgev{
    \Delta z_{i}^\sig(\omega)
  }
  &=
  \tilde \kappa
  \sum^\infty_{
    n=0
  }
  \frac1{(n\!-\!1)!}
  V_{\rmm{wl},i}
  \Big[
    \smallint\! \bgev{h_\mn(k)}
  ,
    \smallint\! \bgev{\Delta z_{i}^{\sig_1}(\omega_1)}
  ,...,
    \smallint\! \bgev{\Delta z_{i}^{\sig_n}(\omega_n)}
    ,
    \Del z^\sig_i(\oma)
  \Big]
  \ .
\end{align}
The left-hand-side comes from the kinetic term of the action and includes the inverse propagator $im\oma^2$.
The right-hand-side comes from the expansion of the interaction terms and imitates a relativistic force $f^\sig(\oma)$.
A simplifying property of the Polyakov action is the linear coupling of the worldline to the graviton.
In general, we would expect multipoint graviton interactions on the worldline.
Such interactions appear when we include spin.
We have inserted the power counting parameter $\tilde\kappa=1$ to track the $\kappa$ scaling of the vertex rules.
In addition, the fluctuation fields have their own dependence on $\kap$.
This additional $\kap$ scaling implies that terms with larger $n$ of the right-hand-side scale with larger powers of $\kap$.

The corresponding expansion and derivation of the Einstein fields equations follow equivalent steps.
Starting from the Polyakov action Eq.~\eqref{WQFT:Polyakov} we get:\checked
\begin{align}\label{WQFT:EFE}
  -i\mP^{\ab\mn} k^2
  \bgev{h_\ab(k)}
  &=
  \sum_{\substack{
      i\in\{1,2\}
      \\
      n=0..\infty
  }}
  \frac{  \tilde \kappa
}{n!}
  V_{\rmm{wl},i}
  \Big[
    h_\mn(k),
    \smallint\! \bgev{\Del z_i^\sigo(\omao)}
    ,...,
    \smallint\! \bgev{\Del z_i^\sign(\oman)}
  \Big]
  \\
  &\qquad
  +
  \sum_{n=0}^\infty
  \frac{
    \tilde \kappa^{n+1}
  }{
    (n\!+\!2)!
  }
  V_{\rm GR}
  \Big[
    h_\mn(k),
    \smallint\! \bgev{h_\mno(k_1)}
    ,...,
    \smallint\! \bgev{h_{\mu_{n+2}\nu_{n+2}}(k_{n+2})}
    \Big]
  \ ,
  \nn
\end{align}
\ese
This is the Einstein field equations for $\ev{h_\mn}$.
The first line corresponds to the energy-momentum tensor of the worldlines and the second line to the non-linear (pseudo) energy-momentum of the gravitational field itself ($\tau_{\rm GR}^\mn$ from Eq.~\ref{WEFT:EHInteraction}).
The two equations~\eqref{WQFT:FEOM} are the classical equations of motion for $\Del z_i^\sig(\oma)$ and $h_\mn(k)$ written in terms of Feynman vertices in momentum space.

It is advantageous to draw these equations diagrammatically.
The geodesic equation, Eq.~\eqref{WQFT:GE}, of the first particle becomes:
\bse\label{WQFT:DEOM}
\begin{align}\label{WQFT:DGE}
  \begin{tikzpicture}[baseline={(current bounding box.center)}]
    \coordinate (c) at (0,0) ;
    \coordinate (out) at (-1.5,0) ;
    \coordinate (w1) at (1.5,0) ;
    \coordinate (w2) at (1.4,.5) ;
    \coordinate (w3) at (1.4,-.5) ;
    \coordinate (g1) at (0,-1.5) ;
    \coordinate (g2) at (0,1.2) ;
    \draw[worldline] (out) -- (c) ;
    \node[anchor=center,circle,fill=white,draw=black] () at (c) {$z_1$};
    \node[anchor=center,circle,fill=white,draw=white,text=white] () at (g1) {$h$};
    \node[anchor=center,circle,fill=white,draw=white,text=white] () at (g2) {$h$};
  \end{tikzpicture}
  \ 
  =
  \ 
  \begin{tikzpicture}[baseline={(current bounding box.center)}]
    \coordinate (c) at (0,0) ;
    \coordinate (dd) at (.8,0) ;
    \coordinate (out) at (-1.5,0) ;
    \coordinate (w1) at (1.5,0) ;
    \coordinate (w2) at (1.4,.5) ;
    \coordinate (w3) at (1.4,-.5) ;
    \coordinate (g1) at (0,-1.5) ;
    \coordinate (g2) at (0,1.2) ;
    \node[anchor=center,circle,fill=white,draw=white,text=white] () at (g1) {$h$};
    \node[anchor=center,circle,fill=white,draw=white,text=white] () at (g2) {$h$};
    \draw[worldline] (c) -- (out) ;
    \draw[snake it] (g1) -- (c) ;
    \draw[worlddot2] (c) -- (dd) ;
    \node[anchor=center,circle,fill=white,draw=black] () at (g1) {$h$};
  \end{tikzpicture}
  \ 
  +
  \ 
  \begin{tikzpicture}[baseline={(current bounding box.center)}]
    \coordinate (c) at (0,0) ;
    \coordinate (out) at (-1.5,0) ;
    \coordinate (w1) at (1.5,0) ;
    \coordinate (w2) at (1.4,.5) ;
    \coordinate (w3) at (1.4,-.5) ;
    \coordinate (g1) at (0,-1.5) ;
    \coordinate (g2) at (0,1.2) ;
    \node[anchor=center,circle,fill=white,draw=white,text=white] () at (g1) {$h$};
    \node[anchor=center,circle,fill=white,draw=white,text=white] () at (g2) {$h$};
    \draw[worldline] (c) -- (out) ;
    \draw[worldline] (c) -- (w1) ;
    \draw[snake it] (g1) -- (c) ;
    \node[anchor=center,circle,fill=white,draw=black] () at (g1) {$h$};
    \node[anchor=center,circle,fill=white,draw=black] () at (w1) {$z_1$};
  \end{tikzpicture}
  \ 
  +
  \frac12
  \ 
  \begin{tikzpicture}[baseline={(current bounding box.center)}]
    \coordinate (c) at (0,0) ;
    \coordinate (out) at (-1.5,0) ;
    \coordinate (w1) at (1.5,0) ;
    \coordinate (w2) at (1.4,.5) ;
    \coordinate (w3) at (1.4,-.5) ;
    \coordinate (g1) at (0,-1.5) ;
    \coordinate (g2) at (0,1.2) ;
    \node[anchor=center,circle,fill=white,draw=white,text=white] () at (g1) {$h$};
    \node[anchor=center,circle,fill=white,draw=white,text=white] () at (g2) {$h$};
    \draw[worldline] (c) -- (out) ;
    \draw[worldline] (c) -- (w2) ;
    \draw[worldline] (c) -- (w3) ;
    \draw[snake it] (g1) -- (c) ;
    \node[anchor=center,circle,fill=white,draw=black] () at (g1) {$h$};
    \node[anchor=center,circle,fill=white,draw=black] () at (w2) {$z_1$};
    \node[anchor=center,circle,fill=white,draw=black] () at (w3) {$z_1$};
  \end{tikzpicture}
  \
  +
  \ 
  .\,.\,.
\end{align}
The blobs labelled by $z_1$ or $h$ represent the one-point functions $\ev{\Del z_1^\sig(\oma)}$ and $\ev{h_\mn(k)}$ respectively.
Only the first three terms of the right-hand-side of Eq.~\eqref{WQFT:GE} are shown.
The ellipses indicate the remaining terms with any number $n>2$ of worldline fluctuations.
The equation of motion for the second particle is obtained by symmetry.

The Einstein field equations, Eq.~\eqref{WQFT:EFE}, have the diagrammatic form:
\begin{align}\label{WQFT:DEFE}
  \begin{tikzpicture}[baseline={(current bounding box.center)}]
    \coordinate (c) at (0,0) ;
    \coordinate (w0) at (1.2,1.2) ;
    \coordinate (d1) at (.6,1.2) ;
    \coordinate (d2) at (1.8,1.2) ;
    \coordinate (w1) at (2.7,1.2) ;
    \coordinate (w2) at (2.6,1.7) ;
    \coordinate (w3) at (2.6,.7) ;
    \coordinate (w4) at (0,-1.7) ;
    \coordinate (w5) at (1.5,1.7) ;
    \coordinate (v) at (1.5,0) ;
    \draw[snake it] (c) -- (v) ;
    \node[anchor=center,circle,fill=white,draw=black] () at (v) {$h$};
    \node[anchor=center,circle,fill=white,draw=white,text=white] () at (w4) {$z_1$};
    \node[anchor=center,circle,fill=white,draw=white,text=white] () at (w5) {$z_1$};
  \end{tikzpicture}
  \ 
  =&
  \
  \begin{tikzpicture}[baseline={(current bounding box.center)}]
    \coordinate (c) at (0,0) ;
    \coordinate (w0) at (1.2,1.2) ;
    \coordinate (d1) at (.6,1.2) ;
    \coordinate (d2) at (1.8,1.2) ;
    \coordinate (w1) at (2.7,1.2) ;
    \coordinate (w2) at (2.6,1.7) ;
    \coordinate (w3) at (2.6,.7) ;
    \coordinate (w4) at (0,-1.7) ;
    \coordinate (w5) at (1.5,1.7) ;
    \coordinate (v) at (1.5,0) ;
    \draw[snake it] (c) -- (w0) ;
    \draw[worlddot2] (w0) -- (d1) ;
    \draw[worlddot2] (w0) -- (d2) ;
    \node[anchor=center,circle,fill=white,draw=white,text=white] () at (w4) {$z_1$};
    \node[anchor=center,circle,fill=white,draw=white,text=white] () at (w5) {$z_1$};
  \end{tikzpicture}
  \ 
  +
  \ 
  \begin{tikzpicture}[baseline={(current bounding box.center)}]
    \coordinate (c) at (0,0) ;
    \coordinate (w0) at (1.2,1.2) ;
    \coordinate (d1) at (.6,1.2) ;
    \coordinate (d2) at (1.8,1.2) ;
    \coordinate (w1) at (2.7,1.2) ;
    \coordinate (w2) at (2.6,1.7) ;
    \coordinate (w3) at (2.6,.7) ;
    \coordinate (w4) at (0,-1.7) ;
    \coordinate (w5) at (1.5,1.7) ;
    \coordinate (v) at (1.5,0) ;
    \draw[snake it] (c) -- (w0) ;
    \draw[worlddot2] (w0) -- (d1) ;
    \draw[worldline] (w0) -- (w1) ;
    \node[anchor=center,circle,fill=white,draw=black] () at (w1) {$z_1$};
    \node[anchor=center,circle,fill=white,draw=white,text=white] () at (w4) {$z_1$};
    \node[anchor=center,circle,fill=white,draw=white,text=white] () at (w5) {$z_1$};
  \end{tikzpicture}
  \ 
  +\ 
  \frac12\!\!
  \begin{tikzpicture}[baseline={(current bounding box.center)}]
    \coordinate (c) at (0,0) ;
    \coordinate (w0) at (1.2,1.2) ;
    \coordinate (d1) at (.6,1.2) ;
    \coordinate (d2) at (1.8,1.2) ;
    \coordinate (w1) at (2.7,1.2) ;
    \coordinate (w2) at (2.6,1.7) ;
    \coordinate (w3) at (2.6,.7) ;
    \coordinate (w4) at (0,-1.7) ;
    \coordinate (w5) at (1.5,1.7) ;
    \coordinate (v) at (1.5,0) ;
    \draw[snake it] (c) -- (w0) ;
    \draw[worlddot2] (w0) -- (d1) ;
    \draw[worldline] (w0) -- (w2) ;
    \draw[worldline] (w0) -- (w3) ;
    \node[anchor=center,circle,fill=white,draw=black] () at (w2) {$z_1$};
    \node[anchor=center,circle,fill=white,draw=black] () at (w3) {$z_1$};
    \node[anchor=center,circle,fill=white,draw=white,text=white] () at (w4) {$z_1$};
  \end{tikzpicture}
    \ 
    +
    \ 
    .\,.\,.
    \nn
    \\[-60pt]
    &+
  \
  \begin{tikzpicture}[baseline={(current bounding box.center)}]
    \coordinate (c) at (0,0) ;
    \coordinate (w0) at (1.2,-1.2) ;
    \coordinate (d1) at (.6,-1.2) ;
    \coordinate (d2) at (1.8,-1.2) ;
    \coordinate (w1) at (2.7,-1.2) ;
    \coordinate (w2) at (2.6,-1.7) ;
    \coordinate (w3) at (2.6,-.7) ;
    \coordinate (w4) at (0,+1.7) ;
    \coordinate (w5) at (1.5,-1.7) ;
    \coordinate (v) at (1.5,0) ;
    \draw[snake it] (c) -- (w0) ;
    \draw[worlddot2] (w0) -- (d1) ;
    \draw[worlddot2] (w0) -- (d2) ;
    \node[anchor=center,circle,fill=white,draw=white,text=white] () at (w4) {$z_1$};
    \node[anchor=center,circle,fill=white,draw=white,text=white] () at (w5) {$z_1$};
  \end{tikzpicture}
  \ 
  +
  \ 
  \begin{tikzpicture}[baseline={(current bounding box.center)}]
    \coordinate (c) at (0,0) ;
    \coordinate (w0) at (1.2,-1.2) ;
    \coordinate (d1) at (.6,-1.2) ;
    \coordinate (d2) at (1.8,-1.2) ;
    \coordinate (w1) at (2.7,-1.2) ;
    \coordinate (w2) at (2.6,-1.7) ;
    \coordinate (w3) at (2.6,-.7) ;
    \coordinate (w4) at (0,+1.7) ;
    \coordinate (w5) at (1.5,-1.7) ;
    \coordinate (v) at (1.5,0) ;
    \draw[snake it] (c) -- (w0) ;
    \draw[worlddot2] (w0) -- (d1) ;
    \draw[worldline] (w0) -- (w1) ;
    \node[anchor=center,circle,fill=white,draw=black] () at (w1) {$z_2$};
    \node[anchor=center,circle,fill=white,draw=white,text=white] () at (w4) {$z_2$};
    \node[anchor=center,circle,fill=white,draw=white,text=white] () at (w5) {$z_2$};
  \end{tikzpicture}
  \ 
  +\ 
  \frac12\!\!
  \begin{tikzpicture}[baseline={(current bounding box.center)}]
    \coordinate (c) at (0,0) ;
    \coordinate (w0) at (1.2,-1.2) ;
    \coordinate (d1) at (.6,-1.2) ;
    \coordinate (d2) at (1.8,-1.2) ;
    \coordinate (w1) at (2.7,-1.2) ;
    \coordinate (w2) at (2.6,-1.7) ;
    \coordinate (w3) at (2.6,-.7) ;
    \coordinate (w4) at (0,+1.7) ;
    \coordinate (w5) at (1.5,-1.7) ;
    \coordinate (v) at (1.5,0) ;
    \draw[snake it] (c) -- (w0) ;
    \draw[worlddot2] (w0) -- (d1) ;
    \draw[worldline] (w0) -- (w2) ;
    \draw[worldline] (w0) -- (w3) ;
    \node[anchor=center,circle,fill=white,draw=black] () at (w2) {$z_2$};
    \node[anchor=center,circle,fill=white,draw=black] () at (w3) {$z_2$};
    \node[anchor=center,circle,fill=white,draw=white,text=white] () at (w4) {$z_2$};
  \end{tikzpicture}
    \ 
    +
    \ 
    .\,.\,.
  \nn
  \\[0pt]
  &+\frac12\
  \begin{tikzpicture}[baseline={(current bounding box.center)}]
    \coordinate (c) at (0,0) ;
    \coordinate (w0) at (1.2,1.2) ;
    \coordinate (d1) at (.6,1.2) ;
    \coordinate (d2) at (1.8,1.2) ;
    \coordinate (w1) at (2.7,1.2) ;
    \coordinate (w2) at (2.6,1.7) ;
    \coordinate (w3) at (2.6,.7) ;
    \coordinate (w4) at (0,-1.7) ;
    \coordinate (w5) at (1.5,1.7) ;
    \coordinate (v) at (1.5,0) ;
    \coordinate (g1) at (2.5,1) ;
    \coordinate (g2) at (2.5,-1) ;
    \coordinate (g3) at (3,0) ;
    \draw[snake it] (c) -- (v) ;
    \draw[snake it] (v) -- (g1) ;
    \draw[snake it] (v) -- (g2) ;
    \node[anchor=center,circle,fill=white,draw=black] () at (g1) {$h$};
    \node[anchor=center,circle,fill=white,draw=black] () at (g2) {$h$};
  \end{tikzpicture}
  \
  +\frac16
  \
    \begin{tikzpicture}[baseline={(current bounding box.center)}]
    \coordinate (c) at (0,0) ;
    \coordinate (w0) at (1.2,1.2) ;
    \coordinate (d1) at (.6,1.2) ;
    \coordinate (d2) at (1.8,1.2) ;
    \coordinate (w1) at (2.7,1.2) ;
    \coordinate (w2) at (2.6,1.7) ;
    \coordinate (w3) at (2.6,.7) ;
    \coordinate (w4) at (0,-1.7) ;
    \coordinate (w5) at (1.5,1.7) ;
    \coordinate (v) at (1.5,0) ;
    \coordinate (g1) at (2.5,1) ;
    \coordinate (g2) at (2.5,-1) ;
    \coordinate (g3) at (3,0) ;
    \draw[snake it] (c) -- (v) ;
    \draw[snake it] (v) -- (g1) ;
    \draw[snake it] (v) -- (g2) ;
    \draw[snake it] (v) -- (g3) ;
    \node[anchor=center,circle,fill=white,draw=black] () at (g1) {$h$};
    \node[anchor=center,circle,fill=white,draw=black] () at (g2) {$h$};
    \node[anchor=center,circle,fill=white,draw=black] () at (g3) {$h$};
    \end{tikzpicture}
    \
    +
    \ .\,.\,.
\end{align}
\ese
The blobs are the same as above and represent the one-point functions.
The first and second line of this equation represent the energy-momentum of the first and second worldline respectively.
The third line is due to gravitational self-interaction.

These equations are easily solved by induction which is done systematically by expanding each one-point function in powers of $\kap$.
We then insert these expansions into the equations of motion and collect equal powers of $\kap$ on each side.
We expand the functions as follows:\checked
\begin{align}\label{TL:FieldExpansions}
  \bgev{
    \Delta z_{i}^\sig(\oma)
    }
  &=
  \sum_{n=1}^\infty
  \bgevs{2n}{
    \Delta z_{i}^{\sigma}(\oma)
  }
  \,
  \tilde \kappa^{2n}
  \ ,
  \\
  \bgev{h_\mn(k)}
  &=
  \sum_{n=1}^\infty
  \bgevs{2n-1}{h_\mn(k)}
  \,
  \tilde \kappa^{2n-1}
  \ .
\end{align}
Here, we use the same power parameter $\tilde \kap=1$ to indicate the scaling of the corresponding expansion terms in $\kap$.
We insert the expansions into the equations of motion Eqs.~\eqref{WQFT:FEOM} and collect equal powers of $\tilde \kap$.
To subleading order for each field we get:\checked
\bse
\label{WQFT:RecursiveDiagrams}
\begin{align}
  -i\mP^{\mn\ab} k^2
  \bgev{h_\ab(k)}_{(1)}
  &=
  \sum_{i\in\{1,2\}}
  V_{\rmm{wl},i}
  \big[
    h_\mn(k)
    \big]
  \ ,
  \\
  im_i\oma^2
  \bgevs{2}{
    \Del z_i^\sig(\oma)
  }
  &=
  V_{\rmm{wl},i}
  \Big[
    \smallint\!\bgevs{1}{h_\mn(k)}
    ,
    \Del z_i^\sig(\oma)
    \Big]
  \ ,
  \\
  -i\mP^{\mn\ab} k^2
  \bgevs{3}{
    h_\ab(k)
  }
  &=
  \frac12
  V_{\rm GR}
  \Big[
    h_\mn(k),
    \smallint\!
    \bgevs{1}{h_\mno(k_1)},
    \smallint\!
    \bgevs{1}{h_\mnt(k_2)}
  \Big]
  \\
  &\qquad+
  \sum_{i\in\{1,2\}}
  V_{\rmm{wl},i}
  \Big[
    h_\mn(k),
    \smallint\!
    \bgevs{2}{
      \Del z_i^\sig(\oma)
      }
    \Big]
  \ ,
  \nn
  \\
  im_i\oma^2
  \bgevs{4}{
    \Del z_i^\sig(\oma)
  }
  &=
  V_{\rmm{wl},i}
  \Big[
    \smallint\!\bgevs{3}{h_\mn(k)}
    ,
    \Del z_i^\sig(\oma)
    \Big]
  \\
  &\qquad+
  V_{\rmm{wl},i}
  \Big[
    \smallint\!\bgevs{1}{h_\mn(k)}
    ,
    \smallint\!
    \bgevs{2}{\Del z_i^\sigo(\omao)}
    ,
    \Del z_i^\sig(\oma)
    \Big]
  \ .
  \nn
\end{align}
\ese
One may draw these relations diagrammatically in similar fashion to Eqs.~\eqref{WQFT:DEOM}.
The Feynman diagrams generated from these recursive relations are all the relevant ones for the computation of the one-point functions.

The solution of the recursive relations requires the inversion of the kinetic operators acting on the one-point functions (e.g. on the left-hand-side of Eqs.~\ref{WQFT:RecursiveDiagrams}).
This inversion gives rise to the respective propagators of the fields.
It is only at this point that the difference between the in-out and in-in formalisms enter.
That is, for the in-out theory we use the time-symmetric propagators Eqs.~\eqref{WQFT:IOP} while for the in-in theory we use retarded propagators Eqs.~\eqref{WQFT:IIP}.

\begin{figure}
  \renewcommand*\thesubfigure{\arabic{subfigure}} 
  \centerline{
    \begin{adjustbox}{minipage=\linewidth,scale=1.1}
      \centering
  \begin{subfigure}{.11\textwidth}
    \centering
    \includegraphics{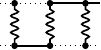}
    \caption{}
    \label{TL:fig:LoopDiagrams0}
  \end{subfigure}
  \begin{subfigure}{.11\textwidth}
    \centering
    \includegraphics{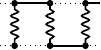}
    \caption{}
  \end{subfigure}
  \begin{subfigure}{.11\textwidth}
    \centering
    \includegraphics{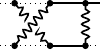}
    \caption{}
  \end{subfigure}
  \begin{subfigure}{.11\textwidth}
    \centering
    \includegraphics{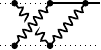}
    \caption{}
  \end{subfigure}
  \begin{subfigure}{.11\textwidth}
    \centering
    \includegraphics{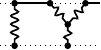}
    \caption{}
  \end{subfigure}
  \begin{subfigure}{.11\textwidth}
    \centering
    \includegraphics{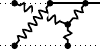}
    \caption{}
  \end{subfigure}
  \begin{subfigure}{.11\textwidth}
    \centering
    \includegraphics{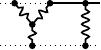}
    \caption{}
  \end{subfigure}
  \begin{subfigure}{.11\textwidth}
    \centering
    \includegraphics{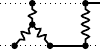}
    \caption{}
  \end{subfigure}
  \begin{subfigure}{.11\textwidth}
    \centering
    \includegraphics{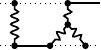}
    \caption{}
  \end{subfigure}
  \begin{subfigure}{.11\textwidth}
    \centering
    \includegraphics{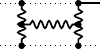}
    \caption{}
  \end{subfigure}
  \begin{subfigure}{.11\textwidth}
    \centering
    \includegraphics{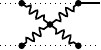}
    \caption{}
  \end{subfigure}
  \begin{subfigure}{.11\textwidth}
    \centering
    \includegraphics{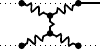}
    \caption{}
    \label{TL:fig:LoopDiagrams21}
  \end{subfigure}
  \begin{subfigure}{.11\textwidth}
    \centering
    \includegraphics{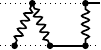}
    \caption{}
  \end{subfigure}
  \begin{subfigure}{.11\textwidth}
    \centering
    \includegraphics{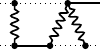}
    \caption{}
  \end{subfigure}
  \begin{subfigure}{.11\textwidth}
    \centering
    \includegraphics{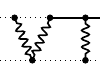}
    \caption{}
    \label{TL:fig:LoopDiagrams3}
  \end{subfigure}
  \begin{subfigure}{.11\textwidth}
    \centering
    \includegraphics{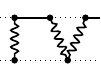}
    \caption{}
  \end{subfigure}
  \begin{subfigure}{.11\textwidth}
    \centering
    \includegraphics{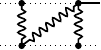}
    \caption{}
  \end{subfigure}
  \begin{subfigure}{.11\textwidth}
    \centering
    \includegraphics{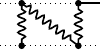}
    \caption{}
    \label{TL:fig:LoopDiagrams15}
  \end{subfigure}
  \begin{subfigure}{.11\textwidth}
    \centering
    \includegraphics{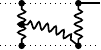}
    \caption{}
  \end{subfigure}
  \begin{subfigure}{.11\textwidth}
    \centering
    \includegraphics{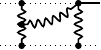}
    \caption{}
  \end{subfigure}
  \begin{subfigure}{.11\textwidth}
    \centering
    \includegraphics{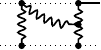}
    \caption{}
  \end{subfigure}
  \begin{subfigure}{.11\textwidth}
    \centering
    \includegraphics{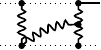}
    \caption{}
  \end{subfigure}
  \begin{subfigure}{.11\textwidth}
    \centering
    \includegraphics{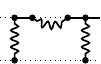}
    \caption{}
    \label{TL:fig:LoopDiagrams22}
  \end{subfigure}
  \begin{subfigure}{.11\textwidth}
    \centering
    \includegraphics{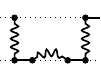}
    \caption{}
  \end{subfigure}
  \begin{subfigure}{.11\textwidth}
    \centering
    \includegraphics{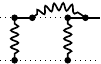}
    \caption{}
  \end{subfigure}
  \begin{subfigure}{.11\textwidth}
    \centering
    \includegraphics{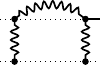}
    \caption{}
  \end{subfigure}
  \begin{subfigure}{.11\textwidth}
    \centering
    \includegraphics{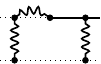}
    \caption{}
  \end{subfigure}
  \begin{subfigure}{.11\textwidth}
    \centering
    \includegraphics{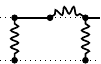}
    \caption{}
  \end{subfigure}
  \begin{subfigure}{.11\textwidth}
    \centering
    \includegraphics{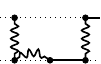}
    \caption{}
  \end{subfigure}
  \begin{subfigure}{.11\textwidth}
    \centering
    \includegraphics{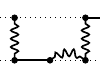}
    \caption{}
  \end{subfigure}
  \begin{subfigure}{.11\textwidth}
    \centering
    \includegraphics{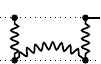}
    \caption{}
  \end{subfigure}
  \begin{subfigure}{.11\textwidth}
    \centering
    \includegraphics{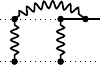}
    \caption{}
  \end{subfigure}
    \end{adjustbox}}
  \caption{
    Diagrams relevant to the comparable mass sector of the worldline observables at the third post-Minkowskian order.
    Every solid line represents either a Grassmann or worldline fluctuation and wiggly lines gravitons.
    For the spinless case only graphs (1)-(12) and (23)-(25) are relevant.
    The graphs (23)-(32) are often referred to as ``mushrooms'' and are zero in the conservative case. 
    This figure is reproduced with minor changes from Ref.~\cite{Jakobsen:2022zsx}.
  }
  \label{WQFT:ComparableGraphs}
\end{figure}

The inclusion of the full $\mO(S^2)$ spinning action to this approach is straightforward except, perhaps, for the anti-commutativity of the Grassmann variables.
Apart from that, the only difference to the recursive relations (the equations of motion) of the Polyakov action is the addition of another field and its corresponding equation of motion.
Thus, in addition to the graviton field $h_\mn(k)$ and the worldline fluctuation $\Del z^\sig(\oma)$ we have the Grassmann fluctuation $\Del \psi^\sig(\oma)$.
Correspondingly, we have their three equations of motion which we solve recursively by inserting perturbative expansions of the one-point functions (Eqs.~\ref{TL:FieldExpansions}).
The field $\Del \bar \psi^\sig(\oma)$ is the Hermitian conjugate of $\Del \psi^\sig(\oma)$ and its equation of motion and solution follows from this relation.
The main feature that the spin adds to the equations of motion is that the worldline part is no longer linear in the graviton field.
Thus, just as in the bulk, any number of gravitons interact on the worldlines.

Let us then finally discuss the anti-commutativity of the Grassmann variables.
The recursive solution of the equations of motion offers a simple approach to keeping track of the signs due to these variables.
The vertex rules in the presence of Grassmann legs anti commute (Eq.~\ref{WQFT:Anticommute}):\checked
\begin{align}
  V_{\rm spin}[...,\Del \psi^\sigo(\omao),...,\Del \psi^\sigt(\omat),...]
  =
  -
  V_{\rm spin}[...,\Del \psi^\sigt(\omat)
    ,...,
    \Del \psi^\sigo(\omao)
    ,...]
  \ .
\end{align}
We will not try to capture this property of the Grassmann vertices with diagrams.
However, the notation $V[...]$ easily captures this anti-commutativity.
When the vertices are integrated into a Grassmann field they no longer pick up a sign as long as the order of fields is respected.
Thus, in that case, the order of the fields in Eq.~\eqref{WQFT:IntegrationNotation} is significant.
We have:\checked
\begin{align}
  V_{\rm spin}[...,\smallint\Del \psi^\sigo(\omao),...,\smallint\Del \psi^\sigt(\omat),...]
  =
  V_{\rm spin}[...,\smallint\Del \psi^\sigt(\omat)
    ,...,
    \smallint\Del \psi^\sigo(\omao)
    ,...]
  \ .
\end{align}
In this case the order of the Grassmann fields does not matter.
With these two rules the signs due to the Grassmann variables in the interaction terms are taken care of.

As a final application of the graph generation discussed in this section, we consider the graphs relevant to the worldline observables at the third post-Minkowskian order which are the topic of \chaps~~\ref{sec:TL} and~\ref{sec:Scattering}.
As we will see below in Sec.~\ref{WQFT:PostMinkowskian} the worldline observables are derived from the on-shell worldline one-point functions.
We thus require all graphs contributing to the worldline one-point functions at the 3PM order.
For that purpose the recursive relations must be solved to one order higher than Eqs.~\eqref{WQFT:RecursiveDiagrams} and spin must be included.
With this approach all relevant graphs are generated and the result is shown in Figs.~\ref{WQFT:ProbeGraphs} and~\ref{WQFT:ComparableGraphs}.
They are drawn with a condensed notation where every solid line could be either a worldline or Grassmann fluctuation.
The full set of graphs is generated by the replacement rule:\checked
\begin{align}
    	\begin{tikzpicture}[baseline={(current bounding box.center)}]
	\coordinate (in) at (-1,0);
	\coordinate (out) at (1,0);
	\draw[worldline] (in) -- (out);
        \node[above] at ($(in)!0.5!(out)$) {$\vphantom{\Del z(\omega)}$} ;
        \node[below] at ($(in)!0.5!(out)$) {$\vphantom{\Del z(\omega)\big)}$} ;
	\end{tikzpicture}
        \to
    	\begin{tikzpicture}[baseline={(current bounding box.center)}]
	\coordinate (in) at (-1,0);
	\coordinate (out) at (1,0);
	\draw[worldline] (in) -- (out);
        \node[above] at ($(in)!0.5!(out)$) {$\Del z(\omega)$} ;
        \node[below] at ($(in)!0.5!(out)$) {$\vphantom{\Del z(\omega)\big)}$} ;
	\end{tikzpicture}
        +
            	\begin{tikzpicture}[baseline={(current bounding box.center)}]
	\coordinate (in) at (-1,0);
	\coordinate (out) at (1,0);
	\draw[worldline] (in) -- (out);
        \node[above] at ($(in)!0.5!(out)$) {$\Del \psi(\omega)$} ;
        \node[below] at ($(in)!0.5!(out)$) {$\vphantom{\Del \psi(\omega)\big)}$} ;
	        \end{tikzpicture}
                \ .
\end{align}
Also, the graphs are assumed to be on-shell so that the external energy is zero, which means some graphs were thrown away that would be non-zero off-shell.

\begin{figure}
    \renewcommand*\thesubfigure{\arabic{subfigure}} 
  \centering
   \begin{subfigure}{\figLa\textwidth}
    \centering
      \begin{tikzpicture}
        \coordinate (ul) at (-1.1,0) ;
        \coordinate (ulp)at (-.6 ,0) ;
        \coordinate (urp)at (.6  ,0) ;
        \coordinate (ur) at (+1.1,0) ;
        \coordinate(ull) at (-1.5,0) ;
        \coordinate(urr) at ( 1.5,0) ;
        \coordinate (dl) at (-1.1,-1) ;
        \coordinate (dm) at (0   ,-1) ;
        \coordinate (dr) at (+1.1,-1) ;
        \coordinate(dll) at (-1.5,-1) ;
        \coordinate(drr) at (1.5 ,-1) ;
        \draw[line width=1pt] (ul) -- (ulp) ;
        \draw[dotted,line width=.8pt] (ur) -- (urp) ;
         \draw[dotted,line width=.8pt] (dl) -- (dm) ;
         \draw[dotted,line width=.8pt] (dm) -- (dr) ;
        \draw[dotted,line width=.8pt] (ulp) -- (urp) ;
        \draw[snake it] (.6,0) arc (0:-180:.6cm) ;
      \end{tikzpicture}
      \caption{
      }
   \end{subfigure}
   \caption{Scaleless self-interaction of the first worldline.}
   \label{WQFT:SI}
\end{figure}
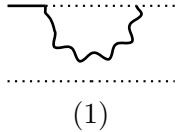

Finally, we note that certain scaleless self-interaction graphs can immediately be set to zero with an example of such a graph given in Fig.~\ref{WQFT:SI}.
In addition to being scaleless that graph is also non-dynamical in the sense that the second worldline is not involved at all and the graph simply corresponds to the first body being acted on by its own Coulomb like potential.
A quick argument for its vanishing observes that the outgoing energy is forced to being zero so that the diagram must be on-shell and must depend only on gauge invariant variables.
The non-dynamical gauge invariant variables are only the masses which do not show up in the integration, so that the integration must be scaleless and vanishes in dimensional regularization.
Such diagrams also show up as subdiagrams in otherwise dynamical graphs and in that case, too, must the entire graph vanish.

\section{Post-Minkowskian Observables from WQFT}
\label{WQFT:PostMinkowskian}
A primary goal of the WQFT is to compute scattering observables.
These are simpler than the full trajectories in part because they do not depend on the gravitational gauge freedom and, instead, are defined on the flat Minkowski background at infinity.
As we will see, observables are directly obtained from the one-point correlation functions by putting them on-shell.
Thus, a primary goal of the WQFT is to compute on-shell one-point functions.
This is in spirit with the scattering amplitudes program of QFT, namely that on-shell amplitudes encode all information about the scattering process in a gauge independent manner.

In the first section~\ref{WQFT:Observables} we derive the relations between on-shell one-point functions and observables.
In addition we introduce the free energy, or eikonal, of the WQFT which is an alternative approach to deriving observables and can be related to the QFT four-point amplitude.
In the second section~\ref{WQFT:Worldline1} we focus on the worldline observables and analyze their expansion in terms of loop integrals.
In the final section~\ref{WQFT:Worldline2} we discuss the computation of worldline observables and the eikonal to the second post-Minkowskian order and quadratic order in spins.

\subsection{Observables from On-Shell Functions and WQFT Eikonal}
\label{WQFT:Observables}
A main objective of the WQFT approach to classical gravitational scattering is the direct computation of gauge invariant observables.
In this section we will express three of the main observables in terms of the one-point correlation functions.
In addition we consider the WQFT eikonal which is equivalent to the on-shell action.
The causal dynamics and, consequently, the causal observables generated by the in-in formalism are of most interest.
However, the dynamics and observables generated by the in-out formalism has certain properties that make them appealing too.
In particular, the eikonal of the in-out theory can be directly related to the four-point scattering amplitude of QFT.

Observables of generic scattering events were considered in Sec.~\ref{WEFT:GS}.
Worldline observables may be defined as the difference between the future and past asymptotic values of the worldline variables as in Eq.~\eqref{WEFT:Kick}.
An independent set of asymptotic worldline (background) variables may be chosen as the momenta $p_i^\sig$, the spin vectors $S_i^\sig$, and the total CoM angular momentum $J^\sig$.
Each of these variables define an observable by considering their total change during scattering.
These are the impulse of each body $\Del p^\sig_i$, the spin kick of each body $\Del S^\sig_i$ and the change in total angular momentum $\Del J^\sig$.
Below we will derive expressions for the impulse and spin kick in terms of the worldline one-point functions.
In contrast, the kick in total angular momentum is not related to the one-point functions in the same direct manner.

Since the (off-shell) worldline one-point functions simply are equivalent to the classical trajectories, it is a problem in classical physics to relate them to observables.
Instead of working with the spin kick we focus on the Grassmann kick from which the spin kick may be derived (Eq.~\ref{WEFT:SpinKicks}).
We start with the impulse and find:\checked
\begin{align}
  \Delta p^\mu_i
  &=
  m_i
  \int_{-\infty}^{\infty}
  d\tau\,
  \bgev{\ddot z_{i}^\sigma(\tau)}
  \nn
  \\
  &=
  -m_i
  \int d\tau
  \int_\omega
  e^{-i\omega\tau}
  \omega^2
  \bgev{z^\sig_i(\omega)}
  \nn
  \\
  &=
  -m_i\,
  \omega^2
  \bgev{
    \Del z_i^\sig(\omega)
  }
  \Big|_{\omega\to0}
  \ .
  \label{WQFT:Impulse}
\end{align}
In the first line we write the impulse as the difference between $m_i \dot z_i(\tau)$ at future and past infinity.
In the second line we transform to momentum space and in the final line the two integrations effectively send $\oma$ to zero.
There we have also replaced the full worldline by its perturbation field as the two time derivatives make the distinction irrelevant.
The same kind of steps work out for the Grassmann kick:\checked
\begin{align}
  \Del \psi^\sig
  &=
  \int_{-\infty}^{\infty}
  d\tau\,
  \bgev{\dot \psi_{i}^\sig(\tau)}
  \nn
  \\
  &=
  -i
  \int d\tau
  \int_\omega
  e^{-i\omega\tau}
  \oma
  \bgev{\psi^\sig_i(\omega)}
  \nn
  \\
  &=
  -i\,
  \omega
  \bgev{
    \Del \psi_i^\sig(\omega)
  }
  \Big|_{\omega\to0}
  \ .
  \label{WQFT:SPINKICK}
\end{align}
Thus in both cases we amputate the external propagator and send the energy to zero.
We may also try to derive an expression for the change in total angular momentum by a similar fashion using its expression in terms of worldline fields Eq.~\eqref{WEFT:TAM}.
Alternatively the change in total angular momentum can be computed from the waveform which is analyzed in Ref.~\cite{Manohar:2022dea}.

The gravitational waves radiated to infinity are described by the waveform Eq.~\eqref{WEFT:Waveform}.
The waveform at infinity in frequency space, $f(k)$, is given by the on-shell momentum space energy-momentum tensor $\tau^\mn(k)$ contracted with polarizations.
This is simply the amputated on-shell graviton field~\cite{Jakobsen:2021smu,Jakobsen:2021lvp}:\checked
\begin{align}\label{eq:waveform}
  f(k)
  &=
  \frac{\kappa
    }{8\pi}
  \eps^\mu\eps^\nu
  k^2
  \bgev{h_\mn(k)}
  \Big|_{
    k^2
    \to
    0
  }
  \ .
\end{align}
The polarizations $\eps^\mu$ satisfy $\eps\cdot k=\eps^2=0$.
The waveform and its time domain expression will be discussed in detail in Sec.~\ref{WF:GK}.

With the three formulas Eqs.~\eqref{WQFT:Impulse}-\eqref{eq:waveform} we have a compact relation between classical scattering observables and on-shell WQFT one-point functions.
The observables are the impulse, the spin kick and the waveform corresponding to  the trajectory fluctuation, the Grassmann fluctuation and the graviton field.
The waveform at $\mO(G^2,S^2)$ and the impulse and spin kick at $\mO(G^3,S^2)$ are the subject of \chaps~\ref{GB} and~\ref{sec:Scattering} respectively.
The on-shell value of connected higher point functions, too, have physical interpretations as observables.
Thus the on-shell value of the connected two-point graviton correlation function describes the propagation of gravitational waves from asymptotic past to future.
This two-point function was computed from worldline effective field theory in Ref.~\cite{Saketh:2022wap} and verified to match WQFT to $\mO(S^2)$.
We note that a simple formula for the kick in angular momentum $\Del J^\mu$ or impact parameter $\Del b^\mu$ would be desirable.
For now, however, the kick in $\Del J^\mu$ may be derived from the waveform using the formalism of Ref.~\cite{Manohar:2022dea}.

In Eqs.~\eqref{WQFT:Impulse}-\eqref{eq:waveform} we did not specify whether the in-in or in-out formalisms were used for computing the observables.
In the rest of this thesis, however, we will mostly be interested in the in-in expectation values.
In that case causal boundary conditions are imposed and the asymptotic variables are defined at past infinity.
Below we will, however, also consider observables derived from the in-out formalism and in that case we denote that explicitly with a subscript.
Those observables do not have causal boundary conditions and asymptotic variables are not defined at past infinity.

Finally, we consider the WQFT free energy, $\chi$.
We will refer to $\chi$ also as the WQFT eikonal and, in fact, it is also equivalent to the on-shell action.
It is simply defined as the logarithm of the partition function (as in Eq.~\ref{WQFT:Connected}) without external sources.
We define:\checked
\bse
\begin{align}
  i
  \chi_{\rm in-out}
  &=
  \hbar
  \log\!\bigg[
  \int
  D[h_\mn,\Del z^\sig_i,\Del \psi^\sig_i]
  e^{i S/\hbar}
  \bigg]
  \bigg|_{\hbar\to0}
  \ ,
  \\
    i
    \chi_{\rm in-in}
    &=
    \hbar\!
    \log\bigg[
      \int
  D[h_{(\pm)\mn},\Del z^\sig_{(\pm)i},\Del \psi^\sig_{(\pm)i}]
  e^{i S_{\rm in-in}/\hbar}
  \bigg]
  \bigg|_{\hbar\to0}
  \ .
\end{align}
\ese
The classical limit simply implies that the WQFT eikonal is the on-shell action:\checked
\bse
\begin{align}
  \chi_{\rm in-out}
  &=
  \inout{S}
  \ ,
  \\
  \chi_{\rm in-in}
  &=
  \inin{S_{\rm in-in}}
  \ .
\end{align}
\ese
These expectation values correspond exactly to the action evaluated on the classical solutions.
In the second case $\chi_{\rm in-in}$ vanishes unless we let the $(-)$ Schwinger-Keldysh background parameters be non-zero.
The eikonal may be evaluated in terms of diagrams by inserting the one-point functions into the action or by considering it, simply, as the sum of connected vacuum diagrams.

The in-out worldline observables can be derived in a simple manner from the in-out eikonal:\checked
\bse
\label{WQFT:ObservablesFromE}
\begin{align}
  \label{WQFT:ImpulseFromE}
  \Del p_{\rmm{(in-out)}i}^\sig
  =
  -
  \frac{
    \pat \chi_\rmm{in-out}}{
    \pat b_{i,\sig}}
  \ ,
  \\
  m_i
  \Del
  \bar \psi_{\rmm{(in-out)}i}^{\sig}
  =
  -i \frac{
    \pat \chi_{\rm in-out}
  }{
    \pat \Psi_{i,\sig}
  }
  \ .
  \label{WQFT:SpinKickFromE}
\end{align}
\ese
Thus, by simple differentiation of the in-out eikonal with respect to the background parameters of the worldlines the corresponding observables are derived.
We put a label (in-out) on the in-out observables to distinguish them from the in-in observables defined with causal boundary conditions that we are usually interested in.

The relations of the observables to the eikonal Eqs.~\eqref{WQFT:ObservablesFromE} are derived using the Euler-Lagrange equations.
Introducing the worldline Lagrangian as $S_{\rmm{spin},i}=\smallint\di\tau L_i$ they read:\checked
\begin{align}
  \frac{\di
  }{
    \di t
  }
  \frac{
    \pat L_i
  }{
    \pat
    \dot z^\sig_i(\tau)
  }
  &=
  \frac{
    \pat L_i
  }{
    \pat
    z^\sig_i(\tau)
  }
  \\
  &=
  \frac{
    \pat L_i
  }{
    \pat
    b^\sig_i
  }
  \nn
\end{align}
In the second line we used the fact that $L_i$ depends on $b_i^\sig$ only through its dependence on $z_i^\sig(\tau)$.
Using the Euler-Lagrange equations and the fact that the eikonal is the on-shell action we derive Eq.~\eqref{WQFT:ImpulseFromE}:\checked
\begin{align}
  \biginout{
    \frac{
      \pat S
    }{
      \pat b_{i,\sig}
    }
  }
  &=
  -\int \di\tau
  \biginout{
    \frac{\pat L_i}{
      \pat b_{i,\sig}}
  }
  \nn
  \\
  &=
  -\int \di \tau
  \frac{\di }{\di \tau}
  \inout{
    \pi_{i}^{\sig}(\tau)
  }
  \nn
  \\
  &=
  -\Del p_{\rmm{(in-out)}i}^{\sig}
\end{align}
In the second step we used the Euler-Lagrange equation with $\pi_i^\sig$ given by:\checked
\begin{align}
  \pi^\sig_i(\tau)
  =
  \frac{
    \partial L_i}{
    \partial \dot z_{i,\sig}(\tau)
  }
  \ .
\end{align}
At past and future infinity where the gravitational field can be neglected this momentum $\pi^\mu_i$ coincides with $p_i^\sig$.
By considering the Euler-Lagrange equations of the Grassmann field we arrive in analogous steps to the relation of the Grassmann kick from the eikonal Eq.~\eqref{WQFT:SpinKickFromE}.
The general pattern is that a derivative with respect to the background parameter of a field results in the kick of the corresponding momentum of that field.
The trajectory fluctuation has two parameters $b^\sig_i$ and $v_i^\sig$ and one may also consider derivatives with respect to the velocity.

The eikonal relations are closely related to the recursive properties of the Feynman vertices Eqs.~\eqref{WQFT:Recursive}.
As a special case they tell us that a partial derivative with respect to $b_i^\sig$ on a worldline vertex pulls out a worldline fluctuation leg with its energy put to zero.
Thus, when we hit a vacuum diagram with a $b_i^\sig$ derivative we pull out an external worldline leg with its energy put to zero at every vertex.
In this way the in-out on-shell one-point function is derived from the vacuum diagrams.
The recursive relation has the same interpretation for the spin kick.
One may use several differentiations with respect to the background parameters in order to pull out additional (on-shell) external legs.

The eikonal depends on the Grassmann variables only through its dependence on the spin tensor.
Thus, a differentiation with respect to the Grassmann variable may be computed using the chain rule.
In this way one can write a formula for the spin kick (of the spin tensor) in terms of differentiation with respect to the spin tensor.
We find \cite{Jakobsen:2021zvh}:\checked
\begin{align}
  m_i \Del S^\mn_{\rmm{(in-out)}i}
  =
  4
  S^{\ \ [\mu}_{i,\alpha}
    \,
  \frac{
    \del
    \chi_{\rm in-out}
  }{
    \del
    S_{i,\nu]\alpha}
  }
  \ .
\end{align}
This formula could, perhaps, also be derived directly using the Euler-Lagrange equations.

In general the eikonal depends only on the background parameters that are gauge invariant with respect to the background symmetries as discussed around Eq.~\eqref{WEFT:WLKin}.
A set of independent gauge invariant background parameters is given by the momenta $p^\sig_i$, the orthogonal spin tensors $S_{i,\bot}^\mn$ (Eq.~\ref{WEFT:SpinO}) and the SUSY impact parameter $\beta^\sig$ (Eq.~\ref{WEFT:Beta}).
The invariance of the eikonal under the background symmetries imply several constraints on the in-out observables: Conservation of total momentum, $P^\mu$, conservation of mass $p_i^2=m_i^2$ and spin length $S_\mn S^\mn$ and conservation of SSC $S^\mn p_\nu$.
We will compute the in-out eikonal to $\mO(G^2,S^2)$ in Sec.~\ref{WQFT:Worldline2} and show how in-in observables to this order may be derived from it.

The in-in eikonal does not have the same kind of simplicity as the in-out eikonal.
Only if we keep the $(-)$ background parameters non-zero do we get a non-vanishing in-in eikonal.
In that case it takes the form \cite{Jakobsen:2022psy}:\checked
\begin{align}
  \chi_{\rm in-in}
  =
  \sum_{i=1,2}
  \Big[
    -b_{(-)i} \cdot \Del p_i
  +
  im_i \bar \psi_{(-)i}
  \cdot \Del \psi_i
  +
  im_i
  \psi_{(-)i}
  \cdot
  \Del \bar \psi_{i}
  \Big]
  +
  .\,.\,.
\end{align}
Here, the ellipses indicates a term linear in $v_{(-)i}^\sig$ and possibly higher order terms in the minus fields.
This eikonal, then, is equivalent to computing $\Del p^\sig_i$ and $\Del \psi^\sig_i$ directly and the minus parameters function only as placeholders.
If at all, this eikonal can still be useful as a bookkeeping variable.

The in-out eikonal has an interesting connection to the QFT amplitudes approach to classical scattering.
Thus, in Ref.~\cite{Mogull:2020sak} it was proposed that the WQFT in-out eikonal is directly related to the classical limit of the QFT $2\to2$ scalar S-matrix.
The exact formula proposed there reads:\checked
\begin{align}
  \label{WQFT:Amplitude}
  e^{i\chi_{\rm in-out}}
  =
  \frac1{4m_1m_2}
  \int_q
  \dd(q\cdot v_1)
  \dd(q\cdot v_2)
  e^{iq\cdot b}
  \ev{\phi_1
    \phi_2
    |S|
    \phi_1
    \phi_2}
  \ .
\end{align}
The relation of the scalar momenta of $\ev{\phi_1 \phi_2 |S| \phi_1 \phi_2}$ and the WQFT background parameters is given in Ref.~\cite{Mogull:2020sak}.
The Eq.~\eqref{WQFT:Amplitude} presents an exciting relation between the WQFT and QFT-amplitudes approaches to classical gravitational scattering.
In Ref.~\cite{Jakobsen:2021zvh} its spinning generalization was verified to $\mO(G^2,S^2)$ by computation of the WQFT in-out eikonal to that order.
This computation is discussed in Sec.~\ref{WQFT:Worldline2}.
A direct comparison at the third post-Minkowskian order has not yet been carried out although some work has been done~\cite{Wang:2022ntx}.

\subsection{Worldline Observables from Classical Loops}
\label{WQFT:Worldline1}
The perturbative post-Minkowskian expansion of the worldline observables enjoys the simplifying property that all non-trivial dependence on the background parameters is limited to the relative Lorentz factor $\gamma$.
Thus, at each order in $G$ and spins we can bootstrap all dependence on the background parameters except for the $\gamma$ dependence.
The $\gamma$ dependence in turn is determined from loop integrals in momentum space.
These integrals thus depend only on a single dimensionless scale $\gamma$ and their computation at each post-Minkowskian order is the current bottleneck of deriving higher orders in $G$.
In this section we will analyze the general structure of the worldline observables and how their computation can be reduced to the evaluation of these single scale loop integrals.
The detailed analysis of the loop integrals appearing at the third post-Minkowskian order is carried out in \chap~\ref{sec:TL}.

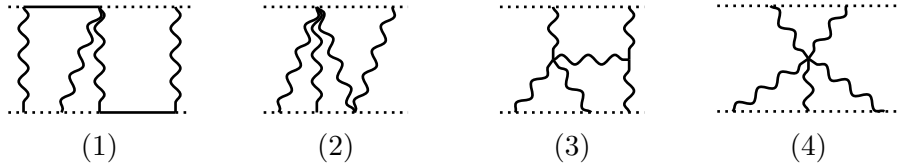
\begin{figure}
  \renewcommand*\thesubfigure{\arabic{subfigure}}
  \centering
  \begin{subfigure}{.18\textwidth}
    \centering
    \begin{tikzpicture}
      \coordinate (ua0)at (-1.2,.7) ;
      \coordinate (ua) at (-1 ,.7) ;
      \coordinate (ub) at (-.5,.7) ;
      \coordinate (uc) at (0  ,.7) ;
      \coordinate (ud) at (.5 ,.7) ;
      \coordinate (ue) at ( 1 ,.7) ;
      \coordinate (ue0)at (1.2,.7) ;
      \coordinate (da0)at (-1.2,-.7) ;
      \coordinate (da) at (-1 ,-.7) ;
      \coordinate (db) at (-.5,-.7) ;
      \coordinate (dc) at (0  ,-.7) ;
      \coordinate (dd) at (.5 ,-.7) ;
      \coordinate (de) at ( 1 ,-.7) ;
      \coordinate (de0)at (1.2 ,-.7) ;
      \draw[worldline] (ua) -- (uc) ;
      \draw[worldline] (dc) -- (de) ;
      \draw[dotted, line width=1pt] (ua0)--(ua) ;
      \draw[dotted, line width=1pt] (da0)--(dc) ;
      \draw[dotted, line width=1pt] (uc)--(ue0) ;
      \draw[dotted, line width=1pt] (de)--(de0) ;
      \draw[snake it] (ua) -- (da) ;
      \draw[snake it] (uc) -- (db) ;
      \draw[snake it] (uc) -- (dc) ;
      \draw[snake it] (ue) -- (de) ;
    \end{tikzpicture}
    \caption{
    }
  \end{subfigure}
  \begin{subfigure}{.18\textwidth}
    \centering
    \begin{tikzpicture}
      \coordinate (ua0)at (-1.2,.7) ;
      \coordinate (ua) at (-1 ,.7) ;
      \coordinate (ub) at (-.5,.7) ;
      \coordinate (uc) at (0  ,.7) ;
      \coordinate (ud) at (.5 ,.7) ;
      \coordinate (ud0) at (.7 ,.7) ;
      \coordinate (ue) at ( 1 ,.7) ;
      \coordinate (ue0)at (1.2,.7) ;
      \coordinate (da0)at (-1.2,-.7) ;
      \coordinate (da) at (-1 ,-.7) ;
      \coordinate (db) at (-.5,-.7) ;
      \coordinate (dc) at (0  ,-.7) ;
      \coordinate (dd) at (.5 ,-.7) ;
      \coordinate (dd0) at (.7 ,-.7) ;
      \coordinate (de) at ( 1 ,-.7) ;
      \coordinate (de0)at (1.2 ,-.7) ;
      \draw[dotted, line width=1pt] (ua0)--(ud0) ;
      \draw[dotted, line width=1pt] (da0)--(dd0) ;
      \draw[snake it] (ub) -- (da) ;
      \draw[snake it] (ub) -- (db) ;
      \draw[snake it] (ub) -- (dc) ;
      \draw[snake it] (ud) -- (dc) ;
    \end{tikzpicture}
    \caption{
    }
  \end{subfigure}
  \begin{subfigure}{.18\textwidth}
    \centering
    \begin{tikzpicture}
      \coordinate (ua0)at (-1.2,.7) ;
      \coordinate (ua) at (-1 ,.7) ;
      \coordinate (ub) at (-.5,.7) ;
      \coordinate (uc) at (0  ,.7) ;
      \coordinate (ud) at (.5 ,.7) ;
      \coordinate (ud0) at (.7 ,.7) ;
      \coordinate (ue) at ( 1 ,.7) ;
      \coordinate (ue0)at (1.2,.7) ;
      \coordinate (da0)at (-1.2,-.7) ;
      \coordinate (da) at (-1 ,-.7) ;
      \coordinate (db) at (-.5,-.7) ;
      \coordinate (dc) at (0  ,-.7) ;
      \coordinate (dd) at (.5 ,-.7) ;
      \coordinate (dd0) at (.7 ,-.7) ;
      \coordinate (de) at ( 1 ,-.7) ;
      \coordinate (de0)at (1.2 ,-.7) ;
      \coordinate (mb) at (-.5,0) ;
      \coordinate (mc) at (0  ,0) ;
      \coordinate (md) at (.5 ,0) ;
      \draw[dotted, line width=1pt] (ua0)--(ud0) ;
      \draw[dotted, line width=1pt] (da0)--(dd0) ;
      \draw[snake it] (ub) -- (mb) ;
      \draw[snake it] (da) -- (mb) ;
      \draw[snake it] (dc) -- (mb) ;
      \draw[snake it] (mb) -- (md) ;
      \draw[snake it] (ud) -- (md) ;
      \draw[snake it] (dd) -- (md) ;
    \end{tikzpicture}
    \caption{
    }
  \end{subfigure}
  \begin{subfigure}{.18\textwidth}
    \centering
    \begin{tikzpicture}
      \coordinate (ua0)at (-1.2,.7) ;
      \coordinate (ua) at (-1 ,.7) ;
      \coordinate (ub) at (-.5,.7) ;
      \coordinate (uc) at (0  ,.7) ;
      \coordinate (ud) at (.5 ,.7) ;
      \coordinate (ud0) at (.7 ,.7) ;
      \coordinate (ue) at ( 1 ,.7) ;
      \coordinate (ue0)at (1.2,.7) ;
      \coordinate (da0)at (-1.2,-.7) ;
      \coordinate (da) at (-1 ,-.7) ;
      \coordinate (db) at (-.5,-.7) ;
      \coordinate (dc) at (0  ,-.7) ;
      \coordinate (dd) at (.5 ,-.7) ;
      \coordinate (dd0) at (.7 ,-.7) ;
      \coordinate (de) at ( 1 ,-.7) ;
      \coordinate (de0)at (1.2 ,-.7) ;
      \coordinate (mb) at (-.5,0) ;
      \coordinate (mc) at (0  ,0) ;
      \coordinate (md) at (.5 ,0) ;
      \draw[dotted, line width=1pt] (ua0)--(ue0) ;
      \draw[dotted,line width=1pt] (da0)--(de0) ;
      \draw[snake it] (ub) -- (mc) ;
      \draw[snake it] (ud) -- (mc) ;
      \draw[snake it] (da) -- (mc) ;
      \draw[snake it] (dc) -- (mc) ;
      \draw[snake it] (de) -- (mc) ;
    \end{tikzpicture}
    \caption{
    }
  \end{subfigure}
  \caption{Four graphs with identical $G$ and mass scaling and number of loop integrations, namely $G^8m_1^2m_2^3$ and three loops.
    First, going from (1) to (2) worldline fluctuations are pinched.
    Then, going from (2) to (3) graviton interactions on the worldlines are pulled into the bulk.
    Finally, going from (3) to (4) all bulk interaction is pinched together into a single multi-graviton vertex.
  }
  \label{WQFT:PEX}
\end{figure}

We will first focus on a generic diagram contributing to either the eikonal or an on-shell one-point worldline function.
We want to determine its mass scaling, its order in $G$ and the number of loop integrations required.
In this case we define the mass scaling of the observables as the overall scaling when they are written in terms of the mass-scaled parameters $v_i^\sig$ and $a^\sig_i$ (or equivalently for arbitrary dimension $d$: $v_i^\sig$ and $\Psi_i^\sig$).
The result is that at the $n$th post-Minkowskian order, a diagram scales as $m_1^{m}m_2^{n+1-m}$ with $1\le m\le n$ and $n-1$ loop integrations are required.

First, we note the following scalings of WQFT Feynman rules.
A worldline interaction of the $i$th body with $m$ worldline fluctuations and $n$ gravitons scales as $m_i \kappa^n$.
A bulk interaction with $n$ gravitons scales as $\kap^{n-2}$.
A worldline propagator of the $i$th body scales as $m^{-1}_i$.
An example of the procedure to be discussed is shown in Fig.~\ref{WQFT:PEX}.

We focus, then, on a generic vacuum WQFT diagram defined in the in-out theory or in-in theory with non-zero $(-)$ parameters.
That is, we focus on diagrams contributing to the WQFT free energy (eikonal) from which any other on-shell $n$-point function can be derived.
In any case, the inclusion of an amputated external line does not change the scalings of $G$ or $m_i$ or the number of loop integrations.
Energy is conserved at every vertex (bulk or worldline) and non-trivial (spacial) loop integrations will only appear from gravitons interacting with the worldlines.
In order to understand the mass and $G$ scalings and the number of loop momenta we therefore first pinch all worldline fluctuations into multipoint graviton contact vertices.
The scalings of vertices and propagators given above ensure that the $G$ and $m_i$ scalings are unchanged under this operation.

We are thus left with vacuum diagrams with propagating gravitons only.
The $G$ and $m_i$ scalings and number of loop integrations are also unchanged if we pull all contact interactions on the worldline into the bulk.
In addition, we can merge all contact interactions in the bulk into a single multi-point bulk interaction.
In this way, we have turned a generic vacuum diagram into an $n$-point graviton vertex interacting with $n$ one-point worldline backgrounds without changing the mass and $G$ scalings or the number of spacial loop integrations.
At this point we may verify the result that was stated above.
Thus, if we take an $(n+1)$-point bulk graviton vertex, it scales as $\kap^{n-1}$.
Together with the $(n+1)$ worldline emission vertices we get $G^n$.
Every worldline emission vertex comes with either $m_1$ or $m_2$ and we get $m_1^{m}m_2^{n+1-m}$ with $1\le m\le n$.
Here, we have neglected the cases $m_1^{n+1}$ and $m_2^{n+1}$ which vanish in dimensional regularization and are not dynamical.
Of the $n+1$ graviton momentum integrations present, one is cancelled by the momentum conservation of the bulk vertex and one integration is simply a Fourier transform (as we will see).
Thus, there are $n-1$ non-trivial loop integrations leftover.

We let $X$ represent either the impulse, the spin kick $\Del S_i^\sig=m_i \Del a_i^\sig$ (or $\Del S_i^\mn$) or the WQFT eikonal.
The post-Minkowskian expansion may then be performed as follows:\checked
\begin{align}
  X
  =
  \sum_{g=1}^\infty
  \Big(
  \frac{GM}{|b|}
  \Big)^g
  X^{(g)}
  \ .
\end{align}
Here, $M$ is the total mass and we use the dimensionless variable $GM/|b|$ as PM expansion parameter.
The above analysis then shows that $X^{(g)}$ may be expanded in masses as follows:\checked
\begin{align}
  \label{WQFT:MassE}
  X^{(g)}
  =
  |b|^n
  \sum_{1\le m\le g}
  \frac{m_1^{m}m_2^{g+1-m}}{M^{g}}
  X^{(g,m)}
\end{align}
The coefficients $X^{(g,m)}$ are independent of mass and $G$ and are computed from $(g-1)$-loop integrals.
The factors $|b|^n$ are inserted to make the expansion coefficients $X^{(g,m)}$ dimensionless.
For the impulse $n=0$ and for the spin kick $\Del S_i^\sig$ we get $n=1$.
In principle the above analysis on mass scalings applied to the (mass scaled) Grassmann kick but the properties are unchanged for the spin kick (which can be verified from their relation Eqs.~\ref{WEFT:SpinKicks}).

The spin dependence of the observables appears in a perturbative expansion in $a^\sig_i$.
This expansion may be carried out independently from the PM expansion.
We define it in terms of spin structures $\mO^{(S)}$:\checked
\begin{align}
  X=
  \sum_\mS
  X^{(\mS)}
  \mO^{(\mS)}
\end{align}
The only spinless structure is $\mO^{(0)}=1$.
At linear order in spins and $d=4$ we have three spin structures for each body which we can define as:\checked
\begin{align}
  \mO^{(1,1,i)}
  =
  \frac{a_i\cdot \hat L}{|b|}
  \ ,
  \qquad
  \mO^{(1,2,i)}
  =
  \frac{a_i\cdot \hat b}{|b|}
  \ ,
  \qquad
  \mO^{(1,3,i)}
  =
  \frac{a_i\cdot (v_1+v_2)}{|b|}
  \ .
\end{align}
The factor of $|b|$ are inserted to make the spin structures dimensionless.
The spin structures at quadratic order in spins can be written in a similar fashion but we will not consider explicit expressions here.
Naturally, the coefficients of the spin expansion $X^{(\mS)}$ are independent of the spin parameters, $a_i^\sig$.

Finally, in the case of vector observables (or higher order tensors) we may expand them on a basis.
In $d=4$ we can choose the four basis elements $v_i^\sig$, $\hat b^\sig$ and $\hat L^\sig$.
We denote the basis elements symbolically as $\mB^{(\beta)}$ and write:\checked
\begin{align}
  X
  =
  \sum_{\beta}
  X^{(\beta)}
  \mB^{(\beta)}
\end{align}
The expansion coefficients, $X^{(\beta)}$, of this expansion are scalars.

The observables can be expanded in $G$, in spins and on a basis simultaneously.
We define coefficients $X^{(g,m;\mS;\beta)}$ by:\checked
\begin{align}\label{WQFT:MP}
  X
  =
  |b|^{n}
  \sum_{g,m,\mS,\beta}
  X^{(g,m;\mS;\beta)}
  \,
  \Big(
  \frac{GM}{|b|}
  \Big)^g
  \,
  \frac{m_1^{m}m_2^{g+1-m}}{M^{g}}
  \mO^{(S)}
  \mB^{(\beta)}
  \ .
\end{align}
The sum on $m$ has the same limits as in Eq.~\eqref{WQFT:MassE} and, again, the first factor $|b|^n$ ensures that the expansion coefficients are dimensionless with $n=0$ for the impulse and $n=1$ for the spin kick.
The coefficients $X^{(g,m;\mS;\beta)}$ are scalars independent of Newton's constant $G$, the masses $m_i$ and the spins $a_i^\mu$.
Since they are dimensionless they must also be independent of $|b|$.
Therefore they are functions only of the relative Lorentz factor $\gam$.
A similar schematic form for the waveform may easily be derived with the same approach since the relevant diagrams of the waveform are simply derived from vacuum diagrams by adding an external graviton.

Finally, then, we have bootstrapped the scattering observables in terms of unknown functions of $\gam$.
These functions are determined from the post-Minkowskian $(g-1)$-loop integrals.
Symmetry constraints of the observables such as the conservation of $p^2_i=m_i^2$ or particle exchange symmetry put corresponding constraints on these functions.
For the in-out eikonal in the spinless case at the $g$th PM order, the counting taking into account symmetries simply gives $g$ or $g-1$ independent functions with $g$ odd or even respectively.
Thus the spinless 1PM and 2PM components of the in-out eikonal each have one non-trivial function and the 3PM and 4PM components each have two functions.

The simple mass scaling (or mass polynomiality) of the scattering observables in Eq.~\eqref{WQFT:MP} was first observed by Damour in Ref.~\cite{Damour:2019lcq}.
In fact, this observation is equivalent to the statement that the observables are functions only of the two combinations $m_i G/|b|$ which are the individual effective expansion parameters of the worldlines in the worldline effective field theory.
A particular consequence of this scaling is the straightforward comparison of the post-Minkowskian series expansion to the corresponding expansion in self-force or mass ratio.
In particular, in the limit $m_1/m_2\ll1$ the motion is geodesic and generally known to all order in $G$.
Thus, at each order $n$ in $G$ the coefficients of $m_1^{n}m_2$ and $m_1m_2^n$ can be directly read off from the known geodesic limit~\cite{Cheung:2020gbf,Damgaard:2022jem,Kol:2021jjc}.
As an example, this implies that the 1PM and 2PM in-out eikonal can be directly matched from geodesic motion.
For the spinless 3PM and 4PM in-out eikonal only one non-trivial function of $\gamma$ remains at each order after matching with geodesic motion.
In fact, these functions are then fully determined from the first self-force order.
Similar results hold for the remaining worldline observables and waveform.

Let us now discuss the appearance of the loop integrals from a generic WQFT diagram.
At the $n$th post-Minkowskian order there are initially $n$ momentum integrations one of them being a Fourier transform and the rest loop integrations.
We denote the $n$ integration momenta by $l_k^\sig$.
Generally, then, there is some numerator with $i$ factors of $l_k^\sig$ and some denominator with $j$ propagators $D_k$.
There are $(n+1)$ energy conserving delta functions from the $(n+1)$ gravitons interacting on the worldline and finally a Fourier factor $e^{-il_n\cdot b}$ coming from the worldline vertex rules.
We write:\checked
\begin{align}
  \label{WQFT:Integral}
  \int_{l_1...l_n}
  e^{-il_n \cdot b}
  \frac{
    l_{m_1}^{\sig_1}
    .\,.\,.
    l_{m_i}^{\sig_i}
  }{
    D_1
    .\,.\,.
    D_j}
  \dd(l_1\cdot v)
  .\,.\,.
  \dd(l_{n-1}\cdot v)
  \dd(l_n\cdot v_1)
  \dd(l_n\cdot v_2)
  \ .
\end{align}
The $n$ loop momenta have been labelled in a specific way such that $l_n^\sig$ appears in the Fourier factor and $l_k$ with $k<n$ appears in each their own delta function and such a labelling is always possible.
The velocities in the delta functions of $l_k$ with $k<n$ have no labels and for every diagram each of those are either $v_1^\mu$ or $v_2^\mu$ depending on the mass sector under consideration.
The denominators $D_k$ are either massless or linear propagators with some $i\eps$ prescription.
The only dependence on external variables is through the delta functions, the Fourier factor and the linear propagators $l_k\cdot v$.

With the labelling in Eq.~\eqref{WQFT:Integral}, the $n$th momentum $l_n^\sig$ has been picked out as the Fourier integral.
It is practical to postpone this Fourier transform and define observables in momentum space ($q$-space).
We usually label the momentum $l_n^\sig$ by $q^\sig$.
We get:\checked
\begin{align}
  \label{WQFT:Integral2}
  \int_q
  e^{-i q\cdot b}
  \dd(q\cdot v_1)
  \dd(q\cdot v_2)
  \int_{l_1...l_{n-1}}
  \frac{
    l_{m_1}^{\sig_1}
    .\,.\,.
    l_{m_i}^{\sig_i}
  }{
    D_1
    .\,.\,.
    D_j}
  \dd(l_1\cdot v)
  .\,.\,.
  \dd(l_{n-1}\cdot v)
  \ .
\end{align}
The inner integral on $l_k$ with $0\le k<n$ is the loop integral.
It depends only on $q^\mu$ and the velocities $v_i^\sig$.
Their scalar products are $v_i^2=1$, $v_i\cdot q=0$, $v_1\cdot v_2=\gam$ and $q^2=-|q|^2$.
The only dimensionful scale is $|q|$ and the dependence on $|q|$ can be determined by dimensional analysis.
The tensor structure is a function of the dimensionless vectors $v_i^\mu$ and $\hat q^\mu$ and the metric $\eta^\mn$.
The non-trivial part of these integrals, then, is their dependence on $\gam$.
When considering the worldline observables in $q$-space, $q^\sig$ plays a similar role to the impact parameter and we will consider it as belonging to the external data.

After the dependence of the loop integral in Eq.~\eqref{WQFT:Integral2} on $|q|$ has been determined from dimensional analysis the Fourier transform takes the following form (with some generic integers $n$ and $m$):\checked
\bse
\label{WQFT:FourierTransform}
\begin{align}
  \int_q
  e^{-i q\cdot b}
  \dd(q\cdot v_1)
  \dd(q\cdot v_2)
  \frac{\hat q^{\mu_1}
    .\,.\,.
    \hat q^{\mu_m}
  }{
    |q|^n
  }
  \ .
\end{align}
Factors of $\hat q^\mu=q^\mu/|q|$ can easily be written as derivatives of the scalar integral ($m=0$) with respect to $b^\mu$.
The scalar integral is well known and given by:\checked
\begin{align}
  \int_q
  e^{-i q\cdot b}
  \dd(q\cdot v_1)
  \dd(q\cdot v_2)
  \frac{1  }{
    |q^\mu|^n
  }
  =
  \frac{1}{    \sqrt{\gam^2-1}
  }
  \frac{1}{
    2^n\,
    \pi^{(d-2)/2}
  }
  \frac{
    \Gam(\frac{d-2-n}{2})
  }{
    \Gam(\frac{n}{2})
  }
  \frac{1}{
    |P_{12}^\mn b_\nu|^{d-2-n}
  }
  \ .  
\end{align}
\ese
It is most easily computed in the orthogonal subspace to $v_i^\mu$ and its dependence on $|b|$ is determined on dimensional grounds.
We inserted the projector $(P_{12}\cdot b)^\sig=b^\sig$ explicitly to remind ourselves of the definition of the impact parameter $\eqref{WEFT:IP}$.

We have now identified the generic structure of the momentum integrals appearing in the post-Minkowskian expansion of the worldline observables.
Their computation to $\mO(G^2)$ will be considered in the next section~\ref{WQFT:Worldline2} and at the third post-Minkowskian order in \chap~\ref{sec:TL}.

\subsection{Worldline Observables at $\mO(G^2,S^2)$ and In-In from In-Out}
\label{WQFT:Worldline2}
The second post-Minkowskian order requires one-loop integration and is from that perspective the first non-trivial order in the perturbative expansion.
In the spinless case the 2PM results were first published by Westpfahl in 1985 Ref.~\cite{Westpfahl:1985tsl}.
With the recent application of quantum field theory to the classical post-Minkowskian expansion, the 2PM results have been rederived in numerous works and with additional properties of the bodies.
In fact, the inclusion of $\mO(S^2)$ effects does not add any significant complexity to the relevant integrals.
The $\mO(G^2,S^2)$ computations presented here are based on Ref.~\cite{Jakobsen:2021zvh}.
They serve as a non-trivial example of the WQFT framework.

We will focus on the computation of the WQFT in-out eikonal $\chi_{\rm in-out}$ to $\mO(G^2,S^2)$.
In fact all worldline observables $\Del p^\mu$, $\Del \psi^\mu$ and the spin kick can be derived from the eikonal to this order.
This includes both the in-out and in-in versions of the observables.
The relation of the in-out observables to $\chi_{\rm in-out}$ was given in Eqs.~\eqref{WQFT:ObservablesFromE} and we will discuss how the in-in observables consequently are derived below.

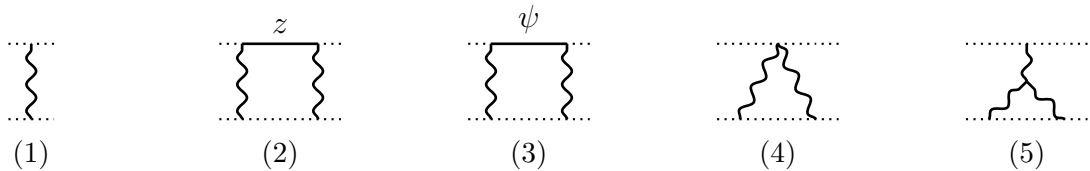
\begin{figure}[h]
  \renewcommand*\thesubfigure{\arabic{subfigure}} 
  \centering
    \begin{subfigure}{.19\textwidth}
    \centering
    \begin{tikzpicture}
      \coordinate (ur) at (+.5  ,0) ;
      \coordinate (ull)at (.2 ,0 ) ;
      \coordinate (urr)at (+.8 ,0 ) ;
      \coordinate (dr) at (+.5  ,-1) ;
      \coordinate (dll)at (.2 ,-1 ) ;
      \coordinate (drr)at (+.8 ,-1 ) ;
      \draw[worlddot] (dll) -- (dr) ;
      \draw[worlddot] (ull) -- (ur) ;
      \draw[worlddot] (urr) -- (ur) ;
      \draw[worlddot] (dr) -- (drr) ;
      \draw[snake it] (dr) -- (ur) ;
    \end{tikzpicture}
    \caption{
    }
  \end{subfigure}
  \begin{subfigure}{.19\textwidth}
    \centering
    \begin{tikzpicture}
      \coordinate (ul) at (-.5  ,0) ;
      \coordinate (um) at (0   ,0) ;
      \coordinate (ur) at (+.5  ,0) ;
      \coordinate (ull)at (-.8 ,0 ) ;
      \coordinate (urr)at (+.8 ,0 ) ;
      \coordinate (dl) at (-.5  ,-1) ;
      \coordinate (dm) at (0   ,-1) ;
      \coordinate (dr) at (+.5  ,-1) ;
      \coordinate (dll)at (-.8 ,-1 ) ;
      \coordinate (drr)at (+.8 ,-1 ) ;
      \draw[worldline] (ul) -- (ur) ;
      \draw[worlddot] (ull) -- (ul) ;
      \draw[worlddot] (urr) -- (ur) ;
      \draw[worlddot] (dll) -- (drr) ;
      \draw[snake it] (dr) -- (ur) ;
      \draw[snake it] (dl) -- (ul) ;
      \node[above] at ($(ul)!0.5!(ur)$) {
        $z$
        };
    \end{tikzpicture}
    \caption{
    }
  \end{subfigure}
  \begin{subfigure}{.19\textwidth}
    \centering
    \begin{tikzpicture}
      \coordinate (ul) at (-.5  ,0) ;
      \coordinate (um) at (0   ,0) ;
      \coordinate (ur) at (+.5  ,0) ;
      \coordinate (ull)at (-.8 ,0 ) ;
      \coordinate (urr)at (+.8 ,0 ) ;
      \coordinate (dl) at (-.5  ,-1) ;
      \coordinate (dm) at (0   ,-1) ;
      \coordinate (dr) at (+.5  ,-1) ;
      \coordinate (dll)at (-.8 ,-1 ) ;
      \coordinate (drr)at (+.8 ,-1 ) ;
      \draw[worldline] (ul) -- (ur) ;
      \draw[worlddot] (ull) -- (ul) ;
      \draw[worlddot] (urr) -- (ur) ;
      \draw[worlddot] (dll) -- (drr) ;
      \draw[snake it] (dr) -- (ur) ;
      \draw[snake it] (dl) -- (ul) ;
      \node[above] at ($(ul)!0.5!(ur)$) {
        $\psi$
        };
    \end{tikzpicture}
    \caption{}
  \end{subfigure}
  \begin{subfigure}{.19\textwidth}
    \centering
    \begin{tikzpicture}
      \coordinate (ul) at (-.5  ,0) ;
      \coordinate (um) at (0   ,0) ;
      \coordinate (ur) at (+.5  ,0) ;
      \coordinate (ull)at (-.8 ,0 ) ;
      \coordinate (urr)at (+.8 ,0 ) ;
      \coordinate (dl) at (-.5  ,-1) ;
      \coordinate (dm) at (0   ,-1) ;
      \coordinate (dr) at (+.5  ,-1) ;
      \coordinate (dll)at (-.8 ,-1 ) ;
      \coordinate (drr)at (+.8 ,-1 ) ;
      \draw[worlddot] (ul) -- (ur) ;
      \draw[worlddot] (ull) -- (ul) ;
      \draw[worlddot] (urr) -- (ur) ;
      \draw[worlddot] (dll) -- (drr) ;
      \draw[snake it] (dr) -- (um) ;
      \draw[snake it] (dl) -- (um) ;
    \end{tikzpicture}
    \caption{
    }
  \end{subfigure}
  \begin{subfigure}{.19\textwidth}
    \centering
    \begin{tikzpicture}
      \coordinate (ul) at (-.5  ,0) ;
      \coordinate (um) at (0   ,0) ;
      \coordinate (ur) at (+.5  ,0) ;
      \coordinate (ull)at (-.8 ,0 ) ;
      \coordinate (urr)at (+.8 ,0 ) ;
      \coordinate (dl) at (-.5  ,-1) ;
      \coordinate (dm) at (0   ,-1) ;
      \coordinate (dr) at (+.5  ,-1) ;
      \coordinate (dll)at (-.8 ,-1 ) ;
      \coordinate (drr)at (+.8 ,-1 ) ;
      \coordinate (mm) at (0,-.5) ;
      \draw[worlddot] (ul) -- (ur) ;
      \draw[worlddot] (ull) -- (ul) ;
      \draw[worlddot] (urr) -- (ur) ;
      \draw[worlddot] (dll) -- (drr) ;
      \draw[snake it] (mm) -- (um) ;
      \draw[snake it] (dr) -- (mm) ;
      \draw[snake it] (dl) -- (mm) ;
    \end{tikzpicture}
    \caption{
    }
  \end{subfigure}
  \caption{
    Graphs contributing to the eikonal, $\chi_{\rm in-out}$, to $\mO(G^2,S^2)$.
    The first graph is the leading order contribution and the four subsequent graphs together with their mirrored ones constitute the subleading order.
  }
  \label{WQFT:EikonalGraphs}
\end{figure}

The relevant graphs for the eikonal, $\chi_{\rm in-out}$ to $\mO(G^2,S^2)$ are shown in Fig.~\ref{WQFT:EikonalGraphs}.
In that figure, there is one graph (1) at leading 1PM order and four graphs (2) - (5) at subleading 2PM order.
In addition to the four 2PM graphs shown there we must add additional four graphs obtained by particle exchange symmetry.
In total, then, there are $1+4+4$ graphs with mass and $G$ scalings $G m_1 m_2$, $G^2 m_1 m_2^2$ and $G^2 m_1^2 m_2$ respectively.
The first two mass and $G$ scalings are the ones shown in Fig.~\ref{WQFT:EikonalGraphs}.

We start by considering the tree level contribution which comes from graph (1) of Fig.~\ref{WQFT:EikonalGraphs}.
We label the exchanged graviton momentum by $q^\mu$ and the integration on that momentum is a Fourier transform due to the two Fourier factors from each worldline vertex rule.
Insertion of Feynman rules leads to the following 1PM expression,\checked
\begin{align}
  \label{WQFT:EE1}
  &
  i\chi_{\rm in-out}^{(1)}=
  \int_q
  \dd(q\cdot v_1)
  \dd(q\cdot v_2)
  \frac{-i\mP^{-1}_{\mn\ab}}{q^2}
  \hat V_{\rmm{spin},1}[h_\mn(q)]
  \hat V_{\rmm{spin},2}[h_\ab(-q)]
  \\\nn
  &\qquad\qquad\qquad\qquad\qquad=
  i\frac{m_1 m_2
  \kappa^2}{4}
  \int_q
  \dd(q\cdot v_1)
  \dd(q\cdot v_2)
  e^{-iq\cdot b}
  \frac{
    \mP^{-1}_{\ab\mn}}{
    q^2}
  v^\alpha
  v^\beta
  v^\mu
  v^\nu
  +
  \mO(S)
  \ ,
\end{align}
where we have printed only the spinless term in the second line.
The Fourier integral on $q^\mu$ is exactly the one discussed in Eqs.~\eqref{WQFT:FourierTransform}.
We note that while the exponent of the Fourier factor is in principle $-iq\cdot(b_2-b_1)$, the two energy-conserving delta functions allow us to rewrite it in terms of the (orthogonal) impact parameter $b^\mu=(P_{12}\cdot(b_2-b_1))^\mu$.
The linear and quadratic orders in spin introduces dependence on $q^\mu$ in the numerator which is rewritten as $b^\mu$ derivatives.

We add terms to quadratic orders in spin to Eq.~\eqref{WQFT:EE1} and insert the Fourier transform.
In this way we get the 1PM contribution to $\chi_{\rm in-out}$ and find:\checked
\begingroup
\allowdisplaybreaks
\begin{subequations}\label{eq:1PMresult}
  \begin{align}
    \left.\chi_{\rm in-out}^{(1)}\right|_{ S_1^0 S_2^0}&=
    \frac{2\pi^{2-\frac{d}2}\Gamma(\frac{d}2-2)((d-2)\gamma^2-1)m_1m_2}
    {(d-2)|b|^{d-4}\sqrt{\gamma^2-1}}\,,\label{eq:1PMnospin}\\
    \left.\chi_{\rm in-out}^{(1)}\right|_{ S_1 S_2^0}&=
    \frac{4\pi^{2-\frac{d}2}\Gamma(\frac{d}2-1)\gamma m_2}{|b|^{d-3}\sqrt{\gamma^2-1}}\,
    {\hat b}\cdot S_1\cdot v_2\,,\\
    \left.\chi_{\rm in-out}^{(1)}\right|_{ S_1 S_2}&=
    \frac{2\pi^{2-\frac{d}2}\Gamma(\frac{d}2-1)}{|b|^{d-2}(\gamma^2-1)^{3/2}}
    \bigg[
          -
    v_2\cdot S_1\cdot S_2\cdot v_1
    \\&\quad
    +
    (\gamma^2-1)
    \Big(
    \gamma\operatorname{tr}( S_1\cdot S_2)
    -
    (d-2)
    (
    {\hat b}\cdot S_1\cdot v_2\,{\hat b}\cdot S_2\cdot v_1
    -
    \gamma {\hat b}\cdot S_1\cdot S_2\cdot {\hat b}
    )
    \Big)
    \bigg]
    \,,\nn\\
    \left.\chi_{\rm in-out}^{(1)}\right|_{ S_1^2 S_2^0}&=
    \frac{2\pi^{2-\frac{d}2}\Gamma(\frac{d}2-1)m_2}{(d-2)|b|^{d-2}(\gamma^2-1)^{3/2}m_1}
    \bigg[
    \big(
    d-1-\gamma^2(d-2)
    \big)
    v_2\cdot S_1\cdot S_1\cdot v_2
    \\
    &\quad
    +(\gamma^2-1)
    \big(
    (d-2)^2
    ({\hat b}\cdot S_1\cdot v_2)^2
    +
    (d-2)
    {\hat b}\cdot S_1\cdot S_1\cdot{\hat b}+\operatorname{tr}(S_1\cdot S_1)
    \big)
    \nn\\&\quad
    -C_{\rmm{E},1}
    \big(
    (d\!-\!2)\gamma^2\!-\!1
    \big)
    \Big(
    (\gamma^2\!-\!1)
    \big(
    (d\!-\!2){\hat b}\cdot S_1\cdot S_1\cdot{\hat b}+\operatorname{tr}(S_1\cdot S_1)
    \big)
    +v_2\cdot S_1\cdot S_1\cdot v_2
    \Big)
    \bigg]
    \,,\nn
\end{align}
\end{subequations}
\endgroup
Each of the four equations display a certain spin sector indicated by the subscript to the vertical lines.
Other spin sectors are obtained by particle exchange symmetry.
The expressions are evaluated in arbitrary dimensions and the covariant gauge is assumed.
However, as the eikonal depends only on gauge invariant asymptotic variables, we may straightforwardly exchange the covariant gauge variables with the SUSY independent variables using Eqs.~\eqref{WEFT:CVP}.
The mass scalings of the terms at linear and quadratic order in spins do not match the spinless $m_1 m_2$.
However, if we would use the mass scaled spin tensors $S_i^\mn/m_i$ instead of $S_i^\mn$ we would recover that scaling in all terms.

Worldline observables in $d$ dimensions at 1PM are derived by simple differentiation:\checked
\bse
\begin{align}
  \Del p^{(1)\mu}_{i,\rm in-in}
  &=
  \Del p^{(1)\mu}_{i,\rm in-out}
  =
  -\frac{
    \pat \chi_{\rm in-out}^{(1)}
  }{
    \pat b_{i,\mu}
  }
  \ ,
  \\
  \Del S^{(1)\mn}_{i,\rm in-in}
  &=
  \Del S^{(1)\mn}_{i,\rm in-out}
  =
  \frac4{m_i}
  S_{i,\alpha}^{\ \ [\mu}
    \ 
    \frac{
      \pat \chi_{\rm in-out}^{(1)}
    }{
      \pat S_{i,\nu]\alpha}
  }
  \ .
\end{align}
\ese
At the 1PM order there is no distinction between the in-in and in-out boundary conditions.
That is so because the only propagator present (that is $1/q^2$) is not sensitive to its $i\eps$-prescription.
When differentiations with respect to the background parameters are carried out care has to be taken with respect to their constraints.
One may instead use the unconstrained variables as discussed in Secs.~\ref{WEFT:GS} and~\ref{sec:Dynamics}.
Thus, one may rewrite the eikonal and observables in terms of unconstrained variables and it is then straightforward to carry out differentiations.

The 2PM contribution comes from graphs (2) - (5) of Fig.~\ref{WQFT:EikonalGraphs} and their mirrored graphs.
We label the two internal graviton momenta connected to the second worldline by $l^\mu$ and $(q^\mu+l^\mu)$.
The worldline energy of the excitations in graphs (2)-(3) become $l\cdot v_1$ and the third graviton momentum of graph (5) is $q^\mu$.
The contribution from graphs (2) - (5) then take the schematic form,\checked
\begin{align}
  \int_q
  \dd(q\cdot v_1)
  \dd(q\cdot v_2)
  e^{-iq\cdot b}
  \sum_{n_1,n_2,n_3}
  \int_{l}
  \dd(l\cdot v_2)
  \frac{\mN_{n_1,n_2,n_3}}{
    D_1^{n_1}
    D_2^{n_2}
    D_3^{n_3}
  }\ ,
\end{align}
where the numerators $\mN_{n_1,n_2,n_3}$ depend on external data (including $q^\mu$) and the loop momenta $l^\mu$.
The denominators are:\checked
\begin{align}
  D_1
  &=l\cdot v_1 \pm i\eps
  \ ,
  \\
  D_2
  &= l^2
  \ ,
  \nn
  \\
  D_3
  &= (l+q)^2
  \ .
  \nn
\end{align}
Power counting of the vertex rules tell us that there will be at most $(2+s)$ internal momenta in $\mN_{n_1,n_2,n_3}$ with $s$ the spin order.
The computation of the numerators only requires simple contractions of the vertex rules.
Some of the vertex rules, however, such as the three point graviton vertex are rather lengthy.
We will not discuss the explicit derivation of the numerators $\mN_{n_1,n_2,n_3}$ here.

The relevant loop integrals at this order is thus different powers of the denominators $D_i$ with factors of $l^\mu$ in the numerator.
Advanced integration techniques will be discussed in more detail in \chap~\ref{sec:TL} including tensor reduction with which loop momenta in the numerator may be eliminated.
As an example of such a rule, a single factor of $l^\mu$ in the numerator is removed with the following relation:\checked
\begin{align}
  \int_{l}
  \dd(l\cdot v_2)
  \frac{l^\mu}{
    D_1^{n_1}
    D_2^{n_2}
    D_3^{n_3}
  }
  =
  \int_{l}
  \dd(l\cdot v_2)
  \frac{1}{
    D_1^{n_1}
    D_2^{n_2}
    D_3^{n_3}
  }
  \bigg[
  \frac{  D_3-D_2-q^2}{2q^2}
  q^\mu
  -
  D_1
  \frac{
    v^\mu_1-\gam v_2^\mu
  }{
    \gam^2-1
  }
  \bigg]
  \ .
\end{align}
Similar relations with any number of factors of $l^\mu$ are easily derived (see e.g. Sec.~\ref{TL:TR}).
The remaining scalar integrals are a standard result in quantum field theory.
Its derivation will be discussed in more detail in Sec.~\ref{sec:TLMushroom}.
The result is:\checked
\begin{align}\label{WQFT:TLI}
  F_{n_1,n_2,n_3}^+
  =
  \int_{l}
  \frac{
      \dd(l\cdot v_2)
}{
    D_1^{n_1}
    D_2^{n_2}
    D_3^{n_3}
  }
  =
    (-i)^{n_1+2n_2+2n_3}
  \frac{
    2^{n_1}
    (4\pi)^{\frac{1-d}{2}}
}{\sqrt{\gam^2-1}^{n_1}}
  \Gam_{n_1,n_2,n_3}(d-1)
  |q|^{d-1-n_1-2(n_2+n_3)}
  \ ,
\end{align}
with the function $\Gam_{n_1n_2n_3}(d)$ given by:\checked
\begin{align}\label{WQFT:GAMF}
  \Gamma_{n_1,n_2,n_3}(d)=
	\frac{\Gamma(n_2+n_3+\frac{n_1}2-\frac{d}2)\Gamma(\frac{n_1}2)}
	{2\Gamma(n_1)\Gamma(n_2)\Gamma(n_3)}
	\frac{\Gamma(\frac{d}2-n_2-\frac{n_1}2)\Gamma(\frac{d}2-n_3-\frac{n_1}2)}
	     {\Gamma(d-n_1-n_2-n_3)}
             \ .
\end{align}
The $i\eps$-prescription of the linear propagator, $D_1$, is taken as positive in Eq.~\eqref{WQFT:TLI}.
It may be computed with the use of Schwinger parameters.
Some techniques and references for the evaluation of loop integrals may be found in Sec.~\ref{sec:TLIntegral}.

Putting all pieces together we get the 2PM contribution at $\mO(S^2)$ which is found to be:\checked
\begingroup
\allowdisplaybreaks
\begin{subequations}\label{eq:2PMresult}
  \begin{align}
      \left.\chi_{\rm in-out}^{(2)}\right|_{ S_1^0 S_2^0}&=
      \frac{3\pi(5\gamma^2-1)(m_1+m_2)m_1m_2}{4|b|\sqrt{\gamma^2-1}}\,,\\
      \left.\chi_{\rm in-out}^{(2)}\right|_{ S_1 S_2^0}&=
      \frac{\pi\gamma(5\gamma^2-3)(4m_1+3m_2)m_2}{4|b|^2(\gamma^2-1)^{3/2}}
      {\hat b}\cdot S_1\cdot v_2\,,\\
      \left.\chi_{\rm in-out}^{(2)}\right|_{ S_1 S_2}&=
      \frac{\pi(m_1+m_2)}{4|b|^3(\gamma^2-1)^{5/2}}
      \Big[(\gamma^2-1)(\gamma(5\gamma^2-3)(2\operatorname{tr}( S_1\cdot S_2)
        +3{\hat b}\cdot S_1\cdot S_2\cdot{\hat b})
        \\
      &\quad
      -
      9(5\gamma^2-1){\hat b}\cdot S_1\cdot v_2\,{\hat b}\cdot S_2\cdot v_1)
      -3(3\gamma^2-1)v_2\cdot S_1\cdot S_2\cdot v_1
      \Big]\,,\nn\\
      \left.\chi_{\rm in-out}^{(2)}\right|_{ S_1^2 S_2^0}&=
      \frac{
        \pi\,m_2
      }{
        64|b|^3(\gamma^2-1)^{5/2}m_1
      }
      \Big[
24(\gamma^2-1)((31\gamma^2-11)m_1+3(5\gamma^2-1)m_2) ({\hat b}\cdot S_1\cdot v_2)^2
        \nn
        \\
      &\quad
        -4(\gamma^2-1)((13\gamma^4-42\gamma^2+21)m_1-4(3\gamma^2-1)m_2)\operatorname{tr}(S_1\cdot S_1)
\\
      &\quad
        -6(\gamma^2-1)((29\gamma^4-66\gamma^2+29)m_1-4(3\gamma^2-1)m_2)  {\hat b}\cdot S_1\cdot S_1\cdot{\hat b}
\nn\\
      &\quad
      -6((49\gamma^4-90\gamma^2+33)m_1+4(5\gamma^4-9\gamma^2+2)m_2)
      v_2\cdot S_1\cdot S_1\cdot v_2\nn\\
      &\quad
      -2C_{E,1}(\gamma^2-1)
      ((125\gamma^4-138\gamma^2+29)m_1+2(45\gamma^4-42\gamma^2+5)m_2)
      \operatorname{tr}(S_1\cdot S_1)\nn\\
      &\quad
      -3C_{E,1}(\gamma^2-1)
      ((155\gamma^4-174\gamma^2+35)m_1+4(30\gamma^4-29\gamma^2+3)m_2)
      {\hat b}\cdot S_1\cdot S_1\cdot{\hat b}\nn\\
      &\quad
      -3C_{E,1}((95\gamma^4-102\gamma^2+23)m_1+4(15\gamma^4-13\gamma^2+2)m_2))
      v_2\cdot S_1\cdot S_1\cdot v_2
      \Big]\,.\nn
  \end{align}
\end{subequations}
\endgroup
Here, we present only its value in four space-time dimensions.
Expressions in arbitrary dimensions are found in the ancillary file to Ref.~\cite{Jakobsen:2021zvh}.
The presentation in terms of different spin sectors is the same as with the 1PM eikonal Eqs.~\eqref{eq:2PMresult}.
Just as with the 1PM terms we recover a similar mass scaling at 2PM for all spin sectors if we use mass scaled spin tensors and the covariant SSC is assumed for all variables (although, again, we may easily re-express the eikonal in terms of SUSY invariant variables).

Observables at second post-Minkowskian order with the in-out boundary conditions are now easily derived:\checked
\bse
\begin{align}
  \Del p^{(2)\mu}_{\rm in-out}
  &=
  -\frac{
    \pat \chi_{\rm in-out}^{(2)}
  }{
    \pat b_{i,\mu}
  }
  \ ,
  \\
  \Del S^{(2)\mn}_{\rm in-out}
  &=
  \frac4{m_i}
  S_{i,\alpha}^{\ \ [\mu}
    \ 
    \frac{
      \pat \chi_{\rm in-out}^{(2)}
    }{
      \pat S_{i,\nu]\alpha}
  }
  \ .
\end{align}
\ese
The in-in observables are no longer equivalent to the in-out ones.
Thus, the $i\eps$-prescription of the internal worldline propagator is now significant while the graviton propagators of the 2PM observables are still insensitive to their $i\eps$-prescription.
We can, however, derive the in-in observables from the in-out ones at this order.
In particular, we find that if we evaluate the in-out observables on averaged background variables, we recover the in-in observables:\checked
\begin{align}
  \label{WQFT:Average}
  \Del X_{\rm in-in}(X_{-\infty})
  =
  \Del X_{\rm in-out}
  \Big(
  \frac12(X_{\infty}+X_{-\infty})
  \Big)
  +
  \mO(G^3)
  \ .
\end{align}
Here, $\Del X$ could be any of the observables $\Del p^\mu$ or $\Del S^\mn$ (or $\Del \psi^\mu$).
The observables depend on the background parameters which are collectively indicated by $X$.
They can be taken as $p_i^\mu$, $S^\mn_i$ and $b^\mu$.
The subscripts $-\infty$ and $\infty$ on $X$ then refers to whether they are defined at past or future infinity (as in Eq.~\ref{WEFT:Kick}).
The variables at future infinity are given in terms of the initial parameters as $X_{\infty}=X_{-\infty}+\Del X_{\rm in-in}(X_{-\infty})$.
We insert the future variables in terms of the initial parameters and omit the minus subscript:\checked
\begin{align}
  \Del X_{\rm in-in}(X)
  =
  \Del X_{\rm in-out}
  \Big(
  X+\frac12 \Del X_{\rm in-in}(X)
  \Big)
  +
  \mO(G^3)
  \ .
\end{align}
The right hand side is now expanded in $G$ and for the 2PM observables we find:\checked
\begin{align}
  \label{WQFT:InFromOut}
  \Del X_{\rm in-in}^{(2)}
  =
  \Del X_{\rm in-out}^{(2)}
  +
  \frac12
  \sum_Y
  \frac{\pat \Del X_{\rm in-out}^{(1)}}{
    \pat Y
  }
  \Del Y^{(1)}_{\rm in-out}
  \ .
\end{align}
Again, $\Del X$ is either the impulse or spin kick and the sum on $Y$ runs over a set of independent background variables such as $p_i^\mu$, $S_i^\mn$ and $b^\mu$.
Also, all variables in Eq.~\eqref{WQFT:InFromOut} are evaluated in terms of the background parameters at past infinity.
This formula gives the 2PM (causal) in-in observables in terms of the in-out observables.
Scattering angles in arbitrary dimensions at 2PM without spin were derived in Ref.~\cite{Cristofoli:2020uzm} and agrees with the spinless versions of the above results.

In principle, we require $\Del b^\mu$ at 1PM order.
We may, however, use the total angular momentum $J^\mu$ as variable instead of $b^\mu$.
That is useful because the kick of $J^\mu$ is zero at 1PM order (it would generally be expected to vanish in the in-out theory).
In that case we only have to include $p_i^\mu$ and $S_i^\mn$ in the sum on $Y$ in Eq.~\eqref{WQFT:InFromOut} and the 2PM in-in observables are fully determined from the 2PM in-out eikonal.
One might also use the total momentum $P^\mu$ and relative momentum $p^\mu$ instead of $p_i^\mu$.
At 1PM the change in $P^\mu$ is also zero so that the sum on $Y$ is even smaller.

We may justify Eq.~\eqref{WQFT:Average} by arguing that the in-out prescription of the worldline fields amounts to specifying background parameters at $\tau=0$.
It is, however, not clear how that formula is generalized to higher orders in $G$.
The 2PM eikonal itself is invariant under shifts of the background parameters:\checked
\begin{align}
  \chi_{\rm in-out}[X+\Del X]
  =
  \chi_{\rm in-out}[X]
  +
  \mO(G^3)
  \ .
\end{align}
Again, one may consider generalizations of this formula to higher orders in $G$.

As discussed around Eq.~\eqref{WQFT:Amplitude}, the WQFT in-out eikonal is directly related to the $2\to2$ QFT scattering amplitude.
The $2\to2$ scattering amplitude was computed to $\mO(G^2,S^2)$ in the papers~\cite{Bern:2020buy,Kosmopoulos:2021zoq} and using the spinning generalization to Eq.~\eqref{WQFT:Amplitude} the eikonal presented here matches exactly the results obtained there.
Note, that in their work, the canonical variables are often used in favor of the covariant ones.
It is, however, simpler to carry out the comparison with their results in terms of covariant variables.
In Refs.~\cite{Bern:2020buy,Kosmopoulos:2021zoq} a formula for obtaining (causal) in-in observables directly from the eikonal (with canonical variables) is presented which is different from the method presented here.
The direct relation of those two approaches is not yet clear.

\chapter{Two-Loop Integrals with Retarded Propagators
}
\label{sec:TL}
The two-loop integrals that are necessary for the computation of worldline observables at the third post-Minkowskian order introduce new complexity and non-trivial dependence on the Lorentz factor $\gam$.
As presented in the introduction Sec.~\ref{INT}, their conservative (potential) contribution was initially derived in Ref.~\cite{Bern:2019nnu} with the remaining radiative contribution first added in Ref.~\cite{Herrmann:2021tct} and today they are generally well understood~\cite{Bern:2019crd,Kalin:2020fhe,Cheung:2020gyp,DiVecchia:2021ndb,Herrmann:2021lqe,Ruf:2021egk,Bjerrum-Bohr:2021vuf,Bjerrum-Bohr:2021din,Brandhuber:2021eyq,DiVecchia:2021bdo,Riva:2021vnj,Mougiakakos:2022sic,Jakobsen:2022zsx,Jakobsen:2022psy,Jakobsen:2022fcj,DiVecchia:2022piu,Heissenberg:2022tsn,Riva:2022fru,FebresCordero:2022jts}.
These two-loop integrals introduce logarithmic dependence of the observables on the Lorentz factor $\gam$ in the form of $\operatorname{arccosh}(\gam)$ and $\log(\frac{\gam+1}{2})$.

The integrals that appear in the WQFT approach are generally simpler than those of the QFT-amplitudes approach because the time-component of loop momenta are explicitly constrained by energy-conserving delta functions.
However with methods like the velocity cuts of Refs.~\cite{Bjerrum-Bohr:2021din,Bjerrum-Bohr:2021wwt} this property of the QFT-amplitudes integrals is made manifest.
Also, the WQFT in-in formalism consistently uses retarded propagators which at the time of Ref.~\cite{Jakobsen:2022psy} was not previously considered for these two-loop integrals (though, naturally, retarded propagators have previously been used in the PM formalism in e.g. Refs.~\cite{Bel:1981be,Westpfahl:1979gu,Saketh:2021sri}).

The computation of the post-Minkowskian integrals is achieved with a streamlined methodology where their $\gam$ dependence is bootstrapped by differential equations and boundary values to those equations are given by the post-Newtonian expansion of the same integrals.
Here, advanced integration techniques such as integration-by-parts relations (IBP relations) \cite{Chetyrkin:1981qh,Laporta:2000dsw}, differential equations \cite{Kotikov:1990kg,Bern:1993kr,Remiddi:1997ny,Gehrmann:1999as,Henn:2013pwa,Henn:2013nsa} and method of regions \cite{Beneke:1997zp,Smirnov:2002pj,Smirnov:2012gma,Becher:2014oda} are essential.

In this chapter we discuss the computation of the two-loop integrals with retarded propagators that appear in the WQFT formalism for the derivation of worldline observables at the third post-Minkowskian order.
In the first Sec.~\ref{sec:TLReduction} we focus on the identification of relevant integral families and their reduction to scalar master integrals.
Of those the comparable mass integrals are the only ones exhibiting non-trivial dependence on $\gam$ and their derivation is the topic of the next Sec.~\ref{sec:TLComparable} with differential equations and the method of regions.
Finally, in the last Sec.~\ref{sec:TLIntegral} we present expressions for the master integrals and discuss the evaluation of boundary (post-Newtonian) integrals.
The content of this chapter is mostly based on Refs.~\cite{Jakobsen:2022psy,Jakobsen:2022fcj}.

We note that in this chapter we will distinguish the infinitesimal propagator pole displacement and the deviation of the space-time dimension $d$ from four both of which are usually denoted by $\eps$.
Thus we define,\checked
\begin{align}
  d=4-2\eps
  \ ,
\end{align}
and use $i\zplus$ to denote the propagator pole displacement with e.g. the retarded prescription given by:\checked
\begin{align}
  D_{\rm R}(k)
  =
  \frac{1}{
    (k^0+i\zplus)^2-\vct{k}^2}
  \ .
\end{align}
Thus, in this relation $\zplus$ is used to represent a positive infinitesimal (real) number which is conventionally, and in the other parts of this thesis, denoted $\eps$.
Using two different variables for $\eps$ and $i\zplus$, however, avoids confusion in this chapter.

\section{Integral Families and Reduction to Master Integrals}
\label{sec:TLReduction}
In this section we describe the reduction of the integration problem of the worldline observables to the integration of a finite set of master integrals using the techniques of tensor reduction and IBP-reduction.
In addition we discuss the additional symmetries of our integrals which further reduce the basis of master integrals.
It is a fascinating fact that multi-loop integral families in dimensional regularization form a vector space~\cite{Frellesvig:2019uqt} and from that perspective we are identifying a suitable basis of (integral) vectors for the worldline observables.

It is interesting that all steps and techniques of this section are insensitive to the $i\zplus$-prescription except the discussion of symmetries.
Thus, for all those parts the results are universal to different choices of propagators including the in-in retarded and the in-out time-symmetric prescriptions.

\subsection{Integral Families and Top Sectors}
\label{TL:IFTS}
Our first task is to identify the relevant loop integrals to the worldline observables at third post-Minkowskian order and at this point we are thus interested only in the parts of the WQFT diagrams that lead to the loop integrals.
Here, the propagator structure is of most importance as this determines the denominators of the resulting loop integrals.
Generally, the vertex rules then introduce loop momenta in the numerator which we will eliminate with tensor reduction in Sec.~\ref{TL:TR}.
In addition we must enforce conservation of four-momentum in the bulk and energy on the worldlines.

\begin{figure}
  \renewcommand*\thesubfigure{\arabic{subfigure}} 
  \centering
  \begin{subfigure}{\figLa\textwidth}
    \centering
    \includegraphics{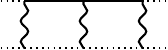}
    \caption{}
  \end{subfigure}
  \begin{subfigure}{\figLa\textwidth}
    \centering
    \includegraphics{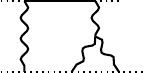}
    \caption{}
  \end{subfigure}
  \begin{subfigure}{\figLa\textwidth}
    \centering
    \includegraphics{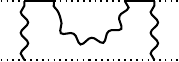}
    \caption{}
  \end{subfigure}
  \begin{subfigure}{\figLa\textwidth}
    \centering
    \includegraphics{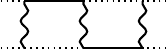}
    \caption{}
  \end{subfigure}
  \begin{subfigure}{\figLa\textwidth}
    \centering
    \includegraphics{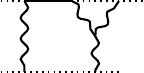}
    \caption{}
  \end{subfigure}
  \begin{subfigure}{\figLa\textwidth}
    \centering
    \includegraphics{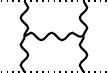}
    \caption{}
  \end{subfigure}
  \caption{
    The six top sectors relevant to the worldline observables at the third post-Minkowskian order.
    All diagrams contributing to the worldline observables (given in Figs.~\ref{WQFT:ComparableGraphs} and~\ref{WQFT:ProbeGraphs}) can be derived from these by insertion of an external line, particle exchange symmetry and/or pinching (and possibly insertion of a trivial graviton propagator).
  }
  \label{TL:fig:TopSectors}
\end{figure}

We are thus interested in determining the most general kind of propagator structure of the WQFT diagrams of the worldline observables at 3PM order.
All relevant diagrams were shown in Figs.~\ref{WQFT:ComparableGraphs} and~\ref{WQFT:ProbeGraphs}.
We ignore both the external line and vertex rules since these only introduce loop momenta in the numerator.
In this way we identify (up to particle exchange symmetry) six top sectors which together describe the most general propagator structure of the graphs.
They are shown in Fig.~\ref{TL:fig:TopSectors} where solid lines denote any worldline fluctuation and wiggly lines gravitons.

All diagrams for the WQFT eikonal can be derived from the top sectors by pinching internal lines and all diagrams for the worldline observables are then subsequently derived by adding an external worldline leg in all possible manners.
As an example graph (5) of Fig.~\ref{WQFT:ComparableGraphs} can be obtained from either of the top sectors (4) or (5) by pinching a worldline propagator or graviton respectively and adding an external line.
A pinched propagator is simply shrunk into a point.
In a few cases a graviton propagator without dependence on the loop momenta has to be introduced.
This is the case with graphs (5) and (6) of Fig.~\ref{WQFT:ProbeGraphs} which both belong to the top sector (2) of Fig.~\ref{TL:fig:TopSectors}.
The top sectors (1) and (2) describe diagrams with mass scaling $m_1m_2^3$ which are the the probe limit graphs Fig.~\ref{WQFT:ProbeGraphs}.
The remaining top sectors (3) - (6) describe comparable mass graphs Fig.~\ref{WQFT:ComparableGraphs} with mass scaling $m_1^2 m_2^2$.

With one small exception, all integrals relevant to the six top sectors are included in the following general integral family:\checked
\begin{align}\label{TL:GeneralFamily}
  &\qquad
  \mI
  ^{(i_1,i_2)\sig_1\sig_2\sig_3}
  _{n_1,n_2,n_3,n_4,n_5,n_6,n_7}
  (|q^\mu|,\gamma,d)
  =
  \int_{l_1l_2}
  \frac{
    \dd(l_1\cdot v_{\bar \imath_1})
    \dd(l_2\cdot v_{\bar \imath_2})\
  }{
    \prod_{j=1}^7
    D_j^{n_j}
  }
  \ ,
  \span
  \span
  \span
  \span
  \\[4pt]
    &
  D_{1} = l_1\cdot v_{i_1} + \sig_1i\zplus
  \ ,
  &
  &
  D_{2} = l_2\cdot v_{i_2} + \sig_2i\zplus
  \ ,
  &&
  D_{3} = (k^0+\sig_3i\zplus)^2-\vct{k}^2
  \ ,
  \nn
  \\
  &
  D_{4} = l_1^2
  \ ,
  &&
  D_{5} = l_2^2
  \ ,
  &&
  \nn
  \\
  &
  D_{6} = (l_1+q)^2
  \ ,
  &&
  D_{7} = (l_2+q)^2
  \ .
  &&
  \nn
\end{align}
The indices $i_j$ take on the values $1$ or $2$ corresponding to the two particles.
The bar on the subscript $\bar\jmath$ of $i_{\bar\jmath}$ swaps the particle indices: $\bar 1=2$ and $\bar 2=1$.
The momentum $k^\mu$ is defined by $k^\mu=l_1^\mu+l_2^\mu+q^\mu$.
The variables $\sig_j$ label the $i\zplus$-prescription of the first three denominators and take on the values $\pm$.
The $i\zplus$-prescription of those denominators is either retarded or advanced according to the signs $+$ or $-$ respectively.
Advanced propagators appear if the labelling of momenta points against the flow of causality.
The $i\zplus$-prescription of the remaining four gravitons does not play a role which we will discuss in more detail in Sec.~\ref{sec:TLComparable}.
On the left-hand-side of Eq.~\eqref{TL:GeneralFamily} we indicated the dependence of the integral on the magnitude of $q^\mu$, the Lorentz factor $\gam$ and the dimension $d$.

The six top sectors of Fig.~\ref{TL:fig:TopSectors} are described by the integral family in Eq.~\eqref{TL:GeneralFamily} with parameters $i_j$ and $(n_j)=(n_1...n_7)$ given by the following values:\checked
\begin{align}
  (1):
  \qquad
  i_1=1\ ,i_2=1\ ,
  \qquad
  (n_j) = (1,1,1,1,1,0,0)
  \ ,
  \\
  (2):
  \qquad
  i_1=1\ ,i_2=1\ ,
  \qquad
  (n_j) = (1,0,1,1,1,1,0)
  \ ,
  \nn  \\
  (3):
  \qquad
  i_1=1\ ,i_2=2\ ,
  \qquad
  (n_j) = (1,0,1,1,0,1,0)
  \ ,
  \nn  \\
  (4):
  \qquad
  i_1=1\ ,i_2=2\ ,
  \qquad
  (n_j) = (1,1,1,1,1,0,0)
  \ ,
  \nn  \\
  (5):
  \qquad
  i_1=1\ ,i_2=2\ ,
  \qquad
  (n_j) = (1,0,1,1,1,1,0)
  \ ,
  \nn\\
  (6):
  \qquad
  i_1=1\ ,i_2=2\ ,
  \qquad
  (n_j) = (0,0,1,1,1,1,1)
  \ .
  \nn
\end{align}
The third top sector, however, cannot in general be described by the integral family Eq.~\eqref{TL:GeneralFamily}.
The two worldline propagators in this top sector have the same energy and it could happen, then, that the same linear propagator appears with different $i\zplus$-prescriptions.
This happens in graph (25) of Fig.~\ref{WQFT:ComparableGraphs} which belongs to the third top sector but is not described by the integral family Eq.~\eqref{TL:GeneralFamily}.
We introduce a second integral family that fully describes the third top sector:\checked
\begin{align}\label{IF:KCovariant}
  &\qquad
  K_{n_1...n_5}^{\sigo\sigt\sigh}
  (|q^\mu|,\gamma,d)
  =
  \int_{l_1l_2}
    \frac{
    \dd(l_1\cdot v_2)
    \dd(l_2\cdot v_1)
  }{
    \prod_{i=1}^5
    D_{(K)i}^{n_i}
    }
    \ ,
    \span
    \span
    \span
    \span
    \\[4pt]
  &
  D_{(K)1} = l_1\cdot v_1 + \sigo i\zplus
  \ ,
  &
  &
  D_{(K)2} = l_1\cdot v_1 + \sigt i\zplus
  \ ,
  &&
  D_{(K)3} = (k^0+\sigh i\zplus)^2-\vct{k}^2
  \ ,
  \nn
  \\
  &
  D_{(K)4} = l_1^2
  \ ,
  &&
  D_{(K)5} = (l_1+q)^2
  \ .
  &&
  \nn
\end{align}
This family has the same worldline propagator appearing with two different $i\zplus$ prescriptions.
We will refer to it as the radiation reaction family.
We note that if we ignore the $i\zplus$ prescriptions the $K$ family is included in the $\mI$ family.

The two integral families labelled by $\mI$ and $K$ of Eqs.~\eqref{TL:GeneralFamily} and~\eqref{IF:KCovariant} fully describe all integrals included in the six top sectors.
All loop integration relevant to the worldline observables at the third post-Minkowskian order is thus included in these integral families when we include the possibility of loop momenta in the numerator which we will analyze below in Sec.~\ref{TL:TR}.

It is practical to split the general integral family $\mI^{(i_1,i_2)\sig_1\sig_2\sig_3}_{n_1,n_2,n_3,n_4,n_5,n_6,n_7}$ into two families defined by the special choices of $(i_1=1,i_2=1)$ and $(i_1=1,i_2=2)$.
We thus define the probe family labelled by $J$ corresponding to $i_1=i_2=1$ by:\checked
\begin{align}\label{IF:JCovariant}
  &\qquad
  J^{\sigma_1\sigma_2}_{n_1...n_7}
  (|q^\mu|,\gamma,d)
  =
  \int_{l_1l_2}
  \frac{
    \dd(l_1\cdot v_2)
    \dd(l_2\cdot v_2)
  }{
    \prod_{i=1}^7
    D_{(J)i}^{n_i}
  }
  \ ,
  \span\span\span\span
  \\[4pt]
  &
  D_{(J)1} = l_1\cdot v_1 + \sigma_1i\zplus
  \ ,
  &
  &
  D_{(J)2} = l_2\cdot v_1 + \sigma_2i\zplus
  \ ,
  &&
  D_{(J)3} = (l_1+l_2+q)^2
  \ ,
  \nn
  \\
  &
  D_{(J)4} = l_1^2
  \ ,
  &&
  D_{(J)5} = l_2^2
  \ ,
  &&
  \nn
  \\
  &
  D_{(J)6} = (l_1+q)^2
  \ ,
  &&
  D_{(J)7} = (l_2+q)^2
  \ .
  &&
  \nn
\end{align}
In this family only the $i\zplus$ prescriptions of the worldline propagators are significant.
The significance of the $i\zplus$ prescriptions of graviton momenta is discussed at the beginning of Sec.~\ref{sec:TLComparable}.
The comparable mass integral family corresponding to $i_1=1$ and $i_2=2$ and labelled by $I$ is defined by:\checked
\begin{align}\label{IF:ICovariant}
  &\qquad
  I^{\sigma_1\sigma_2\sigma_3}_{n_1...n_7}
  (|q^\mu|,\gamma,d)
  =
  \int_{l_1l_2}
  \frac{
    \dd(l_1\cdot v_2)
    \dd(l_2\cdot v_1)
  }{
    \prod_{i=1}^7
    D_{(I)i}^{n_i}
  }
  \ ,
  \span\span\span\span
  \\[4pt]
  &
  D_{(I)1} = l_1\cdot v_1 + \sigma_1i\zplus
  \ ,
  &
  &
  D_{(I)2} = l_2\cdot v_2 + \sigma_2i\zplus
  \ ,
  &&
  D_{(I)3} = (k^0+\sigma_3i\zplus)^2-\vct{k}^2
  \ ,
  \nn
  \\
  &
  D_{(I)4} = l_1^2
  \ ,
  &&
  D_{(I)5} = l_2^2
  \ ,
  &&
  \nn
  \\
  &
  D_{(I)6} = (l_1+q)^2
  \ ,
  &&
  D_{(I)7} = (l_2+q)^2
  \ .
  &&
  \nn
\end{align}
Here, we used the momentum $k^\mu=l_1^\mu+l_2^\mu+q^\mu$ again.

All integral families $I$, $J$ and $K$ generally depend on $|q^\mu|$, $\gam$ and $d$.
The dependence on $|q^\mu|$ is determined by dimensional analysis and thus trivial.
Instead, the dependence on $\gam$ is not a priori constrained.
Only the comparable mass $I$ family, however, depends in a non-trivial way on $\gam$ while both the probe and radiation reaction families $J$ and $K$ have a trivial dependence on $\gam$.
This is easily realized by specializing to a specific frame and integrating out the energy conserving delta-functions.
The resulting expressions are integrals on the $(d-1)$ remaining spacial components of the loop momenta.
We will refer to those as the Euclidean representations of the integral families in contrast to the covariant representations given above.

We define the Euclidean comparable mass integral family by:\checked
\begin{align}\label{IF:IEuclidean}
  &\qquad
  \hat I^{\sig_1\sig_2\sig_3}_{n_1...n_7}(\gam,d)
  =
  \int_{\vct{l}_1\vct{l}_2}  
  \frac{1}{
    \prod_{j=1}^7
    \hat D_{(I)j}^{n_j}
  }
  \ ,
  \span\span\span\span
  \\[4pt]
  &
  \hat D_{(I)1 }= \vct{l}_1\cdot
  \vct{\hat e}
  +\sig_1i\zplus
  \ ,
  &&
  \hat D_{(I)4 }= \vct{l}_1^2
  \ ,
  &&
  \hat D_{(I)6 }= (\vct{l}_1+\vct{\hat q})^2
  \ ,
  \nn
  \\
  &
  \hat D_{(I)2 }= \vct{l}_2\cdot
  \vct{\hat e}
  +\sig_2i\zplus
  \ ,
  &&
  \hat D_{(I)5 }= \vct{l}^2_2
  \ ,
  &&
  \hat D_{(I)7 }= (\vct{l}_2+\vct{\hat q})^2
  \ ,
  \nn
  \\
  &
  \hat D_{(I)3 }=
  (\vct{l}_1+\vct{l}_2+\vct{\hat q})^2
  +
  2(\gamma-1)
  \vct{l}_1\cdot \vct{\hat e}
  \, 
  \vct{l}_2\cdot \vct{\hat e}
  +
  \sig_3
  i\zplus (\vct{l}_2-\vct{l}_1)\cdot \vct{\hat e}
  \ .
  \span\span\span\span
  \nn
\end{align}
In this representation all dependence of the integrand on $\gamma$ is explicit.
We have also factorized out the dependence on $|q^\mu|$ so we are left only with the unit vector $\hat q^\mu=(0,\vct{\hat q})$.
The spacial unit vectors $\vct{\hat e}$ and $\vct{\hat q}$ satisfy $\vct{\hat e}^2=\vct{\hat q}^2=1$ and $\vct{\hat e}\cdot \vct{\hat q}=0$.
We note that this representation of the comparable mass family matches the one used in Ref.~\cite{Bjerrum-Bohr:2021din}.
It may be derived from the covariant integral family Eq.~\eqref{IF:ICovariant} by going to the frame of $v_2$ so that $v_2=(1,\vct{0})$ and $v_1=(\gam,-\gam v \vct{\hat e})$.
In order to arrive at Eq.~\eqref{IF:IEuclidean} one must only scale $\vct{l}_2\cdot \vct{\hat e}\to\gam\vct{l}_2\cdot \vct{\hat e}$.
The $i\zplus$ prescription of this integral corresponds to the retarded prescription of the covariant representation~Eq.~\eqref{IF:ICovariant} and the way for relating the two is briefly discussed below in Eq.~\eqref{WEFT:RP2}.

The Euclidean probe integral family is:\checked
\begin{align}\label{IF:JEuclidean}
  &\qquad
  \hat J^{\sig_1\sig_2}_{n_1...n_7}(d)
  =
  \int_{\vct{l}_1\vct{l}_2}  
  \frac{1}{
    \prod_{j=1}^7
    \hat D_{(J)j}^{n_j}
  }
  \ ,
  \span\span\span\span
  \\[4pt]
  &
  \hat D_{(J)1 }= \vct{l}_1\cdot
  \vct{\hat e}
  +\sig_1i\zplus
  \ ,
  &&
  \hat D_{(J)4 }= \vct{l}_1^2
  \ ,
  &&
  \hat D_{(J)6 }= (\vct{l}_1+\vct{\hat q})^2
  \ ,
  \nn
  \\
  &
  \hat D_{(J)2 }= \vct{l}_2\cdot
  \vct{\hat e}
  +\sig_2i\zplus
  \ ,
  &&
  \hat D_{(J)5 }= \vct{l}^2_2
  \ ,
  &&
  \hat D_{(J)7 }= (\vct{l}_2+\vct{\hat q})^2
  \ ,
  \nn
  \\
  &
  \hat D_{(J)3 }=
  (\vct{l}_1+\vct{l}_2+\vct{\hat q})^2
  \ .
  \span\span\span\span
  \nn
\end{align}
The spacial vectors appearing here are the same as in Eq.~\eqref{IF:IEuclidean}.
This representation is straightforwardly derived in the frame of the second body.

Finally, the Euclidean representation of the radiation reaction integral family is:\checked
\begin{align}\label{IF:KEuclidean}
  &\qquad
  \hat K^{\sigo\sigt\sigh}_{n_1...n_5}(d)
  =
  \int_{\vct{l},\vct{k}}  
  \frac{1}{
    \prod_{j=1}^5
    \hat D_{K,j}^{n_j}
  }
  \ ,
  \span\span
  \\[4pt]
  &
  \hat D_{(K)1 }= \vct{l}\cdot
  \vct{\hat e}
  +
  \sigo i\zplus
  \ ,
  &&
  \hat D_{(K)4}= \vct{l}^2
  \ ,
  \nn
  \\
  &
  \hat D_{(K)2 }= \vct{l}\cdot
  \vct{\hat e}
  +
  \sigt i\zplus
  \ ,
  &&
  \hat D_{(K)5 }=
  (\vct{l}+\vct{\hat q})^2
  \ ,
  \nn
  \\
  &
  \hat D_{(K)3}=
  \vct{k}^2
  -
  (\vct{l}\cdot\vct{\hat e}
  +
  \sigh i\zplus
  )^2
  \ .
  \nn
\end{align}
Again, the spacial unit vectors appearing here are the same as in Eq.~\eqref{IF:IEuclidean}.
The radiation reaction Euclidean representation may be derived from the covariant representation Eq.~\eqref{IF:KCovariant} by going to the frame of the first body with $v_1=(1,\vct{0})$ and $v_2=(\gam,\gam v\vct{\hat e})$.
The vectors $l_1^\mu=(l_1^0,\vct{l})$ and $k^\mu=(k^0,\vct{k})$ of Eq.~\eqref{IF:KCovariant} correspond to $\vct{l}$ and $\vct{k}$ after rescaling $\vct{k}\to \gam v \vct{k}$ and $\vct{l}\cdot \vct{\hat e}\to \gam \vct{l}\cdot\vct{\hat e}$.

The relation of the covariant and Euclidean representations are:\checked
\bse\label{TL:CovariantToEuclidean}
\begin{align}\label{IF:ICovToEuc}
  I^{\sig_1\sig_2\sig_3}_{n_1...n_7}
  (|q|,\gam,d)
  &=
  (-1)^{n+n_1}
  \frac{
    |q|^{2(d-1-n)-n_1-n_2}
  }{
    \sqrt{\gamma^2-1}^{n_1+n_2}
  }
  \hat I^{-\sig_1\sig_2\sig_3}_{n_1...n_7}
  (\gam,d)
  \\
  J_{n_1...n_7}^{\sigo\sigt}(|q|,\gam,d)
  &=
      (-1)^{n}
  \frac{
    |q|^{2(d-1-n)-n_1-n_2}
  }{
    \sqrt{\gam^2-1}^{n_1+n_2}
  }
  \hat J_{n_1...n_7}^{\sigo\sigt}(d)
  \\
  K_{n_1...n_5}^{\sigo\sigt\sigh}
  (|q|,\gam,d)
  &=
  (-1)^n
  \frac{
    |q|^{2(d-1-n)-n_1-n_2}
  }{
    \sqrt{\gam^2-1}^{
      n_1+n_2+2n_3-d+1
    }
  }
  \hat K^{\sigo\sigt\sigh}_{n_1...n_5}(d)
\end{align}
\ese
Here, we defined $n$ by $n=\sum_{i=3...7}n_i$ for the families $I$ and $J$ and $n=\sum_{i=3...5}n_i$ for the family $K$.
The covariant and Euclidean representations of the integral families each have their advantages.
In the end, however, they describe the same integrals and if we have expressions for either version we may easily translate those to results for the other version with the formulas~\eqref{TL:CovariantToEuclidean}.

The $i\zplus$ prescription of Eq.~\eqref{IF:IEuclidean} is found from the covariant representation by rewriting the retarded propagator as,\checked
\begin{align}\label{WEFT:RP2}
  D_{\rm R}(k)
  =
  \frac1{
    k^2+i\zplus k\cdot v
  }
  \ ,
\end{align}
with $v^\mu$ some velocity which could be either of $v_i^\mu$ or $V^\mu$.
This representation of the retarded propagator is similar in appearance to the Feynman prescription.

\subsection{Tensor Reduction}
\label{TL:TR}
The (scalar) integral families $I$, $J$ and $K$ generally appear with loop momenta in the numerator, that is, as tensor integrals.
We will now consider the reduction of tensor integrals to scalar integrals.
The result is that tensor integrals of an integral family can be written as linear combinations of scalar integrals of the same family.
This requires that the denominators of the integral family are complete in a certain sense to be defined.
We will discuss tensor reduction of the comparable mass integral family $I$ which we take as a generic example from which the reduction of the other families can be derived.

We define comparable mass tensor integrals as:\checked
\begin{align}
  I^{\sig_1\sig_2\sig_3}_{n_1...n_7}
  [
  l_1^{\mu_1}...l_1^{\mu_i}
  l_2^{\nu_{1\vphantom{j}}}...l_2^{\nu_j}
  ]
  =
  \int_{l_1l_2}
  \frac{
    \dd(l_1\cdot v_2)
    \dd(l_2\cdot v_1)
  }{
    \prod_{m=1}^7
    D_{(I)m}^{n_m}
  }
  l_1^{\mu_1}
  ...
  l_1^{\mu_i}
  l_2^{\nu_1}
  ...
  l_2^{\nu_j}
\end{align}
It has $i$ factors of $l_1^\mu$ and $j$ factors of $l_2^\mu$ in the numerator.
In addition to the three scalars $|q^\mu|$, $\gam$ and $d$, the tensor integrals depend on the unit vectors $\hat q^\mu$ and $v_i^\mu$ and the metric $\eta^\mn$.
In general, we refer to the unit vectors $\hat q^\mu$ and $v_i^\mu$ and the scalars $|q^\mu|$ and $\gam$ as the external data.

The requirement on the denominators $D_{(I),m}$ is that any scalar product of two loop momenta or of a loop momenta with external data can be written in terms of the denominators.
Having two loop momenta and three external vectors, combinatorics tell us that we need nine denominators.
In our case, the energy conserving delta functions, however, allow us to assume $l_1\cdot v_2=l_2\cdot v_1=0$ which reduces this number to seven.
However, using reverse unitarity (see Sec.~\ref{sec:TLIntegral-by-Parts}) we can also think of the delta functions as denominators so that, indeed, we have nine denominators.

The denominators of the comparable mass integral family are complete in the above sense and a few examples read,\checked
\begin{align}
  \label{TL:TRExample}
  l_1\cdot v_1 = D_{(I)1}
  \ ,
  \qquad
  l_1\cdot q = \frac{
    D_{(I)6}
  -
  D_{(I)4}
  -
  q^2
  }2
  \ ,
  \qquad
  l_1\cdot l_2
  =
  \frac{
    D_{(I)3}
    -D_{(I)6}
    -D_{(I)7}
    +q^2
  }2
  \ ,
\end{align}
which are easily derived.

The completeness of the denominators ensures that the tensor integrals can be written as a sum of scalar integrals of the same integral family.
This fact may be realized inductively by deriving such reductions for low order tensor integrals.
We thus consider tensor integrals with one or two loop momenta in the numerator from which the general pattern for higher order tensor integrals can be derived.

Tensor integrals with a single loop momentum in the numerator must be given by a sum of the three external vectors with some coefficients:\checked
\begin{align}
  \label{TL:TRAnsatz}
  I^{\sig_1\sig_2\sig_3}_{n_1...n_7}
  [
  l_i^{\mu}
  ]
  =
  c_{i,1} v_1^\mu
  +
  c_{i,2} v_2^\mu
  +
  c_{i,3} q^\mu
  \ .
\end{align}
Thus, the integral depends only on the external data and must be expressible in terms of the external vectors.
The three coefficients of this ansatz are determined by dotting both sites of the equation with the same three external vectors.
In this way we get three equations which together determine $c_{i,j}$.
In our case they read:\checked
\begin{align}
  I^{\sig_1\sig_2\sig_3}_{n_1...n_7}
  [
  l_i\cdot q
  ]
  =
  c_{i,3} q^2
  \ ,\quad
  I^{\sig_1\sig_2\sig_3}_{n_1...n_7}
  [
  l_i\cdot v_1
  ]
  =
  c_{i,1} 
  +
  \gam c_{i,2} 
  \ ,\quad
    I^{\sig_1\sig_2\sig_3}_{n_1...n_7}
  [
  l_i\cdot v_2
  ]
  =
  \gam c_{i,1}
  +
  c_{i,2}
  \ .
\end{align}
Here, we generalized the notation with square brackets to simply indicate any numerator of the $I$ integral.
The completeness of our basis of denominators $D_{I,j}$ implies that the left-hand-sides of these equations can be written in terms of scalar integrals of the same integral family.
If we specialize to $l_i=l_1$ we can use the examples Eq.~\eqref{TL:TRExample} to reduce the left-hand-sides:\checked
\begin{align}
  I^{\sig_1\sig_2\sig_3}_{n_1...n_7}
  [
  l_1\cdot q
  ]
  &=
  \frac12
  I^{\sig_1\sig_2\sig_3}_{n_1,n_2,n_3,n_4,n_5,n_6-1,n_7}
  -
  \frac12
  I^{\sig_1\sig_2\sig_3}_{n_1,n_2,n_3,n_4-1,n_5,n_6,n_7}
  -
  \frac12
  q^2
  I^{\sig_1\sig_2\sig_3}_{n_1,n_2,n_3,n_4,n_5,n_6,n_7}
  \ ,
  \nn
  \\
  I^{\sig_1\sig_2\sig_3}_{n_1...n_7}
  [
  l_1\cdot v_1
  ]
  &=
  I^{\sig_1\sig_2\sig_3}_{n_1-1,n_2,n_3,n_4,n_5,n_6,n_7}
  \ ,
  \nn
  \\
  I^{\sig_1\sig_2\sig_3}_{n_1...n_7}
  [
  l_1\cdot v_2
  ]
  &=
  0
  \ .
\end{align}
The coefficients $c_{i,j}$ of our ansatz Eq.~\eqref{TL:TRAnsatz} are now related to scalar integrals through linear equations which are easily inverted.

The tensor integrals with two momenta in the numerator are expanded on all possible outer products of the external vectors and the metric.
Let us consider only the case of two symmetrized loop momenta where the ansatz may be assumed to be symmetric in the two indices.
We get:\checked
\begin{align}
  \label{TL:TRE1}
  I^{\sig_1\sig_2\sig_3}_{n_1...n_7}
  [
  l_i^{(\mu} l_j^{\nu)}
  ]
  =
  c_{1} v_1^\mu v_1^\nu
  +
  2c_{2} v_1^{(\mu} v_2^{\nu)}
  +
  c_{3} v_2^\mu v_2^\nu
  +
  c_{4} q^\mu q^\nu
  +
  2c_{5} q^{(\mu}_{\vphantom{1}} v_1^{\nu)}
  +
  2c_{6} q^{(\mu}_{\vphantom{2}} v_2^{\nu)}
  +
  c_{7} \eta^\mn
  \ .
\end{align}
Here we have seven (new) coefficients $c_{n}$ corresponding to the seven possible tensor structures of the right hand side.
In principle we would need a set of seven coefficients for each combination of loop momenta $i$ and $j$ but we hide that dependence of the coefficients.
Contracting both sides with the same seven tensor structures gives us seven linear equations which determine the seven coefficients.
The reduction of the left-hand-sides to scalar integrals after contraction with one of the tensor structures follow the same steps as with the tensor integrals with one index above.

The reduction of higher order tensor integrals to scalar integrals quickly involve numerous tensor structures and their expressions easily get lengthy.
As a start, we can pick a different basis on which we expand the integrals.
Here it is practical to use $v_{i\bot}^\mu$ from Eqs.~\eqref{WEFT:VBot} instead of $v_i^\mu$.
In terms of these vectors, the ansatz Eq.~\eqref{TL:TRE1} with $l_1^\mu$ becomes:\checked
\begin{align}
  I^{\sig_1\sig_2\sig_3}_{n_1...n_7}
  [
  l_1^{\mu} l_1^{\nu}
  ]
  =
  c_{1} v_{2\bot}^\mu v_{2\bot}^\nu
  +
  c_{2} q^\mu q^\nu
  +
  2c_{3} q^{(\mu}_{\vphantom{2\bot}} v_{2\bot}^{\nu)}
  +
  c_{4} \eta^\mn
  \ .
\end{align}
This amounts to taking advantage of the fact that $l_1^\mu$ is orthogonal to $v_2^\mu$.
Also, it is practical to use the metric,\checked
\begin{align}
  \eta_{(d-3)}^\mn
  =
  P_{12}^\mn
  -
  \hat q^\mu
  \hat q^\nu
  \ ,
\end{align}
which is orthogonal to the subspace of the three external vectors.
With this choice it is much simpler to solve for the coefficients of the ansatz.

The general result of tensor reduction is that all loop integration appearing in the 3PM worldline observables can be reduced to the computation of the scalar integral families defined above in Sec.~\ref{TL:IFTS}.
While both of the integral families $I$ and $J$ are complete in the sense discussed here, the family $K$ is not, but instead it is an iterated integral.
The tensor reduction of the $K$ family may then be performed first on the inner integration and then on the outer.

In the tensor reduction discussed until now we work consistently in $d$ dimensions.
We may also take advantage of the fact that we are only interested in the integrals in $d=4$ dimensions.
Here, we may construct a four dimensional basis from our external vectors by adding $\eps_{\mn\ab} q^\nu v_1^\alpha v_2^\beta$ where the four-dimensional Levi-Civita symbol is defined in the four-dimensional subspace of the $d$-dimensional space.
Loop momenta may then be divided into the four-dimensional part that is expanded on the four-dimensional basis and a part that lives in the $d-4$ extra dimensions.

\subsection{Integration-by-Parts Relations and Symmetries}
\label{sec:TLIntegral-by-Parts}
In this section we discuss integration-by-parts and symmetry relations.
These are linear relations satisfied by the members of the integral families.
Of most importance, perhaps, are the IBP relations with which the infinite family of integrals are reduced to a finite set of independent integrals.
These are referred to as the master integrals and are analogous to a basis of vectors in linear algebra.
A set of master integrals for the $I$ and $J$ families will be presented in the next section~\ref{sec:TLMaster}.
The symmetry relations generally introduce further linear relations between the integrals and consequently reduce the number of master integrals.
The symmetry relations of integrals with retarded $i\zplus$-prescription are different from those with Feynman prescription and one therefore has to be careful with the handling of symmetries.

IBP relations are derived using integration by parts where the boundary terms in dimensional regularization are put to zero.
Let us use the comparable mass integral family as an example.
In order to use the covariant representation of those integrals it is then practical to rewrite the energy-conserving delta functions as cut propagators using reverse unitarity.
The general formula reads:\checked
\begin{align}
  \frac{\dd^{(n)}(\oma)}{(-1)^n n!}
  =
  \frac{i}{
    (\oma+i\zplus)^{n+1}
  }
  -
  \frac{i}{
    (\oma-i\zplus)^{n+1}
  }
\end{align}
The superscript $(n)$ of the delta function indicates differentiations and starting from $n=1$ the general formula is derived by repeated differentiation.
This rewriting of the delta functions allow us to treat the integrand uniformly.
Instead of the I family with delta functions Eq.~\eqref{IF:ICovariant} we consider a family with the same seven denominators together with two additional ones:\checked
\begin{align}
  D_{(I)8}=l_1\cdot v_2+i\zplus
  \ ,\qquad
  D_{(I)9}=l_2\cdot v_1+i\zplus
  \ .
\end{align}
In order to get back to the original family, we cut these two additional denominators.
The fundamental equation, then, which generates IBP relations reads:\checked
\begin{align}
  \label{TL:IBP}
  0=
  \int_{l_1l_2}
  \frac{
    \partial
  }{
    \partial
    l_i^\mu
  }
  y^\mu
  \frac{
    1
  }{
    \prod_{i=1}^9
    D_{(I)i}^{n_i}
  }
  \ .
\end{align}
Here, the vector $y^\mu$ is any of the loop momenta $l_i^\mu$ or external four-vectors $v_i^\mu$ or $q^\mu$.
The partial derivative with respect to the loop momenta hits the denominators $D_{I,i}$ and may hit the four-vector $y^\mu$.

With Eq.~\eqref{TL:IBP} all IBP-relations of the (generalized) I integral family are derived.
They allow us to rewrite arbitrary members of the I integral family in terms of a set of master integrals.
The reduction of the integral family to master integrals can be done with the Laporte algorithm~\cite{Laporta:1996mq,Laporta:2000dsw} and is conveniently done with available packages such as Fire~\cite{Smirnov:2019qkx}, LiteRed~\cite{Lee:2012cn,Lee:2013mka} or Kira~\cite{Maierhofer:2017gsa,Klappert:2020nbg}.
A specific choice of master integrals is discussed below in Sec.~\ref{sec:TLMaster}.
The master integrals may be chosen such that the two additional denominators $D_{I,8}$ and $D_{I,9}$ appear only linearly.
By cuts they are then transformed into delta functions and we arrive at master integrals defined within the original I integral family Eq.~\eqref{IF:ICovariant}.
A convenient property of the IBP-relations is that they are insensitive to the $i\zplus$-prescriptions of the denominators.
Thus, the IBP relations for $I^{\sigo\sigt\sigh}_{n_1...n_7}$ are the same regardless of the superscripts $\sign$.

Next, we discuss symmetries of the integral families which are relabelings of the integrals that match them to other integrals of the same integral family.
The crucial difference in our case to the integrals appearing in the QFT-amplitudes approach is our use of retarded propagators $D_{\rm R}(k)$ instead of the Feynman prescription $D_{\rm F}(k)$.
Here, it is important to take care of the $i\zplus$-prescriptions when flipping the signs of momenta.
While the Feynman prescription is insensitive to this operation $D_{\rm F}(-k)=D_{\rm F}(k)$ the retarded propagator turns into the advanced one $D_{\rm R}(-k)=D_{\rm A}(k)$.
For the worldline propagators one must also take care of possible signs under this operation:\checked
\begin{align}
  \frac{1}{
    -\oma+i \zplus
  }
  =
  -
  \frac{1}{
    \oma-i \zplus}
  \ .
\end{align}
In general, then, the symmetry relations will mix integrals with different $i\zplus$-prescriptions.

The I integral family Eq.~\eqref{IF:ICovariant} enjoys three general symmetry relations:\checked
\begin{align}
  \nn
  &
  \text{shift:}
  &&
  I
  ^{\sigo\sigt\sigh}
  _{n_1,n_2,n_3,n_4,n_5,n_6,n_7}
  =
  I
  ^{\sigo\sigt\sigh}
  _{n_1,n_2,n_3,n_6,n_7,n_4,n_5}
  &&
  (l_i^\mu\to l_i^\mu+q^\mu, q^\mu\to-q^\mu)
  \\\nn
  &
  \text{flip:}
  &&
  I
  ^{\sigo\sigt\sigh}
  _{n_1,n_2,n_3,n_4,n_5,n_6,n_7}
  =
  (-1)^{n_1+n_2}I
  ^{-\sigo-\sigt-\sigh}
  _{n_1,n_2,n_3,n_4,n_5,n_6,n_7}
  &&
  (l_i^\mu\to -l_i^\mu,q^\mu\to-q^\mu)
  \\\label{TL:Syms}
  &
  \text{exchange:}
  &&
  I
  ^{\sigo\sigt\sigh}
  _{n_1,n_2,n_3,n_4,n_5,n_6,n_7}
  =
  I
  ^{\sigt\sigo\sigh}
  _{n_2,n_1,n_3,n_5,n_4,n_7,n_6}
  &&
  (l_1^\mu\leftrightarrow l_2^\mu,v_1^\mu\leftrightarrow v_2^\mu)
\end{align}
We refer to these as the shift, flip and exchange symmetries respectively.
They generally act on both $\sig_i$ and $n_i$.
The IBP-relations result in master integrals that are insensitive to the $i\zplus$ prescriptions and for each such master we thus have up to $2^3$ combinations of prescriptions.
The exact number depends on which of the first three denominators that are present in the given master integral.
Namely, the $i\zplus$ prescription of the denominators $D_i$ with $i$ being 1, 2 or 3 only matters if the respective index $n_i$ is (strictly) positive.
With the inclusion of the symmetry relations above, the master integrals may then be related among themselves and the number of independent master integrals is reduced.
We will refer to an analogous set of symmetries of the $J$ integrals by the same terms, shift, flip and exchange symmetries:\checked
\begin{align}
  &
  \text{shift:}
  &&
  J
  ^{\sigo\sigt}
  _{n_1,n_2,n_3,n_4,n_5,n_6,n_7}
  =
  J
  ^{\sigo\sigt}
  _{n_1,n_2,n_3,n_6,n_7,n_4,n_5}
  &&
  (l_i^\mu\to l_i^\mu+q^\mu, q^\mu\to-q^\mu)
  \nn
  \\
  &
  \text{flip:}
  &&
  J
  ^{\sigo\sigt}
  _{n_1,n_2,n_3,n_4,n_5,n_6,n_7}
  =
  (-1)^{n_1+n_2}J
  ^{-\sigo-\sigt}
  _{n_1,n_2,n_3,n_4,n_5,n_6,n_7}
  &&
  (l_i^\mu\to -l_i^\mu,q^\mu\to-q^\mu)
  \nn
  \\
  &
  \text{exchange:}
  &&
  J
  ^{\sigo\sigt}
  _{n_1,n_2,n_3,n_4,n_5,n_6,n_7}
  =
  J
  ^{\sigt\sigo}
  _{n_2,n_1,n_3,n_5,n_4,n_7,n_6}
  &&
  (l_1^\mu\leftrightarrow l_2^\mu)
  \label{TL:JSyms}
\end{align}
The symmetries Eqs.~\eqref{TL:Syms} and~\eqref{TL:JSyms} could conveniently be defined directly for the generic family $\mI$ of Eq.~\eqref{TL:GeneralFamily}.

The symmetries Eqs.~\eqref{TL:Syms} and~\eqref{TL:JSyms} are generally obeyed by the $I$ and $J$ integral families.
There may however be additional symmetries when the indices of some denominators vanish.
Thus, as an example, if the third index of either family is zero, the integral family factorizes into a product of two one-loop integrals each with independent shift and flip symmetries.

Let us finally discuss two important symmetry relations relevant to the worldline propagators: Cut and partial fraction relations.
We work with a special example and define the integrals $I^{\sigo\sigt}$ and $J^{\sigo\sigt}$ by:\checked
\bse\label{TL:IJ}
\begin{align}
  \label{TL:Iot}
  I^{(\sigo\sigt)}
  &=
  I_{1,1,1,1,1,0,0}^{\sigo\sigt\sigh}
  =
  \int_{l_1l_2}
  \frac{
    \dd(l_1\cdot v_2)\dd(l_2\cdot v_1)
  }{
    (
    l_1\cdot v_1+\sigo i\zplus
    )
    (
    l_2\cdot v_2+\sigo i\zplus
    )
    l_1^2
    l_2^2
    (l_1+l_2+q)^2
  }
  \\
  \label{TL:Jot}
  J^{(\sigo\sigt)}
  &=
  J_{1,1,1,1,1,0,0}^{\sigo\sigt}
  =
  \int_{l_1l_2}
  \frac{\dd(l_1\cdot v_2)\dd(l_2\cdot v_2)}{
    (
    l_1\cdot v_1+\sigo i\zplus
    )
    (
    l_2\cdot v_1+\sigo i\zplus
    )
    l_1^2
    l_2^2
    (l_1+l_2+q)^2
  }
\end{align}
\ese
In the first line, the variable $\sigh$ determines the retarded prescription of the final propagator of that line.
This retarded prescription is not printed explicitly but the following manipulations only assume a consistent prescription for the third graviton.
Both integrals will appear in our basis of master integrals.
The particle exchange symmetry identifies $I^{+-}=I^{-+}$ and $J^{+-}=J^{-+}$ and for the $J$ integrals, the flip symmetry identifies $J^{++}=J^{--}$ (For the $I$ integrals this flip symmetry also affects the active propagator, i.e. $\sigh$).
Naively we thus have five independent integrals.

Cuts are defined by considering the difference between two propagators with opposite $i\zplus$ prescriptions.
As an example, we consider $I^{(++)}-I^{(+-)}$ where the only difference between the two terms is the second worldline propagator.
The cut,\checked
\begin{align}
  \frac1{l_2\cdot v_2+i\zplus}
  -
  \frac1{l_2\cdot v_2-i\zplus}
  =
  -i\dd(l_2\cdot v_2)
  \ ,
\end{align}
introduces a delta function of the energy $l_2\cdot v_2$.
The cut $I^{(++)}-I^{(+-)}$ thus has delta functions of both $l_1\cdot v_2$ and $l_2\cdot v_2$ and must belong to the $J$ family.
In fact, one easily realizes that it is mapped to the exact same cut of the $J$ integrals:\checked
\bse
\begin{align}
  I^{(++)}-I^{(+-)}
  &=
  J^{(++)}-J^{(+-)}
  \ ,
  \\
  I^{(--)}-I^{(+-)}
  &=
  J^{(--)}-J^{(+-)}
  \ .
\end{align}
\ese
The same reasoning goes through for the other cut $I^{(--)}-I^{(+-)}$ resulting in the identity of the second line.
Together with the identity between $J^{(++)}$ and $J^{(--)}$ we see that $I^{(++)}$ and $I^{(--)}$ must be identical too.
The initial five integrals are now reduced to a set of three independent integrals which may taken as $I^{(++)}$, $J^{(++)}$ and $J^{(+-)}$.

Finally, with the use of a partial fraction identity, we reduce the three independent integrals to two.
The partial fraction identity is applied to the $J$ family and reads:\checked
\begin{align}
  \frac1{\omao(\omao+\omat)}
  +
  \frac1{\omat(\omao+\omat)}
  =
  \frac1{\omao\omat}
  \ .
\end{align}
In our case $l_i\cdot v_1$ will play the role of $\omega_i$ and we must take care of their $i\zplus$ prescriptions:\checked
\begin{align}
  \label{TL:PF1}
  \frac1{(\omao+i\zplus)(\omao+\omat+i\zplus)}
  +
  \frac1{(\omat+i\zplus)(\omao+\omat+i\zplus)}
  =
  \frac1{(\omao+i\zplus)(\omat+i\zplus)}
  \ .
\end{align}
Here, we let all of them be positive.
The partial fraction identity is applied by realizing that $J^{(+-)}$ may be written as follows after the relabelling $l_1\to l_1+l_2+q$, $l_2\to-l_2$ and $q\to-q$:\checked
\begin{align}
  J^{(+-)}
  =
  -\int_{l_1l_2}
  \frac{
    \dd(l_1\cdot v_2)\dd(l_2\cdot v_2)
  }{
    (l_1\cdot v_1+l_2\cdot v_1+i\zplus)
    (l_2\cdot v_1+i\zplus)
    l_1^2
    l_2^2
    (l_1+l_2+q)^2
  }
  \ .
\end{align}
The integral $J^{(+-)}$ thus plays the role of the two terms of the left-hand-side of Eq.~\eqref{TL:PF1} and the integral $J^{(++)}$ plays the role of the right-hand-side.
We get the identity:\checked
\begin{align}\label{TL:PF2}
  -2J^{(+-)}=J^{(++)}
  \ .
\end{align}
In fact, the initial two $J$ integrals of Eq.~\eqref{TL:Jot} are fully constrained by this identity together with their cut.
In general, the initial five integrals are now reduced to two independent integrals which can be taken as $I^{(++)}$ and $J^{(++)}$.

\subsection{Master Integrals
}
\label{sec:TLMaster}
The IBP-relations allow us to write all integrals of the integral families in terms of a finite set of master integrals.
In this section we present sets of master integrals for the (retarded) $I$ and $J$ integral families that are sufficient for all integrals appearing in the computation of worldline observables at the third post-Minkowskian order.

As discussed above, the IBP-relations give a set of master integrals for each choice of $i\zplus$.
Naively, we then have $2^3=8$ copies of integrals for the $I$ integral family and $2^2=4$ copies for the $J$ integral family after taking into account $i\zplus$.
The symmetries Eqs.~\eqref{TL:Syms} and~\eqref{TL:JSyms} allow us, though, to relate most of the different $i\zplus$ combinations to each other.
Together, then, with cuts and partial fractions all integrals may be derived from the ones with all $i\zplus$ positive.
In this section, then, we will generally ignore the $i\zplus$ prescriptions.

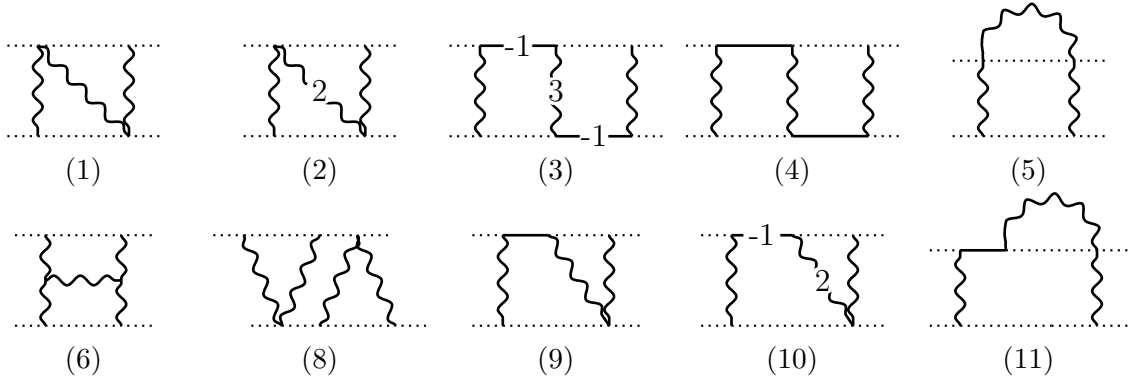
\begin{figure}
  \renewcommand*\thesubfigure{\arabic{subfigure}} 
  \centering
  \begin{subfigure}{.18\textwidth}
    \centering
    \includegraphics{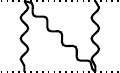}
    \caption{}
  \end{subfigure}
  \begin{subfigure}{.18\textwidth}
    \centering
    \includegraphics{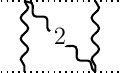}
    \caption{}
  \end{subfigure}
  \begin{subfigure}{.18\textwidth}
    \centering
    \includegraphics[trim=0 .31cm 0cm 0]{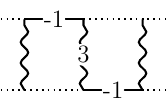}
    \caption{}
  \end{subfigure}
  \begin{subfigure}{.18\textwidth}
    \centering
    \includegraphics{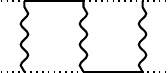}
    \caption{}
  \end{subfigure}
  \begin{subfigure}{.18\textwidth}
    \centering
    \includegraphics{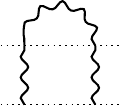}
    \caption{}
  \end{subfigure}
  \begin{subfigure}{.18\textwidth}
    \centering
    \includegraphics{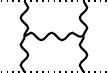}
    \caption{}
  \end{subfigure}
  \begin{subfigure}{.18\textwidth}
    \setcounter{subfigure}{7}
    \centering
    \includegraphics{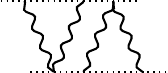}
    \caption{}
  \end{subfigure}
   \begin{subfigure}{.18\textwidth}
    \setcounter{subfigure}{8}
    \centering
      \begin{tikzpicture}
        \coordinate (ul) at (-.8 ,0) ;
        \coordinate (ulp)at (-.2 ,0) ;
        \coordinate (ur) at (+.6 ,0) ;
        \coordinate(ull) at (-1.2  ,0) ;
        \coordinate(urr) at ( 1  ,0) ;
        \coordinate (dl) at (-.8 ,-1.2) ;
        \coordinate (dr) at (+.6 ,-1.2) ;
        \coordinate(dll) at (-1.2  ,-1.2) ;
        \coordinate(drr) at (1   ,-1.2) ;
        \draw[dotted,line width=.8pt] (dl) -- (dr) ;
        \draw[dotted,line width=.8pt] (ulp) -- (ur) ;
        \draw[line width=1pt] (ul) -- (ulp) ;
        \draw[snake it] (ul) -- (dl) ;
        \draw[snake it] (ulp) -- (dr) ;
        \draw[snake it] (ur) -- (dr) ;
        \draw[dotted,line width=.8pt] (ull) -- (ul) ;
        \draw[dotted,line width=.8pt] (urr) -- (ur) ;
        \draw[dotted,line width=.8pt] (dll) -- (dl) ;
        \draw[dotted,line width=.8pt] (drr) -- (dr) ;
      \end{tikzpicture}
      \caption{
      }
  \end{subfigure}
  \begin{subfigure}{.18\textwidth}
    \centering
      \begin{tikzpicture}
        \coordinate (ul) at (-1 ,0) ;
        \coordinate (ulp)at (-.2 ,0) ;
        \coordinate (ur) at (+.6 ,0) ;
        \coordinate(ull) at (-1.4  ,0) ;
        \coordinate(urr) at ( 1  ,0) ;
        \coordinate (dl) at (-1 ,-1.2) ;
        \coordinate (dr) at (+.6 ,-1.2) ;
        \coordinate(dll) at (-1.4  ,-1.2) ;
        \coordinate(drr) at (1   ,-1.2) ;
        \draw[dotted,line width=.8pt] (dl) -- (dr) ;
        \draw[dotted,line width=.8pt] (ulp) -- (ur) ;
        \draw[line width=1pt] (ul) -- (ulp) ;
        \draw[snake it] (ul) -- (dl) ;
        \draw[snake it] (ulp) -- (dr) ;
        \draw[snake it] (ur) -- (dr) ;
        \draw[dotted,line width=.8pt] (ull) -- (ul) ;
        \draw[dotted,line width=.8pt] (urr) -- (ur) ;
        \draw[dotted,line width=.8pt] (dll) -- (dl) ;
        \draw[dotted,line width=.8pt] (drr) -- (dr) ;
        \node at ($(ulp)!0.5!(dr)$) {
          \tikz\draw[white,fill=white] (0,0) circle (1ex);
        };
        \node at ($(ulp)!0.5!(dr)$) {
          2
        };
        \node at ($(ulp)!0.5!(ul)$) {
          \tikz\draw[white,fill=white] (0,0) circle (1.3ex);
        };
        \node at ($(ulp)!0.5!(ul)$) {
          -1
        };
      \end{tikzpicture}
      \caption{
      }
  \end{subfigure}
    \begin{subfigure}{.18\textwidth}
    \centering
      \begin{tikzpicture}
        \coordinate (ul) at (-1.2,0) ;
        \coordinate (ulp)at (-.6 ,0) ;
        \coordinate (urp)at (.6  ,0) ;
        \coordinate (ur) at (.6,0) ;
        \coordinate(ull) at (-1.6,0) ;
        \coordinate(urr) at ( 1,0) ;
        \coordinate (dl) at (-1.2,-1) ;
        \coordinate (dm) at (0   ,-1) ;
        \coordinate (dr) at (+.6,-1) ;
        \coordinate(dll) at (-1.6,-1) ;
        \coordinate(drr) at (1 ,-1) ;
        \draw[line width=1pt] (ul) -- (ulp) ;
        \draw[line width=1pt] (ur) -- (urp) ;
        \draw[dotted,line width=.8pt] (dl) -- (dm) ;
        \draw[dotted,line width=.8pt] (dm) -- (dr) ;
        \draw[dotted,line width=.8pt] (ulp) -- (urp) ;
        \draw[snake it] (dl) -- (ul) ;
        \draw[snake it] (dr) -- (ur) ;
        \draw[snake it] (.6,0) arc (0:180:.6cm) ;
        \draw[dotted,line width=.8pt] (ull) -- (ul) ;
        \draw[dotted,line width=.8pt] (urr) -- (ur) ;
        \draw[dotted,line width=.8pt] (dll) -- (dl) ;
        \draw[dotted,line width=.8pt] (drr) -- (dr) ;
      \end{tikzpicture}
      \caption{
      }
    \end{subfigure}
    \caption{
      Here, 10 out of the total 12 master integrals of the $I$ family are drawn schematically and numbered correspondingly.
    Wiggly and solid lines indicate massless and linear propagators respectively.
    Numbers on propagators indicate their power when it is different from one.
    Routings of advanced and retarded propagators may be indicated by arrows.
    }
    \label{TL:IMastersGraphs}
\end{figure}

The master integrals are conveniently separated into $b$-type and $v$-type integrals defined from their total number of linear propagators.
Thus, integrals with $(n_1+n_2)$ even are referred to as $b$-type and integrals with $(n_1+n_2)$ odd as $v$-type.
The names refer to the fact that in the spinless case $b$-type integrals show up in front of $b^\mu$ in the impulse and $v$-type integrals in front of $v_i^\mu$.
A basis of $b$-type $I$ master integrals is given by:\checked
\bse\label{TL:IMasters}
\begin{align}
  I_{(1)}
  &=
  I_{0,0,1,1,1,0,0}
  \ ,\\\nn
  I_{(2)}
  &=
  I_{0,0,2,1,1,0,0}
  \ ,\\\nn
  I_{(3)}
  &=
  I_{-1,-1,3,1,1,0,0}
  \ ,\\\nn
  I_{(4)}
  &=
  I_{1,1,1,1,1,0,0}
  \ ,\\\nn
  I_{(5)}
  &=
  I_{0,0,1,1,0,1,0}
  \ ,\\\nn
  I_{(6)}
  &=
  I_{0,0,1,1,1,1,1}
  \ ,\\\nn
  I_{(7)}
  &=
  I_{-1,-1,1,1,1,1,1}
  \ ,\\\nn
  I_{(8)}
  &=
  I_{0,0,0,1,1,1,1}
  \ .
\end{align}
In fact, only the fourth master integral is sensitive to the $i\zplus$-prescription of the linear propagators.
A basis of the $v$-type $I$ master integrals is then given by:\checked
\begin{align}
  I_{(9)}
  &=
  I_{1,0,1,1,1,0,0}
  \ ,\\\nn
  I_{(10)}
  &=
  I_{-1,0,2,1,1,0,0}
  \ ,\\\nn
  I_{(11)}
  &=
  I_{1,0,1,1,0,1,0}
  \ ,\\\nn
  I_{(12)}
  &=
  I_{-1,0,1,1,1,1,1}
  \ .
\end{align}
\ese
The $I$ master integrals are drawn schematically in Fig.~\ref{TL:IMastersGraphs}.
For the masters that are sensitive to $i\zplus$ prescriptions we will always assume the graviton propagator to be retarded (i.e. $\sig_3=1$ in Eq.~\ref{IF:ICovariant}) and denote the $i\zplus$ prescriptions of the worldlines by superscripts when relevant i.e. $I_{(11)}^{\sig}=  I_{1,0,1,1,0,1,0}^{\sig\pm+}$ or $I_{(4)}^{\sigo\sigt}= I_{1,1,1,1,1,0,0}^{\sigo\sigt+}$.

\begin{figure}[h]
  \renewcommand*\thesubfigure{\arabic{subfigure}} 
  \centering
  \begin{subfigure}{\figLa\textwidth}
    \centering
    \begin{tikzpicture}
      \coordinate (ul) at (-1  ,0) ;
      \coordinate (um) at (0   ,0) ;
      \coordinate (ur) at (+1  ,0) ;
      \coordinate(ull) at (-1.4,0) ;
      \coordinate(urr) at ( 1.4,0) ;
      \coordinate (dl) at (-1  ,-1.2) ;
      \coordinate (dm) at (0   ,-1.2) ;
      \coordinate (dr) at (+1  ,-1.2) ;
      \coordinate(dll) at (-1.4,-1.2) ;
      \coordinate(drr) at (1.4 ,-1.2) ;
      \draw[dotted,line width=.8pt] (dl) -- (dr) ;
      \draw[dotted,line width=.8pt] (ul) -- (ur) ;
      \draw[snake it] (dl) -- (um) ;
      \draw[snake it] (dm) -- (um) ;
      \draw[snake it] (dr) -- (um) ;
      \draw[dotted,line width=.8pt] (ull) -- (ul) ;
      \draw[dotted,line width=.8pt] (urr) -- (ur) ;
      \draw[dotted,line width=.8pt] (dll) -- (dl) ;
      \draw[dotted,line width=.8pt] (drr) -- (dr) ;
    \end{tikzpicture}
    \caption{
    }
  \end{subfigure}
  \begin{subfigure}{\figLa\textwidth}
    \centering
    \begin{tikzpicture}
      \coordinate (ul) at (-1  ,0) ;
      \coordinate (um) at (0   ,0) ;
      \coordinate (ur) at (+1  ,0) ;
      \coordinate(ull) at (-1.4,0) ;
      \coordinate(urr) at ( 1.4,0) ;
      \coordinate (dl) at (-1  ,-1.2) ;
      \coordinate (dm) at (0   ,-1.2) ;
      \coordinate (dr) at (+1  ,-1.2) ;
      \coordinate(dll) at (-1.4,-1.2) ;
      \coordinate(drr) at (1.4 ,-1.2) ;
      \draw[line width=1pt] (ul) -- (um) ;
      \draw[dotted,line width=.8pt] (dm) -- (dr) ;
      \draw[dotted,line width=.8pt] (dl) -- (dm) ;
      \draw[dotted,line width=.8pt] (um) -- (ur) ;
      \draw[snake it] (dl) -- (ul) ;
      \draw[snake it] (dm) -- (um) ;
      \draw[snake it] (dr) -- (um) ;
      \draw[dotted,line width=.8pt] (ull) -- (ul) ;
      \draw[dotted,line width=.8pt] (urr) -- (ur) ;
      \draw[dotted,line width=.8pt] (dll) -- (dl) ;
      \draw[dotted,line width=.8pt] (drr) -- (dr) ;
    \end{tikzpicture}
    \caption{
    }
  \end{subfigure}
  \begin{subfigure}{\figLa\textwidth}
    \centering
    \begin{tikzpicture}
      \coordinate (ul) at (-1  ,0) ;
      \coordinate (um) at (0   ,0) ;
      \coordinate (ur) at (+1  ,0) ;
      \coordinate(ull) at (-1.4,0) ;
      \coordinate(urr) at ( 1.4,0) ;
      \coordinate (dl) at (-1  ,-1.2) ;
      \coordinate (dm) at (0   ,-1.2) ;
      \coordinate (dr) at (+1  ,-1.2) ;
      \coordinate(dll) at (-1.4,-1.2) ;
      \coordinate(drr) at (1.4 ,-1.2) ;
      \draw[line width=1pt] (ul) -- (um) ;
      \draw[dotted,line width=.8pt] (dm) -- (dr) ;
      \draw[dotted,line width=.8pt] (dl) -- (dm) ;
      \draw[line width=1pt] (um) -- (ur) ;
      \draw[snake it] (dl) -- (ul) ;
      \draw[snake it] (dm) -- (um) ;
      \draw[snake it] (dr) -- (ur) ;
      \draw[dotted,line width=.8pt] (ull) -- (ul) ;
      \draw[dotted,line width=.8pt] (urr) -- (ur) ;
      \draw[dotted,line width=.8pt] (dll) -- (dl) ;
      \draw[dotted,line width=.8pt] (drr) -- (dr) ;
    \end{tikzpicture}
    \caption{
    }
  \end{subfigure}
  \caption{
    The three probe master integrals.
    They should be interpreted in the same manner as the schematically drawn $I$ master integrals in Fig.~\ref{TL:IMastersGraphs}.
  }
  \label{TL:JMastersGraphs}
\end{figure}
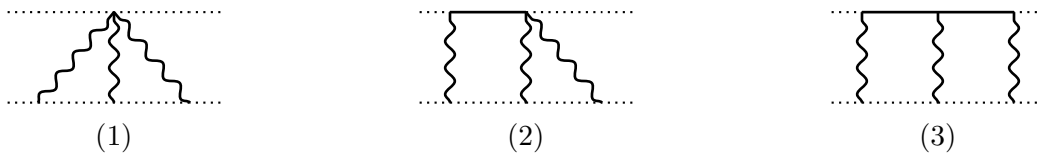

The probe family $J$ is simpler than the comparable mass family $I$ and there are fewer master integrals which may be chosen as the following three integrals:\checked
\begin{align}
  \label{TL:JMasters}
  J_{(1)}
  &=
  J_{0,0,1,1,1,0,0}
  \ ,\\\nn
  J_{(2)}
  &=
  J_{1,0,1,1,1,0,0}
  \ ,\\\nn
  J_{(3)}
  &=
  J_{1,1,1,1,1,0,0}
  \ .
\end{align}
The first and third are $b$-type and the second $v$-type.
They are drawn schematically in Fig.~\ref{TL:JMastersGraphs}.
The $i\zplus$ prescriptions of the worldlines are, again, as with the $I$ masters indicated by superscripts.

\section{Comparable Mass Integrals
}
\label{sec:TLComparable}
The comparable mass integrals are truly post-Minkowskian in the sense that from the post-Newtonian perspective they resum an infinite velocity-expansion into a non-trivial analytic dependence on $\gamma$.
In this section we will analyze these integrals and how their dependence on $\gamma$ can be obtained.
We will consider two methods: Differential equations in Sec.~\ref{sec:TLDifferential} and the method of regions in Sec.~\ref{sec:TLMethod}.

The differential equations allow us to determine all dependence on $\gamma$ through their solution.
The integrals are then functions only of undetermined integration constants.
These are constants with respect to $\gam$ but generally depend on $\eps$.
This method effectively bootstraps the dependence on $\gamma$ without ever computing any integrals.
In a sense, it disentangles the dependence of the integrals on $\gam$ from that on $\eps$.
This disentangling is most clear in the canonical basis where the dependence of the differential equations on $\gam$ and $\eps$ factorizes.

On the other hand, the method of regions allows us to expand the integrals in the post-Newtonian limit where $\gamma-1\to0$.
Such a series expansion is interesting (and non-trivial) in its own right.
Thus, in this way we may approximate the integrals for small velocities or we can get some intuition of what the full resummed expression looks like.
In addition, this method supplements the method of differential equations by allowing us to fix the undetermined integration constants from the solution of the differential equations by matching them to the post-Newtonian limit.

Before moving on we will briefly discuss the fact that only the retarded $i\zplus$ prescription of the ``active graviton'' with momentum $k^\mu=l_1^\mu+l_2^\mu+q^\mu$ is significant in the two-loop integrals.
We first note, that the $i\zplus$ prescription of a (massless) momentum $l^\mu$ is relevant only when the momentum goes on-shell $l^2=0$.
On the other hand, every time a momentum is emitted from a (background) one-graviton worldline vertex its energy with respect to that worldline with velocity $v^\mu_i$ is constrained by energy conservation $\dd(l\cdot v_i)$.
Such a constraint effectively prevents the momentum from going on-shell except if it vanishes identically $l^\mu=0$.
It follows that such graviton momenta generally do not go on-shell and their $i\zplus$ prescription can be ignored.
At 2PM all graviton momenta touch a one-point graviton emission vertex (ignoring the outgoing line) while at 3PM only the active graviton with momentum $k^\mu$ does not touch any such vertices.

\subsection{Differential Equations and Canonical Basis
}
\label{sec:TLDifferential}
The method of differential equations~\cite{Kotikov:1990kg,Bern:1993kr,Remiddi:1997ny,Gehrmann:1999as,Henn:2013pwa,Henn:2013nsa} is a powerful method for bootstrapping dependence of loop integrals on their parameters via differential equations (DEs) in those parameters.
The case of interest to us is particularly simple because our integrals have only a single dimensionless scale which results in ordinary DEs.
In addition, the DEs of the comparable mass integrals may be written in canonical form~\cite{Parra-Martinez:2020dzs,Henn:2013pwa,Dlapa:2022nct} which is a great simplification for the $\eps\to0$ expansion.

The basic idea of this method is that differentiation with respect to $\gamma$ of any master integral must lead to a linear sum of master integrals of the same integral family.
In mathematical notation this translates into the equation:\checked
\begin{align}
  \frac{\pat}{\pat\gam}
  I_{(n)}(\gam,\eps)
  =
  \sum_{m}
  \matrixM'_{nm}(\gam,\eps)
  I_{(m)}(\gam,\eps)
  \ .
\end{align}
Thus, the set of master integrals satisfy a coupled set of first order linear differential equations.
In this expression we indicate only the dependence of the integrals on $\gam$ and $\eps$ as their dependence on $|q^\mu|$ is trivial from dimensional analysis.
Equivalently, one may simply put $|q^\mu|\to1$ and restore it later.

In order to differentiate the covariant representation of the comparable mass integrals Eq.~\eqref{IF:ICovariant} with respect to $\gamma$ we may use the following equation~\cite{Parra-Martinez:2020dzs},\checked
\begin{align}
  \frac{\pat}{\pat\gam}
  =
  \frac{
    v_{1\bot}^\mu
  }{
    \sqrt{\gam^2-1}}
  \frac{\pat}{
    \pat v_1^\mu}
  \ ,
\end{align}
where $v_1^\mu$ is taken to be unconstrained.
Otherwise we can use the Euclidean representation of the integrals Eq.~\eqref{IF:IEuclidean} where dependence on $\gamma$ is explicit.
The fact that the differential of a master integral with respect to $\gam$ can be written as a linear sum of new master integrals follows from the use of tensor reduction and IBP-relations.
With tensor reduction the differential is reduced to a sum of scalar integrals and they are consequently reduced to master integrals.

In our case the system of DEs is further simplified by the choice of a canonical basis.
With this choice the dependence on $\eps$ and $\gam$ of the matrix $\matrixM'_{nm}(\gam,\eps)$ factorizes.
In terms of the canonical basis of integrals $I^{\rm C}_{(n)}$ the DEs have the following simple structure:\checked
\begin{align}
  \frac{\pat}{\pat\gam}
  I^{\rm C}_{(n)}(\gam,\eps)
  =
  \eps
  \sum_m
  \matrixM_{nm}(\gam)
  I^{\rm C}_{(m)}(\gam,\eps)
  \ .
\end{align}
This basis is suitable for an expansion in $\eps$ and the matrix now only depends on $\gam$.

In our case we may choose a canonical basis as in Refs.~\cite{Bjerrum-Bohr:2021din,Jakobsen:2022psy}.
For the $b$-type integrals we find the following canonical basis:\checked
\bse\label{TL:CANB}
\begin{align}
  I^{\rm C}_{(1)}
  &=
  2\eps^2\sqrt{\gam^2-1}
  \,I_{(1)}
  \ ,\\\nn
  I^{\rm C}_{(2)}
  &=
  2\eps\sqrt{\gam^2-1}
  \,I_{(2)}
  \ ,\\\nn
  I^{\rm C}_{(3)}
  &=
  -4
  \,I_{(3)}
  +
  (1+2\eps)\gam
  \,I_{(2)}
  \ ,\\\nn
  I^{\rm C}_{(4)}
  &=
  -\eps^2(\gam^2-1)
  \,I_{(4)}
  \ ,\\\nn
  I^{\rm C}_{(5)}
  &=
  \frac{2(4\eps-1)(2\eps-1)}{
    \sqrt{\gam^2-1}}
  I_{(5)}
  \ ,\\\nn
  I^{\rm C}_{(6)}
  &=
  2\eps^2\sqrt{\gam^2-1}
  \,I_{(6)}
  \ ,\\\nn
  I^{\rm C}_{(7)}
  &=
  -8\eps^2
  \,I_{(7)}
  +4\eps^2\gam
  \,I_{(8)}
  \ ,\\\nn
  I^{\rm C}_{(8)}
  &=
  2\eps^2
  \,I_{(8)}
  \ .
\end{align}
For the $v$-type integrals a canonical basis is given by:\checked
\begin{align}
  I^{\rm C}_{(9)}
  &=
  -2\eps
  \sqrt{\gam^2-1}
  \,I_{(9)}
  \ ,\\\nn
  I^{\rm C}_{(10)}
  &=
  I_{(10)}
  \ ,\\\nn
  I^{\rm C}_{(11)}
  &=
  \frac{1-2\eps}{3}
  I_{(11)}
  \ ,\\\nn
  I^{\rm C}_{(12)}
  &=
  -\eps
  \,I_{(12)}
  \ .
\end{align}
\ese
The canonical basis of integrals may be found with publicly-available packages including epsilon and Fuchsia~\cite{Prausa:2017ltv,Gituliar:2017vzm}.
The canonical integrals are normalized with respect to $\eps$ so that their expansion in $\eps\to0$ starts at $\eps^0$.
The relation between the master integrals $I_{(n)}$ and the canonical $I_{(n)}^{\rm C}$ is a linear transformation which in our case is almost diagonal and can be read off from Eqs.~\eqref{TL:CANB}.

Instead of using $\gam$ as variable it is practical to use $x$ defined by:\checked
\bse
\begin{align}
  \gamma
  &=
  \frac12(\frac1{x}+x)
  \ ,
  \\
  x
  &=
  \gamma
  -
  \sqrt{\gam^2-1}
  \ .
\end{align}
\ese
This variable is useful as it rationalizes the following square root,\checked
\begin{align}
  \sqrt{\gam^2-1}
  =
  \frac12(\frac1{x}-x)
  \ ,
\end{align}
which shows up in the differential equations.
When $\gam$ increases from one to $\infty$ as $1<\gam<\infty$ the variable $x$ decreases from one to zero $1>x>0$.

We gather the master integrals into two vectors corresponding to the $b$ and $v$-type integrals:\checked
\bse
\begin{align}
  \vec{I}_b^{\rmm{C}}(x,\eps)
  &=
  (
  I^{\rmm{C}}_{(1)}(x,\eps)
  ,...,
  I^{\rmm{C}}_{(8)}(x,\eps)
  )
  \ ,
  \\
  \vec{I}_v^{\rmm{C}}
  (x,\eps)
  &=
  (
  I^{\rmm{C}}_{(9)}
  (x,\eps)
  ,...,
  I^{\rmm{C}}_{(12)}
  (x,\eps)
  )
  \ .
\end{align}
\ese
In terms of these vectors, their differential equations are found to take the form:\checked
\bse
\begin{align}
  \frac{\pat}{\pat x}
  \vec{I}^{\rmm{C}}_\alpha
  (x,\eps)
  &=
  \eps
  \matrixM_\alpha(x)
  \cdot
  \vec{I}^{\rmm{C}}_\alpha
  (x,\eps)
  \ ,
  \\
  \matrixM_\alpha(x)
  &=
    \Big(
  \frac{
    \matrixM
    _\alpha^0}{x}
  +
  \frac{
    \matrixM
    _\alpha^+}{1+x}
  -
  \frac{
    \matrixM
    _\alpha^-}{1-x}
  \Big)
  \ ,
\end{align}
\ese
with $\alpha\in\{b,v\}$ and square matrices $\matrixM_\alpha^\sigma$ with $\sigma\in\{\pm,0\}$.
The matrices $\matrixM_\alpha^\sig$ are independent of both $\eps$ and $x$.
The $b$-type matrices are found to be:\checked
\bse
\begin{align}
  \matrixM_b^0=
  \begin{pmatrix}
    -6 & 0  & -1 & 0  & 0  & 0  & 0  & 0  \\
    0  & 2  & -2 & 0  & 0  & 0  & 0  & 0  \\
    12 & 2  & 0  & 0  & 0  & 0  & 0  & 0  \\
    0  & -1 & 0  & 0  & 0  & 0  & 0  & 0  \\
    0  & 0  & 0  & 0  & 2  & 0  & 0  & 0  \\
    0  & 4  & 2  & 0  & 4  & 2  & -2 & 0  \\
    12 & 0  & 0  & 0  & 8  & 2  & -2 & 0  \\
    0  & 0  & 0  & 0  & 0  & 0  & 0  & 0  
  \end{pmatrix}
  \ ,
  \\
  \matrixM_b^\pm=
  \begin{pmatrix}
    6  & 0  & 0  & 0  & 0  & 0  & 0  & 0  \\
    0  & -2 & 0  & 0  & 0  & 0  & 0  & 0  \\
    0  & 0  & 0  & 0  & 0  & 0  & 0  & 0  \\
    0  & 0  & 0  & 0  & 0  & 0  & 0  & 0  \\
    0  & 0  & 0  & 0  & -2 & 0  & 0  & 0  \\
    0  & -4 & 0  & 0  & -4 & -2 & 0  & 0  \\
    0  & 0  & 0  & 0  & 0  & 0  & 2  & \pm4  \\
    0  & 0  & 0  & 0  & 0  & 0  & 0  & 0  
  \end{pmatrix}
  \ .
\end{align}
\ese
The $v$-type matrices, then, are found to have the following expressions:\checked
\begin{align}
  \matrixM_v^0
  &=
  \begin{pmatrix}
    0  & -4 & 0  & 0  \\
    0  & -2 & 0  & 0  \\
    0  & 0  & 2  & 0  \\
    0  & -2 & 0  & 0  
  \end{pmatrix}
  \ ,\ 
  \matrixM_v^+
  =
  \begin{pmatrix}
    0  & 0  & 0  & 0  \\
    0  & 6  & 0  & 0  \\
    0  & 0  & -2 & 0  \\
    0  & 0  & -6 & 2  
  \end{pmatrix}
  \ ,\ 
  \matrixM_v^-
  =
  \begin{pmatrix}
    0  & 0  & 0  & 0  \\
    0  & 2  & 0  & 0  \\
    0  & 0  & 2  & 0  \\
    0  & -4 & -6 & 2  
  \end{pmatrix}
  \ .
\end{align}
The $b$ and $v$-type matrices above are equivalent to the expressions found in Ref.~\cite{Jakobsen:2022psy} (except for the difference in conventions).

A generic solution of the DEs in canonical form may be written down as follows~\cite{Henn:2013pwa}:\checked
\begin{align}
  \vec{I}_\alpha^{\rmm{C}}(x,\eps)
  =
  \mP
  e^{\eps
    \int^x \di x' \matrixM_\alpha(x')}
  \vec{c}_\alpha(\eps)
  \ .
\end{align}
Here, $\mP$ is an ordering operator and $\vec{c}_\alpha(\eps)$ is the integration constant which depends only on the dimension $\eps$.
With the normalization of our canonical basis such that their $\eps$-expansion starts at $\eps^0$ the integration constants may be expanded in a similar fashion:\checked
\begin{align}
  \label{TL:ICs}
  \vec{c}_\alpha(\eps)
  =
  \vec{c}_\alpha^{(0)}
  +
  \eps \vec{c}_\alpha^{(1)}
  +
  \eps^2 \vec{c}_\alpha^{(2)}
  +
  ...
\end{align}
In our application we require at most the linear order in $\eps$ and to this order the solution of the integrals read:\checked
\begin{align}
  \vec{I}_\alpha^{\rm C}(x,\eps)
  &=
  \vec{c}_\alpha^{(0)}
  +
  \eps
  \int^x
  \di x'
  \matrixM_\alpha(x')\cdot \vec{c}_\alpha^{(0)}
  +
  \eps
  \vec{c}^{(1)}_\alpha
  +
  \mO(\eps^2)
  \\
  &=
  \vec{c}_\alpha^{(0)}
  +
  \eps
  \vec{c}^{(1)}_\alpha
  +
  \eps
  \Big(
  \matrixM_\alpha^0
  \log(x)
  +
  \matrixM_\alpha^+
  \log(1+x)
  -
  \matrixM_\alpha^-
  \log(1-x)
  \Big)
  \cdot \vec{c}_\alpha^{(0)}
  +
  \mO(\eps^2)
  \nn
\end{align}
The integration on $x'$ is trivial and leads to the logarithms of the letters $x$, $(x-1)$ and $(x+1)$.
At this point all relevant dependence on $x$ and $\eps$ is explicit and the only remaining task is to fix the constants $\vec{c}_\alpha^{(0)}$ and $\vec{c}_\alpha^{(1)}$.

\subsection{Method of Regions
}
\label{sec:TLMethod}
The method of regions~\cite{Beneke:1997zp,Smirnov:2002pj,Smirnov:2012gma,Becher:2014oda} allows us to expand the comparable mass integrals in the post-Newtonian limit $\gamma\to1$ or, equivalently, $v\to0$.
In particular, we can use this method to derive boundary conditions with which the integration constants $\vec{c}_\alpha(\eps)$ Eq.~\eqref{TL:ICs} can be determined.
More generally, this method can be used to compute post-Newtonian approximations to the post-Minkowskian integrals which can be carried out to any desired order in the velocity.
In this setting, the PN expansion with the method of regions can be carried out on arbitrary members of the integral families or directly on the full integrand of the observables.
The method of regions has been used extensively in the post-Minkowskian expansion~\cite{Cheung:2018wkq,Bern:2021yeh,Herrmann:2021tct,Bern:2022jvn,Dlapa:2022lmu,Jakobsen:2022psy}.

The method of regions instructs us to identify integration regions defined by scalings of the integration variables.
In our current setting, namely the PN limit of PM loop integrals, the integration variables are loop momenta and we must identify different scalings of these momenta that generally depend on the small scale $v$.
Finally, the integral is written as a sum of contributions for each relevant region.
The regions of the post-Minkowskian integrals can be categorized by defining potential and radiative gravitons defined by specific scalings of their momenta.
The respective scalings of a graviton with momentum $l^\mu$ in a given frame is defined by the following relations:\checked
\bse\label{eq:RegionScalings}
\begin{align}
  &\text{Potential:}
  &&
  \hspace*{-1cm}
  l^\mu=(l^0,\vct{l}) \sim (v,1)
  \ ,
  &&
  \\
  &\text{Radiative:}
  &&
    \hspace*{-1cm}
    l^\mu=(l^0,\vct{l}) \sim (v,v)
    \ .
  &&
\end{align}
\ese
Thus, the time component of potential momenta scale as $v$ while the spacial component scale as 1.
Instead, both the time and spacial components of radiative momenta scale as $v$.
The two scalings are also often referred to as potential and radiative modes.
The important distinction between the two scalings is that while potential modes are off-shell, radiative modes go on-shell and carry radiation.
This classification of modes is significant only in the limit $v\to0$.
Thus, if $v\sim1$ the two modes in Eqs.~\eqref{eq:RegionScalings} are indistinguishable.
The potential and radiative modes are well-known in the (post-Newtonian) EFT approach gravity~\cite{Goldberger:2004jt,Galley:2015kus,Almeida:2022jrv}.
In principle one might consider any other scaling of the loop momenta, though, at 3PM we find that these two scalings and corresponding regions are sufficient.

Our application of the method of regions is to expand the comparable mass integrals in the post-Newtonian (PN) limit $v\to0$.
The $I$ integral family Eq.~\eqref{IF:ICovariant} has five massless propagators, $k^2$, $l_i^2$ and $(l_i+q)^2$, with $k^\mu=l_1^\mu+l_2^\mu+q^\mu$.
In the PN limit, all of these but $k^2$ are enforced to have potential scaling due to the energy-conserving delta functions.
Thus, the delta functions constrain the momenta $l_i^\mu$ and $(l_i^\mu+q^\mu)$ to be orthogonal to $v_{\bar\imath}^\mu$ (with $\bar1=2$ and $\bar2=1$).
Then as both velocities behave as $v_i^\mu\sim(1,v)$ in the PN limit, the orthogonal momenta, $l^\mu$, must behave oppositely as potential momenta $l^\mu\sim(v,1)$.
In contrast, there is no combination of velocities to which $k^\mu$ is orthogonal and this momentum can scale both in the potential and radiative region.
It is then the scaling of this graviton which will define the two relevant regions of our integrals which we refer to as potential and radiative corresponding to this scaling.
Thus, in the PN limit the comparable mass integrals are written as a sum of two contributions, potential and radiative.

The potential and radiative scalings~\eqref{eq:RegionScalings} of the four-momenta can be applied directly to the covariant representation of the comparable mass integrals Eq.~\eqref{IF:ICovariant}.
After one of the two scalings have been assigned to $k^\mu$ the integrand is simply Taylor expanded in $v$ and each term in the expansion is integrated by itself.
These integrations are much simpler as the integration no longer depends on $\gam$ and it is in fact given by the $J$ and $K$ integral families.
This approach is very systematic but is, perhaps, at first more abstract because $\gam$-dependence is implicit in the two velocities $v_i^\mu$ and the energy-conserving delta-functions constrain the momenta.
Instead, we will work with the Euclidean representation of the integrals in the rest of this section where the $\gam$-dependence is explicit and the delta functions have been integrated out.
At any rate the covariant and Euclidean representations are simply related through Eq.~\eqref{IF:ICovToEuc}.

We are interested, then, in the PN expansion of the Euclidean representation of the comparable mass integrals Eq.~\eqref{IF:IEuclidean} which reads:\checked
\begin{align}\label{CM:IEuclidean}
  &\qquad
  \hat I^{\sig_1\sig_2\sig_3}_{n_1...n_7}(\gam,d)
  =
  \int_{\vct{l}_1\vct{l}_2}  
  \frac{1}{
    \prod_{j=1}^7
    \hat D_{(I)j}^{n_j}
  }
  \ ,
  \span\span\span\span
  \\[4pt]
  &
  \hat D_{(I)1 }= \vct{l}_1\cdot
  \vct{\hat e}
  +\sig_1i\zplus
  \ ,
  &&
  \hat D_{(I)4 }= \vct{l}_1^2
  \ ,
  &&
  \hat D_{(I)6 }= (\vct{l}_1+\vct{\hat q})^2
  \ ,
  \nn
  \\
  &
  \hat D_{(I)2 }= \vct{l}_2\cdot
  \vct{\hat e}
  +\sig_2i\zplus
  \ ,
  &&
  \hat D_{(I)5 }= \vct{l}^2_2
  \ ,
  &&
  \hat D_{(I)7 }= (\vct{l}_2+\vct{\hat q})^2
  \ ,
  \nn
  \\
  &
  \hat D_{(I)3 }=
  (\vct{l}_1+\vct{l}_2+\vct{\hat q})^2
  +
  2(\gamma-1)
  \vct{l}_1\cdot \vct{\hat e}
  \, 
  \vct{l}_2\cdot \vct{\hat e}
  +
  \sig_3
  i\zplus (\vct{l}_2-\vct{l}_1)\cdot \vct{\hat e}
  \ .
  \span\span\span\span
  \nn
\end{align}
Again, $\vct{\hat e}$ and $\vct{\hat q}$ are spacial unit vectors orthogonal to each other.
The dependence on the Lorentz factor $\gam$ is now present only in the third propagator which is also the one that includes the active momentum $\vct{k}=\vct{l}_1+\vct{l}_2+\vct{q}$.
It is then reasonable that the expansion in regions is sensitive only to this momentum.

The potential and radiative scalings of $k^\mu$ translate into the two scalings $\vct{k}\sim1$ and $\vct{k}\sim v$ of the spacial vector $\vct{k}$.
For the expansion of the Euclidean representation it is natural to use $(\gam-1)$ as expansion parameter and we define $\chi$ by\checked
\begin{align}
  \chi=
  \sqrt{2}\sqrt{\gam-1}
  =
  v
  +
  \frac38 v^3
  +
  ...
\end{align}
The parameter $\chi$ scales with $v$ and is thus a valid PN expansion parameter.
Each scaling of $\vct{k}$ defines a contribution of the full integral which we label by potential and radiative:\checked
\begin{align}\label{TL:SOT}
  \hat I^{\sigo\sigt\sigh}_{\non...\nse}(\chi)
  =
  \hat I^{(\rmm{pot})\sigo\sigt}_{\non...\nse}(\chi^2)
  +
  \chi^{d-1-2n_3}
  \hat I^{(\rmm{rad})\sigo\sigt\sigh}_{\non...\nse}(\chi^2)
  \ .
\end{align}
The factor of $\chi$ in front of the radiative contribution is practical as the remaining radiative contribution $\hat I^{\rmm{rad}\sigo\sigt\sigh}_{\non...\nse}$ then simply is a power series in $\chi^2$.
The potential region has only two superscripts $\sig_i$ since the potential scaling of $\vct{k}$ makes the third index $\sig_3$ superfluous.

Let us first analyze the potential contribution $\hat I^{(\rmm{pot})\sigo\sigt}_{\non...\nse}$.
Here, all momenta shown in Eq.~\eqref{CM:IEuclidean} scale as $1\gg \chi$ and we may simply expand the third denominator as a geometric series:\checked
\begin{align}
  \frac1{\hat D_{(I)3}}
  &=
  \frac1{
      (\vct{l}_1+\vct{l}_2+\vct{q})^2
  +
  \chi^2
  \vct{l}_1\cdot \vct{\hat e}
    \, 
  \vct{l}_2\cdot \vct{\hat e}
    +
    \sig_3  i\zplus (\vct{l}_2-\vct{l}_1)\cdot \vct{\hat e}
  }
  \nn
  \\
  &\to
  \frac{1}{
    (\vct{l}_1+\vct{l}_2+\vct{q})^2
  }
  \sum_{m=0}^{\infty}
  \Big(\!\!
  -\chi^2
  \Big)^m
  \bigg(
  \frac{
    \vct{l}_1\cdot \vct{\hat e}
    \, 
    \vct{l}_2\cdot \vct{\hat e}
  }{
    (\vct{l}_1+\vct{l}_2+\vct{q})^2
  }
  \bigg)^m
\end{align}
In the potential region the third propagator $\hat D_{(I)3}$ is effectively replaced by $(\vct{l}_1+\vct{l_2}+\vct{q})^2$ which is exactly the third propagator of the Euclidean probe family $\hat D_{(J)3}$.
In general, we see that the potential contribution of $\hat I^{(\rmm{pot}),\sigo\sigt}_{\non...\nse}$ is given as an expansion in $\chi^2$ with coefficients given by $\hat J^{\sigo\sigt}_{\non...\nse}$.
A general expression for the potential contribution in terms of the integral family $\hat J$ is given by:\checked
\begin{align}\label{CM:ExactPotential}
  \hat I^{(\rmm{pot})\sigo\sigt}_{\non...\nse}(\chi^2)
  =
  \sum_{m=0}^\infty
  \frac{
    (n_3+m-1)!
  }{
    (n_3-1)! m!
  }
  \big(\!\!
  -\chi^2
  \big)^{m}
  \hat J_{n_1-m,n_2-m,n_3+m,n_4...n_7}^{\sigo\sigt}
  \ .
\end{align}
Here $n_3>0$ with the case of negative or vanishing $n_3$ being trivial.
If desired, the right-hand-side can now be reduced to master integrals with the much simpler IBP-rules of the integral family $\hat J$.

From the perspective of the Euclidean representation the radiative region is slightly more complicated.
Here, $\vct{k}$ scales as $\chi$ and it is advantageous to change variables from $\vct{l}_i$ to $\vct{l}$ and $\vct{k'}$ defined by:\checked
\bse
\begin{align}
  \vct{l}
  &=
  \vct{l}_1
  \ ,
  \\
  \vct{k}'
  &=
  \frac{1}{\chi}
  \vct{k}
  =
  \frac{1}{\chi}
  (\vct{l}_1+\vct{l}_2+\vct{q})
  \ .
\end{align}
\ese
This transformation introduces a Jacobian $\chi^{d-1}$.
The variables $\vct{l}$ and $\vct{k}'$ scales as 1 in the radiative region.
We re-express the denominators $\hat D_{(I)i}$ in terms of the new variables:\checked
\begin{align}
  &
  \hat D_{(I)1} = \vct{l}\cdot
  \vct{\hat e}
  +i\eps\sig_1
  \ ,
  &&
  \hspace*{-4cm}
  \hat D_{(I)4} = \vct{l}^2
  \ ,
  &&
  \hat D_{(I)6} = (\vct{l}+\vct{q})^2
  \ ,
  \nn
  \\
  &
  \hat D_{(I)2} =
  - (\vct{l}
  -
  \chi \vct{k}')\cdot
  \vct{\hat e}
  +i\eps\sig_2
  \ ,
  &&
  \hspace*{-4cm}
  \hat D_{(I)5} =
  (\vct{l}
  +\vct{q}
  -
  \chi\vct{k}'
  )^2
  \ ,
  &&
  \hat D_{(I)7} =
  (\vct{l}-\chi\vct{k}')^2
  \ ,
  \nn
  \\
  &
  \hat D_{(I)3} =
  \chi^2
  \Big(
  \vct{k}'^2
  -
  (\vct{l}\cdot \vct{\hat e}+\sigh i \zplus)^2
  +
  \chi
  \vct{l}
  \cdot
  \vct{\hat e}
  \vct{k}'
  \cdot
  \vct{\hat e}
  \Big)
  \ .
  \label{CM:IRadiative}
\end{align}
We see that the third denominator has an overall factor of $\chi^2$ and this factor together with the Jacobian $\chi^{2(d-1)}$ results in the prefactor that we introduced in front of the radiative contribution $\chi^{d-1-2n_3}$ in Eq.~\eqref{TL:SOT}.
The $\chi\to0$ behavior of each denominator $\hat D_{(I),i}$ is described by the denominators of the $\hat K$ integral family.
Thus, when each denominator is expanded in $\chi$ the relevant integrals will belong to the $\hat K$ integral family.
As several propagators now include $\chi$-dependence the generic expression for $\hat I^{\rmm{(rad)}\sigo\sigt\sigh}_{n_1...n_7}$ in terms of the $\hat K$ integrals is more complicated than the corresponding potential contribution.
At leading order we simply find:\checked
\begin{align}
  \hat I
  ^{(\rmm{rad})\sigo\sigt\sigh}
  _{\non...\nse}
  (\chi^2)
  =
  (-1)^{n_2}
  \hat K^{\sigo-\sigt\sigh}_{n_1,n_2,n_3,n_4+n_7,n_5+n_6}
  +
  \mO(\chi^2)
  \ .
\end{align}
Higher orders may consistently be derived by expanding the relevant denominators in $\chi$.
The series only has even powers of $\chi$ because odd powers come with a numerator that is odd in $\vct{k}'$ while the denominators are even in $\vct{k}'$ so that these terms vanish.


\section{Integral Expressions and Computation of Boundary Integrals}
\label{sec:TLIntegral}
In this section we present results for the master integrals of the $I$ and $J$ integral families and a generic formula for the $K$ integral family.
These expressions are reproduced from Ref.~\cite{Jakobsen:2022psy} although we use a slightly different notation.
The derivation of the expressions for the $J$ and $K$ integrals are then discussed in Secs.~\ref{sec:TLProbe} and~\ref{sec:TLMushroom}.
It is interesting that the only step in which genuine integration is required is for those integrals $J$ and $K$ relevant to the PN limit.
General discussions of loop integration and dimensional regularization can be found in Refs.~\cite{Smirnov:2004ym,Srednicki:2007qs}.

We begin with the comparable mass master integrals and present expressions for the canonical master integrals $I_{(n)}^{\rm C}$.
We use the following variable,\checked
\begin{align}
  \zeta=(4\pi)^{-2+2\eps}
  e^{-2\gamma_E\eps}
  \ ,
\end{align}
with the Euler–Mascheroni constant $\gam_E$.
The $b$-type canonical $I$ integrals are:\checked
\begin{align}
  I^{\rmm{C}}_{(1)}
  &=
  \frac{\eps}{2}
  \log(x)
  +
  \mO(\eps^2)
  \ ,\\\nn
  I^{\rmm{C}}_{(2)}
  &=
  -\frac12
  +\mO(\eps)
  \ ,\\\nn
  I^{\rmm{C}}_{(3)}
  &=
  -\frac12
  +\mO(\eps)
  \ ,\\\nn
  I^{{\rmm{C}},++}_{(4)}
  &=
  \zeta
  \Big[
  -\frac12
  +
  \frac{\eps}{2}
  \log(x)
  +\mO(\eps^2)
  \Big]
  \ ,\\\nn
  I^{{\rmm{C}},+-}_{(4)}
  &=
  \zeta
  \Big[
  1
  +
  \frac{\eps}{2}
  \log(x)
  +\mO(\eps^2)
  \Big]
  \ ,\\\nn
  I^{\rmm{C}}_{(5)}
  &=
  \frac12
  +\mO(\eps)
  \ ,\\\nn
  I^{\rmm{C}}_{(6)}
  &=
  -\eps
  \log(x)
  +\mO(\eps^2)
  \ ,\\\nn
  I^{\rmm{C}}_{(7)}
  &=
  \mO(\eps^2)
  \ ,\\\nn
  I^{\rmm{C}}_{(8)}
  &=
  \mO(\eps^2)
  \ .
\end{align}
In all expressions $\sig_3=+$ and for the fourth master integral the two superscripts indicate the value of $\sig_1$ and $\sig_2$.
In fact, all of these master integrals except the fourth are independent of all $\sig_i$.
Also, we have put $|q|\to1$ in these expressions and their dependence on $|q|$ must be restored from dimensional analysis.
They are provided only to the order in $\eps$ relevant for their application to observables at the third post-Minkowskian order.

The $v$-type $I$ canonical master integrals are:\checked
\begin{align}
  I^{C,+}_{(9)}
  &=
  i\pi\zeta
  \Big[
  \frac{1}{2}
  -
  \eps
  \log(2x)
  +
  \mO(\eps^2)
  \Big]
  \ ,\\\nn
  I^{C,-}_{(9)}
  &=
  i\pi\zeta\Big[
  -\frac{1}{2}
  -\eps
  \log(\frac{x}{2})
  +\mO(\eps^2)
  \Big]
  \ ,\\\nn
  I^{C}_{(10)}
  &=
  i\pi\zeta\Big[
  \frac14
  -\frac{\eps}2
  \Big(
  \log(2x)+\log(1-x)-3\log(1+x)
  \Big)
  +\mO(\eps^2)
  \Big]
  \ ,\\\nn
  I^C_{(11)}
  &=
  i\pi\zeta
  \Big[
  -\frac16
  -\frac{\eps}3
  \Big(
  \log(8x)
  -\log(1-x^2)
  \Big)
  +\mO(\eps^2)
  \Big]
  \ ,\\\nn
  I^C_{(12)}
  &=
  i\pi\zeta
  \Big[
  -\frac{\eps}2
  \Big(
  \log(4x)
  -2\log(1+x)
  \Big)
  +\mO(\eps^2)
  \Big]
  \ .
\end{align}
Again, $\sigh=+$ in all cases and $|q|\to1$.
We note that only the combinations,\checked
\begin{align}
  \operatorname{arccosh}\gam
  &=
  -
  \log(x)
  \ ,
  \\
  \log\frac{\gam+1}2
  &=
  2\log(1+x)-\log(4x)
  \ ,
\end{align}
appear in the observables and it might be advantageous to make that manifest in the choice of master integrals.
Expressions for a similar canonical basis of comparable mass integrals with Feynman $i\zplus$ prescription is found in Ref.~\cite{Bjerrum-Bohr:2021din}.
A particular simple property of the integrals with retarded prescription is that they are (pseudo) real (for $1<\gam<\infty$), that is, they are purely real up to an overall imaginary factor.
Instead, the integrals with Feynman prescription are real only in the region where $-1<\gam<1$~\cite{Herrmann:2021tct}.
This is verified in general by using complex conjugation together with the flip symmetry~\cite{Jakobsen:2022psy}.

The (Euclidean) probe master integrals are given by:\checked
\begin{align}
  \hat J_{(1)}(d)
  &=
  (4\pi)^{-3+2\eps}\frac{\Gamma^3(\frac12-\eps)\Gamma(2\eps)}{\Gamma(\frac32-3\eps)}
  \ ,
  \\\nn
  \hat J_{(2)}^{+}(d)
  &=
  -i
    (4\pi)^{-\frac52+2\eps}
  \frac{\Gamma(\frac12-2\eps)\Gamma^2(\frac12-\eps)\Gamma(-\eps)\Gamma(\frac12+2\eps)}
       {2\Gamma(\frac12-3\eps)\Gamma(1-2\eps)}
       \ ,
       \\\nn
       \hat J^{++}_{(3)}(d)
       =
       -2
       \hat J^{+-}_{(3)}(d)
       &=-  (4\pi)^{-2+2\eps}
       \frac{\Gamma^3(-\eps)\Gamma(1+2\eps)}{3\Gamma(-3\eps)}
       \ .
\end{align}
We recall that $d=4-2\eps$.
The first, second and third probe master integrals have zero, one and two linear propagators respectively and their $i\zplus$ prescriptions are indicated by the superscripts.

Finally, the (Euclidean) mushroom integral family is given by,\checked
\begin{align}
  &\hat K_{n_1n_2n_3n_4n_5}^{+-+}(d)
  =
  (-i)^{n_1-n_2}
  \frac{2^{n_1+n_2+2n_3}}{(8\pi)^{d-1}}
  \frac{
    \Gam(\frac{2n_3-d+1}{2})
  }{
    \Gam(n_3)}
  \\\nn
  &
  \qquad\qquad\qquad\qquad\qquad\qquad\qquad
  \times
  \frac{
    \cos\!\big(
    \frac{\pi}2(
    n_1-n_2-d+1
    )
    \big)}{
    \cos\!\big(
    \frac{\pi}2(
    n_1+n_2-d+1
    )
    \big)}
  \Gam_{n_1+n_2+2n_3-d+1,n_4,n_5}(d-1)
  \ ,
\end{align}
with the factor $\Gam_{n_1,n_2,n_3}(d)$ defined in Eq.~\eqref{WQFT:GAMF}.

\subsection{Probe Integrals}
\label{sec:TLProbe}
In this section, we discuss the computation of the probe master integrals which are $J_{(1)}$, $J_{(2)}^{\sigo}$ and $J_{(3)}^{\sigo\sigt}$.
Crucially, we only have to compute $J_{(1)}$ and the cuts of $J_{(2)}$ and $J_{(3)}$ and all combinations of $i\zplus$ may be obtained from those with symmetries.
The three building blocks (cuts) read:\checked
\bse
\begin{align}
  J_{(1)}
  &=
  \int_{l_1l_2}
  \frac{
    \dd(l_1\cdot v_2)
    \dd(l_2\cdot v_2)
    }{
    l_1^2
    l_2^2
    (l_1+l_2+q)^2
  }
  &&
  =
  -|q|^{2d-8}
  \Del_{(1)}
  \ ,
  \\
  i\big(
  J_{(2)}^+
  -
  J_{(2)}^-
  \big)
  &=
  \int_{l_1l_2}
  \frac{
    \dd(l_1\cdot v_1)
    \dd(l_1\cdot v_2)
    \dd(l_2\cdot v_2)
    }{
    l_1^2
    l_2^2
    (l_1+l_2+q)^2
  }
  &&
  =
  -\frac{|q|^{2d-9}}{\sqrt{\gam^2-1}}
  \Del_{(2)}
  \ ,
  \\
  -2\big(
  J_{(3)}^{++}
  -
  J_{(3)}^{+-}
  \big)
  &=
  \int_{l_1l_2}
  \frac{
    \dd(l_1\cdot v_1)
    \dd(l_1\cdot v_2)
    \dd(l_2\cdot v_1)
    \dd(l_2\cdot v_2)
    }{
    l_1^2
    l_2^2
    (l_1+l_2+q)^2
  }
  &&
  =
  -\frac{
    |q|^{2d-10}
  }{
    \gam^2-1}
  \Del_{(3)}
  \ .
\end{align}
\ese
The three integrals $\Del_{(n)}$ depend only on the dimension $d$ and can be computed with Feynman parametrization.

For the second $J$ master integral the two $i\zplus$ prescription are related by flip symmetry or complex conjugation.
We get:\checked
\begin{align}
  J_{(2)}^\pm
  &=
  \mp\frac{i}{2} \Del_{(2)}
  \ .
\end{align}

For the third master there are naively four $i\zplus$ combinations.
Flip symmetry relates two of those in pairs $J^{++}_{(2)}=J^{--}_{(2)}$ and $J^{+-}_{(2)}=J^{-+}_{(2)}$.
The partial fraction identity $J^{++}_{(2)}=-2J^{+-}_{(2)}$ from Eq.~\eqref{TL:PF2} provides a third relation.
All four combinations may now be written in terms of the cut $\Del_{(3)}$ and for the combinations $(++)$ and $(+-)$ we get:\checked
\begin{align}
  J_{(3)}^{++}
  =
  -2J_{(3)}^{+-}
  =
  -\frac13
  \Del_{(3)}
  \ .
\end{align}

All three cuts may be related to the following generic two-loop integral $\Del(d_1,d_2;a,b,c)$ which in turn may be related to the (following) one-loop integral $\Del(d;a,b)$.
They are defined by:\checked
\bse
\begin{align}
  \Del(d_1,d_2;a,b,c)
  &=
  \int
  \frac{
    \di^{d_1}\vct{l}_1
  }{
    (2\pi)^{d_1}
  }
  \frac{
    \di^{d_2}\vct{l}_2
  }{
    (2\pi)^{d_2}
  }
  \frac{1}{
    |\vct{l}_1|^a
    |\vct{l}_2|^b
    |\vct{l}_1+\vct{l}_2+
    \vct{\hat q}|^c
  }
  \ ,
  \\
  \Del(d_1;a,b)
  &=
  \int
  \frac{
    \di^{d_1}\vct{l}_1
  }{
    (2\pi)^{d_1}
  }
  \frac{1}{
    |\vct{l}_1|^a
    |\vct{l}_1+\vct{\hat q}|^b
  }
  \ .
\end{align}
\ese
For these integrals we assume that $\vct{\hat q}$ lives in the subspace of $\vct{l}_i$ and that the vector of lower dimension, say, $\vct{l}_1$ lives in a subspace of $\vct{l}_2$.

In the iterated integral $\Del(d_1,d_2;a,b,c)$ we may first do the $\vct{l}_2$ integration and afterwards the $\vct{l}_1$ integration which are both of the form $\Del(d;a,b)$ with the result that:\checked
\begin{align}
  \Del(d_1,d_2;a,b,c)
  =
  \Del(d_1;a,b+c-d_2)
  \Del(d_2;b,c)
  \ .
\end{align}
The one-loop integral $\Del(d;a,b)$ is computed with conventional methods~\cite{Smirnov:2004ym,Srednicki:2007qs} and found to be:\checked
\begin{align}
  \Del(d;a,b)
  =
  \frac1{\sqrt{4\pi}^d}
  \frac{
    \Gam(\frac{d-a}2)\Gam(\frac{d-b}2)
  }{
    \Gam(\frac{a}2)\Gam(\frac{b}2)
  }
  \frac{
    \Gam(\frac{a+b-d}{2})
  }{
    \Gam(\frac{2d-a-b}{2})
  }
  \ .
\end{align}
The three building blocks $\Del_n$ with $n=1,2,3$ are expressed in terms of $\Del(d_1,d_2;a,b,c)$ as follows:\checked
\bse
\begin{align}
  \Del_1(d)
  =
  \Del(d-1,d-1;2,2,2)
  \ ,
  \\
  \Del_2(d)
  =
  \Del(d-1,d-2;2,2,2)
  \ ,
  \\
  \Del_3(d)
  =
  \Del(d-2,d-2;2,2,2)
  \ .
\end{align}
\ese

\subsection{Mushroom Integrals}
\label{sec:TLMushroom}
We consider the computation of the radiation reaction integral family.
Discussions of this integral can be found in Refs.~\cite{Galley:2015kus,Herrmann:2021tct,Bjerrum-Bohr:2021vuf} and here we follow Ref.~\cite{Jakobsen:2022psy}.
Its covariant representation is:\checked
\begin{align}\label{TL:KCovariant}
  &\qquad
  K_{n_1...n_5}^{\sigo\sigt\sigh}
  (|q^\mu|,\gamma,d)
  =
  \int_{l_1l_2}
    \frac{
    \dd(l_1\cdot v_2)
    \dd(l_2\cdot v_1)
  }{
    \prod_{i=1}^5
    D_{(K)i}^{n_i}
    }
    \ ,
    \span
    \span
    \span
    \span
    \\[4pt]
  &
  D_{(K)1} = l_1\cdot v_1 + \sigo i\zplus
  \ ,
  &
  &
  D_{(K)2} = l_1\cdot v_1 + \sigt i\zplus
  \ ,
  &&
  D_{(K)3} = (k_0+\sigh i\zplus)^2-\vct{k}^2
  \ ,
  \nn
  \\
  &
  D_{(K)4} = l_1^2
  \ ,
  &&
  D_{(K)5} = (l_1+q)^2
  \ .
  &&
  \nn
\end{align}
Without loss of generality we assume $\sigo=+$, $\sigt=-$ and $\sigh=+$.
The momentum of the active graviton is $k^\mu=l_1^\mu+l_2^\mu+q^\mu$.
It is practical to change variables to $k^\mu$ and $l_1^\mu$ and evaluate $k^0$ in the frame of $v_1$.
The second delta function of Eq.~\eqref{TL:KCovariant} then implies that $k^0=l_1\cdot v_1$.
The third propagator now reads:\checked
\begin{align}
  D_{(K)3}
  =
  (l\cdot v_1+i\zplus)^2
  -
  \vct{k}^2
\end{align}
The integral on $k^\mu$ is a nested one-loop integral given by:\checked
\begin{align}
  \int_k
  \frac{\dd(k\cdot v_1-l_1\cdot v_1)}{
    \big(
    (l\cdot v_1+i\zplus)^2
    -
    \vct{k}^2
    \big)^{n_3}}
  =
  \Big(
  \frac{e^{-i\pi}}{
    4\pi}
  \Big)^{\frac{d-1}{2}}
  \frac{
    \Gam(\frac{2n_3-d+1}{2})
  }{
    \Gam(n_3)}
  (l\cdot v_1 + i\zplus)^{d-1-2n_3}
  \ .
\end{align}
This result may be derived with Schwinger parameters.
The only subtlety is the handling of the (imaginary) mass term.
Here it is practical to rewrite the imaginary mass as\checked
\begin{align}
  (l\cdot v_1+i\zplus)^2-\vct{k}^2
  =
  -
  \Big(
  (\zplus-i l\cdot v_1)^2
  +
  \vct{k}^2
  \Big)
  \ .
\end{align}
This rewriting is useful as $\zplus$ always is assumed positive so that the real part of $(\zplus-i\oma)$ always is positive.

The radiation reaction integral is now reduced to a one-loop integral which reads:\checked
\begin{align}\label{TL:DWL}
  F_{n_1,n_2,n_3,n_4}^{+-}
  =
  \int_{l_1}
  \frac{\dd(l_1\cdot v_2)}{
    (l_1\cdot v_1+i\zplus)^{n_1}
    (l_1\cdot v_1-i\zplus)^{n_2}
    l_1^{2n_3}
    (l_1+q)^{2n_4}
  }
  \ .
\end{align}
Here, we do not restrict $n_1$ and $n_2$ to be integers and they can instead generally be complex numbers.
The integral can be reduced to the one-loop integral encountered at 2PM order by using Schwinger parametrization.
For the worldlines those are introduced as follows:\checked
\begin{align}
  \frac{1}{(\oma\pm i\zplus)^\alpha}
  =
  e^{\mp i\frac{\pi}2 \alpha}
  \frac1{\Gam(\alpha)}
  \int_{-\infty}^{\infty} \di u\,
  \theta(u)
  u^{\alpha-1}
  e^{-u(0\mp i \oma)}
\end{align}
Inserting this expression into Eq.~\eqref{TL:DWL} one finds:\checked
\begin{align}
  F_{n_1,n_2,n_3,n_4}^{+-}
  =
  e^{i\pi n_2}
  \frac{
    \cos\!
    \big(
    \frac{\pi}2
    (n_2-n_1)
    \big)
  }{
    \cos\!
    \big(
    \frac{\pi}2
    (n_2+n_1)
    \big)
  }
  F_{n_1+n_2,n_3,n_4}^+
  \ ,
\end{align}
with the one-loop integral $F_{n_1,n_2,n_3}$ given in Eq.~\eqref{WQFT:TLI}.


\chapter{Gravitational Bremsstrahlung of Spinning Bodies}
\label{GB}
The waveform $f(x)$ describes gravitational waves radiated to asymptotic infinity.
In the present physical setup of the scattering of two compact bodies we may refer to the waveform as gravitational bremsstrahlung in analogy to the electromagnetic case.
The leading order gravitational bremsstrahlung of the two body scattering event starts at the second post-Minkowskian order.
This leading order contribution for spinless bodies was first computed in the 1970s in a series of papers by Kovacs, Thorne and Crowley~\cite{Kovacs1,Crowley:1977us,Kovacs:1977uw,Kovacs:1978eu}.
In particular an explicit expression for the waveform is found in the third paper~\cite{Kovacs:1977uw}.

In this chapter we will reproduce their seminal result and generalize it to include spin effects to quadratic order following the content of the two papers~\cite{Jakobsen:2021smu,Jakobsen:2021lvp}.
The present notation, however, differ in some regards to the one used there.
The main result of that work, namely the leading order $\mO(G^2,S^2)$ waveform, has a rather lengthy expression and it is therefore not easily printed.
In Sec.~\ref{WF:LO} we consider its schematic form and its derivation from WQFT.
For general results we refer to the ancillary file of the original paper~\cite{Jakobsen:2021lvp}.

The gravitational waves carry along energy and angular momentum.
This flux is described by the waveform from which we may derive quantities such as the total emitted energy or angular momentum or the power spectrum.
The radiation of energy and angular momentum is the topic of Sec.~\ref{WF:RE} where we present the leading post-Minkowskian loss of total energy and post-Newtonian expansions of the angular and power spectra of the energy flux.

Finally, in Sec.~\ref{sec:IntegralsWaveform} we consider the momentum space integrals that are relevant for the computation of the waveform.

The (QFT-amplitudes) integrand for the spinless 2PM gravitational bremsstrahlung was first considered in Refs.~\cite{Goldberger:2016iau,Goldberger:2017frp,Luna:2017dtq}.
Its integration in time domain was then first performed in Ref.~\cite{Jakobsen:2021smu} with a corresponding first consideration of its frequency domain version in Ref.~\cite{Mougiakakos:2021ckm}.
The novel results of the $\mO(S^2)$ spinning waveform were then derived in Ref.~\cite{Jakobsen:2021lvp} and later verified in Ref.~\cite{Riva:2022fru} (see also~\cite{Riva:2022uxb}).
Additional work on the leading order gravitational bremsstrahlung include Refs.~\cite{Mougiakakos:2022sic,Riva:2021vnj} and the recent next-to-leading order corrections~\cite{Elkhidir:2023dco,Herderschee:2023fxh,Brandhuber:2023hhy,Georgoudis:2023lgf}.
See also the QFT-amplitudes approaches for deriving waveforms~\cite{Cristofoli:2021vyo,Adamo:2022qci}.

\section{Leading Order Waveform at $\mO(S^2)$}
\label{WF:LO}
In comparison to the worldline observables, the waveform depends on more variables and is in this respect more complicated.
Thus, in addition to the worldline background parameters, it depends on the coordinates at asymptotic infinity which can be taken as the direction of observation and the retarded time.
Already at leading order its expression is very lengthy and it is therefore advantageous to develop an efficient notation to describe it.
The topic of Sec.~\ref{WF:GK} is the analysis of the kinematics and geometry of the waveform and there we will define a covariant notation for the waveform.
It is a task for future work to identify variables with which the waveform might take an even simpler form or that highlight the most important parts of the waveform.

In Sec.~\ref{WF:SF} we present schematic forms for the leading order waveform.
These are useful in order to get an idea of the general structure of the waveform.
Importantly, they tell us in which way the waveform depend on its variables.
In Sec.~\ref{WF:DD} we then consider the derivation of the waveform and its schematic forms from WQFT.

\subsection{Geometry and Kinematics of Waveform}
\label{WF:GK}
The waveform in frequency space, $f(k)$, is given from the one-point graviton function as (Eq.~\ref{eq:waveform}):\checked
\begin{align}
  \label{WF:FD}
  f(k)
  =
  \frac{\kap}{8\pi}
  \eps^\mu
  \eps^\nu
  k^2
  \bgev{h_\mn(k)}
  \Big|_{k^2\to0}
  \ .
\end{align}
It is gauge invariant under general covariant coordinate transformations and describes on-shell gravitons traveling to infinity.
The polarizations $\eps^\mu$ are null vectors $\eps^2=0$ and orthogonal to the wave vector $k\cdot\eps=0$.
The graviton field has spin two which is related to its two polarization vectors.

The waveform in time domain (Eq.~\ref{WEFT:Waveform}) is defined by:\checked
\begin{align}
  \label{WF:TD}
  f_v(x)
  =
  \int_\omega
  e^{-i\omega u_v}
  f(k)
  \ .
\end{align}
Here, $k^\mu$ should be inserted as $k^\mu\to\oma n^\mu_v$.
The time domain waveform is defined with respect to a given frame $v^\mu$ which we indicate by the subscripts $v$ and it is not invariant under boosts of this frame.
The null vector $n^\mu_v$ in the frame $v^\mu$ is given by $n^\mu_v=(1,\vct{\hat{x}})$.
We refer to it as the unit wave vector.
The retarded time is $u_v$ and given by $u_v=t-|\vct{x}|=n_v\cdot x$.

Alternatively, the frequency and time domain waveform may be referred to as the momentum and position space waveform respectively which is consistent with our notation $f(k)$ and $f_v(x)$.
The position space waveform does not depend on all four components of $x^\mu$ but only on the retarded time and the unit vector $\vct{\hat{x}}$.
That is, it is only defined at asymptotic infinity where $|\vct{x}|\to\infty$.

The transformation of the position space waveform and its variables under a change of frame from $v^\mu$ to $v'^\mu$ is given by:\checked
\begin{subequations}\label{WF:VVariables}
\begin{align}
  n_v^\mu
  &=
  \Gamma\,
  n_{v'}^\mu
  \ ,
  \\
  f_v
  &=
  \frac1{\Gamma}\,
  f_{v'}
  \ .
\end{align}
\end{subequations}
The time dilation of the retarded time $u_v=n_v\cdot x$ follows from the transformation of $n_v^\mu$.
The factor $\Gamma$ is:\checked
\begin{align}
  \Gamma=
  n_v\cdot v'
  =
  \frac1{n_{v'}\cdot v}
  \ .
\end{align}
This factor is interpreted as the (local) relative Lorentz factor between the frames $v^\mu$ and $v'^\mu$.
Its value depends on the unit wave vector $n^\mu_v$.
The Eqs.~\eqref{WF:VVariables} are consistent with the definitions of the unit wave vector $n_v\cdot v=n_{v'}\cdot v'=1$.
The position space waveform is given by the leading order behavior of the full metric at asymptotic infinity:\checked
\begin{align}
  \frac12\eps^\mu\eps^\nu h_\mn(x)
  =
  \frac{f_v(x)}{|\vct{x}|}
  +
  \mO
  \Big(
  |\vct{x}|^{-2}
  \Big)
  \ .
\end{align}
This equation is consistent with the length contraction of the waveform under boosts.
The frame, the unit wave vector and the coordinates are related through the formula:\checked
\begin{align}
  x^\mu
  =
  |\vct{x}|n^\mu_v
  +
  u_v v^\mu
  \ .
\end{align}
The transformations Eqs.~\eqref{WF:VVariables} are only valid in the limit $|\vct{x}|\to\infty$.

The polarizations transform under infinitesimal coordinate transformations as:\checked
\begin{align}
  \del \eps^\mu=\alpha n^\mu_v
  \ ,
\end{align}
with infinitesimal parameter $\alpha$.
In a given frame we may choose the transverse traceless gauge which we denote by $\eps_{\rm tt}$ defined by:\checked
\begin{align}
  \eps_\rmm{tt}\cdot v=0
  \ .
\end{align}
In that case we may expand $\eps_\rmm{tt}$ in a helicity basis.
It is defined with respect to spherical coordinates with polar angle $\theta$ and azimuthal angle $\phi$.
Concrete formulas for the explicit choice of these angles are presented below in Eq.~\eqref{WF:BasisFrame}.
From the spherical coordinates, we define a local basis of four vectors $v^\mu$, $n_v^\mu$ and $\hat\theta_v^\mu$ and $\hat \phi_v^\mu$.
The vectors $\hat \theta^\mu_v$ and $\hat \phi^\mu_v$ are the unit vectors of $\theta^\mu_v$ and $\phi^\mu_v$ defined by:\checked
\bse
\begin{align}
  \theta^\mu_v
  &=
  \frac{
    \pat n^\mu_v
  }{
    \pat \theta
  }
  \ ,
  \\
  \phi^\mu_v
  &=
  \frac{
    \pat n^\mu_v
  }{
    \pat \phi
  }
  \ .
\end{align}
\ese
They are orthogonal to the frame $v^\mu$ and thus spacial vectors there.
The spacial vectors $\vct{\hat n}$, $\vct{\hat \theta}$ and $\vct{\hat \phi}$ form an orthonormal (spacial) basis.
The helicity basis for the transverse traceless polarizations is now defined by:\checked
\begin{align}
  \eps^\mu_{tt}
  =
  \hat \theta^\mu_v
  +
  i\hat \phi^\mu_v
  \ .
\end{align}
These polarizations are clearly null vectors and orthogonal to $n^\mu_v$.
The Cartesian plus and cross polarizations, $\eps_+^\mn$ and $\eps_\times^\mn$, are defined by:\checked
\begin{align}
  \eps^\mu_\rmm{tt}
  \eps^\nu_\rmm{tt}
  &=
  \eps_+^\mn
  +
  i
  \eps_\times^\mn
  \\
  &=
  (\hat \theta^\mu\hat \theta^\nu
  -
  \hat \phi^\mu\hat\phi^\nu)
  +
  2i
  \hat \theta^{(\mu}
  \hat \phi^{\nu)}
  \ .\nn
\end{align}
In the (helicity) transverse traceless basis, the waveform is given by:\checked
\begin{align}
  f_{\rm tt}
  =
  f_+ + i f_\times
  \ .
\end{align}
It is also conventional to define $f_\rmm{tt}^\mn$ by:\checked
\begin{align}
  f_\rmm{tt}^\mn
  =
  \operatorname{Re}\big(
  f^*_\rmm{tt}
  \eps^\mu_\rmm{tt}
  \eps^\nu_\rmm{tt}
  \big)
  \ ,
\end{align}
with the star signifying complex conjugation.
This is what is commonly referred to as the transverse traceless waveform.

\subsubsection{Kinematics of Scattering Event}
In addition to the internal variables describing the waveform, the bremsstrahlung of the scattering event will also depend on the initial conditions of the scattering.
It is then natural to define our coordinate system in terms of those initial parameters.
In particular we choose to work in the generic frame $V^\mu$ introduced in Eq.~\ref{WEFT:GF}.
We will assume all quantities to be defined in this frame and ignore the frame subscript on the waveform and other frame dependent quantities.
The frame $V^\mu$ was defined as an arbitrary frame spanned by the two velocities $v_i^\mu$:\checked
\begin{align}
  \label{WF:Arbitrary}
  V^\mu =
  \alpha_1 v_1^\mu
  +
  \alpha_2 v_2^\mu
  \ ,
\end{align}
with constants $\alpha_i$.
This generic frame includes the CoM and rest frames as special cases.

We choose the vectors $V^\mu$, $V_\bot^\mu$, $\hat b^\mu$ and $\hat L^\mu$ as a complete set of orthonormal vectors where the spacial vectors $\vct{\hat V}_\bot$, $\vct{\hat b}$ and $\vct{\hat L}$ form a right-handed system.
We orient the spherical coordinate system along $\vct{\hat V}_\bot$.
In that case:\checked
\bse\label{WF:BasisFrame}
\begin{align}
  n^\mu
  &=
  V^\mu
  +
  V_\bot^\mu
  \cos\theta
  +
  \Big(
  \hat b^\mu
  \cos\phi
  +
  \hat L^\mu
  \sin\phi
  \Big)
  \sin\theta
  \ ,
  \\
  \hat \theta^\mu
  &=
  -
  V_\bot^\mu
  \sin\theta
  +
  \Big(
  \hat b^\mu
  \cos\phi
  +
  \hat L^\mu
  \sin\phi
  \Big)
  \cos\theta
  \ ,
  \\
  \hat\phi^\mu
  &=
  -\hat b^\mu
  \sin\phi
  +
  \hat L^\mu
  \cos\phi
  \ .
\end{align}
\ese

The respective rest frames play an important role in describing the scattering waveform.
We define projectors:\checked
\begin{align}\label{WF:P}
  P_i^\mn=
  \eta^\mn-v_i^\mu v_i^\mu
  \ .
\end{align}
The respective retarded times, $u_i=u_{v_i}$ of the rest frames are:\checked
\begin{align}
  u_i
  =
  \frac{
    n\cdot (x-b_i)
  }{
    n\cdot v_i
  }
  \ ,
\end{align}
which is verified using the transformations Eqs.~\eqref{WF:VVariables} under boosts together with a translation.
We label the initial relative straight-line motion of the two bodies in terms of retarded times by $w^\mu$:\checked
\bse\label{WF:SIP}
\begin{align}
  \label{WF:SIPF}
  w^\mu
  =
  b^\mu
  +
  u_2 v_2^\mu
  -
  u_1 v_1^\mu
  \ .
\end{align}
We refer to this vector, $w^\mu$, as the shifted impact parameter.
This vector and its projections into the respective rest frames of $v_i$  play an important role in the leading order waveform.
The projections are:\checked
\begin{align}
  w_1^\mu
  &=
  (P_1\cdot w)^\mu
  =
  b^\mu
  +
  u_{2}(v_2-\gam v_1)^\mu
  \ ,
  \\
  w_2^\mu
  &=
  (P_2\cdot w)^\mu
  =
  b^\mu
  -
  u_{1}(v_1-\gam v_2)^\mu
  \ .
\end{align}
\ese
The shifted impact parameter is orthogonal to the unit wave vector,\checked
\begin{align}
  w\cdot n=0
  \ ,
\end{align}
which is a consequence of the linear relation between the two retarded times of the bodies.

\subsection{Schematic Form and Wave Memory}
\label{WF:SF}
Let us now consider schematic forms of the leading order post-Minkowskian position space waveform at quadratic order in the spins.
The simple schematic form presented in Ref.~\cite{Jakobsen:2021lvp} is:\checked
\begin{align}
  \label{WF:SF1}
  f^{(2)}(x)
  =
  m_1m_2
  \sum_{\substack{
      s=0,1,2
      \\
      i=1,2
  }}
  \frac1{|w_i|^{2s+1}}
  \bigg(
  \alpha^{(s)}_i
  +
  \frac{
    \beta^{(s)}_i}{
    |w|^{2s+2}
    }
  \bigg)
  \ .
\end{align}
Here, we sum on the two bodies $i=1,2$ and on orders of spin $s=0,1,2$.
The four-vectors appearing in the denominators are the shifted impact parameter $w^\mu$ and its projections into the frames of the two bodies $w_i^\mu$ given in Eqs.~\eqref{WF:SIP}.
The numerators $\alpha_i^{(s)}$ and $\beta^{(s)}_i$ are bilinear in $\eps^\mu$ and polynomial in the retarded times, $u_1$ and $u_2$, and in the scalar products of $v_1^\mu$, $v_2^\mu$, $n^\mu$, $b^\mu$, $\eps^\mu$ and $S_i^\mn$ with the exception of rational functions of $\gamma$ and poles in $n\cdot v_i$.
In addition they depend linearly on the finite size coefficients $C_{\rmm{E},i}$ and the numerators $\alpha_i^{(s)}$ also depend on the square root $|w_i^\mu|$.
They are homogeneous with weight (-1) in the unit wave vector $n^\mu$ which ensures the correct transformation properties under boosts Eq.~\eqref{WF:VVariables}.

The structure of the denominators are:\checked
\bse
\begin{align}
  |w_1|
  &=
  \sqrt{
  |b|^2
  +
  (\gam^2-1)
  u_2^2
  }
  \ ,
  \\
  |w|
  &=
  \sqrt{
    |b|^2
    -
    u_1^2
    -
    u_2^2
    +
    2\gam
    u_1
    u_2
  }
  \ ,
\end{align}
\ese
with $|w_2|$ given by particle exchange symmetry.

The expressions for the numerators $\alpha_i^{(s)}$ and $\beta_i^{(s)}$ are generally quite lengthy.
As an example, however, we focus on the spinless case $s=0$ and let $\vct{\hat n}$ point along $\vct{\hat V_\bot}$, that is, we observe the waveform from the direction of the relative velocity $V_\bot^\mu$ in the frame $V^\mu$.
With the parameterization Eqs~\ref{WF:BasisFrame}) this corresponds to $\theta=0$ and $\phi$ is redundant and mixes the plus and cross polarizations.
For $\eps^\mu$ is implies that $\eps\cdot v_i=0$.
In this special case we find:\checked
\bse\label{WF:SC}
\begin{align}
  \alpha_1^{(0)}
  &=
  0
  \ ,
  \\
  \beta_1^{(0)}
  &=
  -\frac{
    2(2\gam^2-1)(b\cdot\eps)^2}{
    (\gam^2-1) (n\cdot v_1)^2}
  n_\mu
  \Big(
  (P_1\cdot v_2)^\mu
  +
  \frac{\gam^2-1}{|b|^2}
  u_2 (
  u_1 v_1^\mu
  +
  u_2 v_2^\mu
  -
  2\gam u_2 v_1^\mu
  )
  \Big)
  \ .
\end{align}
\ese
This simple case gives some impression of the general structure of these coefficients.
Full expressions can be found in the ancillary file to Ref.~\cite{Jakobsen:2021lvp}.
The definition of the coefficients $\alpha_i^{(s)}$ and $\beta_i^{(s)}$ is not unique as presented in Eq.~\eqref{WF:SF1} together with the polynomial structure of the coefficients discussed below that equation.
Thus, we may define transformations of the coefficients under which parts of them mix together and use this in order to simplify them.

A second, related, schematic formula for the waveform is given by:\checked
\begin{align}\label{eq:WaveformSchematic}
  f^{(2)}(x)
  =
  m_1 m_2
  \sum_{i=1,2}
  \Big[
    \mN_i^{\mu}
  \!\Big(
  \frac{
    \partial
  }{
    \partial b_i^\alpha
  }
  \Big)\,
  \mJ_{i,\mu}
  +
  \mM_i^{\mn}\!
  \Big(
  \frac{
    \partial
  }{
    \partial b_i^\alpha
  }
  \Big)\,
  \mI_{i,\mn}
  \Big]
  \ .
\end{align}
Here, the two numerators $\mN_i^\mu(\pat_{b_i}^\alpha)$ and $\mM_i^\mn(\pat_{b_i}^\alpha)$ act with differentiations on the integrals $\mJ_{i}^\mu$ and $\mI_{i}^\mn$.
These integrals are given by the following expressions:\checked
\begin{subequations}
  \label{eq:ResultsWaveform}
\begin{align}\label{eq:ResultJ}
  \mJ^\mu_1
  &=
  \frac{
    (P_1\cdot v_2)^\mu
  }{
    (\gamma^2-1)
    |w_1|
  }
  -
  \frac{b^\mu}{|b|^2}
  \Big(
  \frac{1}{\sqrt{\gamma^2-1}}
  +
  \frac{u_2}{|w_1|}
  \Big)
  \ ,
  \\
  \label{eq:ResultI}
  \mI^\mn_1
  &=
  \frac{
    \Pi^\mn_1
    v_2\cdot \Pi_1\cdot n
    -
    2
    (\Pi_1\cdot v_2)^{(\mu}
    (\Pi_1\cdot n)^{\nu)}
  }{
    (\gamma^2-1)
    (n\cdot v_1)^2
    |b|^2
    |w|^2
    |w_1|
  }
  \ .
\end{align}
\end{subequations}
The integrals with subscript 2 are given by particle exchange symmetry.
The projectors $\Pi_i^\mn$ are defined by,\checked
\begin{align}\label{WF:PiProjector}
  \Pi^\mn_i
  =
  |w_i|^2
  P_i^\mn
  +
  w_i^\mu
  w_i^\nu
  \ ,
\end{align}
and are orthogonal to $v_i^\mu$ and $w_i^\mu$ so that they project vectors into the two-dimensional subspace orthogonal to those vectors.
When derivatives with respect to $b_i^\mu$ are taken in Eq.~\eqref{eq:WaveformSchematic}, the integrals should be expressed in terms of the unconstrained background parameters.

The numerators $\mN_i^\mu(\pat_{b_i}^\alpha)$ and $\mM_i^\mn(\pat_{b_i}^\alpha)$ are expanded in spins as follows:\checked
\bse\label{WF:NumeratorSE}
\begin{align}
  \label{WF:NSE}
  \mN_i^\mu
  \Big(
  \frac{\pat}{\pat b_i^\alpha}
  \Big)
  =
  \mN_i^{(0)\mu}
  +
  \mN_i^{(1)\mu\alpha}
  \frac{\pat}{\pat b_i^\alpha}
  +
  \mN_i^{(2)\mu\ab}
  \frac{\pat}{\pat b_i^\alpha}
  \frac{\pat}{\pat b_i^\beta}
  \\
  \label{WF:MSE}
  \mM_i^\mn
  \Big(
  \frac{\pat}{\pat b_i^\alpha}
  \Big)
  =
  \mM_i^{(0)\mn}
  +
  \mM_i^{(1)\mn\alpha}
  \frac{\pat}{\pat b_i^\alpha}
  +
  \mM_i^{(2)\mn\ab}
  \frac{\pat}{\pat b_i^\alpha}
  \frac{\pat}{\pat b_i^\beta}
\end{align}
\ese
The numerators $\mN^{(s)\mu}$ and $\mM^{(s)\mn}$ are related to the coefficients $\alpha^{(s)}$ and $\beta^{(s)}$.
They are, however, simpler than those and rather elegant spinless results for numerators defined in an analogous way are given in Ref.~\cite{Jakobsen:2021smu}.
In the same special case as above Eq.~\eqref{WF:SC} they take the more compact form:\checked
\bse
\begin{align}
  \mN^{(0)\mu}_1
  &=
  0
  \\
  \mM^{(0)\mn}_1
  &=
  -2(2\gam^2-1)
  \eps^\mu \eps^\nu
\end{align}
\ese

The coefficients $\alpha^{(s)}_i$ and $\beta^{(s)}_i$ are related to worldline fluctuations and gravitational self-interaction respectively.
The analogous association holds for the numerators $\mN^\mu_i$ and $\mM^\mn_i$.
We will see this explicitly below when we derive the schematic forms from diagrams.
In a theory like electromagnetism without self-interactions in the bulk, the second kind of coefficient (or numerator) would not be there and the waveform is in that case much simpler.
The first schematic form in therms of $\alpha^{(s)}_i$ and $\beta^{(s)}_i$ Eq.~\eqref{WF:SF1} can be derived from the other in terms of $\mN^\mu_i$ and $\mM^\mn_i$ Eq.~\eqref{eq:WaveformSchematic} by noticing how the differentiations associated with the spin powers increase the powers of the denominators $|w^\mu|$ and $|w_i^\mu|$.

As a final example of an explicit result of the waveform we consider the wave memory:\checked
\begin{align}\label{WF:WaveMemory}
  \Delta f
  &= \int_{u=-\infty}^{u=\infty}  d f(u)
  \\
  &=f(u\to\infty)-f(u\to-\infty)
  \ .
  \nn
\end{align}
In the aligned spin case we find:\checked
\begin{align}
  \Del f^{(2)}
  =
  \Bigg(
  1+
  \frac{
    2 v a_+\cdot \hat L
  }{
    |b|(1+v^2)
  }
  +
  \frac{|a_+|^2}{
    |b|^2
  }
  -
  \sum_{i=1,2}
  \frac{C_{E,i} |a_i|^2}{
    |b|^2
    }
  \Bigg)
  \Del f^{(2)}_{S=0}
  \ .
\end{align}
Here $a_{+}^\mu=a_1^\mu+a_2^\mu$ and the aligned spins assumption implies that $a_i^\mu=\pm|a_i^\mu| \hat L^\mu$ with the sign determining whether each spin is aligned or anti-aligned with the orbital angular momentum.
The aligned spin wave memory is simply proportional to the spinless wave memory $\Del f^{(2)}_{S=0}$ given by,\checked
\begin{align}
  \Del f^{(2)}_{S=0}
  =
  m_1m_2
  \frac{
    4(2\gam^2-1)
  }{
    |b|^2 \sqrt{\gam^2-1}
    (n\cdot v_1)^2
  }
  \eps\cdot v_1
  \Big(
  2\eps\cdot b
  \,
  v_1\cdot n
  -
  \eps \cdot v_1
  \,
  b\cdot n
  \Big)
  +
  (1\leftrightarrow2)
  \ ,
\end{align}
where a term with particle labels being exchanged must be added as indicated by the final term $(1\leftrightarrow2)$.
The wave memory will be used to compute the total radiated angular momentum below.

\subsection{Derivation from Diagrams}
\label{WF:DD}
We proceed with a derivation of the leading order waveform and its schematic forms.
The goal, then, is to compute the leading order contributions in $\kap$ to the on-shell one-point graviton function which gives us the leading order of the frequency domain wave form Eq.~\eqref{WF:FD}.
The integrals required in the frequency domain, however, are not easily computed and we focus, instead, on the time domain waveform which is simply a one-dimensional Fourier transform of the frequency domain waveform as in Eq.~\eqref{WF:TD}.
Thus, having knowledge of the time domain waveform, only a one-dimensional Fourier transform is required to get back to the frequency domain waveform.

\begin{figure}
  \renewcommand*\thesubfigure{\arabic{subfigure}} 
  \centering
  \begin{subfigure}{.19\textwidth}
    \centering
    \includegraphics{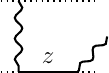}
    \caption{}
  \end{subfigure}
  \begin{subfigure}{.19\textwidth}
    \centering
    \includegraphics{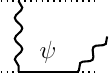}
    \caption{}
  \end{subfigure}
  \begin{subfigure}{.19\textwidth}
    \centering
    \includegraphics{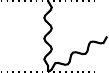}
    \caption{}
  \end{subfigure}
  \begin{subfigure}{.19\textwidth}
    \centering
    \includegraphics{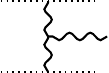}
    \caption{}
  \end{subfigure}
  \caption{
    Relevant WQFT graphs for the leading order waveform.
    The four present graphs must be supplemented by the mirrored versions of the first three graphs.
  }
  \label{WF:LOWaveform}
\end{figure}

The leading order (spinning) one-point graviton function has contributions from seven WQFT diagrams.
Six of those diagrams are related to each other by particle exchange symmetry and we need only focus on four of the seven diagrams.
The four relevant diagrams are shown in Fig.~\ref{WF:LOWaveform} and the remaining three diagrams are obtained by symmetry.
The first three diagrams (1) - (3) of Fig.~\ref{WF:LOWaveform} describe emissions from the worldline and the fourth graph (4) describes emission in the bulk.
We will first analyze the three worldline emission graphs which are simpler than the fourth gravitational self-interaction graph.
In a theory without self-interaction such as electromagnetism we would only have worldline emission graphs.

We focus on the graphs (1) - (3) of Fig.~\ref{WF:LOWaveform} and their time domain waveform Eq.~\eqref{WF:TD}.
The external propagator is amputated, the external momentum $k^\mu$ is on-shell and external indices are contracted with polarizations.
In addition we let $k^\mu=\oma n^\mu$ and integrate on $\oma$ with a Fourier factor.
We label the graviton momentum emitted from the first worldline $q_1^\mu$ and by energy conversation the energy of the worldline fluctuation must be $q_1\cdot v_2$ with both momentum and energy oriented in the direction of causality.
The contribution from the worldline emission graphs may then be written as follows:\checked
\begin{align}
  \label{WF:E1}
  f(x)
  \Big|_{\rm wl,1}
  =
  \int_\omega
  e^{-ik \cdot x}
  \int_{q_1}
  e^{-i q_1\cdot b + ik\cdot b_2}
  \tilde\mN_1
  \frac{
    \dd(q_1\cdot v_1)
  }{
    q_1^2
    (q_1\cdot v_2 + i\eps)^2
  }
  \dd\big((k-q_1)\cdot v_2\big)
  \ .
\end{align}
We denote the contribution of the three worldline emission graphs with the subscript ``wl'' and a ``1'' which refers to the fact that these three graphs are naturally evaluated in the frame of body 1.
The numerator $\tilde\mN_1$ which is related to $\mN_1^\mu$ defined in Eq.~\eqref{eq:WaveformSchematic} is a function of the external kinematics and polynomial in the momenta $q_1^\mu$ and $k^\mu$.
It is computed from composing the relevant WQFT Feynman vertices.
All three worldline emission graphs share the same graviton propagator $q_1^2$ but they have different (or none) worldline propagators $q_1\cdot v_2$.
We may, however, introduce uniform expressions with quadratic worldline propagators by adding a corresponding factor in the numerator which cancels the denominator.
The energy-conserving delta functions and Fourier factor are due to the worldline vertex rules.

As a first simplification we note that $\tilde \mN_1$ is proportional to a factor of $q_1\cdot v_2$.
That is clear for the spin fluctuation (2) and contact interaction (3) of Fig.~\ref{WF:LOWaveform}.
For the worldline fluctuation (1) this is clear after inspection of the worldline vertex rule of the external graviton.
This rule is proportional to the graviton momentum $k^\mu=\oma n^\mu$ and using the delta constraints of Eq.~\eqref{WF:E1} the energy $\oma$ is seen to be proportional to the worldline energy $q_1\cdot v_2$.
We may therefore redefine the numerator in order to make this factor explicit $\tilde \mN_1\to q_1\cdot v_2\tilde \mN_1$.

In addition we note that the $\oma$ integration in Eq.~\eqref{WF:E1} is trivially performed by eliminating one of the delta functions.
Taking both delta constraints into account, the exponent of the Fourier factor rearranges itself into the simple expression $-iw_1\cdot q_1$:\checked
\begin{align}
  -ik\cdot x-ik\cdot b_2+iq_1\cdot b\to -iq_1\cdot w_1
  \ .
\end{align}
This is verified using the relations of Sec.~\ref{WF:GK}.

With these simplifications, the waveform Eq.~\eqref{WF:E1} becomes:\checked
\begin{align}
  \label{WF:E2}
  f(x)
  \Big|_{\rm wl,1}
  =
  \int_{q}
  e^{-i q\cdot w_1}
  \tilde\mN_1
  \frac{
    \dd(q\cdot v_1)
  }{
    q^2
    (q\cdot v_2 + i\eps)
  }
  \ .
\end{align}
Here, we relabelled $q_1^\mu\to q^\mu$.
Power counting of the vertex rules tell us that $\tilde \mN_1$ in Eq.~\eqref{WF:E2} is linear in $q$ with another power of $q$ for each order in the spins.
We may therefore redefine $\tilde\mN_1\to q_\mu\tilde \mN_1^\mu(-iq_\alpha)$.
In addition we factor out $m_1 m_2$ and a factor of $4\pi$ and get:\checked
\begin{align}
  f(x)
  \Big|_{\rm wl,1}
  =
  m_1 m_2\,
  4\pi 
  \int_{q}
  e^{-i q\cdot w_1}
  q_\mu
  \mN_1^\mu(-iq_\alpha)
  \frac{
        \dd(q\cdot v_1)
  }{
    q^2
    (q\cdot v_2 + i\eps)
  }
  \ .
\end{align}
This numerator $\mN_1^\mu(-iq_\alpha)$ is the one present above in Eq.~\eqref{eq:WaveformSchematic}.

We move on to the gravitational self-interaction graph (4) of Fig.~\eqref{WF:LOWaveform}.
We label the internal momenta emitted from the first and second body by $q_1^\mu$ and $q_2^\mu$ respectively.
The external momentum is labelled by $k^\mu=\oma n^\mu$ as above.
The contribution from this diagram then takes the following schematic form:\checked
\begin{align}
  \label{WF:GRE1}
  f(x)
  \Big|_{\rm GR}
  =
  \int_\oma e^{-ik\cdot x}
  \int_{q_1 q_2}
  e^{i(q_1 b_1+q_2 b_2)}
  \,
  \dd^d(q_1+q_2-k)
  \tilde \mM
  \frac{\dd(q_1\cdot v_1)\dd(q_2\cdot v_2)}{q_1^2 q_2^2}
  \ .
\end{align}
Again, $\tilde \mM$ is a numerator depending on the external background parameters and polynomial in the graviton momenta.

The idea is to split this particle symmetric contribution into two individual contributions from each body.
This is achieved using a partial fraction decomposition:\checked
\begin{align}
  \label{WF:PFD}
  \frac1{q_1^2 q_2^2}
  =
  -
  \frac12
  \frac1{
    q_1^2
    \,
    q_1\cdot k}
  -
  \frac12
  \frac1{
    q_2^2
    \,
    q_2\cdot k}
  \ .
\end{align}
This equation is derived from the constraint $k^\mu=q_1^\mu+q_2^\mu$ together with on-shell $k^\mu$, namely $k^2=0$.
The equation may be verified by eliminating $k^\mu$ inserting $2q_1\cdot k=q_1^2-q_2^2$ and $2q_2\cdot k=q_2^2-q_1^2$.
Both sides are then functions only of $q_i^2$ and are seen to be identical.

Insertion of the partial fraction identity Eq.~\eqref{WF:PFD} into the waveform $f(x)|_{\rm GR}$ Eq.~\eqref{WF:GRE1} results in the expression:\checked
\begin{align}
  \label{WF:GRE2}
  f(x)
  \Big|_{\rm GR}
  =&
  -\frac12
  \int_\oma e^{-ik\cdot x}
  \int_{q_1}
  e^{-iq_1 b+i k b_2}
  \tilde \mM
  \frac{\dd(q_1\cdot v_1)}{q_1^2\, q_1\cdot k}
  \dd\big((k-q_1)\cdot v_2\big)
  \\
  &-\frac12
  \int_\oma e^{-ik\cdot x}
  \int_{q_2}
  e^{iq_2 b+i k b_1}
  \
  \tilde \mM
  \frac{\dd(q_2\cdot v_2)}{q_2^2\, q_2\cdot k}
  \dd\big((k-q_2)\cdot v_1\big)
  \ .
  \nn
\end{align}
In the first and second lines we have integrated away $q_2^\mu$ and $q_1^\mu$ respectively using the $d$-dimensional delta constraint $q_1^\mu+q_2^\mu=k^\mu$.
The two lines are obtained from each other by particle exchange symmetry.
The first line has the same kind of energy conserving delta functions and Fourier factor as the contributions from the worldline emissions of the second body.
It is thus practical to split the gravitational contribution into two terms $f(x)|_{\rm GR}=f(x)|_{\rmm{GR},1}+f(x)|_{\rmm{GR},2}$.
The term $f(x)|_{\rmm{GR},1}$ corresponds to the first line of Eq.~\eqref{WF:GRE2} and $f(x)|_{\rmm{GR},2}$ to the second line.

We integrate away the energy, $\oma$, in $f(x)|_{\rmm{GR},1}$ just as above using the delta constraint $k\cdot v_2=\oma n\cdot v_2=q_1\cdot v_2$.
The denominator $q_1\cdot k$ becomes $q_1\cdot n q_1\cdot v_2/n\cdot v_2$.
Just as above, the exponent of the Fourier factor reduces to $q_1\cdot w_1$.
We replace $q_1^\mu$ by $q^\mu$ and find:\checked
\begin{align}
  f(x)
  \Big|_{\rm GR,1}
  &=
  \int_{q}
  e^{-iq\cdot w_1}
  \,
  \tilde \mM_1
  \frac{
    \dd(q\cdot v_1)
  }{
    q^2
    \,
    q\cdot v_2
    \,
    q\cdot n
  }
  \ .
\end{align}
The numerator $\tilde \mM$ was redefined to pick up the factor $n\cdot v_2$ and now depends on the particle label.

As above, power counting of the vertex rules tell us that $\tilde \mM_1$ is proportional to two factors of $q_\mu$ and another factor of $q_\mu$ for each power in spins.
We redefine $\tilde \mM\to q_\mu q_\nu \tilde\mM^\mn(-iq_\alpha)$ and factor out the masses $m_1m_2$ and $4\pi$ and find:\checked
\begin{align}
  f(x)
  \Big|_{\rm GR,1}
  &=
  m_1m_2\,
  4\pi
  \int_{q}
  e^{-iq\cdot w_1}
  \,
  q_\mu q_\nu \mM^\mn(-iq_\alpha)
  \frac{
    \dd(q\cdot v_1)
  }{
    q^2
    \,
    q\cdot v_2
    \,
    q\cdot n
  }
\end{align}
Again, $\mM^\mn(q_\alpha)$ is the same numerator as defined in Eq.~\eqref{eq:WaveformSchematic}.

We can now collect the contributions (1) - (3) and (4) of Fig.~\ref{WF:LOWaveform} into a single expression.
Adding the mirrored graphs of (1) - (3) using particle exchange symmetry, we get the schematic form of the full leading order waveform:\checked
\begin{align}
  \label{WF:Result}
  f(x)
  =
  m_1m_2\,
  4\pi\int_q
  e^{-i q \cdot w_1}
  \frac{
    \dd(q\cdot v_1)}{
    q^2}
  \bigg[
    \frac{
      q_\mu\mN^\mu_1(-iq_\alpha)}{
      q\cdot v_2+i\eps}
    +
    \frac{
      q_\mu q_\nu \mM^\mn_1(-iq_\alpha)
    }{
      q\cdot v_2\, q\cdot n
    }
    \bigg]
  +
  (1\leftrightarrow2)
  \ .
\end{align}
If we re-express the momenta of $\mN_i^\mu(-iq_\alpha)$ and $\mM^\mn_i(-iq_\alpha)$ in terms of derivatives with respect to $w_i^\mu$ or, equivalently, $b_i^\mu$ we get the schematic form Eq.~\eqref{eq:WaveformSchematic}.
The relevant integrals, then, are \checked
\bse
\label{eq:IntegralsWaveform}
\begin{align}
  \mJ^\mu_1
  &=
  4\pi\int_q
  \delta(q\cdot v_1)
  e^{-iq \cdot w_1}
  \frac{
    q^\mu
  }{
    q^2
    (q\cdot v_2+i\eps)
  }
  \ ,
  \\
  \mI^\mn_1
  &=
  4\pi\int_q
  \delta(q\cdot v_1)
  e^{-iq\cdot w_1}
  \frac{
    q^\mu q^\nu
  }{
    q^2
    (q\cdot v_2)
    (q\cdot n)
  }
  \ .
\end{align}
\ese
and integrals with subscript 2 defined by symmetry.
We will compute these in Sec.~\ref{sec:IntegralsWaveform} finding the expressions in Eqs.~\eqref{eq:ResultJ} and~\eqref{eq:ResultI}.

The result Eq.~\eqref{WF:Result} (or the related schematic forms Eqs.~\eqref{WF:SF1} and~\eqref{eq:WaveformSchematic}) gives the leading order post-Minkowskian contribution to the waveform to quadratic order in spins.
We have, however, not discussed the explicit computation of the numerators which are obtained by contracting the relevant vertex rules together.
This process is in principle straightforward and involves only simple algebra and the difficulty, instead, lies in the lengthy expressions of the individual vertex rules with the most lengthy one being in this case the graviton three vertex.
This difficulty is overcome by the use of computer algebra where the laborious contractions are easily automated and the lengthy numerators are computed.
They are presented in the ancillary file to Ref.~\cite{Jakobsen:2021lvp}.
Note, however, that we use slightly different notation and conventions here.

\section{Radiation of Energy and Angular Momentum}
\label{WF:RE}
The waveform describes gravitational waves traveling to infinity sourced by the scattering event.
The waves carry away linear and angular momentum from the scattering system of the two bodies.
Conservation of linear and angular momentum implies that the amount that is carried away by the gravitational waves is balanced by the amount lost by the particle system.
The total linear momentum of that system is $P^\mu$ and the total CoM angular momentum is $J^\mu$.
The change during the scattering event of these two quantities may therefore be computed from the waveform at infinity.
Their kicks in terms of the waveform is given by:\checked
\bse\label{WF:Rad}
\begin{align}
  \label{WF:RadE}
  P^\mu_{\rm rad}
  &=
  \frac1{16\pi G}
  \int \di u \di\Omega
  |\dot f_{\rm tt}|^2 n^\mu
  \ ,
  \\
  J^\mu_{\rm rad}
  &=
  \frac1{16\pi G}
  \eps^{\mn\rs}
  \hat P_\nu
  \int
  \di u
  \di\Omega
  \dot f^\ab_{\rm tt}
  \Big(
  f_{\rmm{tt},\alpha\rho}
  \eta_{\beta\sig}
  -
  \frac12
  x_\rho\pat_\sig
  f_{\rmm{tt},\ab}
  \Big)
  \ .
  \label{WF:RadJ}
\end{align}
\ese
Here, we use the notation $P_{\rm rad}^\mu=-\Del P^\mu$ and $J_{\rm rad}^\mu=-\Del J^\mu$ so that e.g. the corresponding $E_{\rm rad}$ is positive.
The angular integration is $\di\Omega=\sin(\theta)\di\theta\di\phi$ with $0\le \phi\le 2\pi$ and $0\le \theta\le \pi$.
Dots on the waveform denote differentiation with respect to the retarded time and integration on the retarded time is from past to future infinity.
The derivation of these equations is e.g. discussed in Ref.~\cite{Manohar:2022dea}.
In the first Sec.~\ref{WF:RAM} we use the second formula Eq.~\eqref{WF:RadJ} to compute the leading order post-Minkowskian radiated angular momentum to quadratic order in spins.

From the radiation formulas~\eqref{WF:Rad} we can also read off the differential fluxes of linear and angular momentum.
In particular we can read off the energy flux per spherical angle and the power spectrum, i.e. radiated energy per frequency.
While we are not able to compute these in the post-Minkowskian setting, we derive their post-Newtonian expansion in Sec.~\ref{WF:PNE}.

\subsection{Radiated Angular Momentum at Leading Order}
\label{WF:RAM}
The total radiated CoM angular momentum starts at the second post-Minkowskian order and at this order a surprising simplification of the formula for $J_{\rm rad}^\mu$ implies that it can be computed directly from the wave memory and the 1PM static gravitational fields of each body.

\begin{figure}[h]
  \renewcommand*\thesubfigure{\arabic{subfigure}} 
  \centering
  \begin{subfigure}{.19\textwidth}
    \centering
    \begin{tikzpicture}
      \coordinate (dl) at (-.5  ,0) ;
      \coordinate (dm) at (0   ,0) ;
      \coordinate (dr) at (+.5  ,0) ;
      \coordinate (dll)at (-.8 ,0 ) ;
      \coordinate (drr)at (+.8 ,0 ) ;
      \coordinate (ul) at (-.5  ,-1.2) ;
      \coordinate (um) at (0   ,-1.2) ;
      \coordinate (ur) at (+.5  ,-1.2) ;
      \coordinate (ull)at (-.8 ,-1.2 ) ;
      \coordinate (urr)at (+.8 ,-1.2 ) ;
      \coordinate (mrr)at (1,-.6) ;
      \draw[worlddot] (ul) -- (ur) ;
      \draw[worlddot] (ull) -- (ul) ;
      \draw[worlddot] (urr) -- (ur) ;
      \draw[worlddot] (dll) -- (drr) ;
      \draw[snake it] (mrr) -- (um) ;
    \end{tikzpicture}
    \caption{
    }
  \end{subfigure}
  \begin{subfigure}{.19\textwidth}
    \centering
    \begin{tikzpicture}
      \coordinate (dl) at (-.5  ,0) ;
      \coordinate (dm) at (0   ,0) ;
      \coordinate (dr) at (+.5  ,0) ;
      \coordinate (dll)at (-.8 ,0 ) ;
      \coordinate (drr)at (+.8 ,0 ) ;
      \coordinate (ul) at (-.5  ,-1.2) ;
      \coordinate (um) at (0   ,-1.2) ;
      \coordinate (ur) at (+.5  ,-1.2) ;
      \coordinate (ull)at (-.8 ,-1.2 ) ;
      \coordinate (urr)at (+.8 ,-1.2 ) ;
      \coordinate (mrr)at (1,-.6) ;
      \draw[worlddot] (ul) -- (ur) ;
      \draw[worlddot] (ull) -- (ul) ;
      \draw[worlddot] (urr) -- (ur) ;
      \draw[worlddot] (dll) -- (drr) ;
      \draw[snake it] (mrr) -- (dm) ;
    \end{tikzpicture}
    \caption{
    }
  \end{subfigure}
  \caption{
    Static contributions to the waveform at first post-Minkowskian order.
    Their static nature is apparent due to the lack of interactions between the two worldlines.
  }
  \label{WF:StaticWF}
\end{figure}
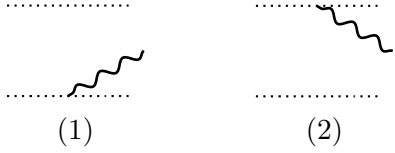

Let us first examine the static 1PM waveform which gets its contribution from the one-graviton emission graphs of each body shown in Fig.~\ref{WF:StaticWF}.
This corresponds to the Coulomb like fields of each of the bodies in straight-line motion.
We find the 1PM frequency space waveform simply as the one-point vertex Eq.~\eqref{WQFT:OnePoint} with external polarizations and on-shell external momentum $k^\mu$.
When we insert the external momentum in terms of the unit wave vector, $k^\mu=\omega n^\mu$, we find that all spin dependence vanishes.
Fourier transform to time domain is trivial.
The results are,\checked
\bse
\begin{align}
  f^{(1)}(k)
  &=
  2\dd(\oma)
  \sum_i
  m_i
  \frac{(\eps\cdot v_i)^2}{
    n\cdot v_i}
  \ ,
  \\
  f^{(1)}(x)
  &=
  2
  \sum_{i=1,2}
  m_i
  \frac{(\eps\cdot v_i)^2}{
  n\cdot v_i}
  \ .
\end{align}
\ese
The 1PM time domain waveform is clearly independent of time.
The independence of spins can be understood from the fact that spin only appears in the metric of a stationary object at subleading orders in $1/|\vct{x}|$.
The static 1PM waveform never appears in the linear momentum flux Eq.~\eqref{WF:RadE} due to the time derivatives on the waveform.

The angular momentum flux is a product of the time derivative of the waveform and spacial derivatives of the waveform.
The first and second factors get their leading order contribution from the dynamical 2PM waveform and the static 1PM waveform respectively.
Inserting the leading order contribution of each factor, we get the an expression for the leading order radiated angular momentum:\checked
\begin{align}
  J^\mu_{\rm rad}
  =
  \frac{G^2}{16\pi}
  \eps^{\mn\rs}
  \hat P_\nu
  \int
  \di\Omega
  \Big(
  f^{(1)}_{\rmm{tt},\alpha\rho}
  \eta_{\beta\sig}
  -
  \frac12
  x_\rho\pat_\sig
  f_{\rmm{tt},\ab}^{(1)}
  \Big)
  \int
  \di u
  \dot f^{(2)\ab}_{\rm tt}
  +
  \mO(G^3)
  \ .
\end{align}
Only the 2PM factor of the waveform depends on time and the time integration yields the leading order wave memory $\Del f_{\rm tt}^{(2)\mn}$ (Eq.~\ref{WF:WaveMemory}).
Thus, the radiated angular momentum is given by spherical integration of the static 1PM waveform and the 2PM wave memory:\checked
\begin{align}
  \label{WF:JRadLO}
  J^{(2)\mu}_{\rm rad}
  =
  \frac{1}{16\pi}
  \eps^{\mn\rs}
  \hat P^\nu
  \int
  \di\Omega\,
  \Big(
  f_{\rmm{tt},\alpha\rho}^{(1)}
  \eta_{\beta\sig}
  -
  \frac12
  x_\rho
  \pat_\sig
  f_{\rmm{tt},\ab}^{(1)}
  \Big)
  \Del f^{(2)\ab}_{\rm tt}
  \ .
\end{align}
This is a surprisingly simple formula for the leading order radiation of angular momentum first derived by Damour in Ref.~\cite{Damour:2020tta}.

The angular integrations required are simple and the only denominators with angular dependence of the integrand of Eq.~\eqref{WF:JRadLO} are powers of $n\cdot v_i$.
We work in the CoM frame using equations of Sec.~\ref{WF:GK} and find:\checked
\begin{align}
  n\cdot v_i
  =
  \frac1{m_i}
  \Big(
  E_i
  +
  (-1)^i
  p_\infty
  \cos(\theta)
  \Big)
  \ .
\end{align}
Products of the denominators $n\cdot v_1$ and $n\cdot v_2$ are turned into sums with partial fraction identities and apart from powers of these simple poles the numerator is polynomial in the sine and cosine of $\theta$ and $\phi$.
Spherical integration on these poles with sines and cosines in the numerator is now simple.
The waveform should be inserted in the transverse traceless gauge and here it is helpful that only the plus component of the 1PM waveform is nonzero,\checked
\begin{align}
  f^{(1)}_{\rm tt}(x)
  =
  f^{(1)}_{+}(x)
  =
  2 p_\infty \sin^2\theta
  \sum_i
  \frac{v_i}{
    1
    +
    (-1)^i
    v_i
    \cos\theta}
  \ ,
\end{align}
with $v_i=E_i/p_\infty$.
We will not consider the explicit computation here.

The result factorizes into the spinless result and a spin dependent factor.
This result was first derived in Ref.~\cite{Jakobsen:2021lvp} and we present it here as found in Ref.~\cite{Jakobsen:2022zsx}:\checked
\begin{align}
  \label{WF:JRadAO}
  J_{\rmm{rad}}^{(2)\mu}
  =
  -\frac{4M^3\nu^2 (2\gam^2-1)
  }{
    |b|\Gamma}
  \mI(\gam)
  \operatorname{Re}\!
  \bigg[
    \zeta^\mu
    \Bigg(
    1+
    \frac{2 v a_+\cdot \zeta}{
      |b|(1+v^2)
    }
    +
    \frac{
      (a_+\cdot \zeta)^2
    }{
      |b|^2
    }
    -
    \sum_{i=1,2}
    \frac{C_{E,i}}{
      |b|^2
    }
    (a_i\cdot \zeta)^2
    \Bigg)
    \bigg]
  \ .
\end{align}
Here, $\operatorname{Re}(X)$ denotes the real part of $X$ and $\zeta^\mu$ is a complex vector:
\begin{align}
  \zeta^\mu = \hat L^\mu + i \hat b^\mu
  \ .
\end{align}
The vectors $a^\mu_\pm$ are the symmetric and antisymmetric sums of the individual Pauli-Lubanski vectors $a^\mu_\pm=a^\mu_1\pm a^\mu_2$.
The prefactor $\mI(\gamma)$ expressed first in terms of $\gam$ then in terms of the relative velocity $v$ is\checked
\begin{align}
  \label{WF:IFunction}
  \mI(\gamma)
  &=
  \frac{8-5\gam^2}{
    3(\gam^2-1)
    }
  -
  \gam\frac{3-2\gam^2}{\sqrt{\gam^2-1}^3}
  \operatorname{arccosh}\gam
  \\\nn
  &=
  -\frac83
  +
  \frac1{v^2}
  +
  \frac{3v^2-1}{v^3}
  \operatorname{arctanh}v
  \ .
\end{align}
This function is positive and monotonically increasing from zero to infinity as $v$ goes from zero to one.
As a power series in $v$ it only has even powers.

In the non-spinning case, the angular momentum only changes magnitude during the scattering, and thus $J^\mu_{\rm rad}$ points in the direction of $J^\mu$.
This is no longer the case with generic spins where the system also looses angular momentum in the direction of the impact parameter.
Remarkably the result Eq.~\eqref{WF:JRadAO} was generalized to all orders in spins for Kerr black hole scattering in Ref.~\cite{Alessio:2022kwv}.
The total radiated angular momentum at the third post-Minkowskian order was computed in Refs.~\cite{Manohar:2022dea,DiVecchia:2022piu,Heissenberg:2022tsn}, though until now without spin effects.

We note that one must take care when considering the loss of angular momentum and its relativistic definition.
Thus as a surprising fact the time components of $\Del J^\mn$ were found to be non-zero already at the first PM order~\cite{Gralla:2021qaf,Gralla:2021eoi}.
These components, however, are projected out in the definition of the CoM Pauli-Lubanski angular momentum $J^\mu$.
Careful considerations of the definition of angular momentum, its radiative loss and the role of zero frequency gravitons (the static waveform) are found in Refs.~\cite{Veneziano:2022zwh,Manohar:2022dea,DiVecchia:2022owy,Heissenberg:2022tsn,Javadinezhad:2022ldc,Cristofoli:2022phh,Riva:2023xxm}.

\subsection{Post-Newtonian Expansion of Energy Fluxes}
\label{WF:PNE}
The radiation of linear momentum starts, in contrast to the angular momentum, at the third post-Minkowskian order.
Its integration offers a greater challenge than the corresponding angular momentum and we will only be able to derive that result after the analysis of the third post-Minkowskian loop integrals in \chap~\ref{sec:TL}.
Here, however, we will consider the post-Newtonian expansion of the waveform and energy fluxes which has a much simpler dependence on the coordinates.

In our post-Newtonian expansion we will find that the schematic forms Eqs.~\eqref{WF:SF1} and~\eqref{eq:WaveformSchematic} are not advantageous and that they create spurious terms that in fact cancel among themselves.
In order to realize why this cancellation happens we must go back to the derivation from diagrams where the graviton self-interaction graph was artificially split into two contributions.
When those two contributions are kept together no spurious terms appear.

We will focus on the total radiated energy as seen from the rest frame of the first body, $P_{\rm rad}\cdot v_1$, which we denote by $E'_{\rm rad}$:\checked
\begin{align}
  E'_{\rm rad}
  =
  P_{\rm rad}\cdot v_1
  \ .
\end{align}
We consider, then, the post-Newtonian expansion of the waveform in the frame of the first body.
In particular, we are interested in the denominators of the schematic form Eq.~\eqref{WF:SF1} which are $|w^\mu|$ and $|w_i^\mu|$.
Their post-Newtonian expansion is the most important one, because the remaining dependence of the waveform on its coordinates ($u_1$, $\theta$, $\phi$) simply is polynomial except for simple poles in the polar angle.

First, let us focus on $|w_2^\mu|$ which is given by:\checked
\begin{align}
  |w_2^\mu|^2
  =
  |b|^2 +
  \gam^2 v^2
  u_1^2
  \ .
\end{align}
This function will be our basic building block in the post-Newtonian limit and should not be expanded in velocity since $u_1$ could be arbitrarily large.
For that reason we define a new time coordinate $\tilde u$ and a rescaled version of $|w_2|$, $\tilde w$:\checked
\bse
\begin{align}
  \tilde u
  &=
  \frac{
    \gam v
    u_1
  }{
    |b|
  }
  \\
  \tilde w
  &=
  \sqrt{1+\tilde u^2}
\end{align}
\ese
The length of $w_2^\mu$ is simply proportional to the dimensionless $\tilde w$ by $|w_2^\mu|=|b| \tilde w$.
The dimensionless $\tilde w$, then, will be the basic building block of the PN expanded waveform.

We move on to the PN limit of $|w_1^\mu|$.
Its full expression is:\checked
\begin{align}
  |w_1^\mu|
  =
  |b|^2 +
  \gam^2 v^2
  u_2^2
  \ .
\end{align}
We must express the retarded time of the second body, $u_2$, in terms of $u_1$ which is the retarded time in our current frame and this may be done using $n\cdot w=0$.
This should then be re-expressed in terms of the reduced retarded time $\tilde u$ and we find:\checked
\begin{align}
  \gam v u_2
  =
  \gam v \frac{
  u_1
  -
  b\cdot n
  }{v_2\cdot n}
  =
  |b|\tilde u
  +
  \mO(v)
  \ .
\end{align}
In the second equality we use that $b\cdot n$ and $v_2\cdot n$ are constant at leading order in $v$.
The leading PN behavior of $\gam v u_2$ is then simply $|b|\tilde u$.
As a consequence, at leading order $|w_1^\mu|$ is also simply proportional to $\tilde w$ as $|w_2^\mu|=|b| \tilde w + \mO(v)$.
This is reasonable as in the PN limit the two projections $w_i^\mu$ must approximately be equal.

Finally, we move on to the shifted impact parameter, $w^\mu$.
Expressed in terms of $\tilde u$ the PN limit of $|w^\mu|$ is found to be:\checked
\begin{align}
  \label{WF:Spurious}
  |w^\mu|^2
  &=
  |b|^2 \tilde w^2
  -
  (n\cdot(b-v_{1\bot}))^2
  +
  \mO(v)
  \ .
\end{align}
This limit is not simply proportional to $\tilde w$ but instead to an expression including angular dependence.
From the naive PN expansion of the schematic forms Eqs.~\eqref{WF:SF1} and~\eqref{eq:WaveformSchematic} it would then seem that two kinds of denominators appear.
Namely, we would expect the simple $\tilde w$ and the more complicated PN limit of Eq.~\eqref{WF:Spurious}.
It is found, however, that the complicated denominators cancel when the sum on particle labels is performed.
In the end, only the simple denominator $|\tilde w|$ appears in the post-Newtonian expansion.

The spurious denominators appear from the expansion of $|w^\mu|$ which in turn comes from the gravitational self-interaction graph (4) of Fig.~\ref{WF:LOWaveform}.
The contribution from that graph was considered in Eq.~\eqref{WF:GRE1} where it was separated into a part for each body and it is this separation that is the cause of the spurious terms.

We go back to the expression for that graph before separating it in two pieces Eq.~\eqref{WF:GRE1} and then simply integrate away $q_2$ and $\oma$ with two of the three delta functions.
We find, then, the following expression:\checked
\bse
\begin{align}
  \label{WF:GREP}
  f(x)\Big|_{\rm GR}
  &=
  \int_{q_1}
  e^{-i q_1\cdot w_1}
  \tilde \mM
  \frac{\dd(q_1\cdot v_1)}{
    q_1^2
    (q_1^2
    +
    q_1\cdot \mL\cdot q_1)
  }
  \\
  \mL^\mn
  &=
  -2\frac{(P_1\cdot v_2)^{(\mu} (P_1\cdot n)^{\nu)}}{v_2\cdot n}
  =
  2\gam v
  \frac{
    \hat v_{1\bot}^{(\mu}
    (P_1\cdot n)^{\nu)}
  }{
    v_2\cdot n
  }
\end{align}
\ese
The second denominator is $q_2^2$ after integrating away the delta functions with the following steps:\checked
\begin{align}
  q_2^2
  \to
  q_1^2-2k\cdot q_1
  \to
  q_1^2
  -
  2
  \frac{
    v_2\cdot P_1\cdot  q_1
  \,
  n\cdot P_1\cdot q_1
  }{v_2\cdot n}
  \ .
\end{align}
The operator $\mL^\mn$ scales with $v$ and the second denominator of Eq.~\eqref{WF:GREP} is expanded as a geometric series in the PN limit:\checked
\begin{align}
  \frac{1}{q_1^2+q_1\cdot \mL\cdot q_1}
  =
  \frac1{q_1^2}
  \sum_n
  (-1)^n
  \bigg(
  \frac{
    q_1\cdot \mL\cdot q_1
  }{
    q_1^2
  }
  \bigg)^{n}
\end{align}
The resulting integrals are Fourier transforms of integer powers of $1/q_1^2$ with some tensor structure.
The only denominators will therefore be integer powers of $1/|w_1^\mu|$.
In fact, the expression Eq.~\eqref{WF:GREP} was directly (PM) integrated in Ref.~\cite{Jakobsen:2021smu} without splitting it up in two parts.

The important conclusion is that in the post-Newtonian expansion of the waveform, the only non-trivial dependence on $\tilde u$ is through powers of $\tilde w^{-1}$ such that at higher PN orders higher powers of $\tilde w^{-1}$ appear.
We will not consider the explicit PN expansion of the waveform here and only present the same (spinless) integrated results for $\di E_{\rm rad}/\di\Omega$ and $\di E_{\rm rad}/\di \omega$ as found in Ref.~\cite{Jakobsen:2021smu}.
The computation of these terms and the inclusion of higher orders in velocity and spins is, however, straightforward following the method outlined here with the only challenge being, again, the rather lengthy expressions involved.

We note that the PN expanded waveform can be transformed to frequency domain in terms of known functions.
The Fourier transform of the first few power of $\tilde w^{-1}$ reads:\checked
\bse
\begin{align}
  \int
  \di
  \tilde u\,
  e^{i\tilde u \tilde \oma}
  \frac1{\tilde w}
  &=
  2
  K_{0}(|\tilde \oma|)
  \ ,
  \\
  \int
  \di
  \tilde u\,
  e^{i\tilde u \tilde \oma}
  \frac1{\tilde w^2}
  &=
  \pi
  e^{-|\tilde \oma|}
  \ ,
  \\
  \int
  \di
  \tilde u\,
  e^{i\tilde u \tilde \oma}
  \frac1{\tilde w^3}
  &=
  2
  |\tilde\oma| K_{1}(|\tilde\oma|)
  \ .
\end{align}
\ese
Here, $K_0(|\tilde\oma|)$ and $K_1(|\tilde\oma|)$ are modified Bessel functions of the second kind.
They appear in the PN expansion of the power spectrum below.
The time integrations relevant to either $P_{\rm rad}$ or $J_{\rm rad}$ Eqs.~\eqref{WF:Rad} are trivial and given by:\checked
\begin{align}
  \int_{-\infty}^\infty
  \di \tilde u
  \frac{\tilde u^{n}}{
    \tilde w^m}
  =
  \frac{1+(-1)^n}{2}
  \frac{
    \Gamma\Big(
    \frac{m-n-1}{2}
    \Big)
    \Gamma\Big(
    \frac{n+1}{2}
    \Big)
  }{
    \Gamma\Big(
    \frac{m}{2}
    \Big)
  }
  \ .
\end{align}
When $n$ is odd the integral is naturally zero.
Also, we require $m-2\ge n\ge 0$ which is always the case in our application.

The energy flux per spherical angle, $\di E_{\rm rad}'/\di \Omega$, and the power spectrum, $\di E'_{\rm rad}/\di \omega$, are defined by:\checked
\bse
\begin{align}
  \frac{\di E'_{\text{rad}}}{\di\Omega}
  &=
  \frac1{16\pi G}
  \int\!\di u\, |\dot{f}_{\rm tt}(x)|^{2}
  \ ,
  \\
  \frac{\di E'_{\text{rad}}}{\di\omega}
  &=
  \frac{1}{8\pi G}
  \int\!\di \Omega
  |\omega f_{\rm tt}(k) |^{2}
  \ .
\end{align}
\ese
In the first and second lines we use the time and frequency waveforms respectively with $k^\mu=n^\mu\oma$.
Using the approach described above, we compute each of these differential observables in the PN expansion without spin to next-to-next-to-leading order (NNLO) in the PN expansion.

For the energy flux per spherical angle we find:\checked
\begin{align}
  &\frac{\di E'_{\text{rad}}}{\di\Omega}
  =
  \frac{G^3m_{1}^{2}m_{2}^{2} v}{512 b^{3}}
  \nn
  \\
  &\ \ \ 
 \times
  \Bigl [
  45(\cos^2\!\theta\cos^2\!\phi+\sin^2\!\phi)^2+
  109\sin^4\!\theta+630\sin^2\!\theta\sin^2\!\phi+
  354\sin^2\!\theta\cos^2\!\theta\cos^2\!\phi
  \nn
  \\
  &\ \ \ \ \ 
  +\frac{v}{2}\cos\theta\Bigl(
    135+30\sin^2\!\theta(44+61\sin^2\!\phi)+
    7\sin^4\!\theta(-200+264\sin^2\!\phi+45\sin^4\!\phi)\Bigr )
    \nn
  \\
  &\ \ \ \ \ 
  +
  \frac{v^{2}}{32}
  \Big(
    7545+
    9\sin^2\!\theta(5714+6627\sin^2\!\phi)+
    \sin^4\!\theta(-128104 + 27084\sin^2\!\phi+24255\sin^4\!\phi)
    \nn
  \\
  &\ \ \ \ \ 
  +
    \sin^6\!\theta(53200 - 70728\sin^2\!\phi - 11790\sin^4\!\phi + 525\sin^6\!\phi)
  \Big)
    +\mathcal{O}(v^{3})\Bigr ]  +\mathcal{O}(G^4)\,.
\end{align}
The linear in $v$ term of this series vanishes in $E'_{\rm rad}$ after spherical integration.

For the power spectrum, we find:\checked
\begin{align}
  &\frac{\di E'_{\text{rad}}}{\addcontentsline{}{}{}\di\omega}
  =
  \frac{16G^3 m_1^2 m_2^2 \tilde\oma^2}{15b^2}
  \bigg[
    12 
  \Big(
  (\frac{1}{3} + \tilde\oma ^2) {K_0}^2+3 \tilde\oma  K_0 K_1+ \left(1+\tilde\oma ^2\right) {K_1}^2
  \Big)
  \nn
  \\
  &\ \ \ \ \ 
  +
  \frac{v^2}{7}
  \Big(
  2 \left(5 \tilde\oma ^4-20 \tilde\oma ^2-64\right) {K_0}^2+4 \tilde\oma  \left(19 \tilde\oma ^2-8\right) K_0 K_1+\left(10 \tilde\oma ^4+3 \tilde\oma ^2+300\right) {K_1}^2
  \Big)
  \nn
  \\
  &\ \ \ \ \ 
  +
  \frac{v^4}{189}
  \Big(
  \left(20 \tilde\oma ^6+301 \tilde\oma ^4-1026 \tilde\oma ^2+1584\right) {K_0}^2+\tilde\oma  \left(120 \tilde\oma ^4-1523 \tilde\oma ^2+4518\right) K_0 K_1
  \nn
  \\
  &\ \ \ \ \ 
  +\left(20 \tilde\oma ^6+371 \tilde\oma ^4+4417 \tilde\oma ^2+13860\right) {K_1}^2
  \Big)+\mO(v^5)
  \bigg]
  +{\mathcal O}(G^4)\,.
\end{align}
Here, we use the dimensionless frequency $\tilde\oma=|b|\oma/v$.
The Bessel functions are $K_0=K_0(|\tilde \oma|)$ and $K_1=K_1(|\tilde \oma|)$.
It would be natural to include spin in the power spectrum and energy flux per spherical angle taking advantage of the $\mO(S^2)$ waveform.

\section{Integrals for Leading Order Waveform}
\label{sec:IntegralsWaveform}
In this section we derive the expressions for the integrals $\mJ^\mu$ and $\mI^\mn$ defined in Eqs.~\eqref{eq:IntegralsWaveform} verifying the expressions Eqs.~\eqref{eq:ResultsWaveform} printed above.
Our derivation follows the one given in Ref.~\cite{Jakobsen:2021lvp}.
We reprint the integrals here:\checked
\begin{subequations}
  \label{eq:IntegralsWaveformB}
\begin{align}
  \mJ^\mu_1
  &=
  4\pi\int_q
  \dd(q\cdot v_1)
  e^{-iq \cdot w_1}
  \frac{
    q^\mu
  }{
    q^2
    (q\cdot P_1\cdot v_2+i\eps)
  }
  \ ,
  \\
  \mI^\mn_1
  &=
  4\pi\int_q
  \dd(q\cdot v_1)
  e^{-iq\cdot w_1}
  \frac{
    q^\mu q^\nu
  }{
    q^2
    (q\cdot P_1\cdot v_2)
    (q\cdot P_1\cdot n)
  }
  \ .
\end{align}
\end{subequations}
We have inserted the projector $P_1^\mn=\eta^\mn-v_1^\mu v_1^\nu$ because the integral is most easily evaluated in the frame of $v_1^\mu$ in terms of the spacial parts of the vectors in that frame.
Its insertion is possible due to the energy conserving delta function.
The second integral $\mI^\mn_1$ is related to the first one $\mJ^\mu_1$ by the addition of the denominator $1/(q\cdot n)$ and another momentum in the numerator.
In that sense the first integral is simpler than the second one and we will consider them in that order.
The $i\eps$-prescriptions of the second integral are chosen as ``principal value''.
In fact, we could also have used retarded $i\eps$ prescriptions in the fraction decomposition Eq.~\ref{WF:PFD} which would have made the two integrals even more similar.

The integrals will be considered explicitly in $d=4$ space-time dimensions which is sensible as the numerators have no dim-reg poles and we are focusing on the (physical) $d=4$ waveform.
It could, however, be interesting to ask whether the waveform could be generalized to arbitrary dimensions.

\subsubsection{The Integral $\mJ^\mu_i$}

The integral $\mJ^\mu$ can be computed by re-expressing the worldline propagator as a time-integral.
This reflects the essential property of the worldline propagators, namely that they correspond to time integration with boundary condition at past infinity.
Mathematically this follows from the Fourier transform (Eq.~\ref{WEFT:WLP}):\checked
\begin{align}
  \frac{1}{
    q\cdot v_2+i\eps
  }
  =
  -i\int_{-\infty}^\infty
  \di\tau\,
  e^{i \tau q\cdot v_2}
  \theta(\tau)
  \ .
\end{align}
When inserted in the expression for $\mJ^\mu_1$ we find:\checked
\begin{align}\label{WF:JEquationA}
  \mJ_1^\mu
  =
  4\pi\int_{-\infty}^\infty \di\tau
  \theta(\tau)
  \int_q
  \dd(q\cdot v_1)
  e^{-iq\cdot(w_1-\tau v_2)}
  \frac{q^\mu}{q^2}
  \ .
\end{align}
Now, the momentum integration on $q^\mu$ has only one dimensionful scale $|w_1^\mu-\tau (P_1\cdot v_2)^\mu|$ and the dependence on this scale is given by dimensional analysis.
This integral is most easily evaluated in the rest frame of $v_1^\mu$ where $v_1^\mu=(1,\vct{0})$.
In this frame the integral reduces to a standard Euclidean Fourier transform.
In covariant notation we find:\checked
\begin{align}\label{WF:JEquationB}
  4\pi\int_q
  \dd(q\cdot v_1)
  e^{-iq\cdot(w_1-\tau v_2)}
  \frac{q^\mu}{q^2}
  =
  -i
  \frac{
    (w_1-\tau P_1\cdot v_2)^\mu
  }{|w_1-\tau P_1\cdot v_2|^3}
  \ .
\end{align}
The following relations,\checked
\begin{align}
  (w_1-\tau P_1\cdot v_2)^\mu
  &=
  b^\mu
  +
  (u_2-\tau) P_1\cdot v_2^\mu
  \ ,
  \\
  |w_1-\tau P_1\cdot v_2|^2
  &=
  |b|^2
  +
  (\gamma^2-1)
  (u_2-\tau)^2
  \ ,
\end{align}
imply that the $\tau$ integration of Eq.~\eqref{WF:JEquationA} is equivalent to a retarded integration on $(u_2-\tau)$ from infinite past to $u_2$.
At any rate the $\tau$ integration in Eq.~\eqref{WF:JEquationA} can now be performed and we find the expression given above in Eq.~\eqref{eq:ResultJ}.

\subsubsection{The Integral $\mI^\mn_i$}

We compute the second integral $\mI^\mn_1$ by taking advantage of the simplifying relation that,\checked
\begin{align}\label{WF:IIntegralA}
  \mI^\mn_1 w_{1\nu}
  =
  i |w_1|
  \frac{
    \pat
  }{
    \pat |w_1|}
  \mI^\mu_1
  =
  0
  \ ,
\end{align}
with $\mI^\mu_1$ given by the same expression as $\mI^\mn_1$ with only one momentum in the numerator:\checked
\begin{align}
  \mI^\mu_1
  &=
  4\pi\int_q
  \dd(q\cdot v_1)
  e^{-iq\cdot w_1}
  \frac{
    q^\mu 
  }{
    q^2
    (q\cdot v_2)
    (q\cdot n)
  }
  \ .
\end{align}
The first equality of Eq.~\eqref{WF:IIntegralA} is seen by realizing that $q\cdot w_1=|w_1| q\cdot \hat w_1$ with the unit vector $\hat w_1^\mu=w_1^\mu/|w_1|$ independent of $|w_1|$.
Thus, differentiation with respect to $|w_1|$ pulls down a factor $q\cdot w_1/|w_1|$.
The next equality of Eq.~\eqref{WF:IIntegralA} follows from dimensional analysis of $I_1^\mu$.
In four space-time dimensions this integral is dimensionless and must then depend only on $\hat w_1^\mu$.
Differentiation with respect to $|w_1|$ thus gives zero.

The integral $\mI^\mn_1$ is thus orthogonal to two four vectors: Trivially to $v_1^\mu$ and as just shown to $w_1^\mu$.
The tensor structure of the integral must thus be expressible in a two-dimensional subspace orthogonal to $v_1^\mu$ and $w_1^\mu$.
We may span that space with the two vectors $(\Pi_1\cdot v_2)^\mu$ and $(\Pi_1\cdot n)^\mu$ with $\Pi_1^\mn$ the projector into the two-dimensional subspace orthogonal to $v_1^\mu$ and $w_1^\mu$ defined above in Eq.~\eqref{WF:PiProjector}.

A two-dimensional symmetric tensor has three independent tensor structures and we may write an ansatz for the $\mI^\mn_1$ integral as:\checked
\begin{align}
  \mI^\mn_1
  =
  \Pi_{1\alpha}^{\mu}\Pi_{1\beta}^{\nu}
  \Big(
  c_{vv}
  v_2^\alpha
  v_2^\beta
  +
  c_{nn}
  n^\alpha
  n^\beta
  +
  2 c_{vn}
  v_2^{(\alpha}
  n^{\beta)}
  \Big)
  \ ,
\end{align}
with some scalars $c_{\sig}$ with $\sig\in\{vv,nn,vn\}$.
We may immediately remove one of these scalars using the symmetry of $\mI^\mn_1$ under the exchange $(P_1\cdot v_2)^\mu\leftrightarrow (P_1\cdot n)^\mu$.
Thus we find that $c_{vv}=c_{nn}$.
In addition we may redefine the basis of our ansatz to match that of Ref.~\cite{Jakobsen:2021lvp} and we get,\checked
\begin{align}
  \mI^\mn_1
  =
  c_1 \Pi_1^\mn
  +
  2
  c_2
  (\Pi_1\cdot v_2)^{(\mu}
  (\Pi_1\cdot n)^{\nu)}
  \ ,
\end{align}
with two new constants $c_1$ and $c_2$.
Starting from either ansatz, the coefficients can be fixed using the two following equations:\checked
\bse
\begin{align}
  n_\mu v_{2\nu} \mI^\mn_1
  &=
  -\frac{1}{|w_1|}
  \ ,
  \\
  n_\mu n_\nu
  \mI^\mn_1
  &=
  \frac{
    v_2\cdot \Pi\cdot n
  }{
    (\gamma^2-1)
    |b|^2|w_1|}
  \ .
\end{align}
\ese
Both of these two integrals are similar to the $\mJ^\mu_1$ integral discussed above.
For the second integral one must only change the retarded prescription to a principal value prescription.
In that case one changes the Heaviside function $\theta(\tau)$ in the Fourier transform Eq.~\eqref{WF:JEquationA} to a sign function $\tau/|\tau|$.


\chapter{Worldline Observables and Hamiltonian at $\mO(G^3,S^2)$}
\label{sec:Scattering}
The third post-Minkowskian contributions to the impulse and spin kick introduce several new properties as compared with their first and second PM orders.
Thus, as discussed in Sec.~\ref{WQFT:Worldline1} those two previous orders may be fixed completely from matching them to the probe limit of probe particle motion in a gravitational Kerr/neutron star background.
The most important new property of the 3PM observables, then, is that they include a new mass sector which is related to the first self-force order which in turn introduces radiative effects to the observables.
At this order we may thus speak of both conservative and radiative contributions with the full (radiative) observables having both contributions.
While the probe limit and corresponding mass sectors generally depend only rationally on the Lorentz factor $\gam$, the first self-force mass sector introduces logarithmic dependence on $\gam$ which in the present approach for deriving the observables shows up in the relevant two-loop integrals analyzed in \chap~\ref{sec:TL}.

Complete conservative, spinning results for the impulse and spin kick at the third post-Minkowskian order were first derived in Ref.~\cite{Jakobsen:2022fcj} to quadratic order in the spins.
Full (radiative) results for the impulse and spin kick at corresponding orders in $G$ and spins were then first presented in Ref.~\cite{Jakobsen:2022zsx} with results for the total radiated four-momentum derived earlier in Ref.~\cite{Riva:2022fru}.
The Hamiltonian at this order was first partially derived in Ref.~\cite{FebresCordero:2022jts} with full supplementing results given in Ref.~\cite{Jakobsen:2022zsx}.

In this chapter we present the $\mO(G^3,S^2)$ results of Refs.~\cite{Jakobsen:2022fcj,Jakobsen:2022zsx}.
The general structure of the spinning observables is similar to the original spinless results though the motion is now non-planar and in addition to the impulse we have the new observable the spin kick.
In the first section~\ref{SPIN:WO} we discuss this structure and consider the definition of scattering angles with spin and radiation.
Finally we show how parts of the full observables can be reconstructed with radiation reaction from lower order PM data.

In the next section~\ref{SPIN:CBM} we focus on the mapping of the unbound observables to bound motion where the main result is the conservative two-body Hamiltonian at $\mO(G^3,S^2)$.
In addition we consider the direct mapping of gauge invariant observables by deriving the bound binding energy to next-to-next-to-leading (NNLO) post-Newtonian order from the aligned spins 3PM scattering angle.

Finally in Sec.~\ref{EXP} we provide explicit results for one of the generalized scattering angles including generic spins and radiation and for the spinning two-body Hamiltonian.
Like in the case of the gravitational bremsstrahlung, the results of this section are generally lengthy and complete results are most conveniently found in the ancillary files to Refs.~\cite{Jakobsen:2022fcj,Jakobsen:2022zsx}.

\section{Worldline Observables with Radiation Reaction and Spin}
\label{SPIN:WO}
In this section we focus on the structure of the full (radiative) expressions for the impulse and spin kick.
We will not consider their explicit derivation in any detail except for the following brief discussion.
The observables are computed from the on-shell one-point functions: $\oma^2 \inin{\Del z(\oma)}|_{\oma\to0}$ and $\oma \inin{\Del \psi(\oma)}|_{\oma\to0}$.
Important ingredients are the spinning WQFT Feynman rules Sec.~\ref{WQFT:FV}, the off-shell recursion for graph generation Sec.~\ref{sec:TLWQFT} and the retarded two-loop integrals \chap~\ref{sec:TL}.
All ingredients are easily put together within the WQFT formalism and the systematic generation of graphs, contraction of Feynman vertices of each graph and tensor and IBP-reduction of the integrals can then be done with computer algebra.

First, in Sec.~\ref{SPIN:Structure} the general structure of the 3PM worldline observables is analyzed.
Next, in Sec.~\ref{SPIN:Angle} we consider generalized definitions of scattering angles including spin and radiation and parameterizations of the observables in terms of those.
Finally, in Sec.~\ref{SPIN:LR} we consider a generalization of the Bini-Damour formula to include spin from which parts of the worldline observables can be reconstructed from lower-order PM data.

\subsection{Structure of the Impulse and Spin Kick
}\label{SPIN:Structure}
Let us analyze the structure of the impulse and spin kick at the third post-Minkowskian order.
We find it practical to split both of the observables into four parts:\checked
\bse
\begin{align}
  \Delta p_i^\mu
  &=
  \Delta p_{i,\rm cons}^{(+)\mu}
  +
  \Delta p_{i,\rm cons}^{(-)\mu}
  +
  \Delta p_{i,\rm rad}^{(+)\mu}
  +
  \Delta p_{i,\rm rad}^{(-)\mu}
  \ ,
  \label{eq:parityP}
  \\
  \Delta S_i^\mu
  &=
  \Delta S_{i,\rm cons}^{(+)\mu}
  +
  \Delta S_{i,\rm cons}^{(-)\mu}
  +
  \Delta S_{i,\rm rad}^{(+)\mu}
  +
  \Delta S_{i,\rm rad}^{(-)\mu}
  \ 
  .
  \label{eq:parityA}
\end{align}
\ese
The subscripts ``cons'' and ``rad'' denote conservative and radiative parts respectively.
At the third post-Minkowskian order this split into conservative and radiative parts is simply made with respect to the two regions potential and radiative.
The superscripts $(\pm)$ are defined by even or odd behavior under the transformation $v_i^\mu\to-v_i^\mu$.

The contribution to each of the four parts of the impulse at the third post-Minkowskian order take the following schematic form:\checked
\bse
\label{SPIN:SchematicImpulse}
\begin{align}
  \Delta p_{1,\rm cons}^{(3;+)\mu}
  &=
  \frac{m_1^2 m_2^2}{|b|^3}
  \bigg[
    \frac{
      \text{arccosh}\gamma
    }{
      \sqrt{\gamma^2-1}
    }
    c_1^{(+)\mu}
    +
    \sum_{n=1}^3
    \Big(
    \frac{m_1}{m_2}
    \Big)^{n-2}
    c_{n+1}^{(+)\mu}
    \bigg]
  \ ,
  \\
  \Delta p_{1,\rm rad}^{(3;+)\mu}
  &=
  \frac{
    m_1^2 m_2^2
  }{
    |b|^{3}
  }\,
  \mI(\gam)\,
  c_5^{(+)\mu}
  \ ,
  \\
  \Delta p_{1,\rm cons}^{(3;-)\mu}
  &=
  \sum_{n=1}^3
  \frac{
    \pi m_1^2 m_2^2
  }{
    |b|^{3}}
  \Big(
  \frac{m_1}{m_2}
  \Big)^{n-2}
  c_{n}^{(-)\mu}
  \ ,
  \\
  \Delta p_{1,\rm rad}^{(3;-)\mu}
  &=
  \frac{
    \pi m_1^2 m_2^2
  }{
    |b|^{3}} 
  \bigg[
    c_{4}^{(-)\mu} +
    \frac{\text{arccosh}\gamma}{\sqrt{\gamma^2-1}}
    c_{5}^{(-)\mu}
    +
    \log\left(\frac{1+\gamma}2\right)
    c_{6}^{(-)\mu}
    \bigg]
  \ .
\end{align}
\ese
The function $\mI(\gam)$ appearing in the second line was defined in Eq.~\eqref{WF:IFunction} and is given by:\checked
\begin{align}
    \mI(\gamma)
  =
  \frac{8-5\gam^2}{
    3(\gam^2-1)
    }
  -
  \gam\frac{3-2\gam^2}{\sqrt{\gam^2-1}^3}
  \operatorname{arccosh}\gam
  \ .
\end{align}
The vectors $c_n^{(\pm)\mu}$ can be expanded in basis vectors and spin structures in terms of the following bases $\rho^{(\pm)}$:\checked
\begin{subequations}\label{eq:BasisVectors}
  \begin{align}
    \rho^{(+)}
    \!
    &=
    \!
    \bigg\{
    \hat{b}^\mu,\,
    \frac{a_i\bdot\hat{L}}{|b|}\hat{b}^\mu,\,
    \frac{a_i\bdot\hat{b}}{|b|}\hat{L}^\mu,\,
    \frac{a_i\bdot a_j}{|b|^2}\hat{b}^\mu,\,
    \frac{a_i\bdot\hat{b}\ a_j\bdot\hat{b}}{|b|^2}\hat{b}^\mu,\,
    \frac{a_i\bdot v_{\bar\imath}\ a_j\bdot v_{\bar\jmath}}{|b|^2}\hat{b}^\mu,\,
    \frac{a_i \bdot\hat{b}\ a_j\bdot \hat{L}}{|b|^2}\hat{L}^\mu,\,
    \frac{a_i\bdot \hat{b}\ a_j\bdot v_{\bar\jmath}}{|b|^2}v_k^\mu
    \bigg\}\,,
    \\
    \rho^{(-)}
    \!
    &=
    \!
    \bigg\{
    v_i^\mu,\,
    \frac{a_i\bdot \hat{L}}{|b|}v_j^\mu,\,
    \frac{a_i\bdot v_{\bar\imath}}{|b|}\hat{L}^\mu,\,
    \frac{a_i\bdot a_j}{|b|^2}v_k^\mu,\,
    \frac{a_i\bdot v_{\bar\imath}\,a_j\bdot v_{\bar\jmath}}{|b|^2}v_k^\mu,\,
    \frac{a_i\bdot\hat{b}\,a_j\bdot\hat{b}}{|b|^2}v_k^\mu,\,
    \frac{a_i\bdot v_{\bar\imath}\,a_j\bdot\hat{L}}{|b|^2}\hat{L}^\mu,\,
    \frac{a_i\bdot v_{\bar\imath}\,a_j\bdot \hat{b}}{|b|^2}\hat{b}^\mu
    \bigg\}\,.
  \end{align}
\end{subequations}
The vectors $c_n^{(+)\mu}$ are expanded in terms of the basis $\rho^{(+)}$ and $c_n^{(-)\mu}$ in terms of $\rho^{(-)}$.
The coefficients of these expansions are then simple polynomials in $\gam$ except for denominators in integer powers of $\sqrt{\gam^2-1}$.
At quadratic order in spins these coefficients also include linear dependence on the finite size coefficients $C_{E,i}$.
The factors of $\pi$ in Eqs.~\eqref{SPIN:SchematicImpulse} ensure that the polynomials in $\gam$ have rational coefficients.
We note that the elements in the bases $\rho^{(\pm)}$ in Eqs.~\eqref{eq:BasisVectors} are ordered in terms of powers in spin and that the basis $\rho^{(-)}$ is simply obtained from $\rho^{(+)}$ by exchanging $\hat b^\mu$ and $v_i^\mu$.

The schematic form of the third post-Minkowskian contribution to the spin kick is analogous to the impulse Eqs.~\eqref{SPIN:SchematicImpulse} and takes the following form:\checked
\bse
\label{SPIN:SchematicSpinKick}
\begin{align}
  \Delta S_{1,\rm cons}^{(3;-)\mu}
  &=
  \frac{m_1^2 m_2^2}{|b|^3}
  \Big[
    \frac{\text{arccosh}\gamma}{\sqrt{\gamma^2-1}}
    \,
    d_1^{(-)\mu}
    +
    \sum_{n=1}^3
    \Big(
    \frac{m_1}{m_2}
    \Big)^{\!n-2}
    d_{n+1}^{(-)\mu}
    \Big]
  \ ,
  \\
  \Delta S_{1,\rm rad}^{(3;-)\mu}
  &=
  \frac{m_1^2 m_2^2}{|b|^{3}}
  \mI(\gam)
  \,d_5^{(-)\mu}
  \ ,
  \\
  \Delta S_{1,\rm cons}^{(3;+)\mu}
  &=
  \sum_{n=1}^3
  \frac{\pi m_1^2 m_2^2}{|b|^{3}}
  \Big(
  \frac{m_1}{m_2}
  \Big)^{n-2}
  d_{n}^{(+)\mu}
  \ ,
  \\
  \Delta S_{1,\rm rad}^{(3;+)\mu}
  &=
  \frac{\pi m_1^2 m_2^2}{|b|^{3}}
  \bigg[
    d_{4}^{(+)\mu}
    +
    \frac{
      \text{arccosh}\gamma
    }{
      \sqrt{\gamma^2-1}
    }
    \,d_{5}^{(+)\mu}
    +
    \log\left(\frac{1+\gamma}2\right)
    \,d_{6}^{(+)\mu}
    \bigg]
  \ .
\end{align}
\ese
The vectors $d^{(\pm)\mu}_n$ can be expanded in terms of bases $\rho'^{(\pm)}$ defined by,
\bse
\begin{align}
  \rho'^{(+)}
  &=
  \bigg\{
  \frac{
    a_1\bdot\hat b}{|b|}
  \hat b^\mu
  \,,
  \frac{
    a_1\bdot v_2}{
    |b|}
  v_i^\mu
  \,,
  \frac{
    a_i\bdot \hat b
    \
    a_j\bdot \hat L}{
    |b|^2}
  \hat b^\mu
  \,,
  \frac{
    a_1\bdot \hat b
    \ 
    a_i\bdot \hat b}{
    |b|^2
  }
  \hat L^\mu
  \,,
  \frac{
    a_1\bdot v_2
    \
    a_i\bdot v_{\bar\imath}}{
    |b|^2}
  \hat L^\mu
  \,,
  \frac{
    a_i\bdot \hat L
    \
    a_j\bdot v_{\bar\jmath}}{
    |b|^2}
  v_k^\mu
  \bigg\}
  \ ,
  \\
  \rho'^{(-)}
  &=
  \bigg\{
  \frac{
    a_1\bdot v_2}{
    |b|}
  \hat b^\mu
  \,,
  \frac{
    a_1\bdot \hat b}{
    |b|}
  v_i^\mu
  \,,
  \frac{
    a_i\bdot \hat L
    \
    a_j\bdot v_{\bar \jmath}}{
    |b|^2}
  \hat b^\mu
  \,,
  \frac{
    a_i\bdot \hat b
    \
    a_j\bdot v_{\bar\jmath}}{
    |b|^2}
  \hat L^\mu
  \,,
  \frac{
    a_i\bdot \hat b
    \
    a_j\bdot \hat L}{
    |b|^2}
  v_k^\mu
  \bigg\}
  \ .
\end{align}
\ese
Thus, $d_n^{(+)\mu}$ may be expanded in terms of the basis $\rho'^{(+)}$ and $d_n^{(-)\mu}$ in terms of $\rho'^{(-)}$.
Again, as discussed below Eq.~\eqref{eq:BasisVectors}, the coefficients in this expansion are simple polynomials in $\gamma$ with the exceptions listed there.
Note that the spin kick of the first body must be proportional to one factor of $a_1^\mu$ and thus only basis elements that contain at least one such copy are relevant.

The schematic forms of the impulse and spin kick Eqs.~\eqref{SPIN:SchematicImpulse} and~\eqref{SPIN:SchematicSpinKick} exhibit all non-trivial dependence of the observables on $\gam$.
The functions appearing there follow the same pattern as the spinless impulse, namely that the conservative parts only include the rapidity $\text{arccosh}\gam$ and that the radiative part involves a $\log(\gam+1)/2$ in addition to the rapidity.
The schematic forms are in agreement with the general form of worldline observables discussed in Sec.~\ref{WQFT:Worldline1} including the simple mass dependence.

At third post-Minkowskian order we define the conservative and radiative parts of the observables as the parts that get contributions from the potential and radiative regions respectively (Sec.~\ref{sec:TLMethod}).
At the first and second post-Minkowskian order there are no radiative parts and the full observables are given by their conservative parts.
At the third post-Minkowskian order a single graviton can go on-shell and the radiative parts of the observables are non-zero.

The superscript $(\pm)$ is defined with respect to odd or even behavior under the transformation $v^\mu\to-v^\mu$:\checked
\begin{align}
  \label{eq:timeReversal}
  \left.
  \Delta X^{(\pm)}
  \right|_{v_i^\mu\to-v_i^\mu}
  &=
  \pm\Delta
  X^{(\pm)}
  \ .
\end{align}
Under this transformation the spin vectors, $a^\mu_i$, and impact parameter, $b^\mu$, are left unchanged as well as the angular momentum $L^\mu$.
The spin tensors, however, $S^\mn_i$ pick up a sign as a consequence of their relation to the velocities and spin vectors.

The four parts are symmetric under another simple operation.
Here, we change the sign of the velocity $v$.
In terms of the Lorentz factor this transformation reads,\checked
\begin{align}
  \sqrt{\gam^2-1}\to-\sqrt{\gam^2-1}
  \ ,
\end{align}
while $\gam$ itself is unchanged.
The rapidity, $\text{arccosh}(\gam)=\text{arctanh}(v)$, flips sign under this operation and the logarithm, $\text{log}(\frac{\gam+1}{2})$, is unchanged.
If we now define a transformation where we simultaneously send $v\to-v$ and $b^\mu\to-b^\mu$ we find that the conservative parts of the impulse are unchanged and that the radiative parts pick up a sign.
Under this transformation the spin vectors and tensors are left unchanged but the angular momentum vector $L^\mu$ picks up a sign (while the unit vector $\hat L^\mu$ is unchanged).
Similar behavior of the spin kick under this transformation distinguishes its conservative and radiative parts.

Alternatively, we may therefore think of the four parts of the observables in Eqs.~\eqref{SPIN:SchematicImpulse} and Eqs.~\eqref{SPIN:SchematicSpinKick} as being defined with respect to their symmetries under the two transformations $v^\mu\to-v^\mu$ and the one considered here $v\to-v$ and $b^\mu\to-b^\mu$.

Finally, we note that the two parts of the impulse $\Del p_{1,\rmm{rad}}^{(3;+)\mu}$ and $\Del p_{1,\rmm{cons}}^{(3;-)\mu}$ are simpler than the remaining parts and likewise for the spin kick $\Del S^{(3;-)\mu}_{1,\rmm{rad}}$ and $\Del S^{(3,+)\mu}_{1,\rmm{cons}}$ are simpler than the two other parts.
In fact, in Sec.~\ref{SPIN:LR} we will see that these (simpler) parts can be straightforwardly inferred from the lower post-Minkowskian orders.
All essential information of the observables at the third post-Minkowskian order is then contained in the four remaining parts $\Del p_{1,\rmm{rad}}^{(3;-)\mu}$, $\Del p_{1,\rmm{cons}}^{(3;+)\mu}$, $\Del S^{(3;+)\mu}_{1,\rmm{rad}}$ and $\Del S^{(3,-)\mu}_{1,\rmm{cons}}$.
This may also be understood as a consequence of the constraints that the kicks satisfy including conservation of mass, spin length and SSC.

From the knowledge of $\Del p_i^\sig$ we may compute the total loss of four-momentum $\Del P^\mu$ which matches the result of Ref.~\cite{Riva:2022fru}.
Our results do not, however, enable us to derive a similar expression for the total loss of angular momentum $\Del J^\mu$ which is, then, the last missing piece at quadratic order in spins and third post-Minkowskian order.
One way to extract this would be from the waveform at infinity as in Ref.~\cite{Manohar:2022dea}.

\subsection{Generalized Scattering Angles
}\label{SPIN:Angle}
In the special case of conservative planar motion the spin kick is zero and so too is the change in the total momentum $P^\mu$.
The only observable of interest, then, is the change in the relative momentum $p^\mu$ and this can be parametrized by a single scattering angle $\phi$ as:\checked
\begin{align}\label{SPIN:CON}
  \Del p^\mu_{\rm cons}
  \Big|_{\rm aligned\ spins}
  =
  p_\infty\Big(
  \hat p^\mu
  (\cos\phi-1)
  +
  \hat b^\mu \sin\phi
  \Big)
  \ .
\end{align}
The angle is a scalar and is in many respects simpler than the full kick.
It is thus advantageous to consider the generalization of such angles to the cases of generic spins and radiative motion.
In particular, the definition of conservative scattering angles for generic spins will prove useful in the computation of the Hamiltonian in Sec.~\ref{SPIN:CBM}.
In fact, we find that the observation that aligned conservative motion can be captured by a single scattering angle generalizes to generic spins.
Thus, the conservative impulse and spin kicks can be parametrized in terms of a single scattering angle defined for generic spins.

As a start we define angles $\theta_i$ and $\phi_i$:\checked
\begin{align}
    \label{eq:genericAngles}
    \sin
    \bigg(
    \frac{\theta_1}{2}
    \bigg)
    &=
    \frac{|\Delta p_{1}|}{2\pin}
    \ ,
    &
    \sin
    \bigg(
    \frac{\theta_2}{2}
    \bigg)
    &=
    \frac{|\Delta p_{2}|}{2\pin}
    \ ,
    \\\nn
    \sin(\phi_1)
    &=
    \frac{\hat{b}\cdot \Delta p_{1}
    }{
      \pin
    }\,,
    &
    \sin(\phi_2)
    &=
    -\frac{\hat{b}\cdot \Delta p_{2}
    }{
      \pin
    }
    \ .
  \end{align}
We define conservative versions of these angles $\theta_{\rm cons}$ and $\phi_{\rm cons}$ by inserting the conservative impulse $\Del p^\mu_{1,\rmm{cons}}=-\Del p^\mu_{2,\rmm{cons}}=\Del p_{\rm cons}^\mu$.
In this case the particle label on the scattering angles is superfluous because of the symmetric (opposite) motion of the two particles.
In the conservative case, these angles have simple physical interpretations as scattering angles.
Thus $\theta_{\rm cons}$ measures the total angle between the incoming and outgoing relative momentum $p^\mu$ in the CoM frame and $\phi_{\rm cons}$ measures the total angle of the trajectory as projected into the initial plane of scattering spanned by $b^\mu$ and $p^\mu$.
However, with the inclusion of radiation the simple physical interpretation of the angles is not clear.

The difference between the two conservative angles $\theta_{\rm cons}$ and $\phi_{\rm cons}$ becomes relevant only when we consider misaligned spins.
Thus, for aligned spins when the motion is planar the two angles are identical.
In fact, at linear order in the spins, the two angles are equivalent even for misaligned spins because their only spin dependence is through the combination $L\cdot a_i$.
In general, for quadratic spins and radiative effects, the different angles generally do not agree.
At the third post-Minkowskian order, however, the aligned spins angles including radiation are independent of the particle label and $\theta_i$ coincides with $\phi_i$.

The angle $\theta_i$ is manifestly SUSY-invariant.
That is, it does not depend on the choice of SSC or the coordinate center of the individual bodies.
Instead, as it stands, the angle $\phi_i$ clearly depends on the choice of SSC because of the impact parameter which is not SUSY-invariant.
A SUSY-invariant definition of $\phi_i$ would be achieved by using the SUSY impact parameter $\beta^\mu$ instead of $b^\mu$.
In the following we will mostly focus on $\theta_i$ because of its SUSY although we could also use the SUSY invariant generalization of $\phi_i$.

In Sec.~\ref{SPIN:CBM} we will use $\theta_\rmm{cons}$ to parametrize the $\mO(G^3,S^2)$ two-body Hamiltonian.
Let us consider its schematic form which we expand in the post-Minkowskian expansion in orders of $G$ and in spins.
The schematic form for the full angle including radiation reads:\checked
\begin{align}
  &
  \frac{\theta_k}{\Gamma}
  =
  \sum_{n=1}^3
  \bigg(
  \frac{G M}{|b|}
  \bigg)^n
  \Bigg[
    \thii{n;0}
    -
    \sum_i \thii{n;1,i} \frac{\hat L\cdot a_i}{|b|}
    \nn
    \\
    &\qquad
    +
    \sum_{i,j}
    \frac{a_i^\mu a_j^\nu}{|b|^2}
    \Big(
    -\thii{n;2,1,i,j} \eta_{\mu\nu}
    +\thii{n;2,2,i,j} \hat b_\mu \hat b_\nu
    +\thii{n;2,3,i,j} v_{\bar \imath,\mu} v_{\bar \jmath,\nu}
    +\thii{n;2,4,i,j}_k \hat b_\mu v_{\bar \jmath,\nu}
    \Big)
    \Bigg]
  \nn
  \\
  &\qquad
  +\mO(S^3,G^4)\ .
    \label{eq:angleCov2}
\end{align}
Here, we use the bar notation $v_{\bar 1}=v_2$ and $v_{\bar 2}=v_1$.
Only the final expansion coefficient of the second line depends on the particle label (and only at 3PM).
This is related to the fact that it is the only term which flips sign under $v_i^\mu\to-v_i^\mu$.
The superscripts of the coefficients $\theta^{(n;A)}$ label the PM order $n$ and spin structures $A$.
The signs of the different terms are chosen so that the spacial scalar products come with a positive sign.
Several coefficients are related under particle exchange symmetry $(1\leftrightarrow2)$.
We define analogous expansion coefficients for the conservative scattering angle $\theta_\rmm{cons}$ which we label $\theta^{(n;A)}_{\rm cons}$.
Expressions for the expansion coefficients, $\theta_{i}^{(n;A)}$ are found in Sec.~\ref{SPIN:ESA}.
In addition, they are found in the supplementary material to Ref.~\cite{Jakobsen:2022zsx}.
They are functions only of $\gam$, $\nu$ and $C_{\rmm{E},i}$.
Their dependence on $\gam$ is polynomial except for the same special functions as in the impulse Eqs.~\eqref{SPIN:SchematicImpulse} and (possibly negative) powers of $\sqrt{\gam^2-1}$.
Their dependence on both $\nu$ and $C_{\rmm{E},i}$ is linear.

As an interesting special case we consider the high energy limit of the angle $\theta_i$.
In this limit we let $\gam\to\infty$ while keeping the total energy $E$ and spin vectors $a_i^\mu$ constant.
In this limit the individual masses $m_i$ tend to zero with a finite mass ratio $\nu$.
At leading order in $\gam^{-1/2}$ and to cubic order in $G$ the angle reads:\checked
\begin{align}
  &
  \theta_i
  =
  4
  \frac{G E}{|b|}
  \bigg[
    1
    +
    \frac{\hat L\cdot a_+}{|b|}
    -
    \frac{
      2
      a_+^2
      +
      3
      \big(
      \hat b\cdot a_+
      \big)^2
    }{2|b|^2}
    +
    \sum_j
    C_{{\rm E},j}
    \frac{a_j^2+2\big(\hat b\cdot a_j
      \big)^2}{|b|^2}
    \bigg]
  +
  \frac{32}{3}
  \bigg(\frac{G E}{|b|}\bigg)^3
  \bigg[
    1
    +
    3\frac{\hat L\cdot a_+}{|b|}
    \nn
    \\&\qquad
    -
    \frac{3}{20}
    \frac{
      41 a_+^2
      +
      a_-^2
      +
      50\big(
      \hat b\cdot a_+
      \big)^2}{|b|^2}
    -\frac{945\pi}{8192}C_{{\rm E},i}
    \frac{
      \hat b\cdot a_i
      \,
      v_{\bar \imath} \cdot a_i
    }{|b|^2}
    +
    \frac65 \sum_j C_{{\rm E},j} \frac{
      2 a_j^2
      +
      5\big(
      \hat b\cdot a_j
      \big)^2}{|b|^2}
    \bigg]
  \nn
  \\
  &\qquad+
  \mO(\gamma^{-1/2},G^4)\ .
\end{align}
Here, we use the variables $a_\pm^\mu=a_1^\mu\pm a_2^\mu$ and $v_{\bar 1}=v_2$ and $v_{\bar 2}=v_1$.
From this expression the leading order high-energy contributions to the expansion coefficients of Eq.~\eqref{eq:angleCov2} can be read off.
We note the explicit dependence on the particle label in the second term of the second line.
Interestingly this term vanishes for Kerr black holes.
Further terms in this series are easily generated from the full expressions.
The role of the high-energy limit played an important role after the initial computation of the conservative 3PM results in Ref.~\cite{Bern:2019nnu} because those conservative results did not have a finite high energy limit (see e.g. Refs.~\cite{Amati:1990xe,Damour:2019lcq,DiVecchia:2020ymx,Bern:2020gjj,Damour:2020tta}).
Only after inclusion of radiative effects is this high-energy limit finite with divergent terms cancelling between the radiative and conservative contributions.

While the scattering angle $\theta_\rmm{cons}$ is SUSY invariant, the expansion coefficients $\theta^{(n;A)}_{\rm cons}$ of Eq.~\eqref{eq:angleCov2} are defined with respect to covariant parameters.
Instead, when we consider the spinning Hamiltonian in Sec.~\ref{SPIN:CBM} we will work in canonical gauge.
It is then useful to define canonical expansion coefficients $\theta^{(n;A)}_{\rm can}$ by:\checked
\begin{align}
  &
  \tc
  =
  \sum_{n=1}^3
  \Big(
  \frac{G M}{|b_{\rm can}|}
  \Big)^n
  \Bigg[
    \thiic{n;0}
    -
    \sum_i \thiic{n;1,i} \frac{\hat L_{\rm can}\cdot a_{i,\rm can}}{|b_{\rm can}|}
    \nn
    \\
    &\qquad
    +
    \!
    \sum_{i,j}
    \frac{a_{i,\rm can}^\mu a_{j,\rm can}^\nu}{|b_{\rm can}|^2}
    \Big(\!\!
    -\thiic{n;2,1,i,j} \eta_{\mu\nu}
    +\thiic{n;2,2,i,j} \hat b_{\rm can,\mu} \hat b_{\rm can,\nu}
    +\thiic{n;2,3,i,j} \hat p_\mu \hat p_\nu
    +\thiic{n;2,4,i,j} \hat b_{\rm can,\mu} \hat p_\nu
    \Big)
    \Bigg]
  \nn
  \\
  &\qquad
  +\mO(S^3,G^4)\ .
  \label{SPIN:Can}
\end{align}
We could also define a canonical expansion of the scattering angle including radiation but that is not necessary for our applications.
The canonical expansion is similar to the covariant one in Eq.~\eqref{eq:angleCov2} though the velocities $v_i^\mu$ have been exchanged with the relative momentum $\hat p^\mu$ and all other variables are in canonical gauge.
Thus, all four vectors appearing in Eq.~\eqref{SPIN:Can} are purely spacial in the CoM frame.
The canonical and covariant expansion coefficients can be related using Eqs.~\eqref{WEFT:CV}-\eqref{WEFT:LCN} of Sec.~\ref{sec:Dynamics}.
We solve for the canonical coefficients in terms of the covariant ones and find:
\begin{align}\label{SPIN:CovToCan}
  \frac1{\Gamma}
  \thiic{n;0}
  &=
  \thiio{n;0}
  \ ,
  \\\nn
  \frac1{\Gamma}
  \thiic{n;1,i}
  &=
  \thiio{n;1,i}
  +
  n
  \frac{\pin}{E_i+m_i}
  \thiio{n;0}
  \ ,
  \\\nn
  \frac1{\Gamma}
  \thiic{n;2,1,i,j}
  &=
  \thiio{n;2,1,i,j}
  +
  \frac{(n+1)\pin}{2(E_j+m_j)}
  \thiio{n;1,i}
  +
  \frac{(n+1)\pin}{2(E_i+m_i)}
  \thiio{n;1,j}
  +
  \frac{n(n+1)\pin^2}{
    2(E_i+m_i)(E_j+m_j)
  }
  \thiio{n;0}
  \ ,
  \\\nn
  \frac1{\Gamma}
  \thiic{n;2,2,i,j}
  &=
  \thiio{n;2,2,i,j}
  -
  \frac{(n+2)\pin}{2(E_j+m_j)}
  \thiio{n;1,i}
  -
  \frac{(n+2)\pin}{2(E_i+m_i)}
  \thiio{n;1,j}
  -
  \frac{n(n+2)\pin^2}{2(E_i+m_i)(E_j+m_j)}
  \thiio{n;0}
  \ ,
  \\
  \frac1{\Gamma}
  \thiic{n;2,3,i,j}
  &=
  (-1)^{i+j} \frac{\pin^2 \Gamma^2}{\mu^2}\thiio{n;2,3,i,j}
  +
  \frac{1-(-1)^{i+j}}{2}
  \frac{E_1E_2-m_1m_2+\pin^2}{m_1m_2}
  \thiio{n;2,1,1,2}
  \nn
  \\\nn
  &\qquad\qquad
  -
  \frac{(n+1)\pin}{2(E_j+m_j)}
  \thiio{n;1,i}
  -
  \frac{(n+1)\pin}{2(E_i+m_i)}
  \thiio{n;1,j}
  -
  \frac{n(n+1)\pin^2}{2(E_i+m_i)(E_j+m_j)}
  \thiio{n;0}
  \ .
\end{align}
While the covariant coefficients depend on $\gam$ and the masses $m_i$ in a simple manner, this is no longer true for the canonical coefficients.
This, in turn, makes the dependence on $\gam$ and $m_i$ of the spinning Hamiltonian very complicated.
The intertangling of center of mass variables and the simple covariant angle coefficients are, however, comparably simple in the above equations~\eqref{SPIN:CovToCan}.

The physical interpretation of the scattering angles $\theta_i$ and $\phi_i$ is most natural for conservative motion and as mentioned, in that case, each of them by themselves also include all dynamical information.
With the inclusion of radiation, however, we are not aware of any single scalar parametrizing the entire motion.
It is still possible, though, to define a set of Lorentz generators which parametrize the full (radiative) motion.
We will consider such a parametrization in the remaining part of this section.

The parametrization of the full radiative observables in terms of generalized angles should explicitly make the conservation of $p^2_i$, $S^2_i$ and $p_i\cdot S_i$ manifest.
This may be achieved using Lorentz transformations which clearly conserves scalars:\checked
\bse
\begin{align}
  &\Delta p_i^\mu = ({\Lambda_i}^\mu_{\ \nu}- \eta^\mu_\nu) p_i^\nu
  \ ,
  \\
  &\Delta S_i^\mu = ({\Lambda_i}^\mu_{\ \nu}- \eta^\mu_\nu) S_i^\nu
   \ .
\end{align}
\ese
The fact that $p_i^\mu$ and $S_i^\mu$ transform with the same Lorentz transformation, ${\Lambda_i}^\mu_{\ \nu}$, ensures that $p_i\cdot S_i$ is conserved.
This parametrization was also discussed in Refs.~\cite{Bini:2017xzy,Vines:2017hyw}.

We find it convenient to write the Lorentz transformation as a product of a boost ${\mB_i}^\mu_{\ \nu}$ and a rotation ${\mR_i}^\mu_{\ \nu}$ as follows:\checked
\begin{align}
  {\Lambda_i}^\mu_{\ \nu}
  =
  (\mB_i\cdot\mR_i)^\mu_{\ \nu}
  \ .
\end{align}
We parametrize the rotation and boost in terms of generators, $R_i^\mu$ and $B_i^\mu$:\checked
\bse
\begin{align}
  &
  {\mB_i}^{\mu\nu} = \exp(2 v_{i}^{[\mu} B_{i}^{\nu]})
  \ ,
  \\
  &
    {\mR_i}^{\mu\nu} = \exp(\eps^{\mu\nu\rho\sigma}v_{i\rho}R_{i\sigma})
    \ .
\end{align}
\ese
The rotation is chosen so that it acts on the velocities $v_i^\mu$ as the identity: ${\mR_i}^{\mu}_{\ \nu}v_{i}^{\nu}=v_{i}^\mu$.
The generators may be chosen to be spacelike vectors orthogonal to $v_i^\mu$ so that $B_i\cdot v_i=R_i\cdot v_i=0$.

With the above parametrization of the Lorentz transformation, the equation for the impulse $\Delta p_i^\mu$ becomes:\checked
\begin{align}
    \frac{\Delta p^\mu_i}{m_i} &= ({\mB_i}^\mu_{\ \nu}-\eta^\mu_\nu) v_i^\nu
  \\
  &=
  v_i^\mu
  (\cosh|B_i| -1)
  +
  \hat B^\mu_i \sinh|B_i|
  \ .
  \nn
\end{align}
This boost clearly describes a generic scattering event with radiation where the final timelike momentum is related to the initial one by some boost.
The equation for $\Delta a^\mu$ becomes:\checked
\begin{align}
  \Delta a^\mu_i
  =
  (
  {\mR_i}_{\ \rho}^\mu  {\mB_i}_{\ \nu}^\rho
  -
  \eta^\mu_\nu
  )a^\nu_i
  \ .
\end{align}
One may understand this as follows: The boost describes how the reference frame of $p_i^\mu$ changes during the scattering event and subtracting this boost, $\Delta a^\mu_i$ is merely a rotation in the frame of $p_i^\mu$.
Thus both the boost and the rotation appears in the formula for $\Delta a^\mu_i$.

Knowledge of $B_i^\mu$ and $R_i^\mu$ thus determines the impulse and spin kick.
They are three vectors in the frame of $v_i^\mu$ and in addition they must satisfy certain constraints under parity so that e.g. $R_i^\mu = L^\mu R_i^L + \mO(S^2)$.
It is not, however, clear whether the two generators $B_i^\mu$ and $R_i^\mu$ are simpler than the impulse and spin kick themselves.
If at all, they are explicit three vectors in the frame of $v_i^\mu$ in contrast to the impulse and spin kick which have non-zero components in all directions.
In that sense the two generators $B_i^\mu$ and $R_i^\mu$ make the three conserved scalars $m_i$, $|a_i^\mu|$ and $a_i\cdot v_i$ manifest.

Finally, we note an elegant generalization of the simple conservative parameterization of the aligned spins impulse Eq.~\eqref{SPIN:CON} to include all radiation effects at third post-Minkowskian order.
Thus, for aligned spins we find the following relation to be satisfied by the impulse $\Del p_1^\mu$:
\begin{align}\label{SPIN:SFIMP}
  \Del p_1^\mu
  =
  p^\mu (\cos\theta-1)
  +
  p_\infty \hat b^\mu \sin\theta
  +
  \frac{\gam v_1^\mu-v_2^\mu}{\gam^2-1}
  \Del P\cdot v_2
  +
  \mO(G^4)
  \ .
\end{align}
Here, the angle $\theta=\theta_1$ is the full radiative scattering angle which for aligned spins is independent of the particle label $\theta_1=\theta_2+\mO(G^4)$ to this order.
The relation Eq.~\eqref{SPIN:SFIMP} elegantly reconstructs the (3PM) aligned spins impulse from the knowledge of the radiative scattering angle and total change in four-momentum and was first observed in the spinless case in Ref.~\cite{Saketh:2021sri}.
It may be derived by observing that the conservation of $p_\infty=|p^\mu|$ is broken only at this order and thus still approximately satisfied.
The degree to which it is not satisfied can only appear in the velocity directions which are also constrained from $|p_i^\mu+\Del p_i^\mu|=m_i$.
A generalization of this formula to also include the case of generic spins and the spin kick would be interesting and likewise its generalization to include 4PM effects.

\subsection{Linear Response
}\label{SPIN:LR}
The full impulse and spin kick, $\Del p_i^\mu$ and $\Del S^\mu_i$, have many terms which are redundant in the sense that they follow from symmetry constraints of the observables.
As an example, for the aligned spins conservative motion the scattering angle incorporate the symmetry that the magnitude of the relative momentum $p^\mu$ is conserved during the scattering and that the motion is planar.
For the spinning full (radiative) variables the situation is more complicated because the motion is non-planar, linear and angular momentum is lost and there are two observables rather than one.
In this section, however, we will see that we can still reconstruct parts of the 3PM observables from lower order post-Minkowskian data in a systematic manner.
Thus, we identify essential parts of the observables from which the full radiative ones may be derived.

The redundant parts of the 3PM observables which may be reconstructed from lower PM data are the two parts of the impulse $\Del p_{1,\rmm{rad}}^{(3;+)\mu}$ and $\Del p_{1,\rmm{cons}}^{(3;-)\mu}$ and the two parts of the spin kick $\Del S^{(3;-)\mu}_{1,\rmm{rad}}$ and $\Del S^{(3,+)\mu}_{1,\rmm{cons}}$.
In particular, we find that the two radiative parts $\Del p_{1,\rmm{rad}}^{(3;+)\mu}$ and $\Del S^{(3;-)\mu}_{1,\rmm{rad}}$ may be reconstructed using a direct generalization of the Bini-Damour formula originally used in Ref.~\cite{Damour:2020tta} for the 3PM scattering angle.
The formulas read:\checked
\bse\label{SPIN:RR}
\begin{align}
  \label{SPIN:BDP}
  \Delta p^{(+)\mu}_{1,\rm rad}
  &=
  -
  \frac12
  \frac{
    \partial\Delta p^\mu_1
  }{
    \partial J^\nu
  }
  \Del J^{\nu}
  +
  \mO(G^4)
  \ ,
  \\
  \Delta S^{(-)\mu}_{1,\rm rad}
  &=
  -\frac12
  \frac{
    \partial\Delta S_1^\mu
  }{
    \partial J^\nu
  }
  \Del J^{\nu}
  +
  \mO(G^4)
  \ .
\end{align}
\ese
In these equations we consider the full vectorial observables and likewise we must consider the full vectorial change in angular momentum $\Del J^\mu$ with its leading 2PM order given in Eq.~\eqref{WF:JRadAO}.
The radiative observables on the left-hand-side start at 3PM and they are reconstructed from the 2PM $\Del J^\mu$ and the 1PM leading order contributions of the impulse and spin kick of the right-hand-side.
In the following we will derive this equation using properties of the retarded and advanced propagators and from that derivation, we also find a method for reconstructing the redundant conservative parts of the impulse.

First, though, as a simple consequence of the linear response formula of the impulse Eq.~\eqref{SPIN:BDP} we derive a response formula for the scattering angle $\theta_i$:\checked
\begin{align}
  \theta_{i,\rmm{rad}}^{(+)}
  =
  \frac12
  \frac{
    \pat \theta_i
  }{
    J^\mu
  }
  \Del J^\mu
  +
  \mO(G^4)
  \ .
\end{align}
Here, the radiative part of the angle at 3PM may simply be defined by $\theta_i=\theta_{\rm cons}+\theta_{i,\rmm{rad}}$ and the $(+)$ superscript is defined with respect to the operation Eq.~\eqref{eq:timeReversal}.
In the spinless case where $\Del J^\mu$ is a scalar and we may ignore the $(+)$ superscript of $\theta_{i,\rmm{rad}}$ this formula reproduces exactly the Bini-Damour formula initially used at 3PM in Ref.~\cite{Damour:2020tta} in order to derive the then unknown radiative scattering angle.
That formula (and the use of radiation-reaction) was first considered in Ref.~\cite{Bini:2012ji} with subsequent work in Refs.~\cite{Damour:2020tta,Bini:2021gat,Manohar:2022dea,Jakobsen:2022zsx,Bini:2022wrq,Alessio:2022kwv,DiVecchia:2021ndb}.
This formula implies a very simple structure of the $(+)$ radiative scattering angle which is seen below with explicit results in Eqs.~\eqref{EXP:RADRAD}.
In the aligned spins Kerr case the 3PM contributions to the radiative scattering angle $\theta_{i,\rm rad}$ was derived to all order in spins in Ref.~\cite{Alessio:2022kwv} where the aligned spins and 3PM make the $(+)$ subscript superfluous.

We derive the linear response relations directly at the third post-Minkowskian order by observing that the conservative and radiative parts of the impulse may be written as follows:\checked
\bse
\label{SPIN:ConsRad}
\begin{align}\label{eq:responsePrelim}
  \Delta p^\mu_{i,\rm cons}
  =
  \frac12
  \Big(
  \Delta p^\mu_{i,\rm in-in}
  (J_{-\infty}^\mu,p_{i,-\infty}^\mu,S_{i,-\infty}^\mu)
  +
  \Delta p^\mu_{i,\rm out-out}
  (J_{\infty}^\mu,p_{i,\infty}^\mu,S_{i,\infty}^\mu)
  \Big)
  \ ,
  \\
  \Delta p^\mu_{i,\rm rad}
  =
  \frac12
  \Big(
  \Delta p^\mu_{i,\rm in-in}
  (J_{-\infty}^\mu,p_{i,-\infty}^\mu,S_{i,-\infty}^\mu)
  -
  \Delta p^\mu_{i,\rm out-out}
  (J_{\infty}^\mu,p_{i,\infty}^\mu,S_{i,\infty}^\mu)
  \Big)
  \ .
\end{align}
\ese
Here the ``out-out'' subscript refers to the same observable computed with $i\eps$-prescriptions opposite to the in-in theory.
That is, all causality points away from the outgoing lines or, equivalently, advanced propagators are used.
The ``in-in'' subscript then simply refers to the usual prescription of retarded propagators and highlights the symmetric or antisymmetric prescription of the conservative or radiative pieces.
The two terms in each line are evaluated in terms of past and future asymptotic variables, $X_{-\infty}$ and $X_{\infty}$, respectively (as in Eq.~\ref{WEFT:Kick}).

The conservative and radiative nature of this decomposition Eqs.~\eqref{SPIN:ConsRad} may be understood intuitively as follows.
At 3PM only one graviton propagator is active and the out-out prescription reverses that propagator.
However, the out-out prescription also reverses all worldline propagators so that the variables are, in a sense, defined at future infinity.
The insertion of future variables in the out-out observable then reverses this effect so that variables of both terms are defined at past infinity.
The conservative part is then given by the symmetric average of retarded and advanced graviton propagators and the radiative part is the antisymmetric sum.
We note, however, that the conservative part does not simply correspond to the use of retarded worldline propagators together with the time symmetric average of retarded and advanced propagators for the gravitons.
It does, however, exactly correspond to the potential region.
Likewise does the radiative part defined in this way correspond exactly to the radiative region.
We may write the conservative and radiative parts of the spin kick exactly as in Eqs.~\eqref{SPIN:ConsRad} by exchanging $\Del p_i^\mu$ with $\Del S_i^\mu$.
Again the conservative part defined in this way corresponds to the potential region and the radiative part to the radiative region.

Next, we relate observables computed with the two different boundary conditions through the simple flipping of the signs of the momenta (or velocities):\checked
\bse\label{eq:timeReverse}
\begin{align}
  \Delta p^\mu_{i,\rm out-out}
  (
  J^\mu,p_i^\mu,S_i^\mu
  )
  &=
  \Delta p^\mu_{i,\rm in-in}
  (
  J^\mu,-p_i^\mu,S_i^\mu
  )\ ,
  \\
  \Delta S_{i,\rm out-out}^{\mu}
  (J^\mu,p_i^\mu,S_i^\mu)
  &=
  -\Delta S_{i,\rm in-in}^{\mu}
  (J^\mu,-p_i^\mu,S_i^\mu)
  \ .
\end{align}
\ese
Intuitively this is a time reversal which flips the signs of timelike vectors and leaves spacelike vectors unchanged.
At the level of the Feynman rules the relation is realized from the fact that the retarded $i\eps$ of a momentum $l^\mu$ is always related to the time component of that momentum and in the frame of $v_i^\mu$ may be written as $(l\cdot v_i+i\eps)^2-\vct{l}^2$.
Flipping $v_i$ now turns retarded gravitons into advanced ones.
The same goes through for the worldline propagators which with conservation of energy are given by the same kind of expression $(l\cdot v_i+i\eps)^2$.
The vertex rules are found to obey the same kind of symmetry when one flips the sign of $S_i^\mn$ additionally which translates to keeping $J^\mu$ and $S_i^\mu$ unchanged.
In fact this relation offers an alternative definition to the $(\pm)$ superscripts defined in Eq.~\eqref{eq:timeReversal}:\checked
\bse
\begin{align}
  \Del p^{(\pm)\mu}_{i}
  (J^\mu,p_i^\mu,S_i^\mu)
  &=
  \frac12\Big(
  \Del p_{i,\rmm{in-in}}^\mu
  (J^\mu,p_i^\mu,S_i^\mu)
  \pm
  \Del p_{i,\rmm{out-out}}^\mu
  (J^\mu,p_i^\mu,S_i^\mu)
  \Big)
  \ ,
  \\
  \Del S^{(\pm)\mu}_{i}
  (J^\mu,p_i^\mu,S_i^\mu)
  &=
  \frac12\Big(
  \Del S_{i,\rmm{in-in}}^\mu
  (J^\mu,p_i^\mu,S_i^\mu)
  \mp
  \Del S_{i,\rmm{out-out}}^\mu
  (J^\mu,p_i^\mu,S_i^\mu)
  \Big)
  \ .
\end{align}
\ese

The equations~\eqref{eq:timeReverse} are now used to rewrite Eqs.~\eqref{SPIN:ConsRad} as follows,\checked
\begin{subequations}~\label{eq:respRels}
  \begin{align}
    \Delta p^\mu_{i,\rm cons}=
    \frac12
    \Big[
      &\Delta p_i^\mu
      \Big(
      J^\mu,p_i^\mu,S_i^\mu
      \Big)
    +
    \Delta p_i^\mu
    \Big(
    J^\mu+\Delta J^\mu,
    -p_i^\mu-\Delta p_i^\mu,
    S_i^\mu+\Delta S_i^\mu
    \Big)
    \Big]
    \ ,\label{eq:consRelP}
    \\
    \Delta p^\mu_{i,\rm rad}=
    \frac12
    \Big[
    &\Delta p_i^\mu
    (
    J^\mu,p_i^\mu,S_i^\mu
    )
    -
    \Delta p_i^\mu
    (
    J^\mu+\Delta J^\mu,
    -p_i^\mu-\Delta p_i^\mu,
    S_i^\mu+\Delta S_i^\mu
    )
    \Big]
    \ ,
    \label{eq:radRelP}
  \end{align}
\end{subequations}
for the impulse, and\checked
\begin{subequations}\label{eq:kickRel}
  \begin{align}
    \Delta S^\mu_{i,\rm cons}
    =
    \frac12
    \Big[
    &\Delta S_i^\mu(
    J^\mu,p_i^\mu,S_i^\mu
    )
    -
    \Delta S_i^\mu
    (
    J^\mu+\Delta J^\mu,-p_i^\mu-\Delta p_i^\mu,S_i^\mu+\Delta S_i^\mu
    )
    \Big]\ ,
    \label{eq:consRelA}
    \\
    \Delta S^\mu_{i,\rm rad}=
    \frac12
    \Big[
    &\Delta S_i^\mu(J^\mu,p_i^\mu,S_i^\mu)
    +
    \Delta S_i^\mu
    (J^\mu+\Delta J^\mu,-p_i^\mu-\Delta p_i^\mu,S_i^\mu+\Delta S_i^\mu
    )
    \Big]
    \ ,
    \label{eq:radRelA}
  \end{align}
\end{subequations}
for the spin kick.

Focusing first on the impulse Eqs.~\eqref{eq:respRels} and considering the formulas to zeroth order in the radiation reaction, we find:\checked
\bse
\begin{align}
  \Delta p^\mu_{i,\rm cons}=
  \frac12
  \Big[
    &\Delta p_i^\mu
    \big(
    J^\mu,p_i^\mu,S_i^\mu
    \big)
    +
    \Delta p_i^\mu
    \big(
    J^\mu,
    -p_i^\mu,
    S_i^\mu
    \big)
    \Big]
  +
  ...
  =
  \Del p^{(+)\mu}_{i,\rmm{cons}}
  +
  ...
  \\
  \Delta p^\mu_{i,\rm rad}=
  \frac12
  \Big[
    &\Delta p_i^\mu
    \big(
    J^\mu,p_i^\mu,S_i^\mu
    \big)
    -
    \Delta p_i^\mu
    \big(
    J^\mu,
    -p_i^\mu,
    S_i^\mu
    \big)
    \Big]
  +
  ...
  =
  \Del p^{(-)\mu}_{i,\rmm{rad}}
  +
  ...
\end{align}
\ese
The dots of each line represent radiation reaction corrections.
We are interested in the cases where the zeroth order vanish and the observables are given only by radiation reaction.
For the conservative and radiative impulse we see that this happens for the $(-)$ and $(+)$ parts respectively.
Following the same logic for the spin kick its $(+)$ conservative and $(-)$ radiative kicks are given entirely by radiation reaction.

The exact relation of $\Del p^{(3;+)\mu}_i$ and $\Del S^{(3;-)\mu}_i$ in terms of radiation reaction to lower PM data may now be worked out with the result Eqs.~\eqref{SPIN:RR}.
The resulting equation for the conservative $\Del p_i^{(m;-)\mu}$ and $\Del S_i^{(m;-)\mu}$ have not been worked into a very simple form.
All the relevant information, however, is in Eqs.~\eqref{eq:radRelP} and~\eqref{eq:radRelA}.
See Ref.~\cite{Jakobsen:2022zsx} for more discussion on this final point.

\section{Conservative Bound Motion from Scattering Observables
}\label{SPIN:CBM}
The mapping of the above unbound results to the case of bound motion is essential for their relevance to the prediction of waveforms observed in gravitational wave observatories.
Currently, the most general approach for carrying out this mapping is the derivation of a two-body Hamiltonian (or, equivalently, interaction potential) which in the case of local-in-time dynamics straightforwardly captures both unbound and bound motion.
In contrast to the fourth PM order, the third PM order observables are still derived from local-in-time interactions and the mapping via a local Hamiltonian to the bound motion works well.

The systematic derivation of (post-Minkowskian) interaction potentials from scattering data was first introduced in Ref.~\cite{Cheung:2018wkq} with a matching calculation which effectively computes the gauge invariant data in terms of the potential.
These relations are then inverted and the initially unknown potential is determined from the scattering data.
Naturally, the potential is gauge dependent and only when a suitable gauge is imposed does it have a unique form.
See also Refs.~\cite{Bjerrum-Bohr:2018xdl,Cristofoli:2019neg} for an alternative approach.
Previous work on computing the potential from quantum field theory and scattering amplitudes include~\cite{Iwasaki:1971vb,Hiida:1972xs,Bjerrum-Bohr:2002gqz,Neill:2013wsa,Bjerrum-Bohr:2013bxa,Vaidya:2014kza,Cachazo:2017jef,Guevara:2017csg}.

In our approach in Sec.~\ref{SPIN:H} we will also determine the spinning two-body Hamiltonian from a matching calculation.
Our inclusion of spin into the post-Minkowskian two-body Hamiltonian follows the formalism initially developed in Ref.~\cite{Bern:2020buy} with additional work in Refs.~\cite{Kosmopoulos:2021zoq,Chen:2021kxt,Bern:2022kto,FebresCordero:2022jts}.
Adding spin to the Hamiltonian introduces new gauge freedom and dynamic variables and in contrast to the simple covariant SSC, the spin variables must be considered in the canonical gauge which spoils the simple mass dependence of the covariant observables.
The Kerr $\mO(G^3,S_i^2)$ part of the Hamiltonian were first derived in Ref.~\cite{FebresCordero:2022jts} with the remaining $\mO(G^3,S_i^2 C_{\rmm{E},i})$ and $\mO(G^3,S_1 S_2)$ parts added in Ref.~\cite{Jakobsen:2022zsx}.

The Hamiltonian is in the present case superior to the gauge invariant mappings of unbound to bound observables~\cite{Kalin:2019rwq,Kalin:2019inp,Cho:2021arx,Saketh:2021sri} because those have not yet been generalized to arbitrarily aligned spins.
In Sec.~\ref{SPIN:GI} we do, however, consider the gauge invariant mapping of the aligned spins scattering angle to the post-Newtonian expansion of the bound binding energy.
The spin terms of the binding energy thus derived are accurate to the fourth post-Newtonian order and match results from the PN literature.

The mapping of radiative unbound observables to the bound system is also possible~\cite{Cho:2021arx,Saketh:2021sri,Manohar:2022dea} but we will not consider that here.
See e.g. Ref.~\cite{Jakobsen:2022zsx} for a discussion of the bound radiative information of the present unbound scattering data.

\subsection{Hamiltonian at $\mO (G^3,S^2)$}\label{SPIN:H}
In this section we derive the conservative Hamiltonian for the two-body system to $\mO(G^3,S^2)$ by a matching calculation to the conservative observables presented above.
In fact, we find that we need only match to the conservative scattering angle $\theta_{\rm cons}$ and the Hamiltonian is thus determined from the knowledge of that angle.

We work in the CoM frame with dynamical spacial vectors $\vct{x}(t)$, $\vct{p}(t)$ and $\vct{S}_i(t)$ satisfying canonical Poisson brackets:\checked
\begin{subequations}
  \begin{align}
    &
    \{\vct{x}^m(t),\vct{p}^n(t)\}_{\rm P.B.}=\delta^{mn}
    \ ,
    \\
    &
    \{\vct{S}^m_i(t),\vct{S}^n_i(t)\}_{\rm P.B.}=\epsilon^{mnk}\vct{S}_i^k(t)
    \ .
  \end{align}
\end{subequations}
The variables $\vct{x}(t)$ and $\vct{p}(t)$ correspond to the relative position and momentum of the two bodies respectively and, importantly, the vectors $\vct{S}_i(t)$ are their spin vectors in canonical gauge Eq.~\eqref{WEFT:NFV}.

We write an ansatz for the Hamiltonian as a sum of the energy of each body together with an interaction potential:
\begin{align}\label{eq:hamiltonian}
  H
  \big(
  \vct{x},\vct{p},\vct{S}_i
  \big)
  \!=\!
  \sqrt{\vct{p}^2+m_1^2}\!+\!\sqrt{\vct{p}^2+m_2^2}
  \!+\!
  V
  \big(
  \vct{x},\vct{p},\vct{S}_i
  \big)\ .
\end{align}
The interaction potential is then expanded in spin structures $\mO^A$:
\begin{align}
  V
  \big(
  \vct{x},\vct{p},\vct{S}_i
  \big)
  &=
  \sum_A
  \mO^A\,
  V^A(\vct{x},\vct{p})
  +
  \mO(S^3)
  \\
  &
  =V^{(0)}
  +
  \sum_i V^{(1,i)} \mO^{(1,i)}
  +
  \sum_{i,j,a}
  V^{(2,a,i,j)} \mO^{(2,a,i,j)}
  +
  \mO(S^3)\ .
  \nn
\end{align}
Here the sums on $i\in\{1,2\}$ and $j\in\{1,2\}$ correspond to the bodies and the sum on $a$ counts the three independent spin structures at quadratic order in the spins.

The ansatz of the Hamiltonian, naturally, is not unique and so, too, is the choice of spin structures not unique.
Here, we will work with the following choice of structures:
  \begin{align}
    \mO^{(0)}
    &=1\,,
    \\\nn
    \mO^{(1,i)}
    &=
    \frac{
      ({\vct{x}}\times{\vct{p}})
      \cdot{\vct{a}}_i
    }{
      |{\vct{x}}|^2
    }\,,
    \\\nn
    \mO^{(2,1,i,j)}
    &=
    \frac{
      {\vct{a}}_i\cdot{\vct{a}}_j
    }{
      |{\vct{x}}|^2
    }\,,
    \\\nn
    \mO^{(2,2,i,j)}
    &=
    \frac{
      {\vct{x}}\cdot{\vct{a}}_i
      {\vct{x}}\cdot{\vct{a}}_j
    }{
      |{\vct{x}}|^4
    }\,,
    \\\nn
    \mO^{(2,3,i,j)}
    &=
    \frac{
      {\vct{p}}\cdot{\vct{a}}_i
      {\vct{p}}\cdot{\vct{a}}_j
    }{
      |{\vct{x}}|^2
    }\,,
  \end{align}
with $\vct{a}_i=\vct{S}_i/m_i$.
The first index of the superscript indicates the spin order and the superscripts $i$ and $j$ the two bodies.
At quadratic order the remaining second superscript counts the three different spin structures.
The above choice of Hamiltonian is sometimes referred to as isotropic gauge because the combination $\vct{x}\cdot \vct{p}$ never appears.
Note that the structures $\mO^{(2,a,1,2)}=\mO^{(2,a,2,1)}$ and we require the same symmetry of the corresponding potentials $V^{(2,a,1,2)}=V^{(2,a,2,1)}$.

The post-Minkowskian expansion of the interaction potential now reads:\checked
\begin{align}\label{SPIN:IP}
  V^A(\vct{x},\vct{p})
  =
  \sum_n
  \left(
  \frac{GM}{|\vct{x}|}
  \right)^n
  \cii{n;A}(\vct{p}^2)\,.
\end{align}
For every PM order and every spin structure we have a coefficient $c^{(n;A)}$ encoding the corresponding contribution to the Hamiltonian, and these coefficients together include all relevant information for our two-body Hamiltonian.
In the post-Newtonian expansion they would be expanded further as a formal expansion $\vct{p}^2$.

Our goal now is to compute the scattering angle $\theta_{\rm cons}$ from the Hamiltonian and thus express it in terms of the Hamiltonian coefficients $c^{(n;A)}$.
This expression for $\theta_{\rm cons}$ is then matched to the expression derived from WQFT in terms of asymptotic background variables and in this manner the coefficients $c^{(n;A)}$ are determined.

Hamilton's equations of motion may be derived using the Poisson brackets and read:\checked
\begin{align}\label{eq:hamEqns}
  \dot{\vct{x}}
  &=
  \frac{
    \partial H
  }{\partial \vct{p}}\ , &
  \dot{\vct{p}}
  &=
  -
  \frac{
    \partial H
  }{\partial \vct{x}}\ , &
  \dot{\vct{S}}_i
  &=
  -\vct{S}_i
  \times
  \frac{
    \partial H
  }{\partial \vct{S}_i}\ .
\end{align}
We solve them perturbatively in $G$ with expansions of the dynamical variables given as follows:\checked
\begin{subequations}\label{eq:expansions}
  \begin{align}
    \vct{x}(t)&=\vct{x}^{(0)}(t)+\sum_{n=1}^3G^n\vct{x}^{(n)}(t)+\mO(G^4)
    \ ,
    \\
    \vct{p}(t)&=\vct{p}^{(0)}+\sum_{n=1}^3G^n\vct{p}^{(n)}(t)+\mO(G^4)
    \ ,
    \\
    \vct{S}_i(t)&=\vct{S}_i^{(0)}+\sum_{n=1}^3G^n\vct{S}_i^{(n)}(t)+\mO(G^4)
    \ .
  \end{align}
\end{subequations}
We do not need to know about the functional dependence of the coefficients, $c^{(n;A)}(\vct{p}^2)$ when solving the equations of motion.
Instead, when deriving the perturbative equations of motion we define derivatives of the coefficients as follows:\checked
\begin{align}
  \frac{\partial^m c^{(n;A)}(\vct{p}^2)}{\partial (\vct{p}^2)^m}
  =
  c^{(n;A;m)}(\vct{p}^2)
  \ .
\end{align}
In the perturbative equations of motion the coefficients $c^{(n;A)}(\vct{p}^2)$ and its derivatives $c^{(n;A;m)}(\vct{p}^2)$ only appear evaluated on the background in terms of $(\vct{p}^{(0)})^2$ which as we will now see equals $p_\infty^2$.

Using retarded boundary conditions in our solution of the Hamiltonian equations of motion imply that the background parameters $\vct{x}^{(0)}(t)$, $\vct{p}^{(0)}$ and $\vct{S}_i^{(0)}$ are defined at asymptotic past and they may be directly related to the asymptotic parameters $b^\mu_{\rm can}$, $p^\mu$ and $S_{i,\rm can}^\mu$:\checked
\begin{align}
  \vct{x}^{(0)}(t)
  &=t \frac{E \vct{p}_\infty}{E_1 E_2} - \vct{b}_{\rm can}\,, &
  \vct{p}^{(0)} &= \vct{p}_\infty \,, &
  \vct{S}_i^{(0)} = \vct{S}_{i,\infty}\,,
\end{align}
where $b^\mu_{\rm can}=(0,\vct{b}_{\rm can})$, $p^\mu=(0,\vct{p}_\infty)$ and $S^\mu_{i,\rm can}=(0,\vct{S}_{i,\infty})$ in the CoM frame.

The impulse and spin kick are given by:\checked
\begin{subequations}
  \begin{align}
    \Delta\mathbf{p}^{(n)}
    &=
    \int_{-\infty}^{\infty} \!\di t\,\,
    \dot{\vct{p}}^{(n)}(t)\,,
    \\
    \Delta\mathbf{S}^{(n)}_{i}
    &=
    \int_{-\infty}^{\infty} \!\di t\,\,
    \dot{\vct{S}}_i^{(n)}(t)\,,
  \end{align}
\end{subequations}
though for the scattering angle $\theta_{\rm cons}$ we need only parts of the impulse.
The fact that the coefficients $c^{(n;A)}$ can be read off only from the knowledge of $\theta_{\rm cons}$ is not immediately clear and future work should explore this in more detail.
Here, it would be interesting to derive the Hamiltonian in a way similar to the EFT approach in Ref.~\cite{Cheung:2018wkq} where the scattering amplitude is derived via Feynman rules from the interaction potential.
Rather than using the scattering angle as we did above, it would then be natural to use the WQFT eikonal as the analogous quantity to the scattering amplitude

The Hamiltonian derived by the matching discussed in this section is presented below in Sec.~\ref{EXP:H} to $\mO(G^2,S^2)$ and $\mO(G^3,S^1)$ with full results found in the ancillary file to Ref.~\cite{Jakobsen:2022zsx}.
The spinning terms of the post-Newtonian expansion of this Hamiltonian reproduces the fourth post-Newtonian order first derived by Levi and Steinhoff in Ref.~\cite{Levi:2015ixa,Levi:2016ofk} (naturally one must first relate the two results through a canonical gauge transformation).
Also, it agrees with the results of Ref.~\cite{FebresCordero:2022jts} where the $S_i^2$ terms of the Hamiltonian without finite size effects ($C_{\rmm{E},i}\to0$) were derived.

\subsection{Gauge Invariant Boundary to Bound Mapping}\label{SPIN:GI}
Here, we consider the gauge invariant mapping of the aligned spins scattering angle to the binding energy of the bound system.
Here, we follow the approach of Refs.~\cite{Antonelli:2020ybz,Kalin:2019inp,Liu:2021zxr}.
General discussions of gauge invariant maps from unbound to bound orbits are found in Refs.~\cite{Kalin:2019rwq,Kalin:2019inp,Cho:2021arx,Saketh:2021sri}.
The specialization to aligned spins is due to these maps being established for that case only.
The overall idea is to derive the bound radial action from the unbound one which is simply related to the unbound scattering angle.
Going from the bound radial action to the binding energy is then a problem of the dynamics of bound orbits.

Throughout this section, then, we will assume the spin vectors to be aligned with the orbital angular momentum and we write:\checked
\begin{align}
  S_i^\mu=S_{i,\rmm{can}}^\mu=m_ia_i^\mu =
  G m_i^2 \chi_i
  \hat L^\mu\ .
\end{align}
In the case of aligned spins the covariant and canonical spin vectors are equivalent.
The final parametrization in terms of the directed spin lengths $\chi_i$ is practical and for Kerr black holes $m_i|\chi_i|$ are the radii of the ring singularities and $0<|\chi_i|<1$ (unless naked singularities are considered).

The magnitude of the angular momentum does still depend on the choice of SSC and we define $\lambda$ as the reduced magnitude of the canonical angular momentum:\checked
\begin{align}
  \lambda &= \frac{|L_{\rm can}|}{G M \mu}=
  \frac{|b|\pin}{GM\mu}
  +
  \frac{
    \mE
  }{
    2
  }
  \Big(
  \chi_+
  +
  \frac{\delta}{\Gamma}
  \chi_-
  \Big)\,,
\end{align}
where $\mE=(E-M)/\mu$ is the reduced binding energy and the definitions of the other variables e.g. can be found in Appendix~\ref{NOT:A}.
The variables $\chi_{\pm}$ are given by,\checked
\bse
\begin{align}
  \chi_\pm &= \frac{m_1\chi_1\pm m_2\chi_2}{M}\,,\\
  \chi^2_{{\rm E},\pm}&=\frac{C_{{\rm E},1}m_1^2\chi_1^2\pm C_{{\rm E},2}m_2^2\chi_2^2}{M^2}\,.
\end{align}
\ese
where we also defined $\chi_{\rmm{E},\pm}^2$ that we will use below.

We proceed to the mapping of the unbound radial action to the bound one.
The (reduced) unbound radial action, $w_{r}$ is related to the scattering angle as follows:\checked
\begin{align}\label{eq:lDeriv}
  2\pi \frac{\partial}{\partial \lambda} w_r(\mE,\lambda,\chi_i)
  =
  -
  (\theta_{\rm cons}(\mE,\lambda,\chi_i)+\pi)
  \ .
\end{align}
Thus, integration with respect to $\lambda$ allows us to derive the unbound radial action from the scattering angle.
The result of Ref.~\cite{Kalin:2019inp} now relates the unbound radial action to the bound one $i_r$ as follows:\checked
\begin{align}\label{eq:radial}
  i_r(\mE,\lambda,\chi_i)
  =
  w_r(\mE,\lambda,\chi_i)
  -
  w_r(\mE,-\lambda;-\chi_i)\ .
\end{align}
Here, the unbound action must be analytically continued from the regime $\lambda>0$ to $\lambda<0$.
Likewise $\mE$ is positive for unbound orbits but negative for bound orbits (or $\gam>1$ for unbound and $1>\gam>0$ for bound).
This continuation is most critical for the 1PM contributions where a $\log(\lambda)$ is present in the unbound radial action.
See e.g. Ref.~\cite{Antonelli:2020ybz} for a discussion of this continuation.
Putting the pieces together the bound radial action in terms of the unbound scattering angle reads,\checked
\begin{align}
  i_r(\mE,\lambda,\chi_i)
  =
  -\lambda
  +
  \frac{2\gamma^2-1}{\sqrt{1-\gamma^2}}
  -
  \frac{1}{\pi}
  \sum_n
  \int\!d\lambda
  \frac{\tilde\theta^{(2n)}}{\lambda^{2n}}\,.
\end{align}
with the post-Minkowskian expansion of the scattering angle,\checked
\begin{align}
  \theta_{\rm cons}=\sum_n\frac{\tilde\theta^{(n)}}{\lambda^n}\,,
\end{align}
with PM parameter $1/\lambda$ and $\tilde \theta^{(n)}$ depending only on the spins $\chi_i$ and $\lambda$ through the ratios $\chi_i/\lambda$ which scales as $G^0$ in the sense that e.g. $|\chi_1|/\lambda=m_2|a_1|/|L_{\rm can}|$.

The first surprising conclusion of this mapping is that only the even PM powers of the scattering angle contribute to the bound radial action.
At first, that is a disappointing conclusion as our results for the 3PM scattering angle seem to be irrelevant.
This effect which was already observed in the spinless case~\cite{Kalin:2019rwq}, however, is partly avoided by the use of the impetus formula.
With that formula the leading post-Newtonian orders of the spinless $\tilde\theta^{(4)}$ are derived with the knowledge only of the lower PM orders $\tilde\theta^{(n)}$ with $n<4$.
With spin also the leading PN contributions to $\tilde\theta^{(6)}$ are constructed in this manner.
Again, this method is spelled out in Ref.~\cite{Antonelli:2020ybz}.

Following the above method we have an expression for the bound radial action $i_r$ to next-to-next-to-leading order (NNLO) in the post-Newtonian expansion for each spin order (i.e. NNLO for the spinless terms and NNLO for the linear-in-spin terms etc.).
The binding energy may now be derived to the corresponding perturbative orders.
First, we specialize to circular orbits:\checked
\begin{align}\label{eq:i}
  i_r(\mE,\lambda,\chi_i)=0\ .
\end{align}
Next, we define the orbital frequency $\Omega$ and the related dimensionless variable $x$ in terms of which we want to express the binding energy:\checked
\begin{align}\label{eq:frequency}
  x^{3/2}=G M \Omega = \frac{d\mE}{d\lambda}
  \ .
\end{align}
The first equation~\eqref{eq:i} allows relates the three variables $\mE$, $\lambda$ and $\chi_i$ to each other.
The next equation~\eqref{eq:frequency} allows us to re-express $\lambda$ in terms of $x$.
Finally, we solve for $\mE$ in terms of $x$ and $\chi_i$ and find:\checked
\begin{align}\label{eq:bindingResult}
    -2\mE
    =
    x
    &\bigg[
      1
      -
      x
      \frac{9+\nu}{12}
      -
      x^2
      \frac{81-57\nu+\nu^2}{24}
      +\cdots
      \bigg]
    \\
    +
    x^{5/2}
    &\bigg[
      \frac{
        7
        \chi_+
        -
        \delta \chi_-
      }{3}
      +
      x
      \frac{
        (99-61\nu)
        \chi_+
        -
        (45-\nu)
        \delta \chi_-
      }{18}
      \nn
      \\
      &\quad
      +
      x^2
      \frac{
        (405-1101\nu+29\nu^2)\chi_+
        -
        (243-165\nu-\nu^2)\delta\chi_-
      }{24}
      +\cdots
      \bigg]
    \nn
    \\
    -x^3&\bigg[
      \chi_+^2
      +
      \frac{5x}{36}
      \Big(
      (5-6\nu)\chi_+^2
      -44\chi_+\chi_-
      -(1+8\nu)\chi_-^2
      \Big)
      \nn
      \\
      &\quad
      +
      \frac{7x^2}{216}
      \Big(
      (198-680\nu+3\nu^2)\chi_+^2
      -2(171-137\nu)\delta\chi_-\chi_+
      +
      (63-251\nu+56\nu^2)
      \chi_-^2
      \Big)
      +
      \cdots
      \bigg]
      \nn
      \\
      -x^3&\bigg[
      \chi_{{\rm E},+}^2
      +\frac{5x}{6}\Big(
      (5-\nu)
      \chi_{{\rm E},+}^2
      -
      2\delta\chi_{{\rm E},-}^2
      \Big)
      \nn
      \\
      &\quad
      +
      \frac{x^2}{72}
      \Big(
      (1125-1025\nu+7\nu^2)\chi_{{\rm E},+}^2
      -
      2\delta(279-70\nu)\chi_{{\rm E},-}^2
      \Big)
      +
      \cdots
      \bigg]\,.
    \nn
\end{align}
The four square brackets collect different powers in spin and at quadratic order the Kerr terms and the finite size terms.
In each case they include terms to next-to-next-to-leading order (NNLO).
The spin variables $\chi_i$ are considered independent of the PN expansion parameter, say $v$, and the variable $x$ scales with $v^2$.
The Newtonian term is the first $x$ and the non-spinning corrections appear to 2PN order, the linear in spin terms to 3.5PN order and the quadratic in spin terms to 4PN order.
They match corresponding PN results (see e.g.~\cite{Cho:2022syn}).

\section{Expressions}\label{EXP}
In this section we print explicitly a few results of the present chapter, namely the $\mO(G^3,S^2)$ contributions to generalized scattering angle $\theta_i$ and the Hamiltonian to $\mO(G^2,S^2)$ and $\mO(G^3,S^1)$.
These are useful for getting an impression of the structure and complexity of the spinning results at this order.
At linear order in spins the complexity grows only slightly due to only one spin structure being added to the scattering angle or Hamiltonian.
This is also related to the fact that the conservative misaligned scattering angle at linear order in spins is identical to the aligned one.
At quadratic orders in spins the expressions become significantly more complicated with six independent spin structures in the Hamiltonian and additionally the finite size corrections.

In the first section~\ref{SPIN:ESA} expressions for the generalized scattering angle are presented and, next, in the second section~\ref{EXP:H} expressions for the conservative Hamiltonian are presented.
All expressions are reproduced from Ref.~\cite{Jakobsen:2022zsx} and full expressions for the observables and the Hamiltonian can be found, too, in the ancillary file to that reference.

\subsection{Generic Spins Scattering Angle}
\label{SPIN:ESA}
Here, we print explicit expressions for the scattering angle $\theta_1$ defined in Eq.~\eqref{eq:genericAngles}.
We print the third post-Minkowskian contribution to each of the expansion coefficients of Eq.~\eqref{eq:angleCov2}.
We focus, first, on the conservative parts of those coefficients which do not depend on the particle label and then afterwards on the radiative contributions.
Coefficients that are not found in the following presentation can be obtained from particle exchange symmetry.
We remind, here, the reader that the scattering angle (and more generally the observables of this chapter) were computed with retarded propagators so that all asymptotic variables are defined at past infinity.

The conservative coefficients to $\mO(S^1)$ are:\checked
\begingroup
\allowdisplaybreaks
{\footnotesize
\begin{align}
  &
  \thiio{3;0}
  =
  \frac{2 \left(64 \gamma
    ^6-120 \gamma ^4+60 \gamma ^2-5\right) \Gamma ^2}{3 \left(\gamma
    ^2-1\right)^3}
  -\frac{8 \gamma  \left(14 \gamma ^2+25\right) \nu }{3 \left(\gamma
    ^2-1\right)}
  -\frac{8 \left(4 \gamma ^4-12 \gamma ^2-3\right) \nu  \ \text{arccosh}\gamma
  }{\left(\gamma ^2-1\right)^{3/2}}\,,
  \\
  &
  \thiio{3;1,1}=
  -\frac{2 \gamma 
    \left(16 \gamma ^4-20 \gamma ^2+5\right) \left(5 \Gamma ^2-\delta
    \right)}{\left(\gamma ^2-1\right)^{5/2}}
  +\frac{4 \left(44 \gamma ^4+100
    \gamma ^2+41\right) \nu }{\left(\gamma ^2-1\right)^{3/2}}
  +\frac{48 \gamma  \left(\gamma ^2-6\right) \left(2 \gamma ^2+1\right) \nu 
    \ \text{arccosh}\gamma}{\left(\gamma ^2-1\right)^2}\,.
\end{align}
}
We turn to the quadratic-in-spins conservative coefficients.
For $\thiio{3;2,1,i,j}$ we find:\checked
\begin{adjustwidth}{-1cm}{-1cm}
{\footnotesize
  \begin{align}
    &
    \thiio{3;2,1,1,1}=
    \Gamma ^2 \left(\frac{4 \left(96
      \gamma ^6-160 \gamma ^4+70 \gamma ^2-5\right)}{\left(\gamma
      ^2-1\right)^3}
    -\frac{4 \left(1772 \gamma ^6-2946 \gamma ^4+1346 \gamma
      ^2-137\right) C_{{\rm E},1}}{35 \left(\gamma ^2-1\right)^3}\right)
    \\
    &\qquad\qquad
    +\delta  \left(
    -\frac{8 \left(4 \gamma ^2-2 \gamma -1\right) \left(4
      \gamma ^2+2 \gamma -1\right)}{\left(\gamma ^2-1\right)^2}
    +
    \frac{8 \left(214 \gamma
      ^4-223 \gamma ^2+44\right) C_{{\rm E},1}}{35 \left(\gamma
      ^2-1\right)^2}
    \right)
    \nn
    \\
    &\qquad\qquad
    -\frac{16 \gamma  \left(148 \gamma ^4+374 \gamma ^2+383\right) \nu }{5
      \left(\gamma ^2-1\right)^2}
    +\frac{8
      \gamma  \left(3244 \gamma ^4+7972 \gamma ^2+4639\right) C_{{\rm E},1} \nu
    }{105 \left(\gamma ^2-1\right)^2}
    \nn
    \\
    &\qquad\qquad
    +\text{arccosh}\gamma
    \left(
    -\frac{192
      \left(\gamma ^6-8 \gamma ^4-7 \gamma ^2-1\right) \nu }{\left(\gamma
      ^2-1\right)^{5/2}}
    +
    \frac{16 \left(8 \gamma ^6-56 \gamma ^4-24 \gamma
      ^2-3\right) C_{{\rm E},1} \nu }{\left(\gamma ^2-1\right)^{5/2}}
    \right)
    \nn
    \\
    &\thiio{3;2,1,1,2}=
    \frac{4 \left(96 \gamma ^6-160 \gamma
      ^4+70 \gamma ^2-5\right) \Gamma ^2}{\left(\gamma ^2-1\right)^3}
    -\frac{32 \gamma  \left(15 \gamma ^4+46 \gamma ^2+47\right) \nu
    }{\left(\gamma ^2-1\right)^2}
    -\frac{48
      \left(4 \gamma ^6-36 \gamma ^4-35 \gamma ^2-5\right) \nu \, \text{arccosh}\gamma
    }{\left(\gamma ^2-1\right)^{5/2}}
\end{align}}
\end{adjustwidth}
For the coefficients $\thiio{3;2,2,i,j}$ we find:\checked
\begin{adjustwidth}{-1cm}{-1cm}
{\footnotesize
  \begin{align}
  &\thiio{3;2,2,1,1}=\frac{4 \gamma  \left(9000 \gamma ^{10}+4404 \gamma
    ^8-2152 \gamma ^6-12152 \gamma ^4+8379 \gamma
    ^2-1479\right) \nu }{15 \left(\gamma ^2-1\right)^3
    \left(2 \gamma ^2-1\right)^2}
  \\&\qquad
  +\pi ^2
  \Bigg(-\frac{\left(5 \gamma ^2-3\right) \left(10
    \gamma ^4-5 \gamma ^2+9\right) \gamma ^2 \delta
  }{128 \left(\gamma ^2-1\right)^2 \left(2 \gamma
    ^2-1\right)^2}+\frac{\left(100 \gamma ^8-160 \gamma
    ^6+193 \gamma ^4-78 \gamma ^2+45\right) \gamma ^2
    \Gamma ^2}{128 \left(\gamma ^2-1\right)^2 \left(2
    \gamma ^2-1\right)^3}
  \nn   \\&\qquad
  -\frac{\left(100 \gamma ^9-160
    \gamma ^7-60 \gamma ^6+193 \gamma ^5-12 \gamma
    ^4-78 \gamma ^3+12 \gamma ^2+45 \gamma -36\right)
    \gamma ^2 \nu }{64 \left(\gamma ^2-1\right)^2
    \left(2 \gamma ^2-1\right)^3}\Bigg)
  +\delta 
  \Bigg(\frac{10 \left(4 \gamma ^2-2 \gamma -1\right)
    \left(4 \gamma ^2+2 \gamma -1\right)}{\left(\gamma
    ^2-1\right)^2}
  \nn   \\&\qquad
  -\frac{2 \left(1192 \gamma ^4-1382
    \gamma ^2+295\right) C_{{\rm E},1}}{35 \left(\gamma
    ^2-1\right)^2}\Bigg)
  -\frac{4 \gamma  \left(10744
    \gamma ^6+13474 \gamma ^4+2665 \gamma
    ^2+9237\right) C_{{\rm E},1} \nu }{105 \left(\gamma
    ^2-1\right)^3}
  \nn   \\&\qquad
  +\Gamma ^2 \left(\frac{2 \left(6568
    \gamma ^6-11114 \gamma ^4+5079 \gamma ^2-463\right)
    C_{{\rm E},1}}{35 \left(\gamma
    ^2-1\right)^3}-\frac{2 \left(960 \gamma ^{10}-2560
    \gamma ^8+2540 \gamma ^6-1150 \gamma ^4+225 \gamma
    ^2-13\right)}{\left(\gamma ^2-1\right)^3 \left(2
    \gamma ^2-1\right)^2}\right)
  \nn   \\&\qquad
  +\text{arccosh}\gamma
  \left(\frac{16 \gamma ^2 \left(60 \gamma ^{10}-600
    \gamma ^8+551 \gamma ^6-63 \gamma ^4-63 \gamma
    ^2+15\right) \nu }{\left(\gamma ^2-1\right)^{7/2}
    \left(2 \gamma ^2-1\right)^2}-\frac{32 \left(8
    \gamma ^8-56 \gamma ^6+26 \gamma ^4-18 \gamma
    ^2-3\right) C_{{\rm E},1} \nu }{\left(\gamma
    ^2-1\right)^{7/2}}\right)
  \nn   \\&
  \thiio{3;2,2,1,2}
  =-\frac{2 \left(960 \gamma ^{10}-2560 \gamma ^8+2540
    \gamma ^6-1150 \gamma ^4+225 \gamma ^2-13\right)
    \Gamma ^2}{\left(\gamma ^2-1\right)^3 \left(2
    \gamma ^2-1\right)^2}
  \\&\qquad
  +\frac{4 \gamma  \left(1800
    \gamma ^{10}+2020 \gamma ^8-424 \gamma ^6-3032
    \gamma ^4+1703 \gamma ^2-231\right) \nu }{3
    \left(\gamma ^2-1\right)^3 \left(2 \gamma
    ^2-1\right)^2}
  \nn\\&\qquad
  +\frac{16 \left(60 \gamma ^{12}-664
    \gamma ^{10}+519 \gamma ^8-31 \gamma ^6-43 \gamma
    ^4+7 \gamma ^2-1\right) \nu\  \text{arccosh}\gamma}{\left(\gamma ^2-1\right)^{7/2} \left(2 \gamma
    ^2-1\right)^2}
  \nn\\&\qquad
  +\pi ^2 \left(\frac{3 \left(\gamma
    ^2+1\right) \left(5 \gamma ^4-4 \gamma ^2+3\right)
    \gamma ^2 \Gamma ^2}{32 \left(\gamma ^2-1\right)^2
    \left(2 \gamma ^2-1\right)^3}+\frac{\left(100
    \gamma ^8-60 \gamma ^7-160 \gamma ^6-12 \gamma
    ^5+193 \gamma ^4+12 \gamma ^3-78 \gamma ^2-36
    \gamma +45\right) \gamma ^2 \nu }{64 \left(\gamma
    ^2-1\right)^2 \left(2 \gamma ^2-1\right)^3}\right)
  \nn
\end{align}}
\end{adjustwidth}
Then, for the coefficients $\thiio{3;2,3,i,j}$ we find:\checked
\begin{adjustwidth}{-1cm}{-1cm}
{\footnotesize
  \begin{align}
  &\thiio{3;2,3,1,1}=\frac{\gamma  \left(18624 \gamma ^8+24848 \gamma
    ^6-45192 \gamma ^4-58631 \gamma ^2+36351\right) \nu
  }{15 \left(\gamma ^2-1\right)^4 \left(2 \gamma
    ^2-1\right)}
  +\delta  \Bigg(\frac{704 \gamma ^6-880
    \gamma ^4+312 \gamma ^2-23}{2 \left(\gamma
    ^2-1\right)^3 \left(2 \gamma ^2-1\right)}
  \\&\qquad
  -\frac{2
    \left(1376 \gamma ^4-1294 \gamma ^2+233\right)
    C_{{\rm E},1}}{35 \left(\gamma
    ^2-1\right)^3}\Bigg)-\frac{4 \gamma  \left(8720
    \gamma ^6+14894 \gamma ^4-22663 \gamma
    ^2-37071\right) C_{{\rm E},1} \nu }{105 \left(\gamma
    ^2-1\right)^4}
  \nn\\&\qquad
  +\Gamma ^2 \left(\frac{2 \left(4064
    \gamma ^6-6562 \gamma ^4+2997 \gamma ^2-359\right)
    C_{{\rm E},1}}{35 \left(\gamma
    ^2-1\right)^4}-\frac{1728 \gamma ^8-3664 \gamma
    ^6+2584 \gamma ^4-673 \gamma ^2+41}{2 \left(\gamma
    ^2-1\right)^4 \left(2 \gamma
    ^2-1\right)}\right)
  \nn\\&\qquad
  +\text{arccosh}\gamma
  \left(\frac{64 \left(3 \gamma ^8-35 \gamma ^6+9
    \gamma ^4+42 \gamma ^2+6\right) \nu }{\left(\gamma
    ^2-1\right)^{9/2}}-\frac{16 \left(8 \gamma ^8-80
    \gamma ^6+44 \gamma ^4+99 \gamma ^2+15\right)
    C_{{\rm E},1} \nu }{\left(\gamma
    ^2-1\right)^{9/2}}\right)
  \nn\\&
  \thiio{3;2,3,1,2}=\frac{4 \gamma  \left(96 \gamma ^6-148 \gamma ^4+55
    \gamma ^2-1\right) \Gamma ^2}{\left(\gamma
    ^2-1\right)^4}-\frac{16 \gamma  \left(12 \gamma
    ^8-136 \gamma ^6-21 \gamma ^4+210 \gamma
    ^2+88\right) \nu\ \text{arccosh}\gamma}{\left(\gamma ^2-1\right)^{9/2}}
  \\&\qquad
  -\frac{\left(2880
    \gamma ^{10}+7712 \gamma ^8-5664 \gamma ^6-22688
    \gamma ^4+9219 \gamma ^2+1197\right) \nu }{3
    \left(\gamma ^2-1\right)^4 \left(2 \gamma
    ^2-1\right)}
  \nn
\end{align}}
\end{adjustwidth}
Finally, the coefficients $\thiio{3;2,4,i,j}$ are:\checked
\begin{adjustwidth}{-1cm}{-1cm}
{\footnotesize
  \begin{align}
  &\thiio{3;2,4,1,1}=\pi  \Bigg(-\frac{\left(80 \gamma ^9-144 \gamma ^7-12
    \gamma ^6+42 \gamma ^5+33 \gamma ^4+32 \gamma ^3-21
    \gamma ^2-18 \gamma +6\right) \nu }{4 \left(\gamma
    ^2-1\right)^{5/2} \left(2 \gamma
    ^2-1\right)^2}-\frac{3 \left(30 \gamma ^3-15 \gamma
    ^2-6 \gamma +1\right) C_{{\rm E},1} \nu }{4
    \left(\gamma ^2-1\right)^{3/2}}
  \\&\qquad
  +\Gamma ^2
  \left(\frac{80 \gamma ^8-104 \gamma ^6+12 \gamma
    ^5+20 \gamma ^4-15 \gamma ^3+18 \gamma ^2+9 \gamma
    -6}{8 \left(\gamma ^2-1\right)^{5/2} \left(2 \gamma
    ^2-1\right)^2}+\frac{3 \left(30 \gamma ^4+15 \gamma
    ^3-21 \gamma ^2-\gamma +3\right) C_{{\rm E},1}}{8
    \left(\gamma ^2-1\right)^{5/2}}\right)
  \nn\\&\qquad
  +\delta 
  \left(-\frac{80 \gamma ^8-104 \gamma ^6-12 \gamma
    ^5+20 \gamma ^4+15 \gamma ^3+18 \gamma ^2-9 \gamma
    -6}{8 \left(\gamma ^2-1\right)^{5/2} \left(2 \gamma
    ^2-1\right)^2}-\frac{3 \left(30 \gamma ^4-15 \gamma
    ^3-21 \gamma ^2+\gamma +3\right) C_{{\rm E},1}}{8
    \left(\gamma ^2-1\right)^{5/2}}\right)\Bigg)
  \nn\\
  &\thiio{3;2,4,1,2}=\pi  \Bigg(-\frac{\left(20 \gamma ^7+12 \gamma ^6-21
    \gamma ^5-24 \gamma ^4+4 \gamma ^3+15 \gamma ^2+3
    \gamma -3\right) \Gamma ^2}{4 \left(\gamma
    ^2-1\right)^{5/2} \left(2 \gamma
    ^2-1\right)^2}
  \\&\qquad
  +\frac{\left(20 \gamma ^7-12 \gamma
    ^6-21 \gamma ^5+24 \gamma ^4+4 \gamma ^3-15 \gamma
    ^2+3 \gamma +3\right) \delta }{4 \left(\gamma
    ^2-1\right)^{5/2} \left(2 \gamma
    ^2-1\right)^2}
  \nn\\&\qquad
  +\frac{\left(120 \gamma ^8-56 \gamma
    ^7-194 \gamma ^6+48 \gamma ^5+124 \gamma ^4-\gamma
    ^3-36 \gamma ^2-9 \gamma +6\right) \nu }{4
    \left(\gamma ^2-1\right)^{5/2} \left(2 \gamma
    ^2-1\right)^2}\Bigg)
  \nn
\end{align}}
\end{adjustwidth}
We turn, then, to the radiative contributions.
First, we print the radiative expansion coefficients that are independent of the particle label and follow from radiation reaction response:
{\small\bse\label{EXP:RADRAD}\begin{align}
  \thiir{3;0}&=
  \frac{4 \left(1-2 \gamma ^2\right)^2 \nu
  }{\left(\gamma ^2-1\right)^{3/2}}
  \mathcal{I}(v)
  \\
  \thiir{3;1,1}&=
  -\frac{24 \gamma  \left(2 \gamma ^2-1\right) \nu
  }{\gamma ^2-1}
  \mathcal{I}(v)
  \\
  \thiir{3;2,1,1,1}&=
  -\frac{16 \nu  \left(\gamma ^4 (4
    C_{{\rm E},1}-6)+\gamma ^2 (6-4
    C_{{\rm E},1})+C_{{\rm E},1}-1\right)}{\left(\gamma
    ^2-1\right)^{3/2}}
  \mathcal{I}(v)
  \\
  \thiir{3;2,1,1,2}&=
  \frac{16 \left(6 \gamma ^4-6 \gamma ^2+1\right) \nu
  }{\left(\gamma ^2-1\right)^{3/2}}
  \mathcal{I}(v)
  \\
  \thiir{3;2,2,1,1}&=
  \frac{8 \nu  \left(\gamma ^4 (16
    C_{{\rm E},1}-15)+\gamma ^2 (15-16 C_{{\rm E},1})+4
    (C_{{\rm E},1}-1)\right)}{\left(\gamma
    ^2-1\right)^{3/2}}
  \mathcal{I}(v)
  \\
  \thiir{3;2,2,1,2}&=
  -\frac{8 \left(15 \gamma ^4-15 \gamma ^2+4\right) \nu
  }{\left(\gamma ^2-1\right)^{3/2}}
  \mathcal{I}(v)
  \\
  \thiir{3;2,3,1,1}&=
  \frac{16 \nu  \left(\gamma ^4 (4
    C_{{\rm E},1}-6)+\gamma ^2 (6-4
    C_{{\rm E},1})+C_{{\rm E},1}-1\right)}{\left(\gamma
    ^2-1\right)^{5/2}}
  \mathcal{I}(v)
  \\
  \thiir{3;2,3,1,2}&=
  \frac{16 \gamma  \left(6 \gamma ^4-6 \gamma
    ^2+1\right) \nu }{\left(\gamma ^2-1\right)^{5/2}}
  \mathcal{I}(v)
\end{align}\ese}
The final radiative coefficients $\theta^{3;2,4,i,j}_{k,\rm rad}$ which depend on the particle label $k$ are:\checked
\begin{adjustwidth}{-1.5cm}{-1.5cm}
{\footnotesize
  \begin{align}
  &\theta^{(3;2,4,1,1)}_{1,\rm rad}=
  \frac{3 \pi  \gamma  \left(28 \gamma ^8-80 \gamma
    ^6+41 \gamma ^4+34 \gamma ^2-15\right) \nu\ \text{arccosh}\gamma}{32 \left(\gamma ^2-1\right)^{7/2}
    \left(2 \gamma ^2-1\right)}
  \\&\qquad
  -\frac{3 \pi  \left(14
    \gamma ^7+14 \gamma ^6+101 \gamma ^5-323 \gamma
    ^4+236 \gamma ^3-60 \gamma ^2-23 \gamma +25\right)
    \nu  \log \left(\frac{\gamma +1}{2}\right)}{16
    (\gamma -1)^2 (\gamma +1)^3 \left(2 \gamma
    ^2-1\right)}
  \nn\\&\qquad
  +\frac{\pi  \left(280 \gamma ^{10}+470
    \gamma ^9+1930 \gamma ^8+5397 \gamma ^7-88373
    \gamma ^6+244865 \gamma ^5-273845 \gamma ^4+73199
    \gamma ^3+112249 \gamma ^2-102411 \gamma
    +25759\right) \nu }{160 (\gamma -1)^3 (\gamma +1)^5
    \left(2 \gamma ^2-1\right)}
  \nn\\&\qquad
  +C_{{\rm E},1}
  \Bigg(\frac{3 \pi  \gamma  \left(98 \gamma ^6-239
    \gamma ^4+164 \gamma ^2-39\right) \nu \ \text{arccosh}\gamma}{64 \left(\gamma
    ^2-1\right)^{7/2}}-\frac{3 \pi  \left(49 \gamma
    ^5+169 \gamma ^4-306 \gamma ^3+150 \gamma ^2-63
    \gamma +33\right) \nu  \log \left(\frac{\gamma
      +1}{2}\right)}{32 (\gamma -1)^2 (\gamma
    +1)^3}
  \nn\\&\qquad
  -\frac{\pi  \left(1575 \gamma ^9-810 \gamma
    ^8+1020 \gamma ^7+4140 \gamma ^6+26442 \gamma
    ^5-196568 \gamma ^4+442476 \gamma ^3-521244 \gamma
    ^2+321447 \gamma -80398\right) \nu }{640 (\gamma
    -1)^3 (\gamma +1)^5}\Bigg)
  \nn
  \\
  &\theta^{(3;2,4,2,2)}_{1,\rm rad}=
  -\frac{3 \pi  \left(2 \gamma ^4-13 \gamma ^2+15\right)
    \gamma ^2 \nu\  \text{arccosh}\gamma}{16
    \left(\gamma ^2-1\right)^{7/2} \left(2 \gamma
    ^2-1\right)}+\frac{3 \pi  \left(-110 \gamma ^5+259
    \gamma ^4-186 \gamma ^3+8 \gamma ^2+40 \gamma
    -19\right) \nu  \log \left(\frac{\gamma
      +1}{2}\right)}{8 (\gamma -1)^2 (\gamma +1)^3
    \left(2 \gamma ^2-1\right)}
  \\&\qquad
  +\frac{\pi  \left(-2070
    \gamma ^9+2740 \gamma ^8+20777 \gamma ^7-153646
    \gamma ^6+408765 \gamma ^5-463000 \gamma ^4+119039
    \gamma ^3+196158 \gamma ^2-168991 \gamma
    +40708\right) \nu }{160 (\gamma -1)^3 (\gamma +1)^5
    \left(2 \gamma ^2-1\right)}
  \nn\\&\qquad
  +C_{{\rm E},2}
  \Bigg(-\frac{3 \pi  \left(30 \gamma ^4-59 \gamma
    ^2+21\right) \gamma ^2 \nu \ \text{arccosh}\gamma}{32 \left(\gamma ^2-1\right)^{7/2}}-\frac{3 \pi 
    \left(65 \gamma ^4-190 \gamma ^3+156 \gamma ^2-74
    \gamma +27\right) \nu  \log \left(\frac{\gamma
      +1}{2}\right)}{16 (\gamma -1)^2 (\gamma
    +1)^3}
  \nn\\&\qquad
  +\frac{\pi  \left(-1075 \gamma ^7+980 \gamma
    ^6-6738 \gamma ^5+73384 \gamma ^4-198749 \gamma
    ^3+236162 \gamma ^2-135438 \gamma +30994\right) \nu
  }{160 (\gamma -1)^3 (\gamma +1)^5}\Bigg)
  \nn
  \\
  &\theta^{(3;2,4,1,2)}_{1,\rm rad}=
  -\frac{3 \pi  \left(8 \gamma ^6-22 \gamma ^4+9 \gamma
    ^2+9\right) \gamma ^2 \nu \ \text{arccosh}\gamma}{16
    \left(\gamma ^2-1\right)^{7/2} \left(2 \gamma
    ^2-1\right)}+\frac{3 \pi  \left(17 \gamma ^6-63
    \gamma ^5+107 \gamma ^4-81 \gamma ^3+8 \gamma ^2+16
    \gamma -8\right) \nu  \log \left(\frac{\gamma
      +1}{2}\right)}{4 (\gamma -1)^2 (\gamma +1)^3
    \left(2 \gamma ^2-1\right)}
  \nn\\&\qquad
  +\frac{\pi  \left(-630
    \gamma ^{10}-480 \gamma ^9+4481 \gamma ^8+11476
    \gamma ^7-88357 \gamma ^6+195970 \gamma ^5-191553
    \gamma ^4+34776 \gamma ^3+87891 \gamma ^2-69710
    \gamma +16328\right) \nu }{64 (\gamma -1)^3 (\gamma
    +1)^5 \left(2 \gamma ^2-1\right)}
  \\
  &\theta^{(3;2,4,2,1)}_{1,\rm rad}=
  \frac{3 \pi  \gamma  \left(4 \gamma ^6-24 \gamma ^4+33
    \gamma ^2-9\right) \nu \ \text{arccosh}\gamma}{16
    \left(\gamma ^2-1\right)^{7/2} \left(2 \gamma
    ^2-1\right)}+\frac{3 \pi  \left(20 \gamma ^6-76
    \gamma ^5+120 \gamma ^4-73 \gamma ^3-\gamma ^2+25
    \gamma -11\right) \nu  \log \left(\frac{\gamma
      +1}{2}\right)}{4 (\gamma -1)^2 (\gamma +1)^3
    \left(2 \gamma ^2-1\right)}
  \nn\\&\qquad
  +\frac{\pi  \left(-98
    \gamma ^9+1488 \gamma ^8+9431 \gamma ^7-75972
    \gamma ^6+188037 \gamma ^5-200914 \gamma ^4+48417
    \gamma ^3+85464 \gamma ^2-73947 \gamma
    +17902\right) \nu }{64 (\gamma -1)^3 (\gamma +1)^5
    \left(2 \gamma ^2-1\right)}
\end{align}}
\end{adjustwidth}
\endgroup

\subsection{Conservative Hamiltonian}\label{EXP:H}
Here we present expressions for the Hamiltonian coefficients $c^{(A;n)}(\vct{p}^2)$ of Eq.~\eqref{SPIN:IP} up to $\mO(G^2,S^2)$ and $\mO(G^3,S^1)$.
They are presented as evaluated on $p_\infty$, that is $c^{(A;n)}(p_\infty^2)$, as this is most natural from the matching computation.
The coefficients $c^{(A;n)}(\vct{p}^2)$ are then simply obtained by the substitution $p_\infty\to|\vct{p}|$.
In order to carry that substitution out all dependence on $\gam$ must be rewritten in terms of $p_\infty$ as:
\begin{align}
  \gam = \frac{p_1\cdot p_2}{m_1m_2}
  =
  \frac{
    \sqrt{m_1^2+p_\infty^2}
    \sqrt{m_2^2+p_\infty^2}
    +
    p_\infty^2}{m_1m_2}
  \ .
\end{align}
In addition, we use a differential operator $\mD[X]$ defined by,
\begin{align}
  \mD[X] = \frac{
    \pat (p_\infty X)}{
    \pat p_\infty}
  \ ,
\end{align}
with $X$ some function of $p_\infty$ and the variable,
\begin{align}
  \xi = \frac{E_1 E_2}{E^2}
  \ .
\end{align}
We note that the Hamiltonian coefficients $c^{(n;A)}(p_\infty^2)$ (and likewise the angle coefficients $\theta_{\rm can}^{(n;A)}$ are considered, here, as functions of $p_\infty$ and the constant masses $m_i$ and finite size coefficients $C_{\rmm{E},i}$.
The canonical angle coefficients $\theta_{\rm can}^{(n;A)}$ were given in terms of the covariant angle coefficients $\theta_{\rm cons}^{(n;A)}$ in Eq.~\eqref{SPIN:CovToCan} with the covariant angle coefficients printed above in Sec.~\ref{SPIN:ESA}.

We present first the coefficients to $\mO(G^3,S^1)$ starting with 1PM:
\begingroup
\allowdisplaybreaks
\begin{align}
  \cii{1;0}(p_\infty^2)
  &=
  -\frac{\pin^2}{2 E \xi}
  \thiic{1;0}
  \ ,
  \\
  \cii{1;1,i}(p_\infty^2)
  &=
  -\frac{\pin}{2 E\xi}
  \thiic{1;1,i}
  \ .
  \nn
\end{align}
Next, at 2PM order we get:\checked
  \begin{align}
    \cii{2;0}(p_\infty^2)
    &=
    -\frac{\pin^2}{\pi E\xi}
    \thiic{2;0}
    +
    \frac{1}{8\pin}
    \mD\Big[
      \frac{\pin^3}{E\xi}
      \big(
      \thiic{1;0}
      \big)^2
      \Big]
    \ ,
    \\
    \cii{2;1,i}(p_\infty^2)
    &=
    -\frac{\pin}{\pi E\xi}
    \thiic{2;1,i}
    +
    \frac{1}{4\pin^3}
    \mD\Big[
      \frac{\pin^4}{E\xi}
      \thiic{1;0}\thiic{1;1,i}
      \Big]
    \ .
    \nn
  \end{align}
Finally, at 3PM we get:
\begin{adjustwidth}{-1.5cm}{-1.5cm}
  \begin{align}
    \cii{3;0}(p_\infty^2)
    &=
    -\frac{\pin^2}{4E\xi}
    \thiic{3;0}
    +
    \frac1{2\pi\pin^2}
    \mD\Big[
      \frac{\pin^4}{E\xi}
      \thiic{1;0}
      \thiic{2;0}
      \Big]
    -
    \frac1{48\pin^2}
    \mD^2\Big[
      \frac{\pin^4}{E\xi}
      (\thii{1;0})^3
      \Big]
    \ ,
    \\
    \cii{3;1,i}(p_\infty^2)
    &=
    -\frac{\pin}{4E\xi}
    \thiic{3;1,i}
    +
    \frac1{2\pi\pin^4}
    \mD\Big[\frac{\pin^5}{E\xi}
      \big(
      \thiic{1;0}
      \thiic{2;1,i}
      +
      \thiic{2;0}
      \thiic{1;1,i}
      \big)
      \Big]
    -
    \frac1{16\pin^4}
    \mD^2\Big[
      \frac{\pin^5}{E\xi}
      (\thiic{1;0})^2
      \thiic{1;1,i}
      \Big]
    \ .\nn
  \end{align}
\end{adjustwidth}
The general pattern of the linear-in-spin results is very suggestive and mimics the non-spinning results.

The quadratic-in-spin coefficients do not exhibit the same simplicity as the above linear-in-spin results.
Their 1PM contributions are:\checked
    \begin{align}
      \cii{1;2,1,i,j}(p_\infty^2)
      &=
      -\frac{\pin^2}{4E\xi}
      \thiic{1;2,1,i,j}
      \ ,
      \\\nn
      \cii{1;2,2,i,j}(p_\infty^2)
      &=
      -\frac{3\pin^2}{8E\xi}
      \thiic{1;2,2,i,j}
      +
      \frac{3\pin^2}{16E\xi}
      \frac{
        \thiic{1;1,i}\thiic{1;1,j}
      }{
        \thiic{1;0}
      }
      \ ,
      \\\nn
      \cii{1;2,3,i,j}(p_\infty^2)
      &=
      -\frac{1}{4E\xi}
      \big(
      \thiic{1;2,3,i,j}
      -
      \frac12 \thiic{1;2,2,i,j}
      \big)
      -
      \frac{1}{16E\xi}
      \frac{
        \thiic{1;1,i}\thiic{1;1,j}
      }{
        \thiic{1;0}
      }
      \ .
    \end{align}
    Next, their 2PM contributions read:
  \begin{adjustwidth}{-1.5cm}{-1.5cm}
    \begin{align}
      \cii{2;2,1,i,j}(p_\infty^2)
      &=
      -\frac{2\pin^2}{3\pi E\xi}
      \thiic{2;2,1,i,j}
      +
      \frac{1}{32\pin^3}
      \mD\Big[
        \frac{\pin^5}{E\xi}
        \Big(
        3\thiic{1;1,i}\thiic{1;1,j}
        +
        4\thiic{1;0}\thiic{1;2,1,i,j}
        \Big)
        \Big]
      \\
      \cii{2;2,2,i,j}(p_\infty^2)
      &=
      -\frac{\pin^2}{E\xi}
      \bigg(
      \frac{8\thiic{2;2,2,i,j}}{9\pi}
      +\frac{\thiic{1;0}\thiic{1;2,2,i,j}}{8}
      -\frac{\thiic{1;1,i}\thiic{1;1,j}}{16}
      +
      \frac{\pin}{16}
      \Big(
      \frac{\thiic{1;1,i}}{m_i}
      +
      \frac{\thiic{1;2,j}}{m_j}
      \Big)
      \thiic{1;2,2,i,j}
      \bigg)
      \nn
      \\
      &\qquad
      +
      \frac{1}{32\pin^3}
      \mD\Big[
        \frac{\pin^5}{E\xi}
        \Big(
        6\thiic{1;0}\thiic{1;2,2,i,j}
        -
        7\thiic{1;1,i}\thiic{1;1,j}
        \Big)
        \Big]
      -\frac{4\pin^2}{9\pi E\xi}
      \frac{
        \thiic{2;0}\thiic{1;1,i}\thiic{1;1,j}
      }{
        \big(\thiic{1;0}\big)^2
      }
      \nn
      \\
      &\qquad
      +\frac{\pin^2}{E\xi\thiic{1;0}}
      \bigg(
      \frac{\pin}{32}
      \Big(
      \frac{\thiic{1;1,i}}{m_i}
      +
      \frac{\thiic{1;1,j}}{m_j}
      \Big)
      \thiic{1;1,i}\thiic{1;1,j}
      +\frac2{9\pi}
      \big(
      \thiic{1;1,i}\thiic{2;1,j}+
      \thiic{2;1,i}\thiic{1;1,j}
      \big)
      \bigg)
      \nn
      \\
      \cii{2;2,3,i,j}(p_\infty^2)
      &=
      -\frac{1}{E\xi}
      \bigg(
      \frac{
        2  \thiic{2;2,3,i,j}
      }{3\pi}
      -
      \frac{
        2 \thiic{2;2,2,i,j}
      }{9\pi}
      -
      \frac{
        \thiic{1;0}\thiic{1;2,2,i,j}
      }{8}
      -
      \frac{\pin}{16}
      \Big(
      \frac{\thiic{1;1,i}}{m_i}
      +
      \frac{\thiic{1;2,j}}{m_j}
      \Big)
      \thiic{1;2,2,i,j}
      \bigg)
      \nn
      \\
      &\qquad
      +
      \frac{\thiic{1;1,i}\thiic{1;1,j}}{16}
      -
      \frac{1}{32\pin^5}
      \mD\Big[
        \frac{\pin^5}{E\xi}
        \Big(
        2\thiic{1;0}\thiic{1;2,2,i,j}
        +
        \thiic{1;1,i}\thiic{1;1,j}
        -
        4\thiic{1;0}\thiic{1;2,3,i,j}
        \Big)
        \Big]
      \nn
      \\
      &\qquad
      -\frac{1}{E\xi\thiic{1;0}}
      \bigg(
      \frac{\pin}{32}
      \Big(
      \frac{\thiic{1;1,i}}{m_i}
      +
      \frac{\thiic{1;1,j}}{m_j}
      \Big)
      \thiic{1;1,i}\thiic{1;1,j}
      +\frac1{18\pi}
      \big(
      \thiic{1;1,i}\thiic{2;1,j}+
      \thiic{2;1,i}\thiic{1;1,j}
      \big)
      \bigg)
      \nn
      \\
      &\qquad
      +
      \frac{1}{9\pi E\xi}
      \frac{
        \thiic{2;0}\thiic{1;1,i}\thiic{1;1,j}
      }{
        \big(\thiic{1;0}\big)^2
      }
      \nn
    \end{align}
  \end{adjustwidth}
The expressions for the 3PM quadratic-in-spin coefficients are lengthy and instead found in the ancillary file to Ref.~\cite{Jakobsen:2022zsx}.
\endgroup

\chapter{Conclusion and Outlook}
\label{DIS}
In this work the worldline quantum field theory formalism was presented with the main goal of solving the classical equations of motion of point-like particles and their interaction with gravity in the post-Minkowskian expansion of unbound scattering events and thus deriving gauge invariant observables of that system.
The presentation of the WQFT was comprehensive with basic objects and techniques of the formalism discussed in detail in \chap~\ref{WQFT} including the in-in formalism, Feynman rules, graph generation, observables from on-shell one-point functions and the generic structure of (post-Minkowskian) worldline observables.
The justification and consistent matching of extended compact (astrophysical) bodies to worldline point-like particles is a problem of effective field theory considered in \chap~\ref{WEFT} and forms the basics of relating WQFT results to physical processes.
A particular invention along with the WQFT formalism is the SUSY description of spinning point-like particles with anti-commuting Grassmann variables and their gravitational interaction analyzed in Sec.~\ref{sec:Spin}.
The main perturbative results are the $\mO(G^2,S^2)$ gravitational bremsstrahlung presented in \chap~\ref{GB} and the $\mO(G^3,S^2)$ conservative and radiative worldline observables, the impulse and spin kick, presented in \chap~\ref{sec:Scattering} which all are novel outcomes of WQFT not yet fully verified by other means.
The latter results required the computation of the post-Minkowskian two-loop integrals with retarded propagators analyzed in \chap~\ref{sec:TL} which at the time of Ref.~\cite{Jakobsen:2022psy} was done previously only with Feynman propagators.

Let us first consider, then, what challenges and obstacles lie ahead for the extension of the above post-Minkowskian results to higher orders in $G$ and spin or other finite size effects of the bodies.
Here, it is interesting that the inclusion of spin to the waveform and worldline observables does not change the basis of required master integrals and we would generally expect that for any (perturbative) effective couplings in the worldline action.
The challenge of including spin and other effects is then not related to integration but, instead, to constructing the integrand.
Assuming for a moment that a relevant (effective) worldline action is present which describes the required effects of the bodies the challenge, then, is to turn that action into an integrand or, most conveniently, map it directly to the coefficients of the (same Polyakov point-particle) master integrals.
In the most basic application, the WQFT facilitates that mapping of the action to the integrand by solving the classical equations of motion diagrammatically.
While more sophisticated inventions for optimizing this step exist, the basic approach of WQFT works well at the third PM order and is expected to do so too at the next fourth PM order or, perhaps, at even higher orders.
Such optimizations would be the derivation of simpler gravitational Feynman rules~\cite{Cheung:2017kzx} or on-shell methods developed for the WQFT in analogy with the QFT-amplitudes approach~\cite{Elvang:2013cua}.

The current obstacle, then, for including higher orders of spin into the observables at the third PM order is rather the construction of a worldline action to that order in spins (or other effects).
Much work~\cite{Saketh:2022xjb,Khalil:2022sii}, however, is being done for including new effects into the worldline description of compact bodies, and spinning terms of the action are known to all spin orders linear in the curvature~\cite{Levi:2015msa}.
It only remains to translate those results into a form that is suitable for the WQFT approach which in the case of spin can be facilitated with the Grassmann variables.
We note, however, that the WQFT approach can also be applied to the more traditional approaches of describing spin.

The problem of going to the next fourth PM order is, as mentioned in the introduction, in principle solved for the spinless Polyakov action~\cite{Bern:2021dqo,Bern:2021yeh,Bern:2022jvn,Dlapa:2021npj,Dlapa:2021vgp,Dlapa:2022lmu,Bjerrum-Bohr:2022ows}.
Here, however, those integrals remain to be analyzed in some greater depth by the community.
In principle, though, the inclusion of spin to those results is now a matter of assembling the required integrand.
In fact, the same applies to the recent next-to-leading-order bremsstrahlung results~\cite{Elkhidir:2023dco,Herderschee:2023fxh,Brandhuber:2023hhy,Georgoudis:2023lgf}.
The integrals required at the fourth PM order are significantly more complicated than those of the third order with complete elliptic functions appearing in the observables.
The PM multi-loop integrals are, however, simpler than many of their (quantum) counterparts due to their single scale nature and spacial loop momenta in contrast to space-time momenta and the next fifth PM order four-loop integrals are an exciting prospect for future research.
Here, other means of solving or approximating the integrals may be advantageous such as numerical approaches~\cite{Jinno:2022sbr} or their post-Newtonian expansion.
In contrast to the general use of Feynman propagators of QFT-amplitudes approaches, the WQFT uses retarded propagators which directly imposes causal boundary conditions of the derived observables.

The post-Minkowskian expansion effectively resums the velocity expansion of the post-Newtonian expansion.
Generally, the goal of resumming in different parameters is fascinating from a theoretical perspective, and often too, from a practical point of view.
Here, the self-force expansion~\cite{Pound:2021qin} is essential for describing small-mass-ratio binaries and from the perspective of the post-Minkowskian expansion it resums at each order in the small mass all orders in $G$.
See e.g. Refs.~\cite{Adamo:2021rfq,Adamo:2022rmp} for the QFT-amplitudes approach on curved backgrounds.
In particular the first self-force correction includes both the third and fourth PM orders.
Perhaps, the WQFT and more generally quantum field theory can bring new inventions to the field of self-force including the focus on gauge invariant observables, integration techniques and dimensional regularization.
More generally a fascinating fundamental question is the description of black hole dynamics and whether there exists an effective worldline action describing their motion (in a weak external field) to all orders in their parameters.
Here, resummations in spin are interesting~\cite{Vines:2017hyw,Guevara:2018wpp,Guevara:2019fsj,Chung:2018kqs,Chung:2019duq,Arkani-Hamed:2019ymq,Arkani-Hamed:2017jhn,Guevara:2020xjx} as such an action must generally describe the Kerr black hole and therefore should include its spin to all orders.
Another relevant process in this relation is the dissipative effects of horizon absorption~\cite{Saketh:2022xjb,Goldberger:2005cd}.

A different kind of resummation of the perturbative PM results is offered by the effective-one-body (EOB) formalism.
Here, it is an exciting prospective to include the post-Minkowskian data into the EOB waveform models~\cite{Damour:2016gwp,Damour:2017zjx,Antonelli:2019ytb,Damgaard:2021rnk,Khalil:2022ylj,Damour:2022ybd,Antonelli:2020ybz,Hoogeveen:2023bqa}.
In order, however, to use the PM data for the description of bound orbits it first has to be mapped from the unbound observables.
This mapping works well for the conservative Hamiltonian at the third PM order but has problems with the non-local in time tail effects at the fourth order~\cite{Cho:2021arx}.
In general, if the goal is to derive post-Newtonian physics from the post-Minkowskian expansion, the rather demanding conclusion is that for the $n$th post-Newtonian Hamiltonian one requires the $(n+1)$th post-Minkowskian Hamiltonian.
Thus, the available 4PN data requires the next order fifth PM three-loop data.
It seems most reasonable then, that the PM data should supplement the PN data rather than replacing it.
This, generally, also fits well with the EOB model which mixes possibly different perturbative results into a single formalism.
An interesting development of the WQFT might be to solve directly the bound system by expanding the worldline fields about Kepler orbits and the gravitational field around the Newtonian potential.

An interesting insight of the present work~\cite{Jakobsen:2022fcj} is the parametrization of the conservative worldline observables in terms of a generic spins scattering angle.
Here, it was achieved through the same parametrization of the conservative Hamiltonian and the derivation of this Hamiltonian was the initial main goal.
For that reason, also, this parametrization of the worldline observables was not analyzed in any depth in this work and is thus a promising project for future research.
The goal, thus, is to parametrize the conservative worldline observables and the Hamiltonian in terms of some scalar (and, possibly, its derivatives) which here was taken as a generic spins scattering angle.
Another candidate, though, would be the WQFT eikonal or on-shell action.
Such a description from a scalar is a powerful way to encode the (non-planar) vectorial observables the impulse and spin kick.
These ideas are similar, also, to the way observables are being derived from the scalar scattering amplitudes.

The role of the WQFT in simplifying the classical limit of the QFT-amplitudes approach was not a main focus of this thesis, though it is a fascinating connection of quantum field theory to its classical limit and not yet fully verified at the third PM order.
In fact, worldline quantum field theory can be advantageous for computing quantum amplitudes~\cite{Ahmadiniaz:2020wlm,Ahmadiniaz:2021gsd,Ahmadiniaz:2022yam} where it may for example resum several Feynman diagrams.
From this perspective it would be interesting to compute $\hbar$ corrections to the gravitational scattering amplitudes~\cite{Bjerrum-Bohr:2002gqz,Bjerrum-Bohr:2013bxa,Donoghue:1994dn} of the QFT-amplitudes approach from the WQFT.
Here, the inclusion of spin effects would be natural and the generalization of the originally spinless dressed propagator of Ref.~\cite{Mogull:2020sak} to including spin.
It is in any case a consistency check to verify that the effective descriptions of compact bodies of the worldline EFT and QFT-amplitudes approaches are equivalent (see e.g. Ref.~\cite{Kim:2023drc} for a possible resolution to a recent discrepancy between the two approaches).

Generally, the new era of gravitational wave astronomy is an inspiring time for theoretical physics and tests of general relativity.
The required theoretical predictions of gravitational and astrophysical phenomena offer many tasks for theoretical physics and in particular the understanding of coalescences of compact bodies for the prediction of their waveforms.
Here, worldline effective field theory offers a systematic description of compact bodies in terms of worldlines and the WQFT a methodical approach to perturbatively solving their equations of motion.
Together with the several other approaches for deriving binary dynamics, the continued effort of the community offers an exciting future for gravitational wave physics.

\newpage
Note added:
Since writing this thesis, several new results on classical gravitational scattering have been published that are of interest for the topics of this thesis~\cite{Dlapa:2023hsl,Mandal:2023lgy,Bautista:2023szu,Gonzo:2023goe,Barack:2023oqp,Aoude:2023vdk,Jones:2023tgz,Bern:2023ccb,Bhattacharyya:2023kbh,Jakobsen:2023ndj,Bianchi:2023lrg,Damgaard:2023vnx,DiVecchia:2023frv,Adamo:2023cfp,Druart:2023mba,Damgaard:2023ttc,Rettegno:2023ghr,Aoude:2023fdm,Saketh:2023bul,Bini:2023mdz,Mandal:2023hqa}.
In Refs.~\cite{Dlapa:2023hsl,Barack:2023oqp,Jakobsen:2023ndj,Damgaard:2023ttc} further work on worldline observables at the fourth post-Minkowskian order was done where in particular Ref.~\cite{Jakobsen:2023ndj} (including the present author) presented new conservative results at $\mO(G^4,S^1)$ and Ref.~\cite{Damgaard:2023ttc} with a QFT-amplitudes approach presented a first independent check of the spinless dissipative (4PM) results of Ref.~\cite{Dlapa:2022lmu}.
In Ref.~\cite{Gonzo:2023goe} unbound-to-bound mappings for generic spins were considered.
In Refs.~\cite{Barack:2023oqp,Saketh:2023bul,Mandal:2023hqa} divergences and the need for renormalization in the (classical) effective field theory of compact bodies appeared.
In Ref.~\cite{Bern:2023ccb} first steps towards the fifth post-Minkowskian order were taken by analyzing the potential region of the simpler system of electromagnetism.
In Ref.~\cite{Damgaard:2023vnx} the relation between amplitudes-QFT and worldline formalisms was analyzed in detail.
In Ref.~\cite{Rettegno:2023ghr} new numerical simulations of the scattering of Kerr black holes and comparison with effective-one-body approaches based on the post-Minkowskian data were presented.

\appendix
\chapter{Notation and Conventions}
\label{NOT}
In this Appendix we have collected our conventions for position and momentum space in Sec.~\ref{NOT:P} and for spinning asymptotic worldline states in Sec.~\ref{NOT:A} because these definitions play a role in most chapters of this thesis.
For other conventions of e.g. Feynman rules or integrals the reader must look in the respective chapters of the thesis.

We use the mostly minus metric with space-time dimension $d$ and the speed of light $c$ generally put to unity $c=1$.
Symmetric brackets and antisymmetric square brackets on indices are defined as follows:
\bse\label{A:Brackets}
\begin{align}
  X_{(\mu}Y_{\nu)}
  =
  \frac12
  \big(
  X_\mu Y_\nu
  +
  X_\nu Y_\mu
  \big)
  \ ,
  \\
  X_{[\mu}Y_{\nu]}
  =
  \frac12
  \big(
  X_\mu Y_\nu
  -
  X_\nu Y_\mu
  \big)
  \ ,
\end{align}
\ese
with in the present case $X_\mu$ and $Y_\mu$ two arbitrary (possibly anti-commuting) vectors.
For any timelike or spacelike four vector $Z^\mu$ we define:
\bse\label{A:VA}
\begin{align}
  |Z^\mu|&=|Z|=\sqrt{|Z^2|}
  \ ,
  \\
  \hat Z^\mu
  &=
  \frac{Z^\mu}{|Z|}
  \ .
\end{align}
\ese

\section{Position and Momentum Space}\label{NOT:P}
Our conventions for ($d$-dimensional) momentum and position space read:\checked
\bse\label{WEFT:MomentumConvention}
\begin{align}
  h_\mn(k)
  &=
  \int \di^d x
  \,
  e^{ik\cdot x}
  \,
  h_\mn(x)
  \ ,
  \qquad
    h_\mn(x)
    =
  \int_k
  \,
  e^{-ik\cdot x}
  \,
  h_\mn(k)
  \ ,
\end{align}
with integration measure:\checked
\begin{align}
  \int_k
  =
  \int
  \frac{
    \di^d k
  }{
    (2\pi)^{d}
  }
  \ .
\end{align}
These conventions apply generally to all space-time fields including the special case of the graviton field shown here.
We generally indicate whether the field is considered in position or momentum space by its argument: $h_\mn(x)$ or $h_\mn(k)$.
Thus, considered as ordinary functions $h_\mn(k)$ and $h_\mn(x)$ are not the same function.
Functional derivatives in position and momentum space are defined as:\checked
\begin{align}
  \frac{\del h_\mn(x)}{\del h_\ab(y)}
  =
  \delta_{\mu}^{(\alpha}
  \delta^{\beta)}_{\nu}\,
  \del^d(x-y)
  \ ,
  \qquad
  \frac{\del h_\mn(k)}{
    \del h_\ab(k')}
  =
  \del_{\mu}^{(\alpha}
  \del^{\beta)}_{\nu}\,
  \dd^d(k-k')
  \ ,
\end{align}
with the barred delta function:\checked
\begin{align}
  \dd^d(k)=(2\pi)^d \del^d(k)
  \ .
\end{align}
For worldline fields $w^\sig(\tau)$ we define time and energy (frequency) domains by the one-dimensional versions of the above formulas:
\begin{align}
  w^\sig(\oma)
  =
  \int\di\tau\, e^{i\oma \tau}\,
  w^\sig(\tau)
  \ ,
  \qquad
  w^\sig(\tau)
  =
  \int_\oma
  e^{-i\oma\tau}\,
  w^\sig(\oma)
  \ ,
\end{align}
with
\begin{align}
  \int_\oma
  =
  \int_{-\infty}^{\infty}
  \frac{\di\oma}{2\pi}
  \ ,
\end{align}
and functional derivatives:
\begin{align}
  \frac{\del w^\sig(\tau)}{\del w^\rho(\tau')}
  =
  \del^{\sig}_{\rho}
  \del(\tau-\tau')
  \ ,
  \qquad
  \frac{\del w^\sig(\oma)}{
    \del w^\rho(\oma')}
  =
  \del_{\rho}^{\sig}
  \dd(\oma-\oma')
  \ ,
\end{align}
with the barred delta function:
\begin{align}
  \dd(\oma)=2\pi\del(\oma)
  \ .
\end{align}
\ese

\section{Asymptotic Worldline Variables}\label{NOT:A}
The conventions and definitions of asymptotic worldline states were discussed (without spin) in Sec.~\ref{WEFT:GS} and (with spin) in Sec.~\ref{sec:Dynamics} where more details can be found.

The asymptotic worldline fluctuations are parametrized as follows:
\begin{align}
  z_i^\sig(\tau)
  &=
  b_i^\sig+\tau v^\sig_i
  \ ,
  \qquad
  \psi_i^\sig(\tau)
  =
  \Psi_i^\sig
  \ ,
  \qquad
  \mS_i^\mn(\tau)
  =
  S^\mn_i
  \ ,
\end{align}
with proper time $\tau$ and velocity $v_i^2=1$.

A basic set of gauge invariant variables (under SUSY and translations) are:
\begin{align}
  p_i^\sig
  &=
  m_i v_i^\sig
  \ ,
  \\
  S^\mn_{i,\bot}
  &=
  S_i^\ab
  P_{i,\alpha}^\mu
  P_{i,\beta}^\nu
  \ ,
  \\
  \beta^\sig
  &=
  P_{12}^{\sig\rho}
  (
  \beta_2-\beta_1
  )_\rho
  \ ,
\end{align}
with projectors,
\begin{align}
  P_i^\mn &= \eta^\mn - v_i^\mu v_i^\nu\ ,
  \\
  P_{12}^\mn &=
  \eta^\mn
  +
  \frac{
    v_1^\mu v_1^\nu
    +
    v_2^\mu v_2^\nu
    -
    2\gam v_1^{(\mu}v_2^{\nu)}
  }{\gam^2-1}
  \ ,
\end{align}
and individual SUSY parameters,
\begin{align}
  \beta^\sig_i
  =
  b_i^\sig
  +
  \frac{ S_i^{\sig\rho}
    v_{i,\rho}
  }{m_i}
  \ .
\end{align}
Generally, the covariant gauge is assumed in which,
\begin{align}
  b^\sig_{\rm cov}
  &=
  P_{12}^{\sig\rho}
  (b_{2,\rm cov}-b_{1,\rm cov})_\rho
  =
  \beta^\sig
  \ ,
  \\
  S^\mn_{i,\rm cov}
  &=
  S_{i,\bot}^\mn
  \ ,
\end{align}
and often the subscript ``cov'' is left out.

The relative Lorentz factor $\gam$ and velocity $v$ are:
\begin{align}
  \gam
  &=
  v_1\cdot v_2 = \frac1{\sqrt{1-v^2}}
  \ ,
  \qquad
  v
  =
  \frac{\sqrt{\gam^2-1}}{\gam}
  \ ,
\end{align}
with the useful relation $\gam v=\sqrt{\gam^2-1}$.

The total mass $M$, reduced mass $\mu$ and symmetric mass ratio $\nu$ are:
\begin{align}
  M&=m_1+m_2
  \ ,
  \qquad
  \mu
  =\frac{m_1 m_2}{M}
  \ ,
  \qquad
  \nu
  =
  \frac{\mu}{M}
  \ .
\end{align}
In addition we have the relative mass difference,
\begin{align}
  \del = \frac{m_2-m_1}{M}
  =
  \sqrt{1-4\nu}
  \ ,
\end{align}
where the last equality assumes $m_2>m_1$.

The total (CoM) momentum $P^\mu$ and relative (CoM) momentum $p^\mu$ are given by:
\begin{align}
  P^\mu &= p_1^\mu+p_2^\mu
  \ ,
  \\
  p^\mu
  &=
  (\eta^\mu_\nu
  -
  \hat P^\mu
  \hat P_\nu
  )
  p_1^\nu
  =
  -
  (\eta^\mu_\nu
  -
  \hat P^\mu
  \hat P_\nu
  )
  p_2^\nu
  \nn
  \\
  &=
  \frac{m_1 m_2}{E^2}
  \Big(
  (\gam m_1+m_2)
 v_1^\mu
  -
  (\gam m_2+m_1)
  v_2^\mu
  \Big)
  \ .
\end{align}
From those we define the total energy $E$, reduced energy $\Gam$ and (absolute) relative momentum:
\begin{align}
  E
  &=
  M\Gam
  =
  |P^\mu|
  =
  \sqrt{m_1^2+m_2^2+2\gam m_1 m_2}
  =
  M\sqrt{1+2\nu(\gam-1)}
  \ ,
  \\
  p_\infty
  &=
  |p^\mu|
  =
  \gam v
  \frac{
    m_1m_2}{
    E}
  \ .
\end{align}
In addition we have the individual CoM energies $E_i$, Lorentz factors $\gam_i$ and velocities $v_i$:
\begin{align}
  E_i = P\cdot v_i
  \ ,\qquad
  \gam_i = \frac{E_i}{m_i}
  \ ,\qquad
  v_i = \frac{p_\infty}{E_i}
  \ ,
\end{align}
so that in the CoM frame:
\begin{align}
  p_i^\mu = E_i \hat P^\mu - (-1)^i p_\infty \hat p^\mu
  \ ,\qquad
  v_i^\mu = \gam_i
  \big(
    \hat P^\mu - (-1)^i v_i \hat p^\mu
    \big)
\end{align}

The orbital $L_i^\mn$ and total $J^\mn_i$ angular momentum of each body are,
\begin{align}
  L_i^\mn = 2 b_i^{[\mu} p_i^{\nu]}
  \ ,\qquad
  J_i^\mn = L_i^\mn + S_i^\mn
  \ ,
\end{align}
with the intrinsic angular momentum $S_i^\mn$.
The orbital $L^\mn$ and total $J^\mn$ angular momentum of the system of particles are:
\begin{align}
  L^\mn = \sum_i L_i^\mn
  \ ,\qquad
  J^\mn = \sum_i J_i^\mn
  \ .
\end{align}

Pauli-Lubanski vectors are defined by projecting out the dependence on coordinate centers:
\begin{align}
  L^\mu = \frac12
  \eps^{\mu}_{\ \nu\ab}
  \hat P^\nu L^\ab
  \ ,
  \qquad
  J^\mu
   = \frac12
  \eps^{\mu}_{\ \nu\ab}
  \hat P^\nu J^\ab
  \ .
\end{align}
For the intrinsic spin tensors the covariant $S_i^\mu$ and canonical $S_{i,\rm can}^\mu$ spin vectors are:
\begin{align}
  S_i^\mu
  =
  m_i a_i^\mu
  =
  \frac12
  \eps^\mu_{\ \nu\ab}
  v_i^\nu
  S^\ab_i
  \ ,
  \qquad
  S_{i,\rm can}^\mu
  =
  \frac12
  \eps^\mu_{\ \nu\ab}
  V^\nu
  S_{i,\rm can}^\ab
  \ ,
\end{align}
with the mass-reduced canonical spin vector $a_i^\mu$ and the generic frame $V^\nu$ which in particular includes as a special case $\hat P^\nu$.

\newpage
\addcontentsline{toc}{chapter}{Bibliography}
\bibliographystyle{JHEP}
\bibliography{/Users/gustav/physics/local/per_folders/Papers/Papers_v0/refs.bib}

\end{document}